\documentclass[twocolumn]{aastex63}

\usepackage{graphicx}   
\usepackage{amsmath}    
\usepackage{amssymb}    
\usepackage{gensymb}
\usepackage{multirow}
\usepackage{float}
\usepackage{xcolor}
\usepackage{multirow}

\newcounter{note}
\let\oldmarginpar\marginpar
\renewcommand\marginpar[1]{\-\oldmarginpar[\raggedleft\footnotesize #1]%
{\raggedright\footnotesize #1}}

\definecolor{Dred}{rgb}{0.312,0.070,0.070}

\newcommand{\Cov}{ \mathop{ \rm Cov }\nolimits }

\received{15 August 2020}
\revised{17 November 2020}
\accepted{26 November 2020}
\submitjournal{AJ}

\shorttitle{VLBA North Polar Cap Survey}
\shortauthors{Popkov et al.}

\begin{document}

\title{Parsec-scale properties of steep and flat spectrum extragalactic radio sources from a VLBA survey of a complete north polar cap sample}

\correspondingauthor{Alexander Popkov}
\email{popkov.av@mipt.ru}

\author[0000-0002-0739-700X]{A.~V.~Popkov}
\affiliation{Moscow Institute of Physics and Technology, Institutsky per. 9, Dolgoprudny 141700, Russia}
\affiliation{Astro Space Center of Lebedev Physical Institute, Profsoyuznaya 84/32, 117997 Moscow, Russia}

\author[0000-0001-9303-3263]{Y.~Y.~Kovalev}
\affiliation{Astro Space Center of Lebedev Physical Institute, Profsoyuznaya 84/32, 117997 Moscow, Russia}
\affiliation{Moscow Institute of Physics and Technology, Institutsky per. 9, Dolgoprudny 141700, Russia}
\affiliation{Max-Planck-Institut f\"ur Radioastronomie, Auf dem H\"ugel 69, 53121 Bonn, Germany}

\author[0000-0001-9737-9667]{L.~Y.~Petrov}
\affiliation{NASA Goddard Space Flight Center, Greenbelt, MD 20771, USA}

\author[0000-0002-8017-5665]{Yu.~A.~Kovalev}
\affiliation{Astro Space Center of Lebedev Physical Institute, Profsoyuznaya 84/32, 117997 Moscow, Russia}

\begin{abstract}

We observed with the VLBA at 2.3 and 8.6~GHz a complete flux-density limited sample of 482 radio sources with declination $>+75\degree$ brighter than 200~mJy at 1.4~GHz drawn from the NVSS catalog. 
34\% of the sources show parsec-scale emission above the flux density detection limit of 30~mJy; their accurate positions and parsec-scale structure parameters are determined. 
Among all the sources detected at least at the shortest VLBA baselines, the majority, or 72\%, has a steep single-dish spectrum.
The fraction of the sources with a detectable parsec-scale structure is above 95\% among the flat-spectrum and close to 25\% among the steep-spectrum objects. 
We identified 82 compact steep-spectrum source candidates, which make up 17\% of the sample; most of them are reported for the first time. 
The compactness and the brightness temperature of the sources in our sample show a positive correlation with single-dish and VLBA spectral indices.
All the sources with a significant 8~GHz variability were detected by the VLBA snapshot observations, which independently confirmed their compactness.
We demonstrated that 54\% of the sources detected by the VLBA at 2.3~GHz in our sample have a steep VLBA spectrum. The compact radio emission of these sources is likely dominated by optically thin jets or mini-lobes, not by an opaque jet core.
These results show that future VLBI surveys aimed to search for new sources with parsec-scale structure should include not only flat-spectrum sources, but also steep-spectrum ones in order to reach an acceptable level of completeness.

\end{abstract}


\section{Introduction}
\label{sec:intro}

Over the last decades, more than a dozen of large-area surveys of active galactic
nuclei (AGN) have been carried out using very long baseline interferometry
(VLBI). These surveys include, among others, \citet{1985AJ.....90.1599P}, the series of
the VLBA Calibrator Surveys (\citealt{2002ApJS..141...13B}, \citealt{2003AJ....126.2562F}, \citealt{2005AJ....129.1163P, 2006AJ....131.1872P}, \citealt{2007AJ....133.1236K},
\citealt{2008AJ....136..580P}, \citealt{2016AJ....151..154G}), the VLBA Imaging and Polarimetry Survey  \citep{2007ApJ...658..203H}, the LBA Calibrator
Surveys \citep{2011MNRAS.414.2528P, 2019MNRAS.485...88P}, the VLBA
Galactic Plane Survey \citep{2011AJ....142...35P}.
Due the the narrow field of view, typically, 1--10~arcseconds, VLBI surveys usually follow up the objects detected with low resolution
connected-element interferometry or single-dish observations. 
It was found that if to take a flux-density limited sample of flat-spectrum extragalactic sources and follow them up with VLBI, most of them will be detected (e.g., \citealt{r:pr81}, \citealt{1985AJ.....90.1599P}, \citealt{1996ApJS..107...37T}, \citealt{2005AJ....129.1163P, 2006AJ....131.1872P}, \citealt{2007AJ....133.1236K}), up to 90\%.
The reason for that is the presence of the opaque VLBI core which typically has a flat spectrum and dominates the total emission of flat-spectrum AGNs \citep[e.g.,][]{2005AJ....130.2473K,2012A&A...544A..34P,2014AJ....147..143H}.
We classify a source spectrum as flat if its
flux denisty dependence on frequency $\nu$ is described by
the power law $S(\nu) \propto \nu^{+\alpha}$ with the spectral
index $\alpha \geq -0.5$.
Since the goal of the surveys was to provide a dense grid of calibrators
using the minimum resources, the source selection algorithm was tuned to
maximize the number of detections. Therefore, these surveys targeted almost
exclusively on flat-spectrum sources.
As a result, VLBI catalogs have a heavy bias towards flat-spectrum sources.
Their dominance in VLBI catalogs prompted researchers to consider them to belong to a special class: compact AGNs.

But then a question arises about the nature of steep spectrum
sources ($\alpha<-0.5$) that make up about 90\% 
of extragalactic radio sources brighter than 200~mJy at 1.4~GHz
\citep{2007ARep...51..343M}.
Are they different? Do steep-spectrum sources belong to the
same population as flat-spectrum sources, or do they form a distinctive population?
Steep-spectrum sources remain poorly studied at parsec scales.
Experience from the surveys mentioned above has shown that they are often heavily resolved by VLBI.
At the same time, a number of compact steep-spectrum (CSS) sources have been found and studied by VLBI \citep[e.g.,][]{2006A&A...449..985M, 2007A&A...469..437K, 2013MNRAS.433..147D, 2018MNRAS.477..578C}.
\cite{1998PASP..110..493O} defines CSS sources as those
with steep spectra in centimeter range and sizes $\leq 20$~kpc.
How many extragalactic steep-spectrum sources have VLBI-compact structures?
More generally, what fraction of the whole AGN population has observable parsec-scale jets? Therefore, how many sources are missed by VLBI surveys limited to flat-spectrum targets?

To address these questions, one needs a VLBI survey of a statistically complete, unbiased sample. To date, 
very few such surveys have been made. 
\citet{2005ApJ...618..635G} and \citet{2009A&A...505..509L} presented the
results of the VLBI observations of the Bologna Complete Sample (BCS) of 76 objects
selected from the low-frequency B2 catalog (408~MHz) and 3CR catalog (178~MHz). Their
sample is flux-density limited with a completeness of 80\% but includes only sources
with redshift $z<0.1$. With a sufficiently low detection limit (5~mJy at 5~GHz),
they detected using VLBI 96\% of the observed sources, which means that most of the sources have compact radio nuclei.
The combination of Pearson-Readhead
and Caltech-Jodrell Bank VLBI surveys (PR+CJ1; \citealt{r:pr81, 1988ApJ...328..114P}, \citealt{1995ApJS...98....1P}, \citealt{1995ApJS...98...33T}, \citealt{1995ApJS...99..297X})
formed another statistically complete sample of 200 objects with total flux density at an intermediate frequency of 5~GHz $S_\mathrm{5GHz}>0.7$~Jy, covering 20\% of the sky. Two-thirds of the sources have been
detected and imaged with VLBI at $1.6$ and $5$~GHz. Among 65 sources of this sample with $S_\mathrm{5GHz}>1.3$~Jy, the authors found 
10 compact sources with a steep spectrum. 
\citet{2013MNRAS.434..956C} analyzed
the distribution of the ratio of visibility amplitudes at long (4.5~km) and short (200~m) ATCA baselines and its relation with the spectral index for 5539 sources from the AT20G catalog \citep{r:at20g} at 20~GHz. Among these sources, 27\% are steep
spectrum objects. For most flat-spectrum sources, the visibility amplitude
decreases by less than 15\% between these baselines, which indicates that their
angular size is less than 0.15 arcsec. These authors also found that the share of sources
with size less than 1~kpc and with a steep spectrum in the $1-4.8$~GHz range
is about 11\% in their sample. 

The mJIVE-20 survey \citep{2014AJ....147...14D} and the deep VLBI surveys of the COSMOS \citep{2017A&A...607A.132H} and GOODS-N \citep{2018A&A...619A..48R} fields utilized phase-referencing technique and covered small fields, 200, 2, and 0.5 square degrees, respectively. They observed all known AGNs at 1.4~GHz down to the surface brightness of 1 and 0.1~mJy/beam, respectively. There was no bias to flat-spectrum sources in these surveys. At the same time, they lack bright sources due to the small size of the field.

All these works highlighted the necessity to have a systematic study of parsec-scale
properties of the entire population of radio sources, not only a subsample of
flat-spectrum objects. To address this need, we observed a large, complete, total flux density limited sample with the VLBA and analyzed the relations between the parsec-scale structure and the total simultaneous broadband radio spectra.
This research has several goals: (i) to determine the share of compact objects among flat-spectrum and steep-spectrum radio sources; (ii) to determine the fraction of the VLBI-detected sources in a total-flux-density limited sample selected at 1.4~GHz; (iii) to investigate parsec-scale properties of the VLBI-detected sources with a steep total spectrum, in particular, to understand, whether they are mostly flat-spectrum cores of extended steep-spectrum sources or CSS sources.
A systematic study of these problems is needed not only for understanding the physics of the AGN population and characterization of CSS sources, but also for planning future VLBI observations and constructing VLBI-selected complete samples. The latter is critical for many applications, including the modern multi-messenger neutrino-AGN VLBI studies \citep{2020ApJ...894..101P}.

The paper is structured as follows. In \autoref{sec:sample}, we define the sample and describe the VLBA observations. In \autoref{sec:processing} and \autoref{sec:processing:params}, we dwell on the VLBA data calibration and analysis procedure. We present the results of our survey in \autoref{sec:results}. Our RATAN-600 monitoring program, as well as the sources spectra taken from the literature, are described in \autoref{sec:r600}. We show the results of our joint analysis of the VLBA data and the continuum radio spectra in \autoref{sec:analysis}. We discuss the results in \autoref{sec:discus} and summarize them in \autoref{sec:concl}.

\section{Observing Sample and VLBA Observations}
\label{sec:sample}

For our observations, we selected a sample of sources from the NVSS catalog \citep{1998AJ....115.1693C}, which meet the following criteria:
(i) flux density $S_\mathrm{NVSS}\geq 200$~mJy at 1.4~GHz (NVSS frequency),
(ii) declination $\geq +75\degree$.
The specific choice of the sky area and the flux density cutoff was dictated by the amount of VLBA observing time and recording bitrate we were able to secure. The north polar cap area was chosen for two reasons: it is always observable at all VLBA stations, and there are published broadband radio spectra for all the sources with $S_\mathrm{NVSS}\geq 200$~mJy within this area (see \autoref{sec:r600}).
There are 502 NVSS sources that satisfy these conditions, including two sources with  $S_\mathrm{NVSS}=199.9$~mJy. They are listed in \autoref{tab:sample}. 
We named our program the VLBA North Polar Cap Survey (NPCS). 
Many studies \citep[e.g.,][and the references therein]{2013ApJ...768...37C, 2016ApJ...831..168K, 2016A&ARv..24...13P} showed that AGNs dominate over star-forming galaxies among extragalactic radio sources with centimeter flux density higher than about 1~mJy. We assume, therefore, that all the sources in our sample are AGNs since they are much stronger.

Only for a small fraction of the sources from our sample optical identifications are known. We searched for them in the NASA/IPAC Extragalactic Database (NED, \citealt{1998asal.confE...6M})\footnote{\url{https://ned.ipac.caltech.edu/}}. The optical types are available for 38 sources. There are 7 objects of the BL Lac type, 13 QSO, and 18 radio galaxies, of which 6 sources belong to Seyfert 1 type and 6 sources -- to the Seyfert 2 type. There are the redshift values in the NED for 41 sources from our sample. They are distributed from 0.003 to 3.4 with a median of about 0.6.

\begin{figure}
    \centering
    \includegraphics[width=\columnwidth]{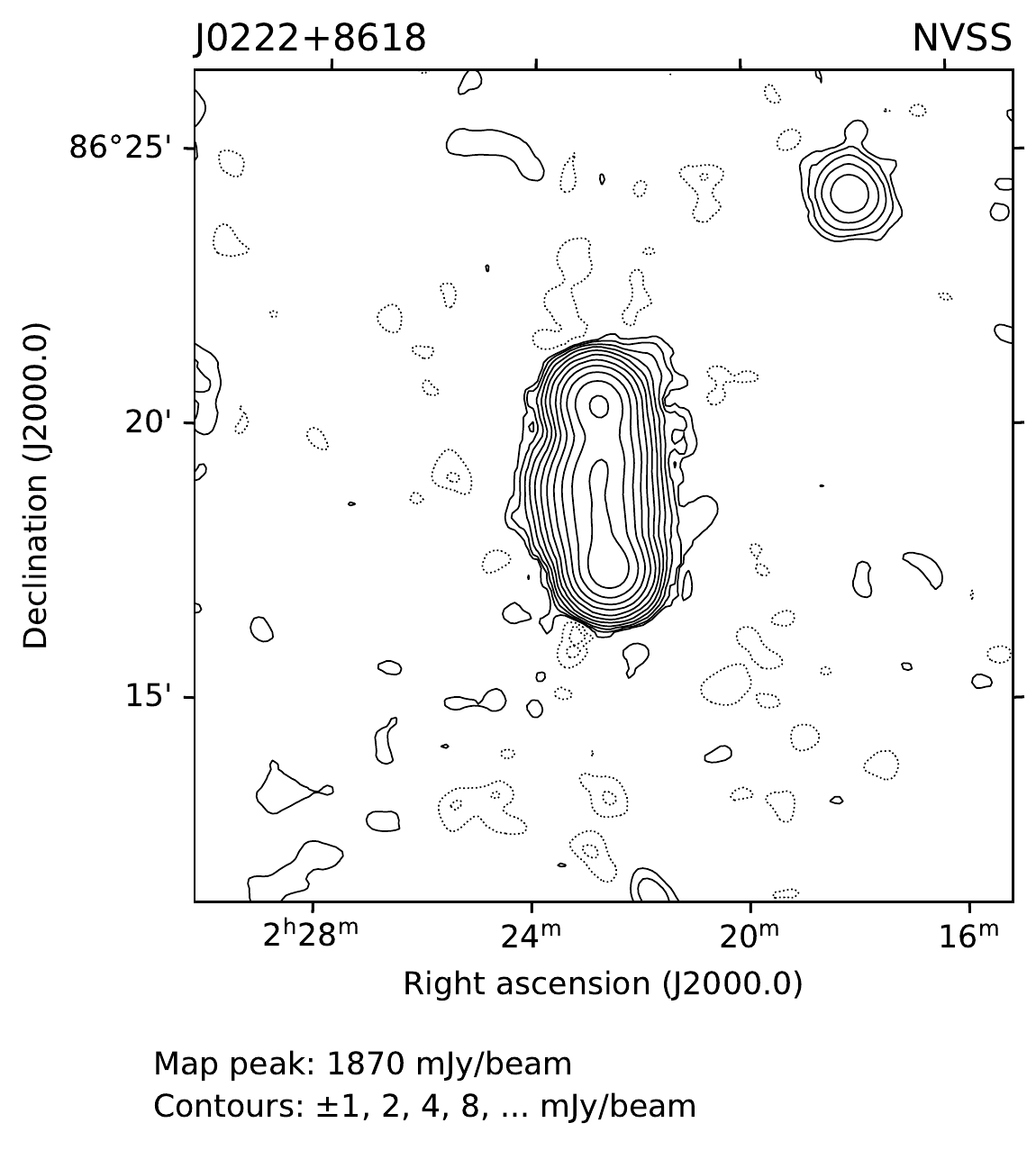}
    \caption{NVSS map for the complex source J0222+8618. The contours are plotted starting at 1~mJy/beam intensity level with an increment of 2. The maps for all 22 complex sources are available as Figure Set~1 in the online version of the article.
    \label{fig:nvss_complex}}
\end{figure}

Some sources from this sample form pairs and groups. Namely, 36 sources have at least one other source from the sample at a distance less than 4~arcmin. At the same time, if 502 sources of the sample were randomly scattered over 703 square degrees of the considered area of the celestial sphere, we would expect only 5 sources with such close neighbors. This is evidence that most of close pairs and groups are not apparently close separate sources, but are large extended sources resolved by the VLA to two or more components. The value of the distance limit of 4~arcmin was found empirically; it seems to be optimal to distinguish extended resolved sources from individual apparently close sources in our sample. For this reason, we consider pairs/groups formed by these 36 sources as single sources and label each pair/group by one joint name. We also consider the source NVSS J204209+751226 as a component of the complex source J2042+7508, because the NVSS map shows that it is indeed a component of this giant radio galaxy, although its distance to the closest neighbour is slightly larger than 4~arcmin. In total, there are 17 complex objects in the sample, consisting of 37 NVSS sources. Therefore, the actual number of sources in the sample is 482. The complex sources are marked in \autoref{tab:sample} by the label `\textit{a}.'
Searching for the NVSS sources forming our sample in the NED
and subsequent visual inspection of the NVSS images showed that 5 additional NVSS sources may be resolved components of extended complex galaxies. Other components of these complex sources are below our flux density cutoff of 200~mJy. These NVSS sources are marked in \autoref{tab:sample} by the label `\textit{b}.'
In Figure Set~\ref{fig:nvss_complex}, we show the NVSS maps for all 22 complex sources with at least one component belonging to our sample. The FITS images were obtained using the NVSS Postage Stamp Server\footnote{\url{https://www.cv.nrao.edu/nvss/postage.shtml}}.

For several sources from our sample, the difference between the coordinates of the compact feature detected by the VLBA (see \autoref{tab:coord}) and of the centroid of the NVSS image results in the difference in the source names formed from these coordinates. These sources are marked in \autoref{tab:sample} by the label `\textit{c}.'

\begin{deluxetable}{lcccc}
    \tablecaption{
    The observing sample and the VLBA detection results.
    \label{tab:sample}
    }
    \tablehead{\colhead{NVSS Name} & \colhead{J2000 Name} & \colhead{B1950 Name} & \colhead{Det2} & \colhead{Det8}}
    \colnumbers
    \startdata
    J001311+774846 & J0013+7748 & 0010+775 & Y & N \\
    J033021+763323 & J0330+7633 & 0324+763 & N & Y \\
    J170524+775559\tablenotemark{c} & J1705+7756 & 1707+779 & Y & Y \\
    J190653+810010\tablenotemark{a} & J1906+8100 & 1911+809 & N & N \\
    J190731+810008\tablenotemark{a} & J1906+8100 & 1911+809 & N & N \\
    \enddata
    \tablecomments{Columns are as follows: (1) -- Source name in the NVSS catalog; (2) -- J2000 object name; (3) -- B1950 object name; (4) -- `Y' if a source was detected at 2.3~GHz in our VLBA survey and `N' otherwise; (5) -- the same as (4) for 8.6~GHz. This table is published in its entirety in the machine-readable format. A portion is shown here for guidance regarding its form and content.}
    \tablenotetext{a}{The NVSS source is a component of an extended complex object. For all the components of each complex object, the same J2000 and B1950 names are given. These names correspond to the compact source detected by the VLBA, or, if the object was not detected, to the brightest of the NVSS components.}
    \tablenotetext{b}{The NVSS source is a component of an extended complex object, but the other components of this complex object are too weak to be included into our sample.}
    \tablenotetext{c}{The sources for which our J2000 name differs from the shortened NVSS name due to the coordinate correction; see \autoref{sec:sample} for details.}
\end{deluxetable}

We observed this sample with the VLBA in three 24-hour observing sessions: on 14, 16, and 23 February 2006 (project code BK130). The telescopes were pointed to each of the 502 original NVSS sources. Each source was observed for about 8~minutes simultaneously in two frequency bands: 2.3~GHz (S~band) and 8.6~GHz (X~band), with single (right circular) polarization. Four 8~MHz-wide intermediate frequencies (IF) were allocated at each band. They cover 148~MHz and 498~MHz at 2.3 and 8.6~GHz bands, respectively. With 1-bit sampling, the data bit rate was 64~Mbit~s$^{-1}$ in each band, and 128~Mbit~s$^{-1}$ in total. In addition to the target sources, we also observed tropospheric calibration sources in eight blocks each day evenly distributed in time; each block consisted of four-to-five sources with different elevations; the calibration source scan duration was 90 seconds. Three sources from the sample (J0017+8135, J1058+8114, J1153+8058) also served as calibrators. They were observed in two-to-three 90-second scans each day of the observations in addition to a 8-minute scan in one of the days.


\section{VLBA Data Processing}
\label{sec:processing}

The data were correlated at the VLBA correlator at the NRAO Array Operations Center in Socorro. The correlator integration time was 0.5~s, and there were 64 spectral channels of width 125~kHz within each IF. This relatively high temporal and spectral resolution was necessary for fringe fitting with poorly known a~priori coordinates, taken for the most of the target sources from the NVSS catalog.

The data a~priori calibration was made independently with two software packages: AIPS \citep{2003ASSL..285..109G} and PIMA \citep{2011AJ....142...35P}. These packages have different important advantages. 
Antenna-based fringe fitting is well implemented in AIPS. Additionally, the AIPS a~priori amplitude calibration and band-pass normalization are extensively tested to deliver no significant amplitude bias.
In turn, PIMA is capable of finding joint baseline-based fringe-fitting solutions for frequency channels widespread around the band. It also precisely estimates the noise level of fringe solutions and the probability of a false detection.
We explored the agreement between the results delivered by the two totally independent software packages.
In both packages, all usual steps of calibration of the VLBI data were made,  including (i) data flagging, (ii) a~priori amplitude calibration, (iii) phase calibration using pulse-cal signal, (iv) fringe fitting, (v) complex band-pass calibration, and (vi) global antenna gain corrections derived from the self-calibration of the strongest sources.


\subsection{A priori calibration and detection filter in PIMA}

PIMA independently processes data collected from a given scan
at a given baseline and a given band, hereafter called an observation.
The procedure of fringe fitting determines the phase delay rate, the group
delay, and the group delay rate using the spectrum of the cross-correlation
function (also known as visibility data) across the band.
See \citet{2011AJ....142...35P} for details of this procedure implementation.
The point-like source model is implicitly assumed in the procedure.
The fringe fitting results were used in two ways. First,
the group delays were used for our astrometric analysis. Second,
the group delays, the phase delay rates, and the group delay rates
were applied to the visibilities, and after that the visibilities were averaged
over frequency within each individual IF and over time.

Upon the fringe fitting completion, the data were exported to the NASA VLBI analysis software VTD/pSolve\footnote{\url{http://astrogeo.org/vtd/}}$^,$\footnote{\url{http://astrogeo.org/psolve/}}.
That software implements a robust algorithm for estimation of source
coordinates, atmospheric path delay in zenith direction, and a clock
function for all the stations but the one taken as the reference
using X-band and S-band group delays in a presence of a high
number of outliers. The procedure is described in full details
by \citet{2020arXiv200809243P}. The robustness is achieved by exploiting our knowledge of the statistics of both detections and non-detections, and the a~priori probability of detection derived
from the empirical signal-to-noise ratio (SNR) distribution. 
This approach was also used by the \textit{RadioAstron} AGN survey \citep{2020AdSpR..65..705K}.
The residuals of the group delays of detected observations have Gaussian distribution with the second moment 0.03 and 0.15~ns for X and S bands respectively, while the residuals of non-detected observations have a uniform distribution within the fringe search window [$-4000, +4000$]~ns. The procedure works both as an estimator of source positions and as a filter of observations
where no interferometric signal was detected. Observations with residuals exceeding 4.5 weighted root mean squares $\sigma$ over all post-fit residuals at S band and 4.0$\sigma$ at X band are discarded as outliers.

A source is considered detected if the number of its observations used in the solution, i.e., not suppressed, is at least 3. Estimation of the right ascension and the declination takes two degrees of freedom. If only two observations of a given source passed the filter, the residuals will be zero for any group delays. If two observations of a given source at S-band were detected, and therefore, their residuals obey the Gaussian distributions with a given second moment, while the third observation was not detected, the probability that its group delay by chance has post-fit residuals less than 4.5$\sigma$ and as a result, was not discarded as an outlier is $4.5\times0.15/4000 = 1.7\times10^{-4}$. Similar calculations for X-band yield the probability of $4\times0.03/4000 = 3\times10^{-5}$. Since there are three possible combinations of two detected observations and one non-detected for a set of three observations, the probability that at least one of three observations is not detected is three times higher, namely $5\times10^{-4}$ for S-band and $9\times10^{-5}$ for X-band. This makes an astrometric solution a very powerful filter.

Since the fringe-fitting procedure processed the observations independently, the fringe reference times for the observations of a given scan were slightly different for different baselines, and the closure of the group delay and the phase delay rate was not preserved. Therefore, the baseline-dependent
group delays, phase delay rates, and group delay rates were
converted to station-based quantities referred to the common
scan reference time using least squares and taking one of the
stations as a reference. The baseline-dependent group delays were
used for astrometric analysis. The station-based group delays and rates were applied to the visibility data, which were then averaged in time and frequency.
The averaged visibilities were used for correlated flux density measurements and source structure modeling.

\subsection{A priori calibration in AIPS and comparison of the results}

In AIPS, we followed the commonly used procedure of VLBA data calibration. In the global fringe fitting, separate solutions were found for each IF; the minimum signal-to-noise ratio for a detection was set to 4. Note that in AIPS, the SNR is defined in a slightly different manner than in PIMA; the PIMA SNR is a $\sqrt{\pi/2}$ times smaller than the AIPS SNR. Such a low AIPS SNR threshold was chosen because there are many sources near or below the detection limit in our complete sample. A source was treated as detected and the data for it were used in subsequent analysis, only if the detection was confirmed by the robust PIMA detection procedure.

The processing of the same dataset of 3 days of the VLBA observations in AIPS and PIMA allowed us to compare their outcomes. For strong sources, both packages yield practically the same result. Considering only observations with the calibrated visibility amplitude greater than 1~Jy, we calculated the median ratio of the visibility amplitudes calibrated in AIPS to the amplitudes calibrated in PIMA. It varies from 94.8\% to 99.8\% for different daily segments and bands. The difference between the calibrated amplitudes of two packages, therefore, does not exceed the typical amplitude calibration uncertainty of VLBI survey data of 5-10\%. For weak sources close to the detection limit, PIMA is more sensitive. AIPS has a known limitation: it cannot process more than one IF unless they are contiguous. Since the frequency allocation of our data was not contiguous due to astrometry calibration requirements, we had to process the data with AIPS using each IF independently, setting APARM(5)=0 in the FRING task. Therefore, we lose sensitivity with respect to PIMA that uses all IFs of a given band for a joint solution.

Taking all this into account, we used the data processed in AIPS for imaging and subsequent analysis of those sources, for which hybrid imaging was robust (see \autoref{sec:processing:imaging}). For other sources (mostly weak and/or strongly resolved), we used the data processed in PIMA.

\subsection{Absolute astrometry}

Observations of the VLBA North Polar Cap Survey were also used for
absolute astrometry. They were processed in a similar way as the VLBA Calibrator Surveys \citep[e.g.,][]{2008AJ....136..580P}. We
refer a reader to that publication which discusses a general
approach and here we dwell upon the technique that is specific for
the analysis of this campaign. 

All dual-band geodetic VLBI data from 24-h observing sessions from 1980.04.01 through 2020.03.09, in total 6498 experiments, and three observing sessions of this survey were processed in three least square runs. The first run used both X- and S-band data from this survey, the second run used only X-band data, and the third run used only S-band data. The number of the detected sources from
the survey used in these solutions is 108, 117, and 157,  respectively.
The estimated parameters are split into three categories: global parameters such as station positions, station velocities, and source coordinates; session-wide parameters, such as pole coordinates, UT1 angle, their time derivatives, and nutation angle offsets; and segment-wide parameters, such as clock function and atmospheric path delay in the zenith direction.
The segment-wide parameters are modeled with a B-spline with a time span of 1~hr.

For accounting systematic errors, we computed weights in the following way:
\begin{equation}
    \label{e:e1}
    w = \frac{1}{k \cdot \sqrt{\sigma_g^2 + a^2 + b^2(e)}},
\end{equation}
where $\sigma_g$ is the group delay uncertainty, $k$ is the multiplicative
factor, $a$ is the elevation-independent additive weight correction, and $b$
is the elevation-dependent weight correction. We used $k=1.3$ based on the
analysis of the VLBI-Gaia offset \citep{r:gaia4}. For processing dual-band
observations, we used $b(e)= \beta \sqrt{\tau(e_1)_{\rm atm,1}^2 + \tau(e_2)_{\rm atm,2}^2}$,
where $\tau(e_i)_{\rm i,atm}$ is the atmospheric path delay at the $i$th
station. We used $\beta=0.02$ in our work. We made trial runs using all geodetic
experiments, and they showed that this choice provides the minimum baseline
length repeatability.

For processing single-band observations, we computed the ionospheric delay
using the Total Electron Contents (TEC) maps from the analysis of the Global Navigation
Satellite System (GNSS) observations. Specifically, we used the CODE TEC time
series \citep{r:schaer99}\footnote{Available at
\href{ftp://ftp.aiub.unibe.ch/CODE}{ftp://ftp.aiub.unibe.ch/CODE}} with
a resolution of $5\degree \times 2.5\degree \times 2$~hr. However, the TEC maps
account only partially for the ionospheric path delay due to the coarseness of
their spatial and temporarily resolution. In order to account for the
contribution of residual ionosphere-driven errors, we used the same approach
as we used for processing single-band Long Baseline Array (LBA) observations
\citep{r:lcs2}. We computed variances of the mismodeled contribution of the
ionosphere to group delay in zenith direction for both stations of
a baseline,  $\Cov_{11}$ and $\Cov_{22}$, as well as their covariances
$\Cov_{12}$. Then, for each observation, we computed the predicted rms
of the mismodeled ionospheric contribution as
\begin{equation}
\begin{split}
    \label{e:e3}
    b^2_{\rm iono}(e) = & \gamma \left( \Cov_{11}^2 M_1^2(e) \right. \\
    & - 2 \Cov_{12} M_1(e) M_2(e) \\
    & + \left. \Cov_{22}^2\, M_2^2(e) \right),
\end{split}
\end{equation}
where $M_1(e)$ and $M_2(e)$ are the mapping functions of the ionospheric
path delay. We used $\gamma=0.5$ in our analysis and added $b^2_{\rm iono}(e)$
to $b^2(e)$ when processed single-band observations.
The additive parameter $a$ was found by an iterative procedure that makes
the ratio of the weighted sum of post-fit residuals to their mathematical expectation close to unity.

We analyzed a dataset of 86 target sources with more than 10 detections in the NPCS campaign at both S- and X-bands. We calculated the position differences from the S-band solution with respect to the X/S solution and similarly for the X-band.
The position differences normalized
over the single-band position uncertainties fit to the Gaussian distribution
over right ascensions and declination with the zero mean and the 2$^\text{nd}$
moment $0.5$ for X-band positions, $0.8$ for S-band right ascensions, and
have a positive bias of $+10$~mas and 2$^\text{nd}$ moment $1.0$ for S-band
declinations. Since the 2$^\text{nd}$ moment of the distribution of the normalized
differences does not exceed one, we conclude that the formal uncertainties
correctly account for ionosphere-driven systematic errors.
The declination bias in the S-band positions was applied in the catalog.

Although a number of sources were observed in other campaigns,
we present here the positions derived only from the
observations of the NPCS campaign (see \autoref{tab:coord}). We also detected four additional sources, which do not belong to our observing sample but lie close to some of the target sources. They were also used in the astrometric solution, and their coordinates are also given in \autoref{tab:coord}. The modern positions of the sources can be found in the Radio Fundamental
Catalog (RFC)\footnote{Available at \href{http://astrogeo.org/rfc}{http://astrogeo.org/rfc}, maintained by Leonid Petrov.}, which is updated
on a quarterly basis.

\begin{deluxetable*}{cccRRRl}
    \tablecaption{
    J2000.0 coordinates measured for all the sources detected in the VLBA NPCS survey, including 162 sources from the target sample and 4 additional sources.
    \label{tab:coord}
    }
    \tablehead{\colhead{Name} & \colhead{RA} & \colhead{Dec} & \colhead{RA error} & \colhead{Dec error} & \colhead{Correlation} & \colhead{Bands}}
    \colnumbers
    \startdata
    J0009+7724 & 00:09:43.09202 & +77:24:42.00489 &    6.80 &  3.30 & -0.322 & XS \\
    J0009+7603 & 00:09:48.23409 & +76:03:18.16477 &   14.84 & 10.27 & -0.498 & XS \\
    J0013+7748 & 00:13:11.70860 & +77:48:46.67715 &  131.78 & 49.91 &  0.556 & S \\
    J0017+8135 & 00:17:08.47491 & +81:35:08.13646 &    1.17 &  0.21 & -0.023 & XS \\
    J0038+8447 & 00:38:11.86078 & +84:47:27.15210 &  194.02 & 19.50 &  0.748 & XS \\
    \enddata
    \tablecomments{Columns are as follows: (1) -- J2000 name; (2) -- Right ascension for the epoch J2000.0; (3) -- Declination for the epoch J2000.0; (4) -- Error in right ascension in milliarcseconds (mas); (5) -- Error in declination in mas; (6) -- Correlation between the errors in right ascension and declination; (7) -- Frequency bands in which a source was detected and which were used for the astrometric solution: S -- 2.3~GHz, X -- 8.6~GHz. This table is available in its entirety in machine-readable form. First five lines are shown here for guidance regarding its form and content.}
\end{deluxetable*}


\subsection{Hybrid imaging and issues with self-calibration of weak or resolved sources}
\label{sec:processing:imaging}

The next stage of data processing was hybrid imaging
in Difmap \citep{1994BAAS...26..987S, 1997ASPC..125...77S}. Before it, bad data points were flagged manually after visual inspection. Hybrid imaging was done by our automatic script, based on the approach developed by \citet{1994AAS...185.0808P}.

A key part of hybrid imaging is self-calibration. However, the data can be self-calibrated not for all the sources in our project. First of all, there are many sources detected at few baselines that do not form a quadrangle or even a triangle and, consequently, self-calibration for them is impossible. Another significant difficulty is that the phase self-calibration of noisy data, which is the case of a significant fraction of sources in our sample, may lead to a significant artificial increase of the average visibility amplitude and may even generate a spurious source from pure noise \citep{1988IAUS..129..509W, 2008A&A...480..289M}. The stability and the correctness of the phase self-calibration depend mainly on the signal-to-noise ratio of the visibilities and the $uv$-coverage. One can increase the SNR of solutions by increasing the solution time interval. However, if the solution interval is too long, the averaging becomes incoherent, and the resulting amplitude is underestimated \citep[e.g.,][]{2010A&A...515A..53M}. Therefore, one needs to decide which sources are appropriate for hybrid imaging, and find the optimal solution interval depending on the data quality. This problem is well-known; see, e.g., \citet{1999ASPC..180..187C}, section~5.3.

\begin{figure}
    \centering
    \includegraphics[width=\columnwidth]{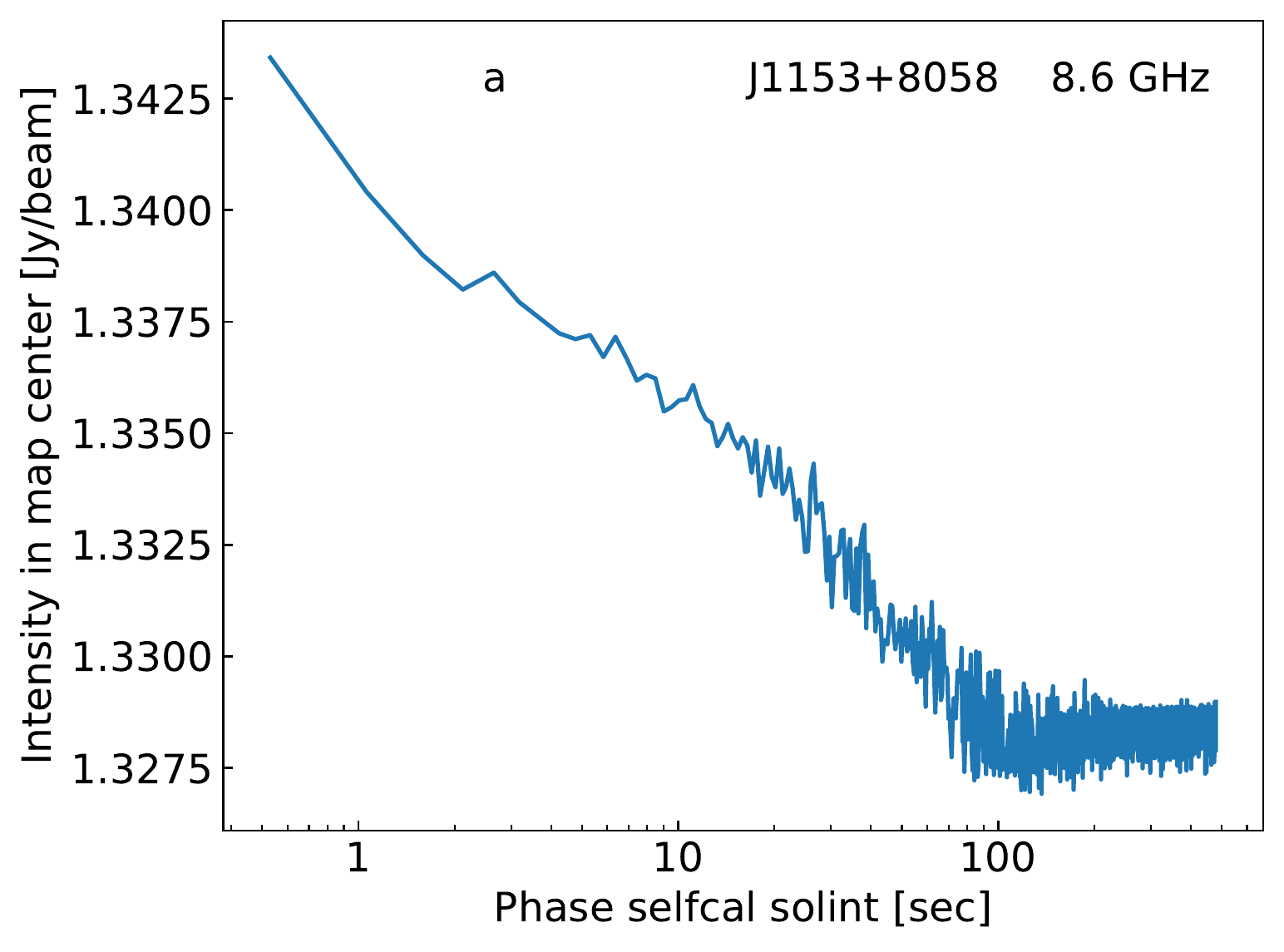}
    \includegraphics[width=\columnwidth]{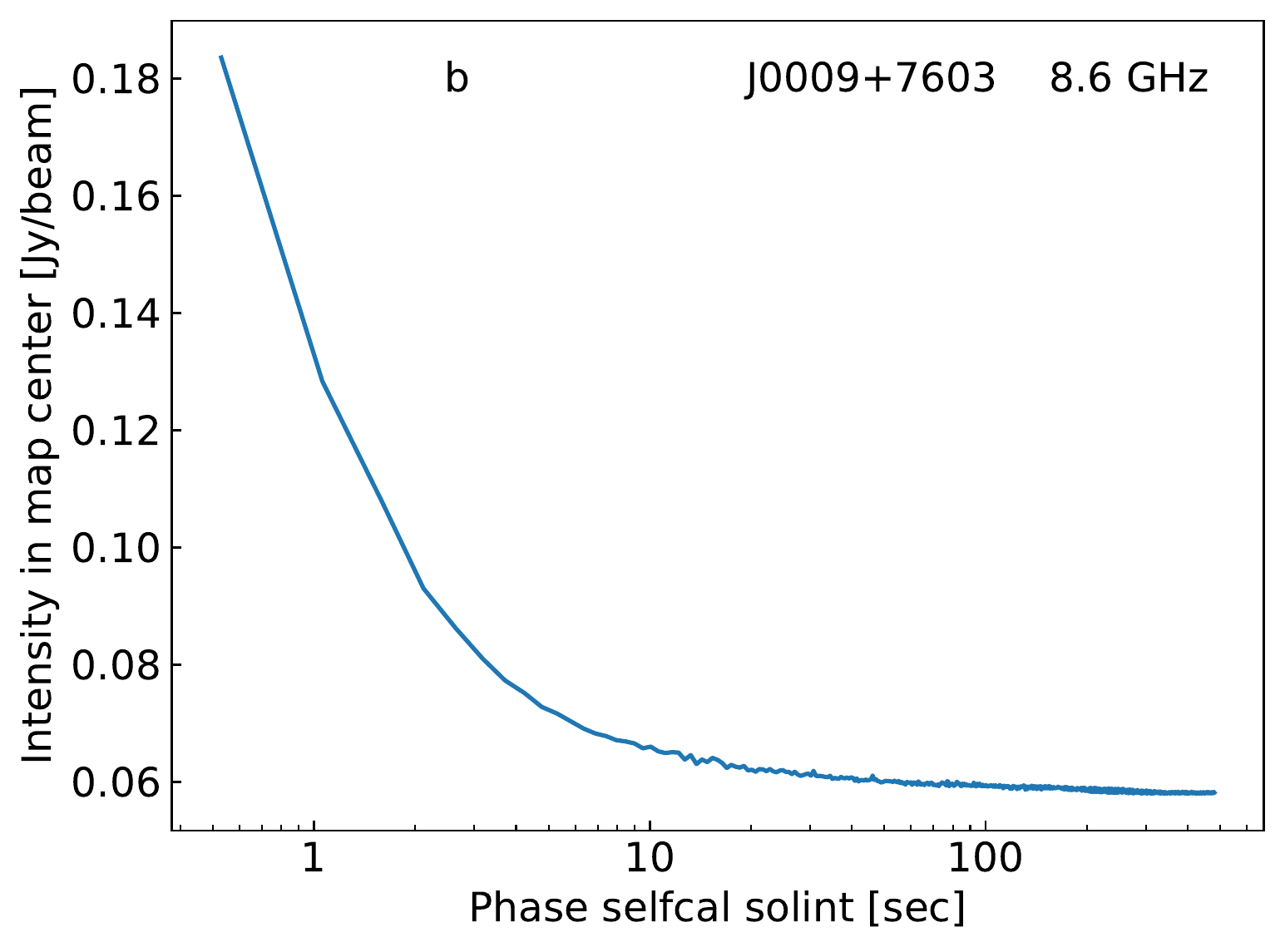}
    \includegraphics[width=\columnwidth]{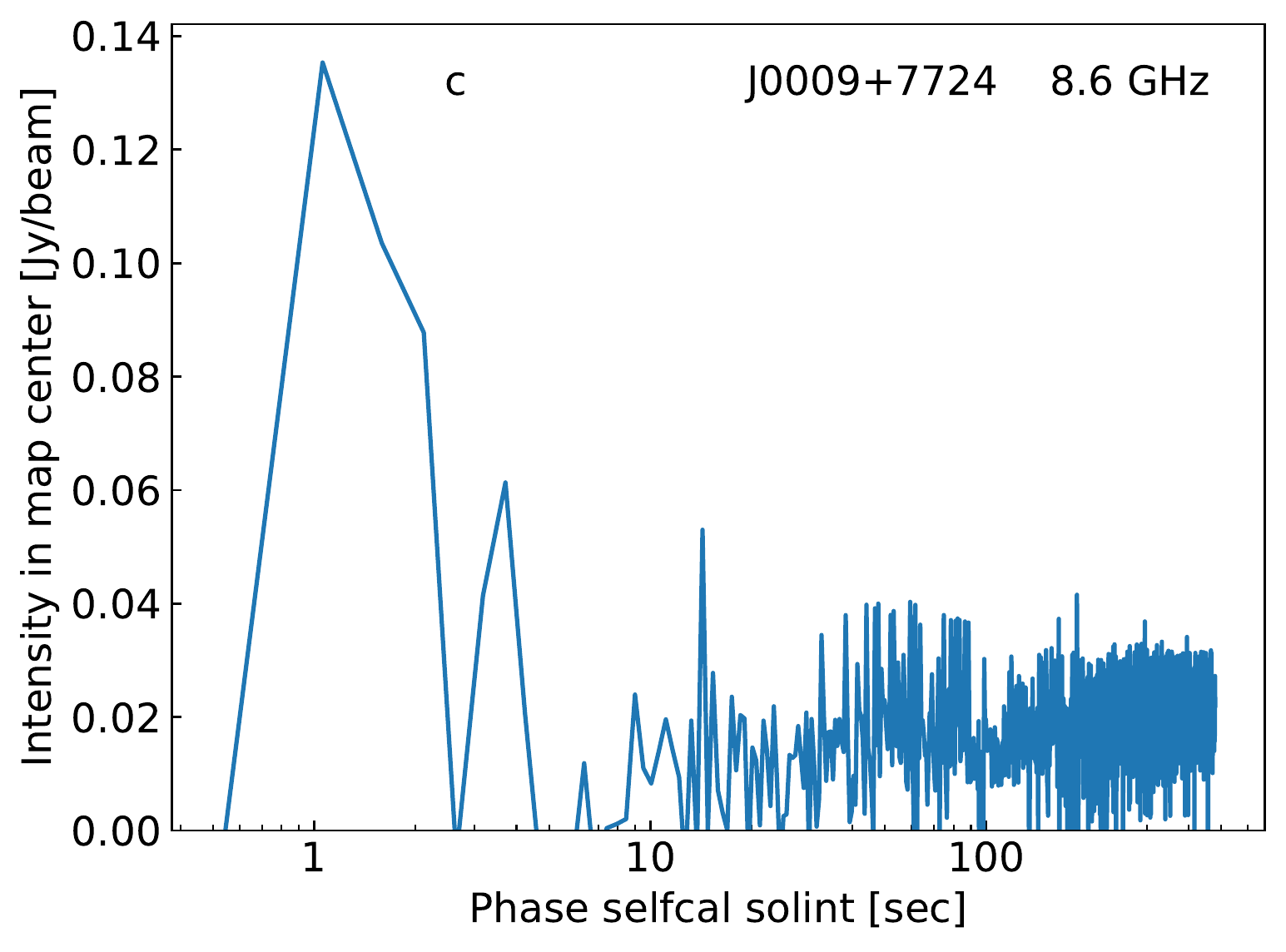}
    \caption{
    The relation between the intensity in the central pixel of the map, produced by automatic hybrid imaging in Difmap, and the phase self-calibration solution interval used in the procedure. (a) Strong source. (b) Weak source. (c) Very weak source. These plots are shown for the data at 8.6~GHz; at 2.3~GHz the situation is similar. See the discussion in \autoref{sec:processing:imaging}.
    \label{fig:solint}
    }
\end{figure}

To test the applicability of phase self-calibration, we performed automatic hybrid imaging of each source detected at six or more baselines with different phase self-calibration solution intervals from the correlator integration time (0.5~s for this experiment) to the whole scan duration (about 8~min). We analyzed the dependence of different parameters of the resulting maps on the phase self-calibration solution interval. We found that the most informative parameter is the relation between the phase self-calibration solution interval and the intensity in the central pixel of the map. The intensity in the central pixel reflects not only the changes in the peak intensity of the map but also the peak shifts. The examples of such relations are shown in \autoref{fig:solint}.

The shape of the curves like those shown in \autoref{fig:solint} varies significantly from source to source. However, the sources may be roughly divided into three groups. For strong sources, with a correlated flux density much higher than the detection limit (panel (a) in \autoref{fig:solint}), the effects discussed above are negligible. Self-calibration works perfectly, correcting visibility phase fluctuations and, therefore, increasing the peak intensity for about 1\%, when short solution intervals are used. For weak sources (panel (b)), the situation is radically different. The phase self-calibration with short, about one second, solution intervals creates a partly fake signal from the noise. As a result, the intensity in the map center is several times higher than that of after self-calibration with solution intervals of about 10~s and longer. In the case of a very weak source (panel (c)), hybrid mapping is completely unstable. Note, however, that for the short solution interval, in the case (c), there appears a relatively bright compact source in the center of the map, which is inconsistent with the data before self-calibration. These examples show that the phase self-calibration of the VLBI data must be applied carefully when dealing with weak and/or very resolved sources.

We inspected by eye the plots similar to that in \autoref{fig:solint} at both frequency bands for all the sources detected at baselines formed by four or more antennas, as well as the resulting maps and calibrated visibilities. We have chosen for hybrid imaging the sources for which this procedure is stable and robust. Whether the phase self-calibration is applicable for a source depends on many factors, including the source structure. There are two main characteristics of the visibility data quality of our snapshot observations: the number of independent points in the $uv$ plane, i.e.\ the number of baselines at which a source is detected, and the median visibility SNR. Here we call the ratio of the amplitude of the visibility at a given baseline coherently averaged over time and frequency to its statistical error as a visibility SNR. Our analysis showed that the hybrid imaging is unstable for the sources with a median visibility $\mathrm{SNR}<6$ and detections at less than 15 baselines. If in our observations a source is detected only at 15 baselines or less, while the total number of VLBA baselines is 45, this means that the source is either strongly resolved, or very weak with the flux density near the detection limit. We found that the hybrid imaging is also unstable for some sources with better data. For this reason, we decided manually whether to perform hybrid imaging of a source or not. For the sources which poor data amount and/or quality prevents their imaging, we performed only the flux density estimation and modeling using the a~priori calibrated visibility amplitudes (see \autoref{sec:processing:params}).

For the sources we found suitable for hybrid imaging, we applied a phase self-calibration solution interval of 8~s to avoid the conversion of noise into a signal in the process of phase self-calibration. A longer self-calibration solution interval limits corrections which compensate short-term phase variations in the atmosphere and might result in a partial incoherence and a loss of the visibility amplitude. 

In process of hybrid imaging, for most of the imaged sources, we made also three iterations of amplitude self-calibration: the first two with one gain correction (per antenna per IF) for the whole scan time, and the third one with a 2-minute solution interval. However, for highly resolved sources, this approach leads to problems. If the correlated flux density sharply decreases with the baseline length at short baselines so that at most baselines it is several times lower than at the shortest one, the CLEAN model cannot fit the data at the shortest baselines. The $uv$ coverage is too poor to allow for a robust model reconstruction. Manipulations with visibility weighting and tapering do not solve the problem. In such situations, the amplitude self-calibration ``corrects'' visibilities to make them closer to a wrong CLEAN model. Therefore, it significantly changes (usually reduces) visibility amplitude at short spacings. For this reason, we do not apply the amplitude self-calibration to sources for which the ratio of the largest time- and frequency-averaged correlated flux density to the median of averaged correlated flux densities over all baselines is $\geq 2.5$. There are 21 such sources at 2.3~GHz and 6 sources at 8.6~GHz.

We set a map pixel size of 0.6~mas at 2.3~GHz and 0.15~mas at 8.6~GHz. The default number of pixels was 1024$\times$1024. We also made maps of wider fields using the similar procedure in Difmap to search for outlying components. For the sources in which such components were found, the number of pixels was increased to 2048$\times$2048 or 4096$\times$4096 (the latter for 8.6~GHz only). The bandwidth smearing causes the intensity loss of more than 10\% at the distances larger than about 450 mas from the map center in our project. Taking into account that the linear size of the CLEAN map is two times smaller than that of the grid it uses, all our maps cover the area within the radius of 450 mas, therefore, bandwidth smearing is not an issue.


\section{Parameters of the sources derived from the visibility data}
\label{sec:processing:params}

From the calibrated visibility data, we obtained a number of parameters that characterize the parsec-scale source structure, in addition to the coordinates and maps of the sources.
For the sources for which hybrid imaging was done (\autoref{sec:processing:imaging}), we used self-calibrated visibility data. For the rest sources, we used the data after a~priori calibration only.

First of all, we measured the total flux density of a source $S_\mathrm{vlba}$ at parsec scales, which we also call the VLBA flux density.
We estimated it as the maximum of time- and frequency-averaged correlated flux densities over baseline projections shorter than 10\% of the longest VLBA baseline of about 8600~km. If the correlated flux density of a source close to the detection limit changes with the baseline projection non-monotonically (e.g., for visibility beatings in a double source), the source may be detected only at baseline projections longer than 10\% of the longest baseline. In such cases, we took the maximum correlated flux density among all the baselines.

Secondly, we calculated the median correlated flux density at the baseline projections longer than 70\% of the longest VLBA baseline. We call it unresolved flux density $S_\mathrm{unres}$ because it comes from the features of the source structure practically unresolved by the VLBA. Note, however, that since the $uv$-coverage of our VLBA snapshots is non-isotropic, the measured value of the unresolved flux density depends not only on the source structure, but also on the array orientation with respect to a source.

The VLBA flux density for the weakest source detected at 2.3~GHz is $29$~mJy, and for the weakest source detected at 8.6~GHz, it is $27$~mJy. We conclude that the detection limit of our survey is around 30~mJy for both frequency bands, in agreement with the expected baseline sensitivity of the VLBA for the bandwidth and the integration time of our observations. 
If there are no detections at long baselines, we put an upper limit on the sources's unresolved flux density. Similarly, if a source is not detected at all, we put an upper limit on its VLBA flux density. In both cases, the upper limit is equal to the detection limit of our observations. 

The single-dish observations (\autoref{sec:r600}) provided us with the total
flux density denoted as $S_{\rm sd}$ from the whole source,
including its extended periphery. The ratios of the flux densities from different spatial scales ($S_\mathrm{unres}$, $S_\mathrm{vlba}$, and $S_\mathrm{sd}$) characterize the source compactness. Let us estimate the largest angular scale $\theta_\mathrm{max}$, for which our VLBA observations are sensitive. The shortest VLBA baseline has a length $D_\mathrm{max}=236$~km. Because we observed circumpolar sources, 
the baseline projections did not differ considerably from their actual length. For 2.3~GHz (wavelength $\lambda=13$~cm), $\theta_\mathrm{max} \sim \lambda/D \approx 6\times10^{-7} \mathrm{rad} \approx 0.1''$. For the median redshift of the sources in our sample $z_\mathrm{med}\approx 0.6$ (calculated among the sources with the redshift given in the NED database), it translates to the linear projected size of about 800~pc. Similar calculations for 8.6~GHz ($\lambda=3.6$~cm) yield a size of about 200~pc. That means that the radiation we observe with the VLBA comes from regions with a characteristic projected size of hundreds of parsec or smaller. At the same time, the extended extragalactic sources have sizes up to megaparsec, and the whole source contributes to the total flux density observed by single-dish telescopes. Therefore, the parameter
\begin{equation}
    C^\mathrm{vlba}_\mathrm{sd} = \frac{S_\mathrm{vlba}}{S_\mathrm{sd}}
\end{equation}
indicates the source compactness at kiloparsec scales, and we call this ratio a kiloparsec-scale compactness parameter. Another ratio,
\begin{equation}
    C^\mathrm{unres}_\mathrm{vlba} = \frac{S_\mathrm{unres}}{S_\mathrm{vlba}},
\end{equation}
may be called a parsec-scale compactness parameter because it indicates what fraction of the VLBA flux density comes from the unresolved parsec-scale core.
When the VLBA did not detect a source and, hence, only the upper limit on $S_\mathrm{vlba}$ is known, then for $C^\mathrm{vlba}_\mathrm{sd}$, we also can derive an upper limit. A similar situation is for $C^\mathrm{unres}_\mathrm{vlba}$ in the case when no detections are available at long projected spacings.

To estimate the size $\theta$ and the brightness temperature $T_\mathrm{b}$ of the main feature of a source, we modeled the source structure in the visibility plane. We used Difmap to fit models of one or two circular Gaussian components to the naturally weighted, self-calibrated complex visibilities.
Such simple models were used because of the limited amount of the observational data -- only one VLBA scan for each source. Modeling the VLBI data by two circular Gaussians was proven to provide robust results for the dominant feature \citep[e.g.,][]{2005AJ....130.2473K,2015MNRAS.452.4274P}. We decided which number of components to use by the following method. At first, we fitted one circular Gaussian component and subtracted it from the data. If the peak of the residual map was greater than six times the map root mean square, we added the second component and re-fitted the whole model.

For weak or extended sources which were not self-calibrated (see \autoref{sec:processing:imaging}), fitting a model to the complex visibilities may lead to erroneous results owing to incorrect phase values. However, some of these sources have robust detections at a large enough number of baselines to fit a model of one circular Gaussian to the visibility amplitudes only. We did it using the maximum likelihood method with the Rician error distribution.

For practically all the sources modeled by two Gaussians, we consider the ``main'' the component located closer to the map center, which is typically the position of the intensity peak. However, for several sources detected at both 2.3 and 8.6~GHz, different components dominate in two bands. In such a case, we consider as ``main'' the component with a flatter spectral index \citep{2014AJ....147..143H} assuming we align them properly. The main component may represent physically different structures, depending on the source morphology -- a jet core, a mini-lobe, or a compact feature of some extended structure. The flux density of the main Gaussian component $S_\mathrm{gauss}$ is, as expected, comparable for most sources to the total VLBA correlated flux density $S_\mathrm{vlba}$, estimated from the visibility amplitudes at short baselines as described above. The median of the ratio of $S_\mathrm{gauss}$ to $S_\mathrm{vlba}$ is about 0.8 for both frequencies. For strongly resolved sources, $S_\mathrm{gauss}$ is slightly higher than $S_\mathrm{vlba}$ due to the sharp decrease of the correlated flux density with the increase of the baseline. 
There are sources for which the decision on the main component identification is based on the spectral index value; their main component at 2.3~GHz accounts for less than half of $S_\mathrm{vlba}$.

Using the flux density of the main component $S_\mathrm{gauss}$ and its full width at half-maximum (FWHM) $\theta$, we calculated its brightness temperature in the observer's frame \citep[e.g.,][]{2005AJ....130.2473K}:
\begin{equation}
\label{eq:tb}
T_\mathrm{b} = \frac{2 \ln 2}{\pi k_\mathrm{B}} \frac{c^{2}S_\mathrm{gauss} }{\nu^{2}\theta^{2}},
\end{equation}
where $k_\mathrm{B}$ is the Boltzmann constant, $c$ is the speed of light, and $\nu$ is the observing frequency.
Fitting a model of a circular Gaussian to visibilities of an unresolved source results in an artificially small or zero-size $\theta$ which does not allow us to estimate $T_\mathrm{b}$ of the component. To avoid this, we calculated the resolution limit $\theta_\mathrm{lim}$ for the maps of the modeled sources following \cite{2005astro.ph..3225L} and \cite{2005AJ....130.2473K}:
\begin{equation}
\theta_\mathrm{lim}=b_\mathrm{maj}\sqrt{\frac{4 \ln 2}{\pi}\ln \left(\frac{\mathrm{SNR}_\mathrm{map}}{\mathrm{SNR}_\mathrm{map}-1}\right)}\,,
\end{equation}
where $b_\mathrm{maj}$ is the major axis FWHM of the beam and SNR$_\mathrm{map}$ is the signal-to-noise ratio in the image plane in the area occupied by the main component; see \cite{2005AJ....130.2473K} for details. If $\theta < \theta_\mathrm{lim}$, we used $\theta_\mathrm{lim}$ as the size upper limit and calculated the brightness temperature lower limit $T_\mathrm{b,lim}$, substituting $\theta_\mathrm{lim}$ in eq.~(\ref{eq:tb}).

Finally, one can calculate the parsec-scale spectral index $\alpha_\mathrm{vlba}$ from the VLBA flux densities at two frequencies. For the sources detected only in one band, we provide limits on $\alpha_\mathrm{vlba}$.
We note that for some sources, there is a problem that leads to biases in $\alpha_\mathrm{vlba}$ due to the so-called ``partial resolution.'' 
The shortest baseline at 2.3~GHz corresponds to the spatial
frequency a factor of 3.7 lower than at 8.6~GHz. If the correlated flux density of a source drops significantly between these spatial frequencies, the evaluated spectral index appears to be steeper than the real one. The shorter wavelength of a given interferometer is not sensitive to the emission of extended regions observable at the longer wavelength. This effect is not strong for core-dominated sources since the AGN core size is relatively small and directly proportional to the wavelength \citep{1979ApJ...232...34B}. However, this effect can be significant for many sources in our sample.


\begin{figure}
    \centering
    \includegraphics[width=\columnwidth]{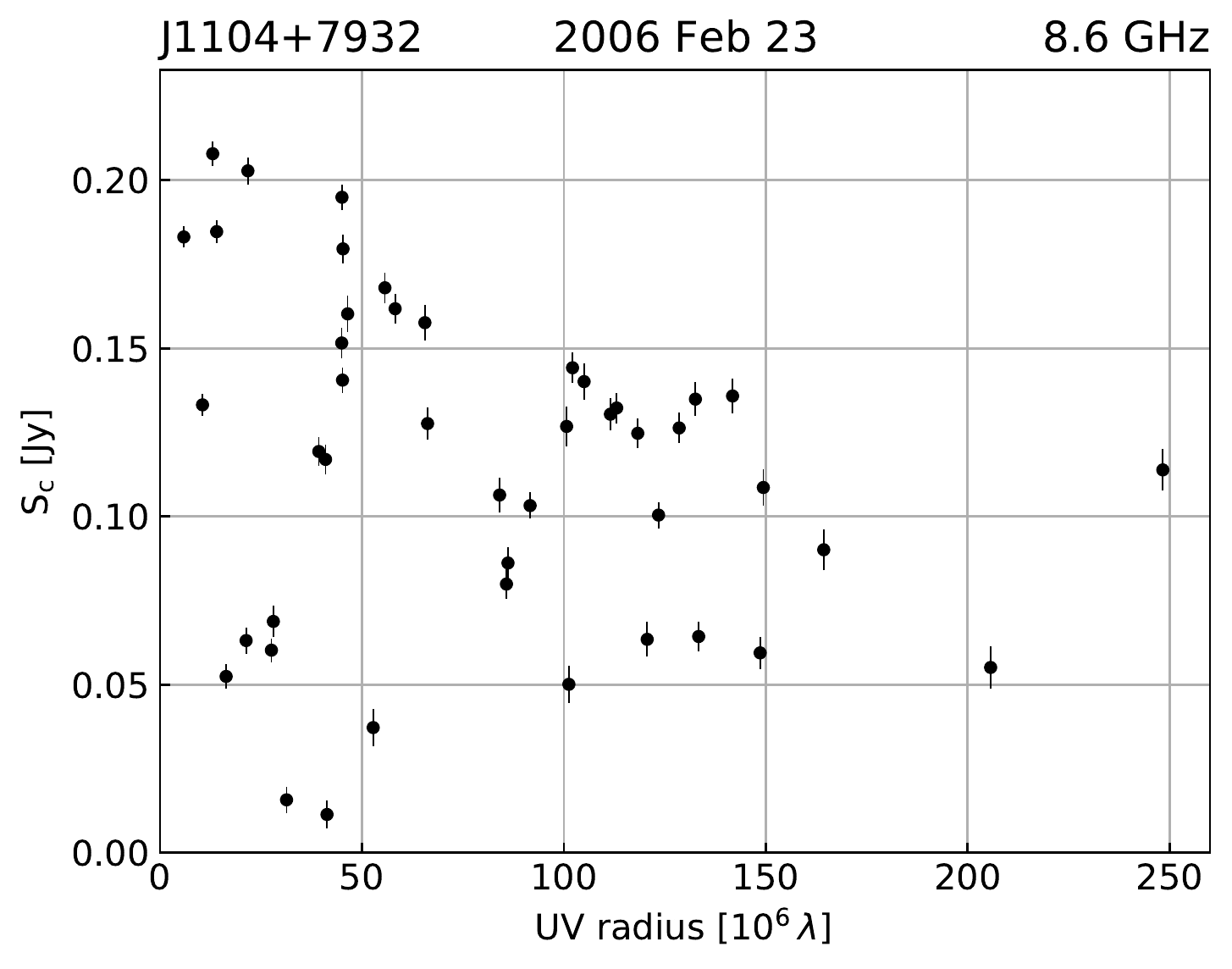}
    \caption{Correlated flux density averaged over time and IFs versus the $uv$ radius for the source J1104+7932 at 8.6~GHz. Similar plots for all the detected sources at 2.3 and/or 8.6~GHz (285 plots) are available online as Figure Set~\ref{fig:radplots}. In the cases when the data for a source were calibrated in AIPS and then underwent hybrid imaging in Difmap with both amplitude and phase self-calibration, they are plotted as filled circles. In the cases when the processing was the same except no amplitude self-calibration was made, the data are plotted as filled triangles. In the cases when no self-calibration was made for the source and the data calibrated in PIMA were used, they are plotted as open circles.}
    \label{fig:radplots}
\end{figure}

\begin{figure}
    \centering
    \includegraphics[width=\columnwidth]{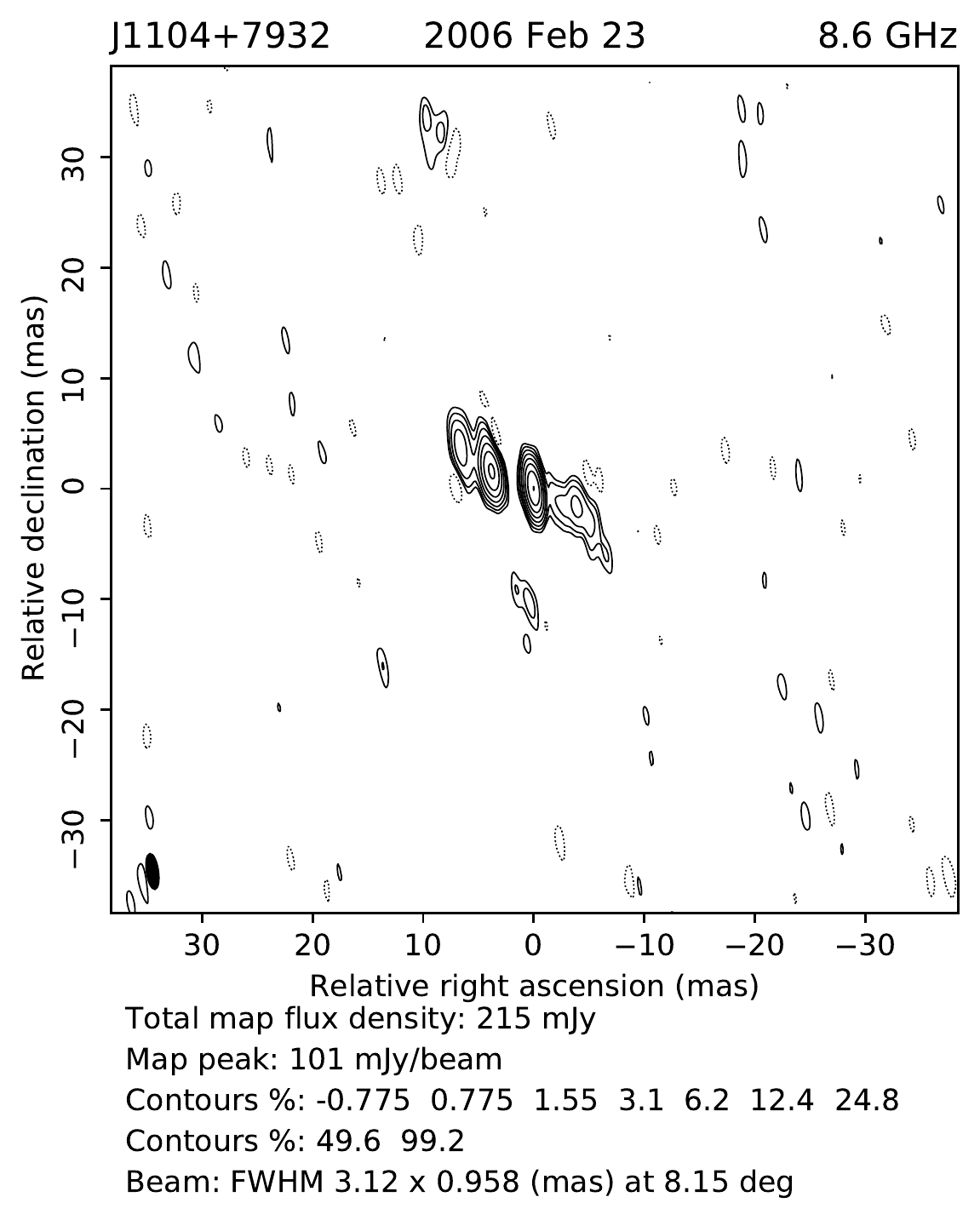}
    \caption{CLEAN map for the source J1104+7932 at 8.6~GHz. The intensity is shown by contours: solid lines are the positive contours, dotted lines are the negative contours. The contour levels in percents of the map peak are specified below the map, as well as the total flux density of the CLEAN model of the source and the intensity of the map peak. First contours correspond to the map noise level~$\times\,3$; each subsequent contour marks intensity increase by the factor of two. The CLEAN beam at the half-maximum level is shown in the map lower left corner as a black ellipse; its major and minor axes and position angle are specified below the map. All the maps obtained as a result of our survey at both 2.3 and 8.6~GHz (168 images) are available online as Figure Set~\ref{fig:maps}.
    \label{fig:maps}}
\end{figure}

\section{VLBA Survey Results}
\label{sec:results}

\begin{deluxetable*}{ccCCCCCCC}
    \tablecaption{
    The VLBA flux density and the compactness parameters of the detected sources.
    \label{tab:vlbaflux}
    }
    \tablehead{
    \colhead{Name} & \colhead{Flag} & \colhead{$S_\mathrm{vlba,2.3}$} & \colhead{$S_\mathrm{unres,2.3}$} & \colhead{$C^\mathrm{unres}_\mathrm{vlba,2.3}$} & \colhead{$S_\mathrm{vlba,8.6}$} & \colhead{$S_\mathrm{unres,8.6}$} & \colhead{$C^\mathrm{unres}_\mathrm{vlba,8.6}$} & \colhead{$\alpha_\mathrm{vlba}$}}
    \colnumbers
    \startdata
    J0009+7724 & \nodata & 55\pm7 & \nodata & <0.54 & 51\pm7 & 31\pm6 & 0.61\pm0.12 & -0.06\pm0.14 \\
    J0009+7603 & \nodata & 49\pm7 & \nodata & <0.62 & 74\pm9 & \nodata & <0.40 & 0.32\pm0.14 \\
    J0013+7748 & CSS & 403\pm41 & \nodata & <0.07 & \nodata & \nodata & \nodata & <-1.97 \\
    J0017+8135 & \nodata & 861\pm94 & 549\pm56 & 0.64\pm0.03 & 1170\pm119 & 699\pm77 & 0.60\pm0.03 & 0.23\pm0.11 \\
    J0038+8447 & CSS & 229\pm23 & \nodata & <0.13 & 40\pm6 & \nodata & <0.75 & -1.32\pm0.13 \\ 
    \enddata
    \tablecomments{Columns are as follows: (1) -- J2000 source name; (2) -- flag: ``CSS'' for compact steep-spectrum source candidates; ``ADD'' for the additional sources not belonging to the target sample detected close to some VLBA pointings; (3) -- VLBA flux density at 2.3~GHz; (4) -- unresolved flux density at 2.3~GHz; (5) -- parsec-scale compactness parameter (ratio of column (4) to column (3)) at 2.3~GHz; (6) -- VLBA flux density at 8.6~GHz; (7) -- unresolved flux density at 8.6~GHz; (8) -- parsec-scale compactness parameter at 8.6~GHz; (9) -- VLBA spectral index. Flux density values are given in mJy.
    This table is available in its entirety in machine-readable form online.}
\end{deluxetable*}

\begin{deluxetable*}{cCCCCCCCCC}
    \tablecaption{The results of the circular Gaussian model fitting to the visibilities.
    \label{tab:modeling}
    }
    \tablehead{\colhead{Name} & \colhead{Model(2.3GHz)} & \colhead{$S_\mathrm{gauss,2.3}$} & \colhead{$\theta_{2.3}$} & \colhead{$T_\mathrm{b,2.3}$} & \colhead{Model(8.6GHz)} & \colhead{$S_\mathrm{gauss,8.6}$} & \colhead{$\theta_{8.6}$} & \colhead{$T_\mathrm{b,8.6}$} & \colhead{$k$}}
    \colnumbers
    \startdata
    J0009+7724 & \nodata & \nodata & \nodata & \nodata & 3 & 41\pm4 & 0.29 & 8.0\times10^{9} & \nodata \\
    J0009+7603 & 3 & 38\pm4 & 1.4 & 4.5\times10^{9} & 1 & 70\pm7 & 0.37 & 8.3\times10^{9} & 1.01 \\
    J0013+7748 & 3 & 464\pm46 & 33 & 1.0\times10^{8} & \nodata & \nodata & \nodata & \nodata & \nodata \\
    J0017+8135 & 2 & 642\pm64 & 0.87 & 2.0\times10^{11} & 2 & 676\pm69 & 0.19 & 3.2\times10^{11} & 1.16 \\
    J0038+8447 & 1 & 154\pm16 & 9.5 & 3.9\times10^{8} & \nodata & \nodata & \nodata & \nodata & \nodata \\
    \enddata
    \tablecomments{Columns are as follows: (1) -- J2000 source name; (2) -- model type for 2.3~GHz; the types are: 1 -- one circular Gaussian, 2 -- two circular Gaussians, 3 -- one circular Gaussian fitted to visibility amplitudes only; (3) -- flux density of the main component at 2.3~GHz (mJy); (4) -- FWHM of the main component at 2.3~GHz (mas); (5) -- brightness temperature of the main component at 2.3~GHz (K); (6) -- model type for 8.6~GHz; (7) -- flux density of the main component at 8.6~GHz (mJy); (8) -- FWHM of the main component at 8.6~GHz (mas); (9) --  brightness temperature of the main component at 8.6~GHz (K); (10) -- negative slope of the angular size -- frequency dependence for the main component. This table is available in its entirety in machine-readable form.}
\end{deluxetable*}

As a result of this survey, we have detected 162 target sources at any band, 153 sources at
2.3 GHz, 116 sources at 8.6 GHz, and 107 sources at both frequencies. Thus, 32\% of the sample of 482 objects have been detected at 2.3~GHz and 24\% at 8.6~GHz.
The detected sources are those that have a compact feature
stronger than the detection limit of the survey, 30 mJy,
at the VLBA spatial frequencies, corresponding to angular sizes $\lesssim0.1''$, or linear sizes less than several hundreds of parsec (see \autoref{sec:processing:params}). Therefore, our fraction of the detected sources is an estimate of the fraction of the sources that have such compact features among all the sources with the NVSS flux density higher than 200~mJy. 
Under a simplifying assumption that the probability of a source detection in the observed sample does not depend on its total flux density, the probability distribution of this fraction is defined by only two parameters: the total number of sources in the studied sample and the number of the detected sources. It allows us to roughly estimate the $1\sigma$ confidence intervals for this fraction as [30\%; 34\%] at 2.3~GHz and [22\%; 26\%] at 8.6~GHz, using the approach from \citet{2011PASA...28..128C}, utilizing the quantiles of the beta distribution. This estimate of the confidence intervals is very coarse, since the total flux densities and variability of sources obviously affect the probability distribution of the fraction of sources with strong compact features. To derive this distribution rigorously, a thorough investigation is needed that goes beyond the scope of this work.

The plots of the averaged correlated flux density versus projected baseline ($uv$ radius) for all the detected sources are given in Figure Set~\ref{fig:radplots}. We have determined coordinates of all the detected sources
with accuracies in a range of 1--100~mas (\autoref{tab:coord}). We have restored maps for 94 sources at 2.3~GHz and 62 sources at 8.6~GHz. All the maps and their parameters are given in Figure Set~\ref{fig:maps}. Three sources of the sample (J0017+8135, J1058+8114, and J1153+8058) were also observed as calibrators in all three days of the program. For them, we present in Figure Set~\ref{fig:radplots} the plots for each day of our observations separately. We have also restored their maps for each day of our observations.  

We have measured the total correlated VLBA flux density for all the detected sources, as described in \autoref{sec:processing:params}. When we had enough data for a source, we also estimated its unresolved flux density, parsec-scale compactness parameter, and VLBA spectral index. These quantities are given in \autoref{tab:vlbaflux}. We fitted simple models, described in \autoref{sec:processing:params}, to visibilities of 132 sources at 2.3~GHz and 80 sources at 8.6~GHz and obtained the parameters of their dominant components.
The size of dominant components was measured in both bands for 60 sources; for them, we calculated the power-law index $k$ of the size-frequency dependence $\theta\propto\nu^{-k}$. These results are given in \autoref{tab:modeling}. Using VLBA and single-dish flux densities, we also calculated the kiloparsec-scale compactness parameters $C^\mathrm{vlba}_\mathrm{sd}$ or their upper limits, which are given in \autoref{tab:spectral}.

Within this survey framework, we also detected several sources that do not belong to our observing sample but lie close to some of our target sources. We also give the obtained parameters of these sources in \autoref{tab:coord} and \autoref{tab:vlbaflux}, marking them with a flag ``ADD'' in the latter. We do not use them in the analysis in the subsequent sections, since they do not belong to our complete flux density limited sample.

The UVFITS files containing the calibrated visibilities for all the sources detected within this project, the FITS images of all the imaged sources, and the data used for the figures are available
online at \url{http://astrogeo.org/npcs}.


\section{Total continuum radio spectra}
\label{sec:r600}

Another type of data we used are the broadband single-dish radio spectra, from which we get total flux densities, spectral indices, and variability amplitudes. We used published spectra as well as the data from our RATAN-600 observational program. As a result, we have single-dish spectra of all the sample sources at one or several epochs.

For all the sources of our sample, except two closest to the North Celestial Pole, there are published quasi-simultaneous broadband radio spectra \citep{2007ARep...51..343M} observed with the RATAN-600 -- a transit-mode ring radio telescope in the Special Astrophysical Observatory of the Russian Academy of Sciences, located in the North Caucasus \citep{1979S&T....57..324K}. The sources were observed at the upper culmination with the Southern sector of the telescope at 1.1, 2.3, 4.8, 7.7, and 11.2~GHz; for about one-third of the sources, flux densities at 21.7~GHz were also measured. The observations were carried out in April--August 2005, i.e., from ten to six months prior to our VLBA observations. The typical accuracy of the flux density measurements is 5-10\,\%. The telescope has a strongly elongated beam. An essential feature of the RATAN-600 spectra is that they are quasi-simultaneous, i.e., the measurements at all frequencies are made consequently within several minutes.
See for details \cite{1999A&AS..139..545K,1999ARep...43..631B}.

Our analysis used these spectra for all the sources, except the most extended, which are larger than the RATAN-600 beam. The RATAN-600 does not have the 8.6~GHz receiver, so the flux density at this frequency was estimated from an interpolation. We calculated the single-dish spectral index $\alpha_\mathrm{sd}$ by fitting a power law to the RATAN-600 flux densities at 2.3, 4.8, and 7.7~GHz. We excluded the measurements that have relative errors larger than 50\%.

For the rest 21 sources, including two closest to the North Celestial Pole and 19 partially resolved by the RATAN-600, we collected all the published non-simultaneous flux density measurements in the studied frequency range from the CATS database\footnote{\url{https://www.sao.ru/cats/}} \citep{2005BSAO...58..118V} and, neglecting the sources variability and assuming that the only reason for flux density differences is the partial resolution of the sources, we fitted the upper envelopes of the collected spectra by a power-law to obtain a single-dish spectral index and flux densities. 

The resulting single-dish flux densities and spectral indices are given in \autoref{tab:spectral}.

\begin{figure}
        \includegraphics[width=\columnwidth]{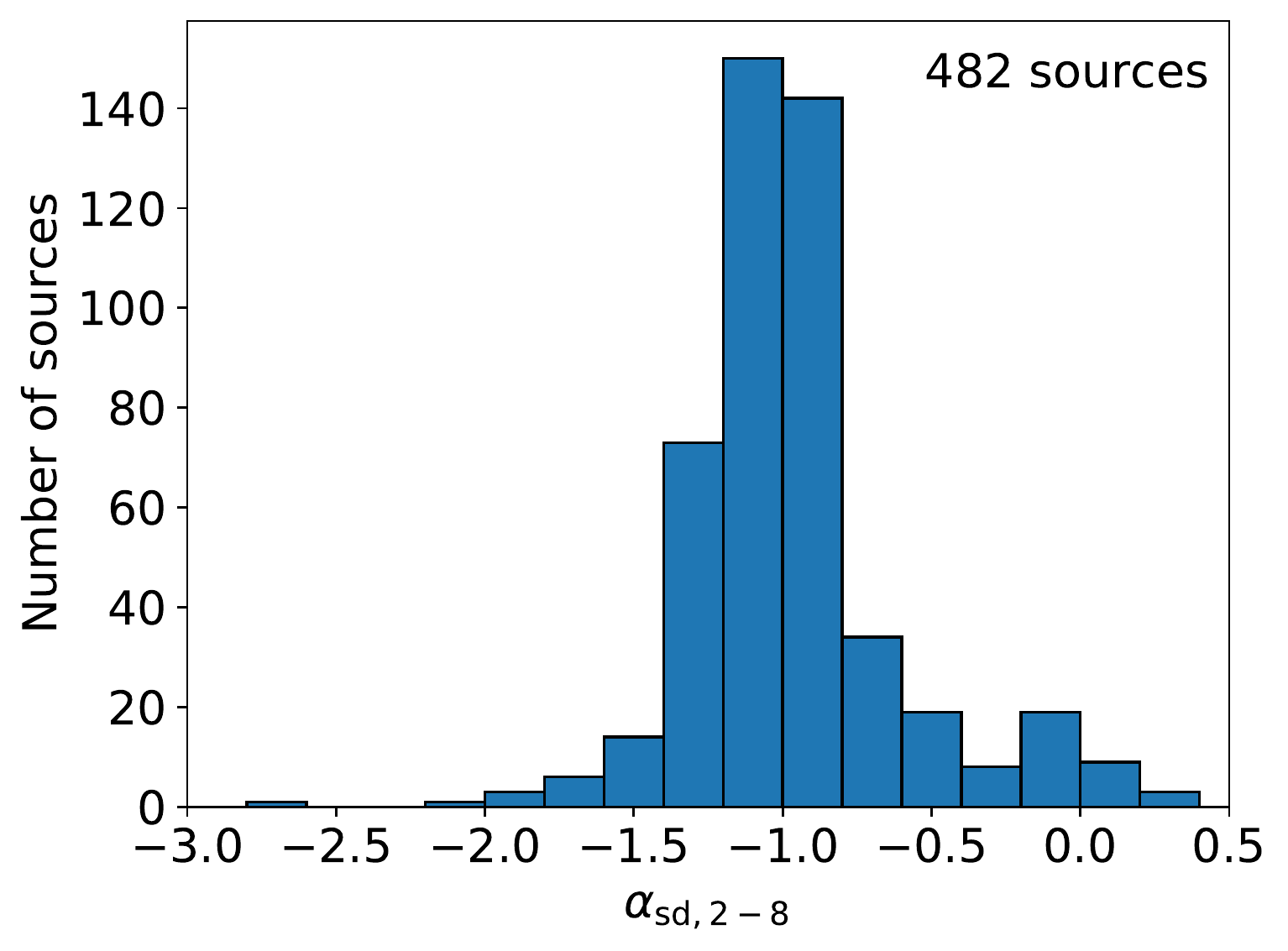}
    \caption{
    Single-dish 2-8~GHz spectral index distribution for the sources of the sample (see \autoref{sec:r600} for details on the spectral index calculation procedure).
    \label{fig:a_hist}
    }
\end{figure}

\begin{deluxetable*}{cCCCCCCC}
    \tablecaption{
    The parameters of total (single-dish) spectra and the kiloparsec-scale compactness for all the sources of our sample.
    \label{tab:spectral}
    }
    \tablehead{
    \colhead{Name} & \colhead{$S_\mathrm{sd,2.3}$} & \colhead{$S_\mathrm{sd,8.6}$} & \colhead{$\alpha_\mathrm{sd}$} & \colhead{$\Delta S_\mathrm{sd,8}$} & \colhead{$V_\mathrm{8}$} & \colhead{$C^\mathrm{vlba}_\mathrm{sd,2.3}$} & \colhead{$C^\mathrm{vlba}_\mathrm{sd,8.6}$}
    }
    \decimalcolnumbers
    \startdata
    J0000+8123 & 164\pm56 & 51\pm11 & -0.86\pm0.32 & \nodata & \nodata & <0.18 & <0.59 \\
    J0005+8135 & 148\pm22 & 31\pm8 & -1.15\pm0.19 & \nodata & \nodata & <0.20 & <0.96 \\
    J0008+8426 & 181\pm27 & 62\pm6 & -0.86\pm0.13 & \nodata & \nodata & <0.17 & <0.48 \\
    J0009+7724 & 475\pm30 & 180\pm12 & -0.74\pm0.08 & 13\pm14 & 0.04\pm0.04 & 0.12\pm0.02 & 0.28\pm0.04 \\
    J0009+7603 & 177\pm10 & 110\pm7 & -0.39\pm0.07 & \nodata & \nodata & 0.28\pm0.04 &  0.68\pm0.09 \\
    \enddata
    \tablecomments{Columns are as follows: (1) -- J2000 source name; (2) -- Single-dish flux density at 2.3~GHz; (3) -- Single-dish flux density at 8.6~GHz; (4) -- Single-dish spectral index in 2-8~GHz range; (5) -- Variability amplitude of the single-dish flux density at 8~GHz; (5) -- Variability index at 8~GHz; (6) -- Kiloparsec-scale compactness parameter defined as the ratio of the VLBA flux density to the single-dish flux density at 2.3~GHz, or its upper limit if a source was not detected by VLBA; (7) -- Kiloparsec-scale compactness parameter at 8.6~GHz. The unit of the flux density is mJy. This table is available in its entirety in machine-readable form.}
\end{deluxetable*}

We classified the shape of the spectra as follows. Five sources (J0626+8202, J0726+7911, J1044+8054, J1823+7938, and J1935+8130) exhibit a peak in their spectra; they were identified as candidates to Gigahertz-peaked spectrum (GPS) sources by \citet{2011ARep...55..187M}. The $S\propto\nu^{+\alpha}$ approximation is not suitable for their spectra; however, for uniformity, we calculate their spectral indices in the same way as for the other sources. We divided the spectra of the other sources into two classes: steep (spectral index $\alpha_\mathrm{sd}<-0.5$) and flat ($\alpha_\mathrm{sd}\ge-0.5$). The spectral index distribution is presented in \autoref{fig:a_hist}. The steep-spectrum sources account for 90\% of our sample selected at 1.4~GHz, flat-spectrum ones --- for 9\%, and peaked spectrum ones --- for 1\%. For optically thin synchrotron sources, there is a steepening of the spectra above few GHz due to synchrotron cooling \citep[e.g.,][]{1991ApJ...383..554C}. However, it does not bias the flat/steep spectrum classification.

To investigate the variability of the sources, we supplemented the spectra from \citet{2007ARep...51..343M} by observations at other epochs. The spectra of 171 sources from the sample with $S_\mathrm{NVSS}\geq 400$~mJy were measured also at the RATAN-600 six years earlier \citep{2001A&A...370...78M}. 
Furthermore, 37 sources from our sample were observed by \cite{2013MNRAS.435.2793R} at 5, 8, 20, and 30~GHz. 

Additionally, we used the data from our RATAN-600 AGN monitoring program. The program's description and its results may be found in \citet{1999A&AS..139..545K, 2000PASJ...52.1027K, 2002PASA...19...83K,2020AdSpR..65..745K}, \citet{2020ApJ...894..101P}. In the framework of this program, more than 4000 compact sources were observed at least once, and about 700 sources were monitored for more than a decade.
The total broadband 1-22~GHz spectra of more than 95\% of the sources were classified by five main types: steep, inverted, super-flat, with a maximum or minimum, and a variable type. They were decomposed into two main spectral components: the first represents the compact jet dominant at higher radio frequencies and the second one represents 
an extended  magnetoshere around the jet including  far jet region and lobes dominant at frequencies of about 1 GHz and lower.
The decomposition of the spectra by two main components has to be valid also for the AGNs in our sample with a detectable parsec-scale structure. However, we note that the steep-spectrum emission at frequencies higher than 1 GHz is attributed, at least partly, to structures of less than about 1~kpc in size, see for details \autoref{sec:analysis}.
Among the sources of our sample, 50 were observed in the framework of our program from 1998 to 2013 with different number of sets (from 1 to 19 epochs) using the combination of the Flat reflector and the Southern sector of the RATAN-600. 

Using all these data, we calculated the variability amplitude of the single-dish flux density for the sources observed at two or more epochs:
\begin{equation}
    \label{eq:varamp}
    \Delta S_\mathrm{sd} = (S-\sigma_{S})_\mathrm{max} - (S+\sigma_{S})_\mathrm{min},
\end{equation}
where $S$ and $\sigma_{S}$ are the single-dish flux density and its error at a given frequency, and maximum and minimum are calculated over all the epochs.
We also calculated the variability index following \citet{1992ApJ...399...16A}:
\begin{equation}
    \label{eq:varind}
    V = \frac{(S-\sigma_{S})_\mathrm{max} - (S+\sigma_{S})_\mathrm{min}} {(S-\sigma_{S})_\mathrm{max} + (S+\sigma_{S})_\mathrm{min}},
\end{equation}
If the values of the variability amplitude and the variability index, calculated according to this formulas, were negative, we set their values to zero. We did not use the flux density measurements of our flux density calibrators in the calculation of the variability parameters. We filtered out measurements with relative errors greater than 1/3 and then inspected the spectra by eye to exclude outliers. The RATAN-600 observations at 2~GHz are often corrupted by man-made radio interference, thus we did not use variability parameters at this frequency in the analysis. Using the RATAN-600 flux densities at 7.7~GHz from all these programs together with the flux densities from \citet{2013MNRAS.435.2793R} at 8.3~GHz, when they were available, we calculated the 8~GHz variability amplitude $\Delta S_\mathrm{sd,8}$ and variability index $V_{8}$ for 167 sources and presented them in \autoref{tab:spectral}.


\begin{table*}
\centering
\caption{Statistics of the VLBA detections of the sources in the complete NVSS flux-density limited sample.}
\label{tab:det}
\begin{tabular}{cccccc}
    \hline
    \hline
    Spectral type & \# sources & \# detected 2.3~GHz & \% detected 2.3~GHz & \# detected 8.6~GHz & \% detected 8.6~GHz \\ 
    (1) & (2) & (3) & (4) & (5) & (6) \\
    \hline
    Flat & 42 & 41 & 98\% & 40 & 95\% \\
    Steep & 435 & 107 & 25\% & 71 & 16\% \\
    Peaked & 5 & 5 & 100\% & 5 & 100\% \\
    \hline
    All & 482 & 153 & 32\% & 116 & 24\% \\
    \hline
\end{tabular}
\tablecomments{Columns are as follows: (1) -- type of the continuum single-dish radio spectrum; (2) --  number of the sources of a given spectral type in the sample; (3) -- number of the sources of a given spectral type detected at 2.3~GHz; (4) -- percent of the detected sources at 2.3~GHz with respect to the number of the sources of a given spectral type in the sample; (5) and (6) -- the same as (3) and (4) for the detections at 8.6~GHz.}
\end{table*}

\section{Analysis}
\label{sec:analysis}

\begin{figure*}
    \centering
        \includegraphics[width=0.49\linewidth]{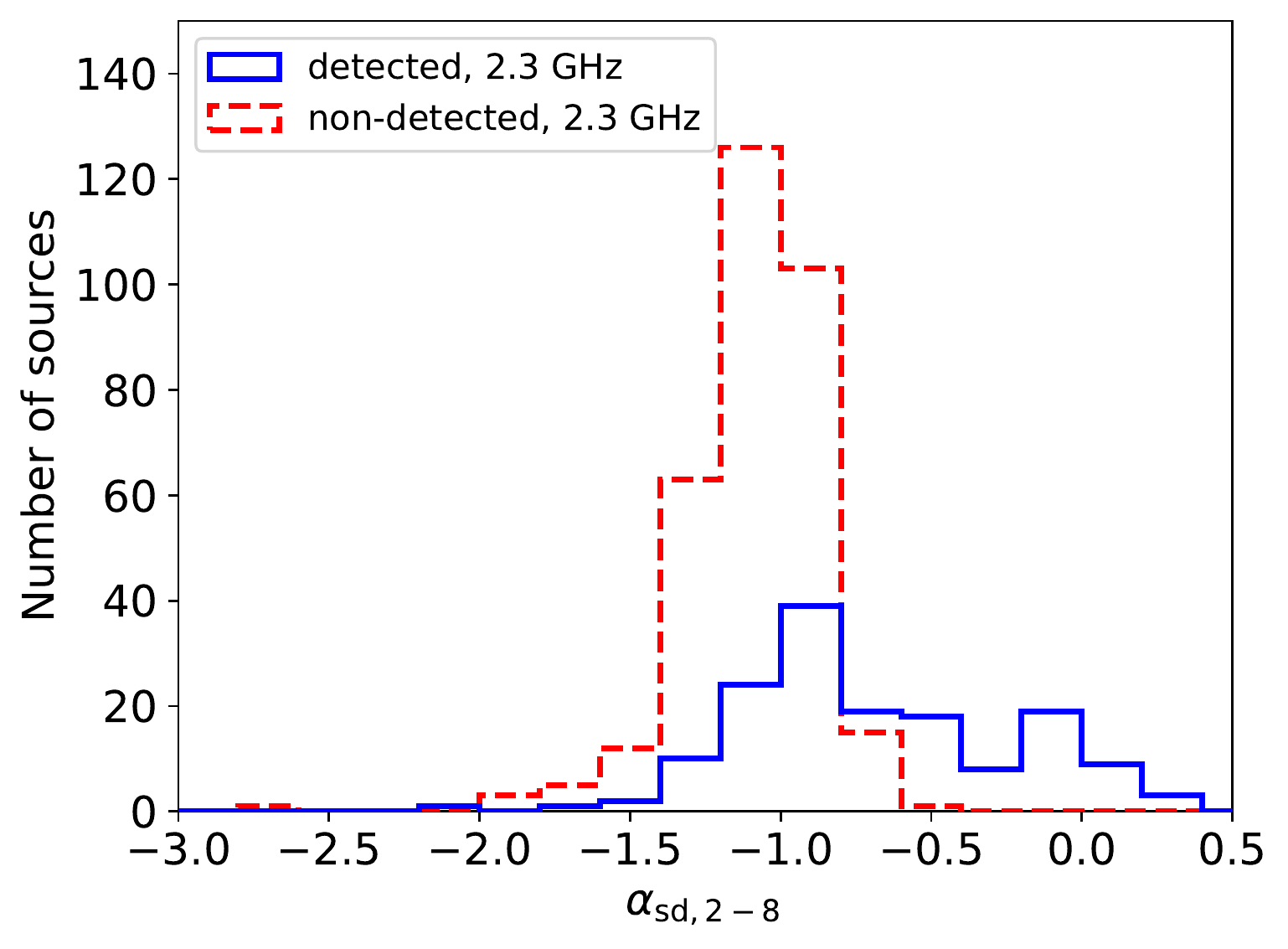}
        \includegraphics[width=0.49\linewidth]{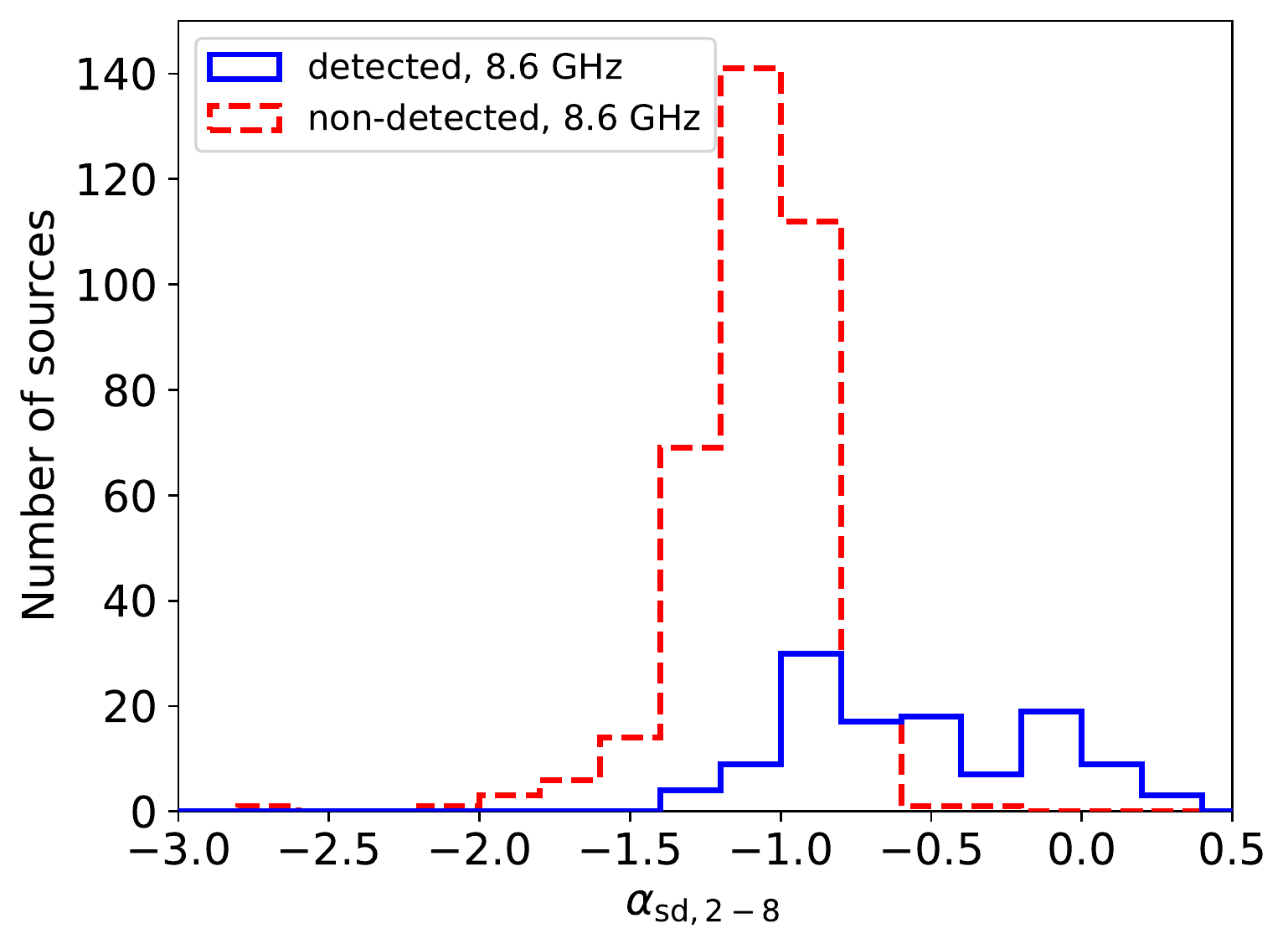}
        \caption{
    Number of the sources in the complete sample detected (solid line) and not detected by the VLBA (dashed line) vs. source single-dish 2-8~GHz spectral index. {\it Left:} Detections at 2.3~GHz. {\it Right:} Detections at 8.6~GHz.
    \label{fig:det}
    }
\end{figure*}

In the previous sections we described our data on the parsec-scale structure and total continuum radio spectra of the sources in our complete sample. Here we present their joint analysis.

\subsection{Relation between VLBA detection and spectral index}
\label{sec:analysis:det}

\autoref{fig:det} shows the distribution of the number of the detected sources at both VLBA frequencies with respect to the single-dish 2-8~GHz spectral index. The statistics of detections of the sources with different spectrum shapes are summarized in \autoref{tab:det}. 

Of flat-spectrum sources, 98\% are detected at 2.3~GHz and 95\% at 8.6~GHz. That is, our observations of a complete sample verify the common assumption that flat-spectrum sources are compact. 
At the same time, there is a significant number of detected steep-spectrum sources. The fraction of the detected objects among the steep-spectrum sources is not very high (25\% at 2.3~GHz and 16\% at 8.6~GHz). However, for our flux-density limited sample selected at 1.4~GHz, the detected sources with a steep single-dish spectrum outnumber all the flat-spectrum sources. We detected 116 sources with steep single-dish spectra at least in one band, which is 27\% of all the sample's steep-spectrum sources. 
As mentioned in \autoref{sec:r600}, there are several Gigahertz-peaked spectrum sources in our sample, which form a separate class. All of them were detected in both bands.


\subsection{Demographics of the complete sample}
\label{sec:analysis:classes}

\begin{figure*}
    \centering
    \includegraphics[width=0.32\linewidth]{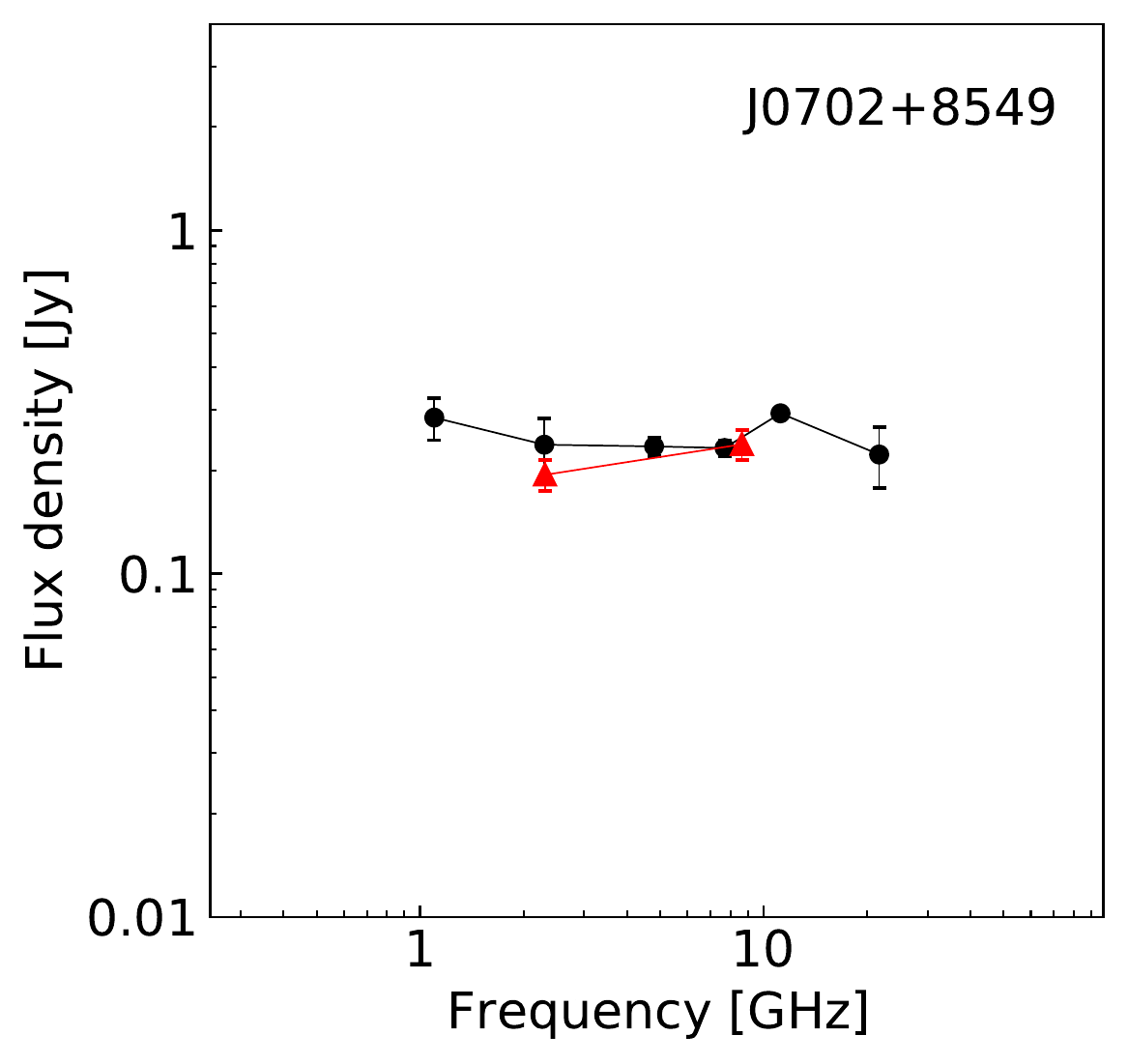}
    \includegraphics[width=0.32\linewidth]{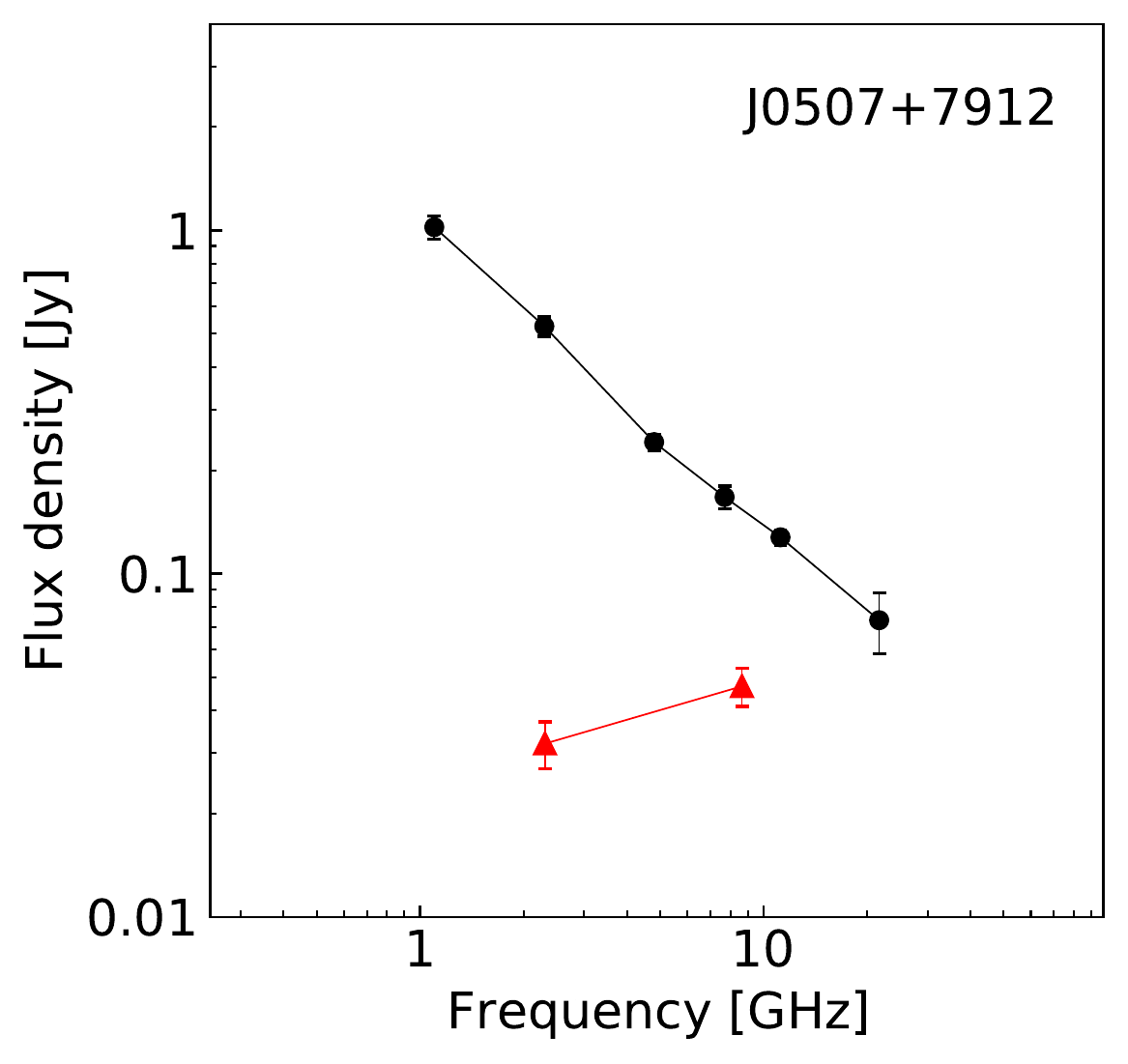}
    \includegraphics[width=0.32\linewidth]{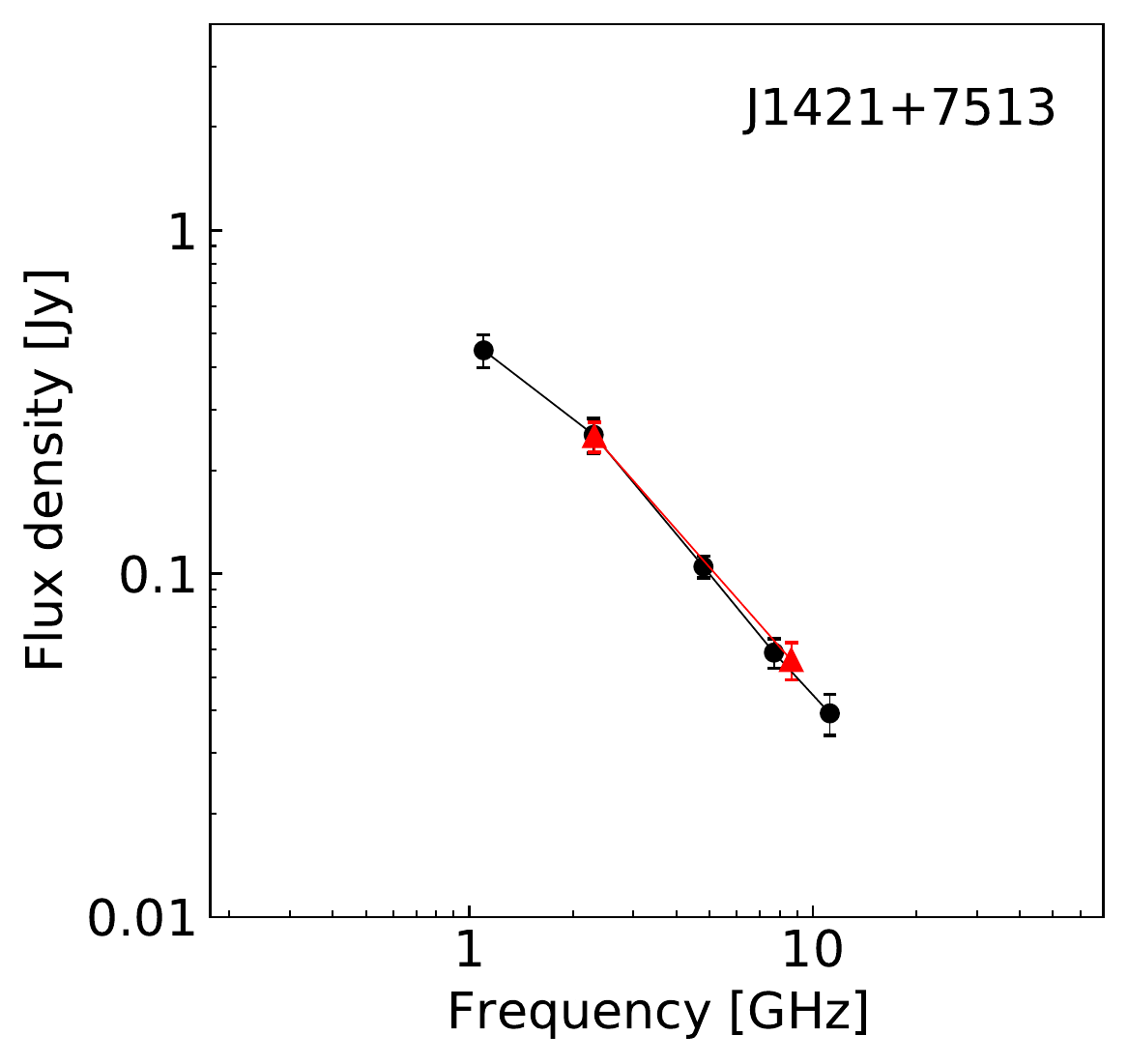}
    \caption{Single-dish and VLBA broadband spectra of all the sources of our sample. The whole Figure Set~\ref{fig:sp_tot_vlba} (482 plots) is available online. The spectra of three sources are shown here as examples. The source name is specified at the top right of each plot. The single-dish spectra are plotted black, the VLBA spectra are plotted by the red color. In cases of the VLBA non-detection, the upper limits on the VLBA flux density are shown by red arrows.
    \label{fig:sp_tot_vlba}}
\end{figure*}

Figure Set~\ref{fig:sp_tot_vlba} shows the single-dish and VLBA broadband spectra for all the sources of the sample. For the sources not detected by the VLBA, upper limits of the VLBA flux density are plotted by arrows.

\begin{figure}
    \centering
    \includegraphics[width=\columnwidth]{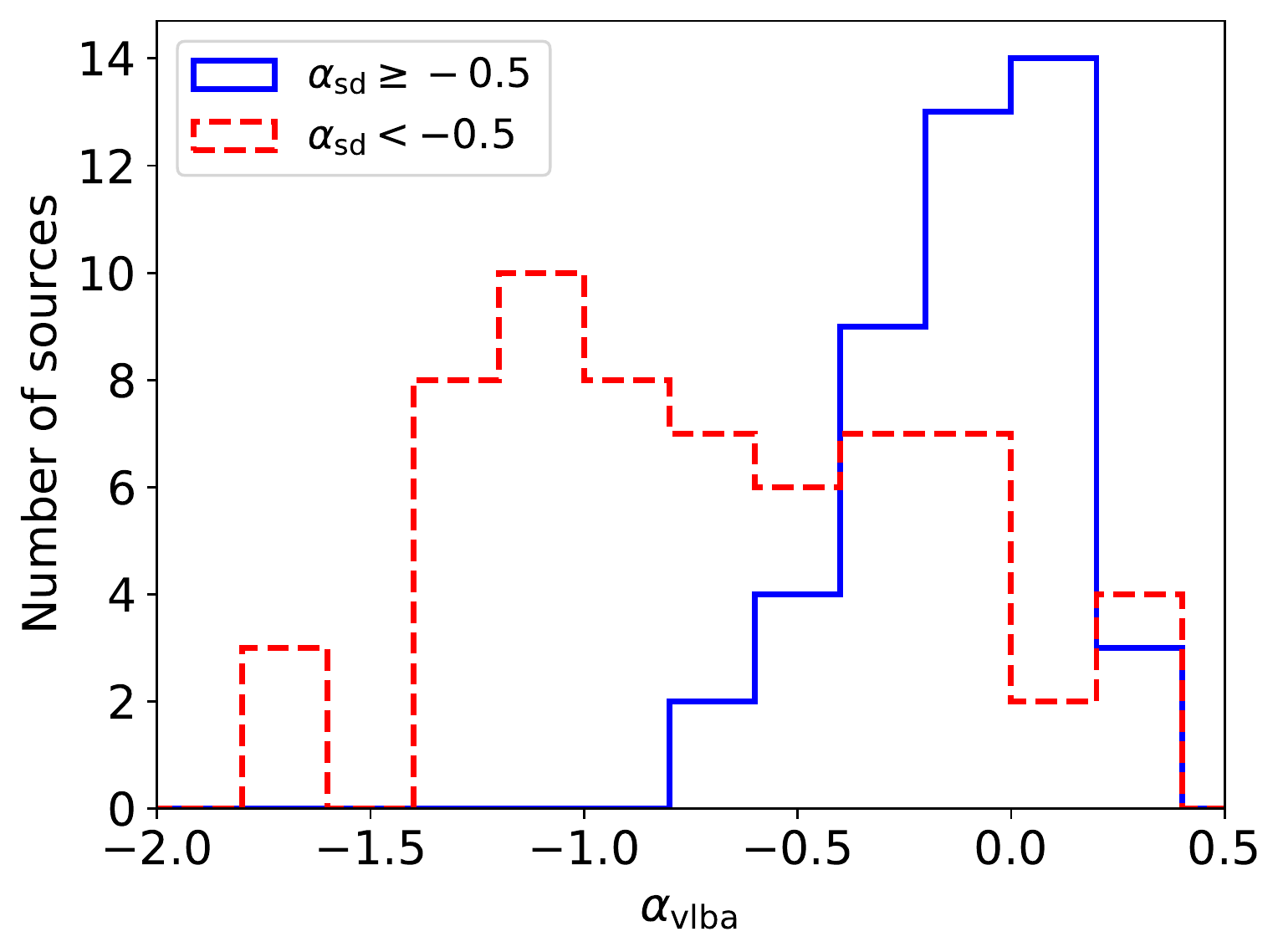}
    \caption{VLBA spectral index histogram for the sources detected in both bands in our survey. The distributions are plotted separately for the sources with a flat single-dish spectrum (solid blue line) and a steep single-dish spectrum (dashed red line). 
    \label{fig:avlba_hist}}
\end{figure}

We can roughly divide steep-spectrum sources into three subclasses:\\
1. Sources with steep single-dish spectra, but flat VLBA spectra. We call them {\it flat-spectrum cores of extended steep-spectrum sources}. These sources necessarily have rather low kiloparsec-scale compactness, because their total emission is dominated by extended kiloparsec-scale jets and lobes. There are 30 such sources in our sample: 7\% of the sources with a steep single-dish spectrum and 6\% of the whole complete sample.\\
2. Sources with both steep single-dish and VLBA spectra. There are 82 such sources in our sample: 19\% of steep-spectrum sources and 17\% of the whole sample.\\
3. Extended steep-spectrum sources. This subclass includes all steep-spectrum sources with no detectable parsec-scale structure.

Both the theory \citep[e.g.,][]{1979ApJ...232...34B} and the observations \citep[e.g.,][]{2014AJ....147..143H} of AGN demonstrate that the emission with a flat spectrum is generated in the inner part of relativistic jets close to the central engine. This region is often called the core. It is bright, compact, and located in the place where the synchrotron optical depth is about unity. In our sample, there are 67 sources with $\alpha_\mathrm{vlba} \geq -0.5$, of which 37 sources have a flat single-dish spectra and 30 sources have a steep single-dish spectra (type 1 above); here, we do not count peaked spectrum sources. 
Therefore, we observe the opaque core in 14\% of the sources in our sample. 

If the VLBA spectrum is steep, then some optically thin compact structures dominate the parsec-scale emission rather than the jet core. The objects of the second subclass are most likely compact steep-spectrum (CSS) sources. A number of them exhibit compact double or compact symmetric morphology in our images, which is typical for CSS sources \citep[see, e.g.,][]{1998PASP..110..493O}.
We refer to these 82 sources as ``CSS candidates,'' because for most of them, we cannot robustly determine the morphology and, thus, cannot say what is detected: a CSS source or just the most compact feature of a hot spot in an extended source. At the same time, 51 of these candidates have $C^\mathrm{vlba}_\mathrm{sd}>0.5$ at 2.3~GHz, which indicates that the VLBA detects their features which dominate their integral single-dish flux density. Such sources most likely are CSS. Note that the number of CSS candidates is larger than the number of all flat-spectrum sources in the sample. All 82 CSS candidates are marked by flag ``CSS'' in \autoref{tab:vlbaflux}. There are several well-known CSS sources among them, e.g., 0403+768 (J0410+7656), 3C~303.1 (J1443+7707), 3C~305.1 (J1447+7656), and 2342+821 (J2344+8226) \citep{1990A&A...231..333F}. Most of them, however, are reported to be VLBI-compact for the first time.

An illustration of the composition of the subpopulation of compact sources within the parent complete NPCS sample is given in \autoref{fig:avlba_hist}. There are two histograms in the figure, showing the distributions of the VLBA spectral index for the sources with a flat and a steep single-dish spectrum. Not surprisingly, most of the flat-spectrum sources have a flat spectrum also at parsec scales. For the sources with a steep single-dish spectrum, the broad distribution of $\alpha_\mathrm{vlba}$ includes sources from both classes 1 and 2. Since CSS sources are more numerous than flat-spectrum cores of extended steep-spectrum sources, the distribution peaks at $\alpha_\mathrm{vlba} \approx -1$.


\subsection{Correlations between parsec-scale structure parameters and spectral index}
\label{sec:analysis:correl}

\begin{deluxetable*}{CRRLRRLRRLRRL}
    \tabletypesize{\footnotesize}
    \tablecaption{
    Kendall rank correlation statistics for different pairs of parameters, characterizing the sources structure and spectra.
    \label{tab:correlation}
    }
    \tablehead{
    & \multicolumn{6}{c}{S band (2.3 GHz)} & \multicolumn{6}{c}{X band (8.6 GHz)} \\
    \colhead{Quantities} & \multicolumn{3}{c}{Values only} & \multicolumn{3}{c}{Values and limits} & \multicolumn{3}{c}{Values only} & \multicolumn{3}{c}{Values and limits} \\
    & \colhead{$\tau$} & \colhead{$N$} & \colhead{$p$} & \colhead{$\tau$} & \colhead{$N$} & \colhead{$p$} & \colhead{$\tau$} & \colhead{$N$} & \colhead{$p$} & \colhead{$\tau$} & \colhead{$N$} & \colhead{$p$}
    }
    \startdata
    C^\mathrm{vlba}_\mathrm{sd}, \alpha_\mathrm{sd} & 0.36\pm 0.05 & 152 & 5\times 10^{-11} & 0.27\pm 0.02 & 481 & 5\times 10^{-30} & 0.37\pm 0.06 & 116 & 7\times 10^{-9} & 0.20\pm 0.02 & 478 & 4\times 10^{-31} \\
    $C^\mathrm{unres}_\mathrm{vlba}$, $\alpha_\mathrm{sd}$ & $0.27\pm 0.08$ & $82$ & $3\times 10^{-4}$ & $0.31\pm 0.04$ & $153$ & $3\times 10^{-13}$ & $0.02\pm 0.09$ & $58$ & $0.84$ & $0.16\pm 0.04$ & $116$ & $1\times 10^{-4}$ \\
    $C^\mathrm{unres}_\mathrm{vlba}$, $\alpha_\mathrm{vlba}$ & $0.37\pm 0.08$ & $82$ & $1\times 10^{-6}$ & $0.33\pm 0.04$ & $153$ & $6\times 10^{-17}$ & $0.29\pm 0.09$ & $57$ & $1\times 10^{-3}$ & $0.20\pm 0.04$ & $116$ & $4\times 10^{-7}$ \\
    $\theta$, $\alpha_\mathrm{sd}$ & $-0.52\pm 0.06$ & $124$ & $2\times 10^{-17}$ & $-0.47\pm 0.06$ & $132$ & $1\times 10^{-15}$ & $-0.40\pm 0.08$ & $67$ & $2\times 10^{-6}$ & $-0.37\pm 0.07$ & $80$ & $9\times 10^{-7}$ \\
    $\theta$, $\alpha_\mathrm{vlba}$ & $-0.49\pm 0.07$ & $89$ & $2\times 10^{-11}$ & $-0.53\pm 0.06$ & $132$ & $3\times 10^{-21}$ & $-0.47\pm0.08$ & $67$ & $2\times 10^{-8}$ & $-0.41\pm 0.07$ & $80$ & $5\times 10^{-8}$ \\
    $T_\mathrm{b}$, $\alpha_\mathrm{sd}$ & $0.56\pm0.06$ & $124$ & $6\times10^{-20}$ & $0.51\pm0.06$ & $132$ & $1\times10^{-18}$ & $0.59\pm0.08$ & $67$ & $1\times10^{-12}$ & $0.52\pm0.07$ & $80$ & $2\times10^{-12}$ \\
    $T_\mathrm{b}$, $\alpha_\mathrm{vlba}$ & $0.44\pm0.07$ & $89$ & $2\times10^{-9}$ & $0.51\pm0.06$ & $132$ & $2\times10^{-20}$ & $0.56\pm0.08$ & $67$ & $3\times10^{-11}$ & $0.48\pm0.07$ & $80$ & $7\times10^{-11}$ \\
    \enddata
    \tablecomments{\small The meaning of the symblos is: $C^\mathrm{vlba}_\mathrm{sd}$ -- kiloparsec-scale compactness parameter, $C^\mathrm{unres}_\mathrm{vlba}$ -- parsec-scale compactness parameter, $\theta$ -- angular size of the main compact feature, $T_\mathrm{b}$ -- its brightness temperature, $\alpha_\mathrm{sd}$ -- single-dish 2-8~GHz spectral index, $\alpha_\mathrm{vlba}$ -- VLBA 2-8~GHz spectral index;
    $\tau$ -- Kendall correlation coefficient, $N$ -- the number of sources for which both quantities in pair are determined, $p$ -- probability that the correlation occurred by chance. Statistics is given for two frequency bands and calculated in two ways: using only measured values as well as values and limits. See \autoref{sec:analysis:correl} for details.
}
\end{deluxetable*}

We investigated correlations between the properties of the sources at VLBI spatial scales and their spectral index. In \autoref{tab:correlation}, we list the Kendall correlation coefficients $\tau$ and the probabilities $p$ that the correlation occurred by chance for different pairs of quantities which were defined above. We also indicate the numbers of the sources for which both given quantities are known for each pair. 
Since for many sources, we know upper or lower limits on some parameters instead of their values (see \autoref{sec:processing:params} and \autoref{sec:results}), we calculated the correlation coefficients in two ways: using only measured values and using both measured and censored values (upper and lower limits).

We used the version of the Kendall correlation coefficient known as ``tau-b'' \citep{Kendall1945}, defined for two variables $x_i$ and $y_i$ ($i$ from 1 to $N$) as:
 \begin{equation}
     \tau = \frac{\sum\limits_{j=1}^{N} \sum\limits_{i=1}^{j-1} a_{ij}b_{ij}}{\sqrt{(n_0-n_1)(n_0-n_2)}}
 \end{equation}
where $n_0 = N(N-1)/2$, $n_1$ and $n_2$ are the numbers of ties in $x$ and $y$ quantity, correspondingly. In case when $x_i$ and $y_i$ may be either measured values or upper or lower limits, $a_{ij} = -1$, if $x_{i}$ is definitely greater than $x_{j}$; $a_{ij} = 0$, if $x_{i}=x_{j}$ or the comparison of $x_{i}$ and $x_{j}$ is uncertain; and $a_{ij} = 1$, if $x_{i}$ is definitely less than $x_{j}$; $b_{ij}$ is defined similarly for $y$ \citep{BHK1974, 1986ApJ...306..490I, 1996MNRAS.278..919A}. We calculated the $p$ values and the errors of $\tau$, using the expression for the variance of the Kendall correlation statistic from \cite{1986ApJ...306..490I}.

\begin{figure*}
    \centering
        \includegraphics[width=0.49\linewidth]{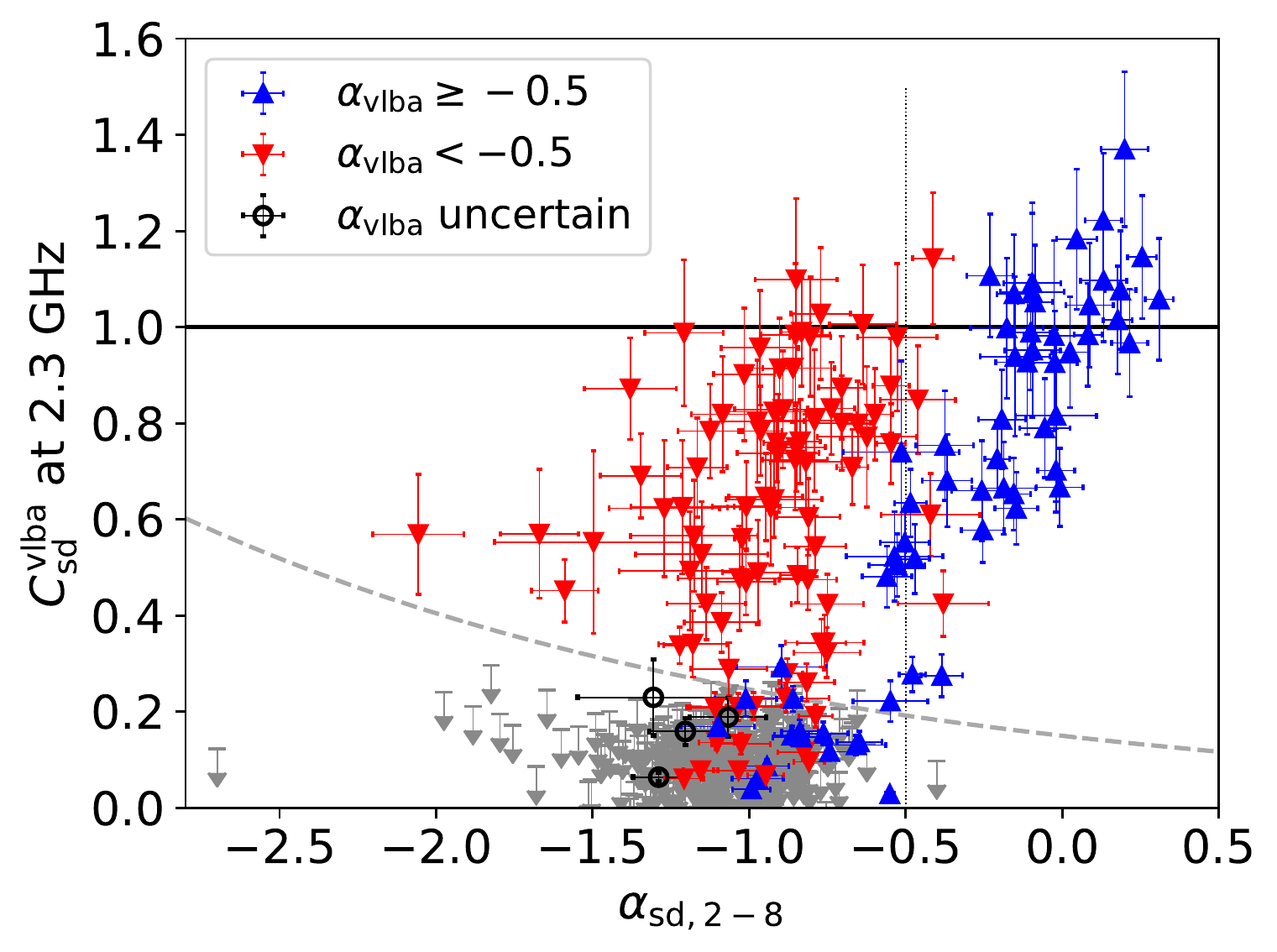}
        \includegraphics[width=0.49\linewidth]{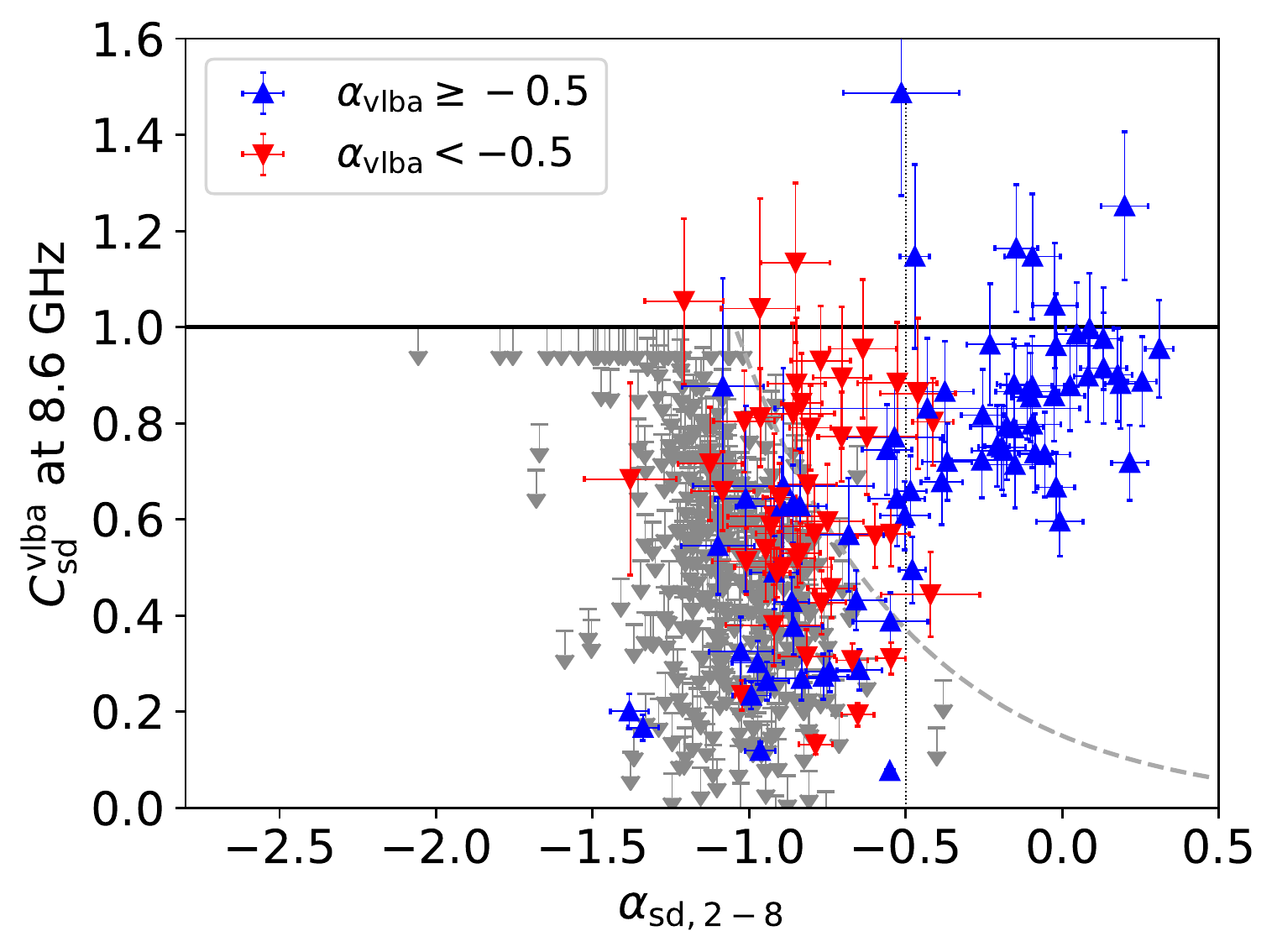}
    \caption{
    Relation between the kiloparsec-scale compactness parameter $C^\mathrm{vlba}_\mathrm{sd}$, defined as the ratio of the VLBA flux density to the total (single-dish) flux density, at two VLBA frequencies and the single-dish 2-8~GHz spectral index $\alpha_\mathrm{sd,2-8}$ . The marker style encodes the type of the VLBA spectrum (see the legend). The upper limits for the sources not detected by the VLBA are marked with gray arrows. The solid line marks the limiting compactness value of~1.0. The dashed line marks the upper envelope of the area, in which non-detected sources could lie (see text for details). The vertical dotted line at $\alpha_\mathrm{sd}=-0.5$ is the border between steep- and flat-spectrum sources in our terminology.
    \label{fig:vlba2ratan}
    }
\end{figure*}

\autoref{fig:vlba2ratan} shows the kiloparsec-scale compactness parameter plotted as a function of the single-dish spectral index. Some sources show non-physical values $C^\mathrm{vlba}_\mathrm{sd}>1$. Besides the measurement errors, it is caused by the sources variability since VLBA and single-dish observations were not simultaneous. Since the sample consists of sources with  single-dish flux density at 1.4~GHz higher than 200~mJy, the largest possible kiloparsec-scale compactness for a non-detected source at frequency $\nu$ with a single-dish spectral index $\alpha_\mathrm{sd}$ equals to (detection~limit)~$/~ [200~\mathrm{mJy}~(\nu/1.4~\mathrm{GHz})^{\alpha_\mathrm{sd}}]$. This upper envelope is plotted with  a grey dashed line.

The kiloparsec-scale compactness (\autoref{fig:vlba2ratan}) correlates with the single-dish spectral index. The corresponding $p$-values in \autoref{tab:correlation} are very low, especially when the upper limits are taken into account. 

In about 90\% of the flat-spectrum sources, the emission detected by the VLBA  from regions of hundreds parsec or less accounts for more than half of the single-dish flux density both at 2.3 and 8.6~GHz. Different steep-spectrum sources have a kiloparsec-scale compactness parameter practically from zero to one. In \autoref{fig:vlba2ratan}, we mark the type of the VLBA spectrum: flat ($\alpha_\mathrm{vlba} \ge -0.5$) or steep ($\alpha_\mathrm{vlba} < -0.5$). Several sources which VLBA spectral type cannot be determined are also indicated.
It helps to visually differentiate CSS candidates and flat-spectrum cores of steep-spectrum extended sources.

\begin{figure*}
    \centering
        \includegraphics[width=0.49\linewidth]{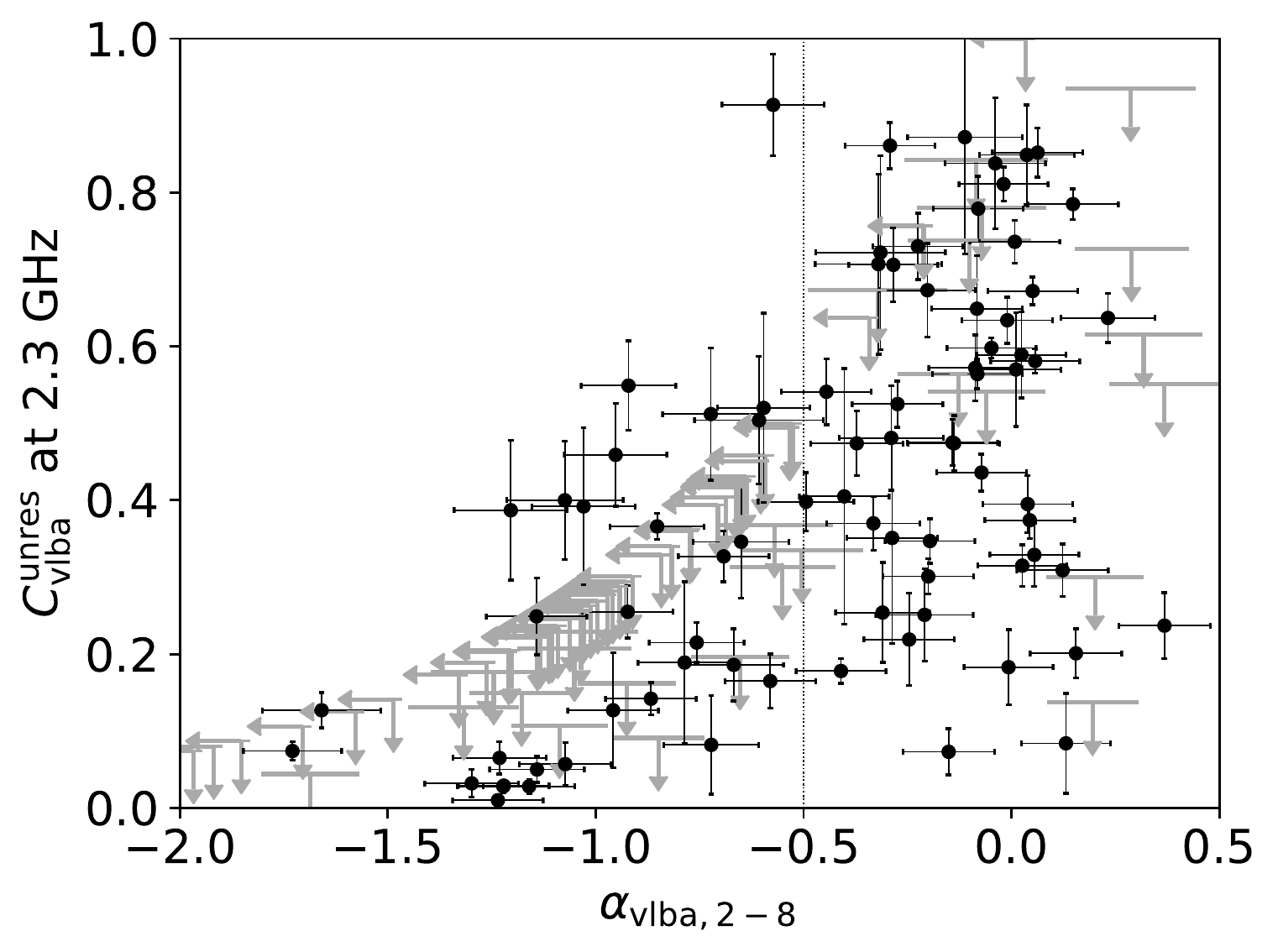}
        \includegraphics[width=0.49\linewidth]{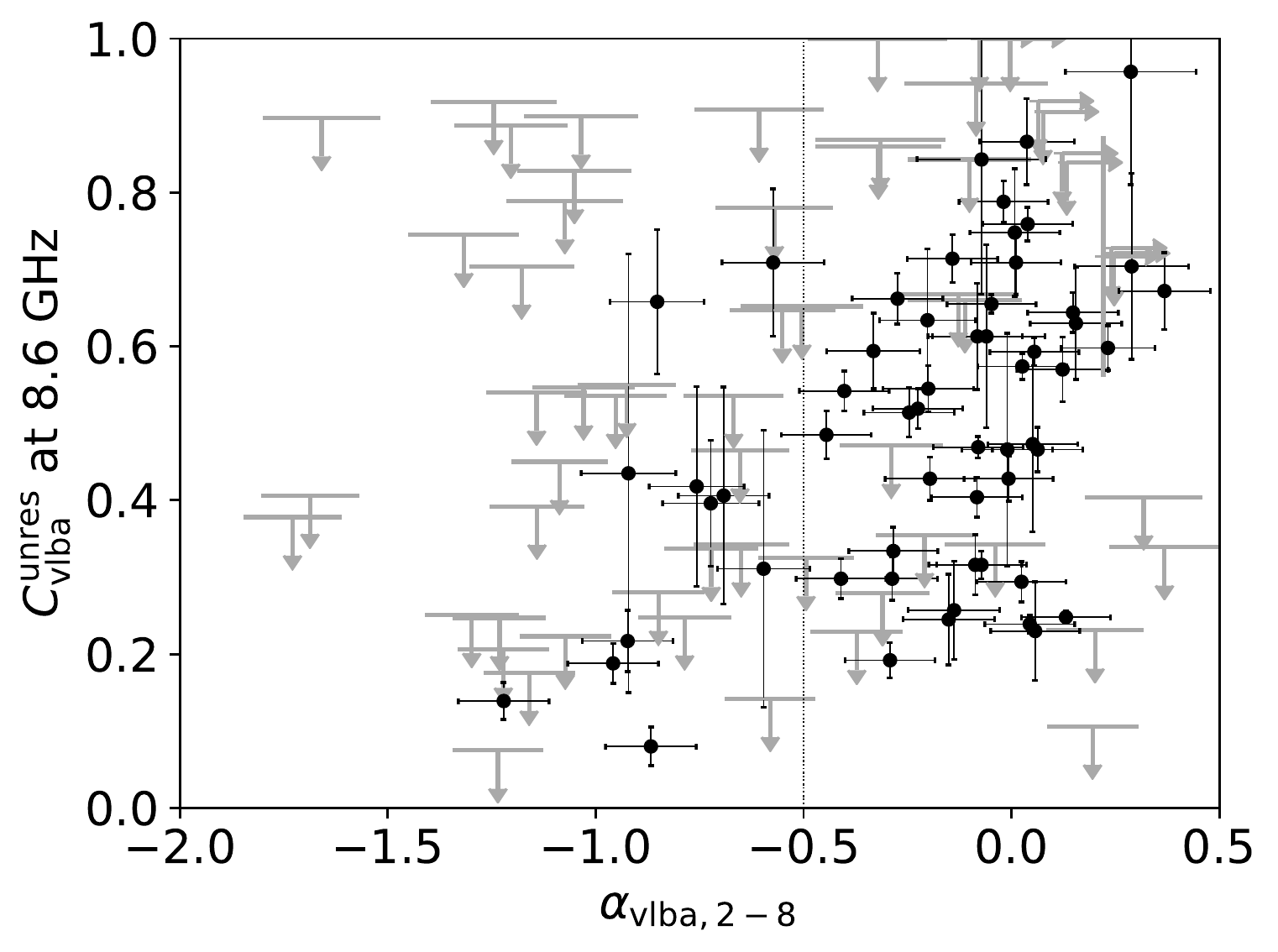}
    \caption{
    Relation between the parsec-scale compactness parameter $C^\mathrm{unres}_\mathrm{vlba}$, defined as the ratio of the unresolved flux density to the VLBA flux density, for two VLBA frequencies and the VLBA spectral index $\alpha_\mathrm{vlba,2-8}$ . The upper limits of compactness for the sources with the VLBA detections only at short baselines and/or upper or lower limits of the VLBA spectral index are marked with gray arrows. Note that less compact sources tend to have steeper spectra.
    \label{fig:unres2vlba_avlba}
    }
\end{figure*}

The parsec-scale compactness parameter $C^\mathrm{unres}_\mathrm{vlba}$ has a stronger correlation with $\alpha_\mathrm{vlba}$ than with $\alpha_\mathrm{sd}$. This is a reasonable result: the compactness at a given spatial scale correlates with the spectral index at the same scale. In \autoref{fig:unres2vlba_avlba}, the parsec-scale compactness is plotted versus the VLBA spectral index. One can see that the CSS source candidates typically have low parsec-scale compactness although some of them show high upper limits especially at 8~GHz which partially complicates the analysis. The sources with $\alpha_\mathrm{vlba}\ge-0.5$ have $C^\mathrm{unres}_\mathrm{vlba}$ in the range from 0 to 1. This scatter is partly due to the sparseness of the $uv$-coverage of our snapshot observations. Some sources have a beam ellipse eccenticity as low as 0.2. When such a narrow beam is directed perpendicular to a jet, a source is resolved stronger than if the beam ellipse were directed along the jet axis.

\begin{figure*}
    \centering
        \includegraphics[width=0.49\linewidth]{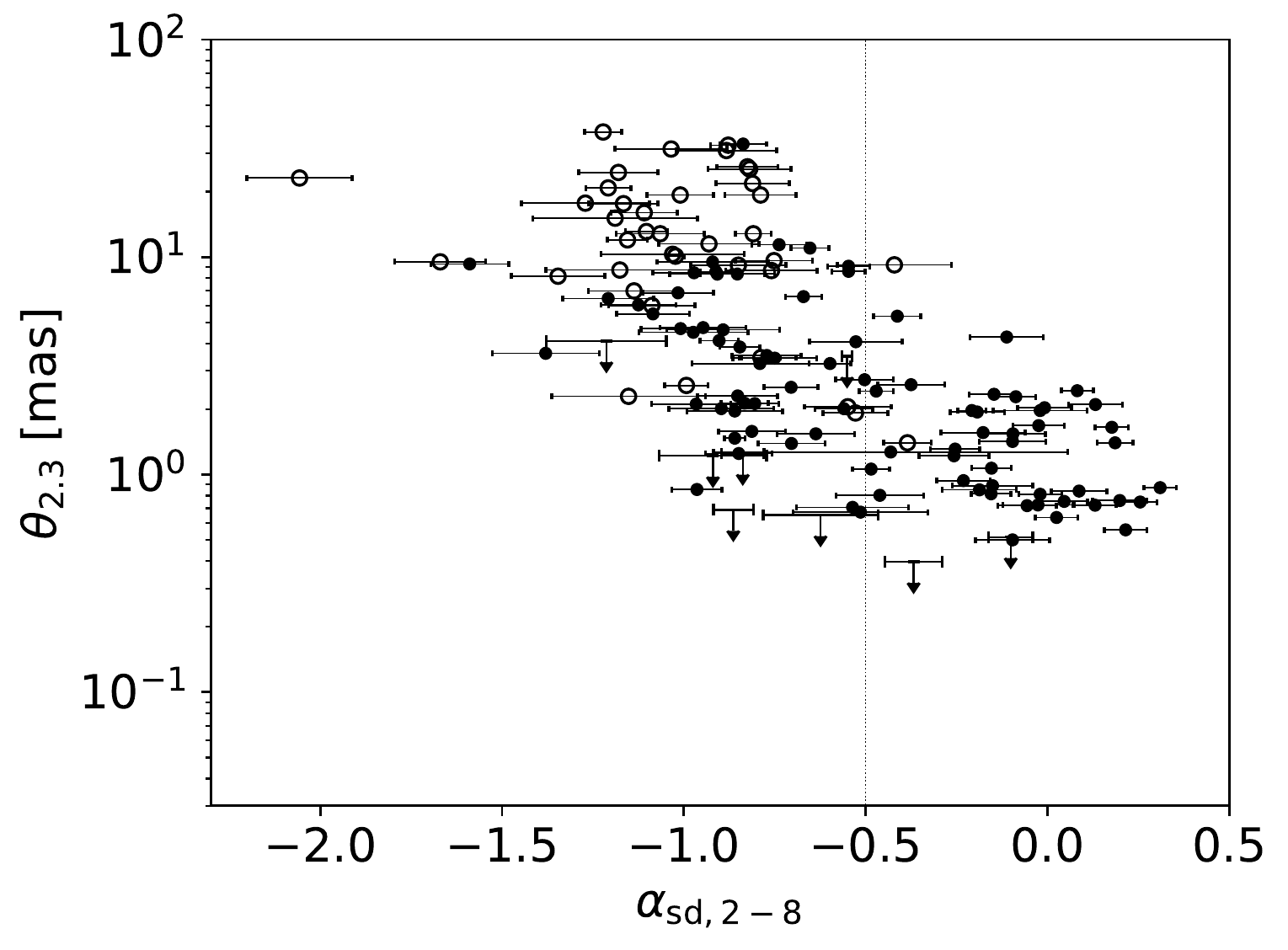}
        \includegraphics[width=0.49\linewidth]{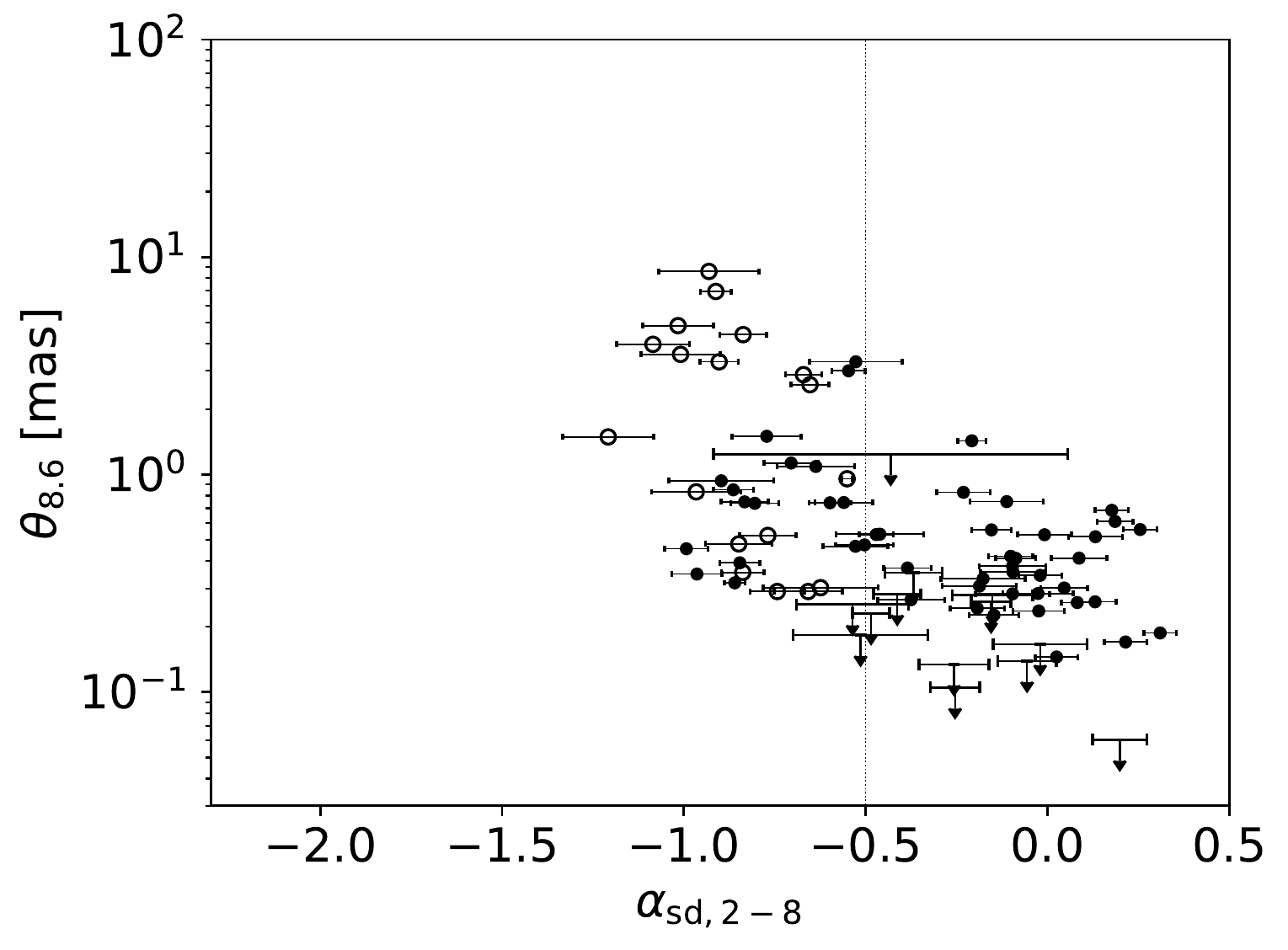}
    \caption{
    Angular size of the main fitted circular Gaussian component of the sources at 2.3~GHz \textit{(left)} and 8.6~GHz \textit{(right)} vs.\ single-dish 2-8~GHz spectral index. The upper limits are plotted for the unresolved sources. We marked the sizes of components fitted to complex visibilities with filled circles, and those fitted to visibility amplitudes only are marked with open circles (see \autoref{sec:processing:params} for details). The vertical dotted line divides steep- and flat-spectrum sources.
    \label{fig:size}
    }
\end{figure*}

The relation between the angular size of the sources and their single-dish spectral index is shown in \autoref{fig:size}. 
Note that we could determine sizes not for all the sources of our sample and not for all the VLBA-detected sources but only for those detected at a large enough number of baselines with an acceptable SNR.
The upper limits are given for unresolved sources. 
The formal errors of the fitted parameters are relatively small, about a few percents. We note however that they are model-dependent, the formal errors should be treated with a caution.
The quantities show a correlation: the flatter is the integral spectrum, the smaller is the dominating feature of the compact structure of a source. In \autoref{tab:correlation}, the correlation coefficients are given for the angular size $\theta$ and the spectral indices $\alpha_\mathrm{sd}$ and $\alpha_\mathrm{vlba}$.

\begin{figure*}
    \centering
        \includegraphics[width=0.49\linewidth]{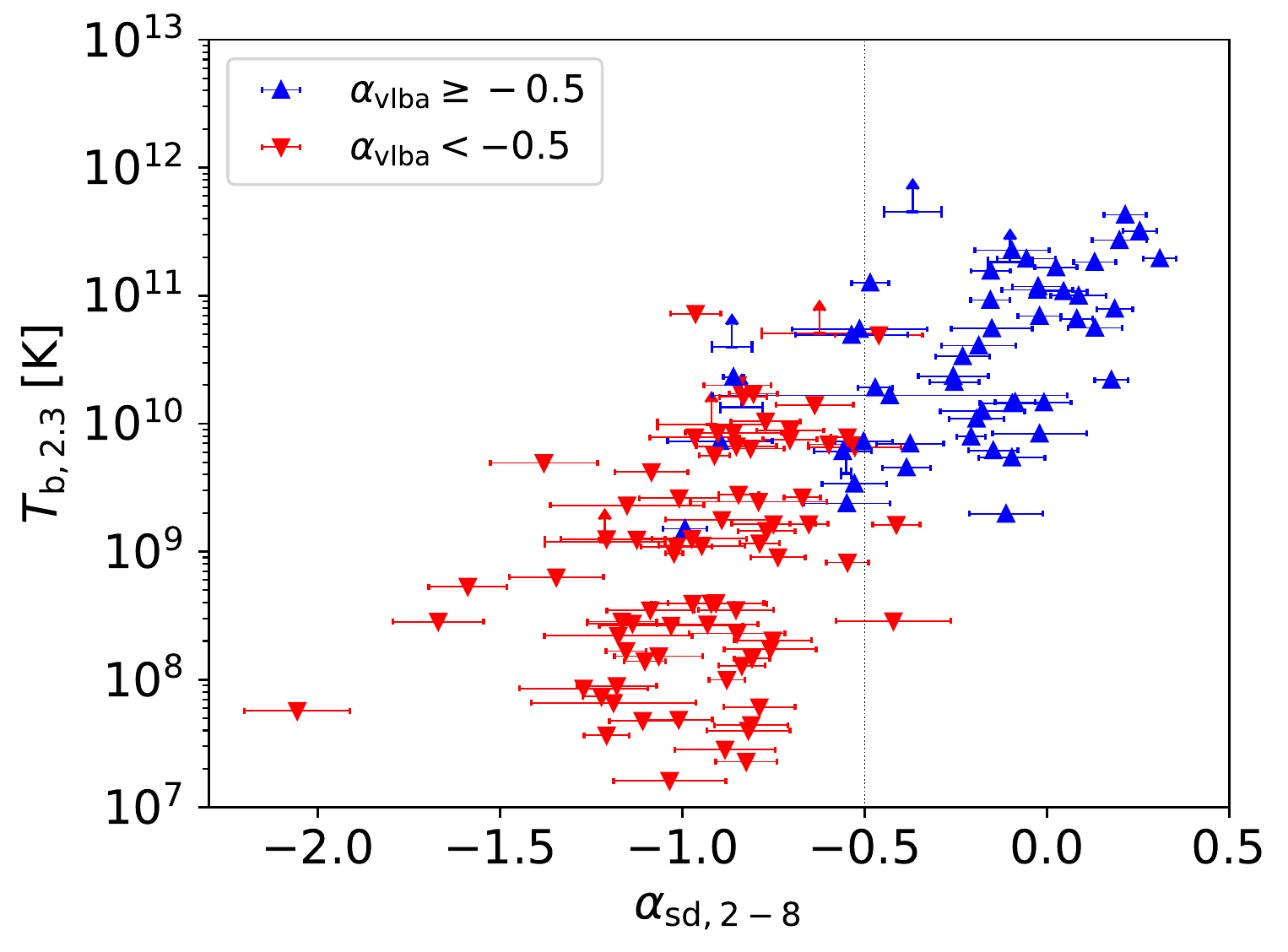}
        \includegraphics[width=0.49\linewidth]{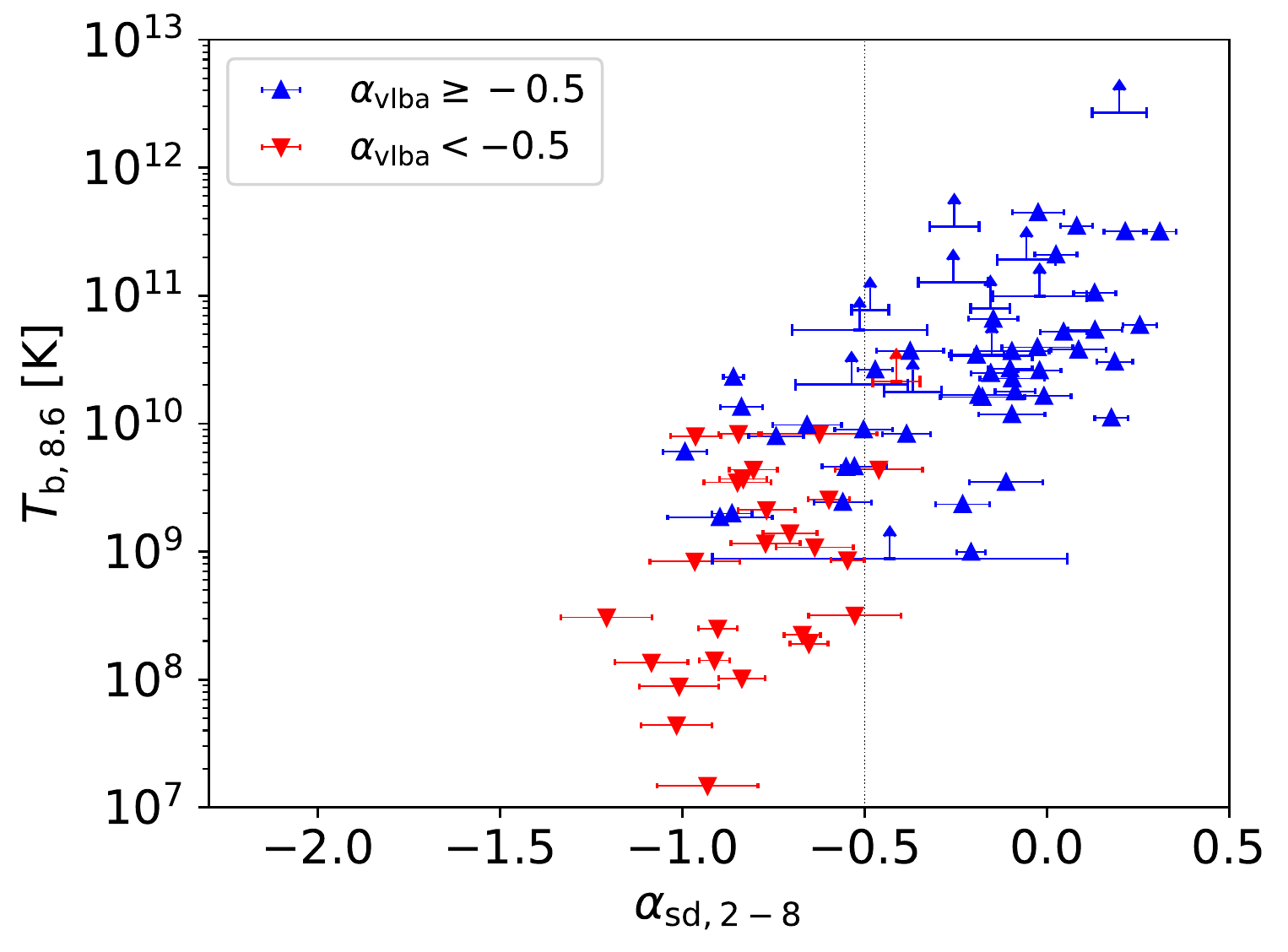}
    \caption{
    Observer's frame brightness temperature of the main fitted circular Gaussian component of the sources at 2.3~GHz \textit{(left)} and 8.6~GHz \textit{(right)} as a function of the single-dish 2-8~GHz spectral index. Symbols are plotted as follows: up-pointing triangles are the sources with a flat VLBA spectrum, down-pointing triangles are the sources with a steep VLBA spectrum.
    \label{fig:tb}
    }
\end{figure*}

\autoref{tab:correlation} also shows that the brightness temperature has a significant correlation with the spectral index. \autoref{fig:tb} shows $T_\mathrm{b}$ versus $\alpha_\mathrm{sd}$. The brightness temperature values are model-dependent, and we estimate their accuracy to be of an order of~2.
Typically, the steep-spectrum sources have one-two orders of magnitude lower $T_\mathrm{b}$ than the flat-spectrum ones. However, in \autoref{sec:analysis:classes}, we show that the sources with a steep single-dish spectrum are very different at parsec scales. 
\autoref{fig:tb} shows that the flat-spectrum cores of steep-spectrum extended sources have practically the same brightness temperature as their ``cousins'' with flat single-dish spectra. However, the correlation between the brightness temperature and the spectral index is still present even if we consider only the sources with flat VLBA spectra. In the sources with flat VLBA spectra, but steep single-dish spectra, the jet is likely directed at a larger angle to the line of sight than in flat-spectrum sources, which is the reason why the core has lower $T_\mathrm{b}$ values and does not dominate the total emission. An example of such sources is J1842+7946 (3C~390.3). It has a steep single-dish spectrum and a flat VLBA spectrum. 
At both frequnecies, its kiloparsec-scale compactness is less than 0.1, and its brightness temperature is about $5\times10^{9}$~K. \citet{2010MNRAS.408.1982L} calculated that the jet of this source is directed at $48\degree$ to the line of sight. It confirms that the properties of the flat-spectrum cores of extended sources can be explained, at least partly, by a large jet viewing angle.

The CSS candidates have one-two orders of magnitude lower average $T_\mathrm{b}$ values than the sources with $\alpha_\mathrm{vlba} \geq -0.5$. Together with their steep VLBA spectrum, it indicates that the optically thick jet core does not dominate their emission. 
VLBI observations with sensitivity and $uv$-coverage better than ours show that the dominating structures of most CSS sources are mini-lobes or jets \citep{2006A&A...449..985M, 2006A&A...450..945K, 2013MNRAS.433..147D, 2007A&A...469..437K}. Our results do not contradict that.


\subsection{Relation between radio flux density variability and parsec-scale structure}
\label{sec:analysis:var}

We investigated the relation between variability of extragalactic radio sources, inferred from multi-epoch single-dish flux density measurements, and their VLBA structure. 
Our variability data are neither complete nor uniform because single-dish spectra of only a fraction of the sources were observed more than once, and different sources have different numbers of observation epochs. However, keeping this in mind, we are still able to draw useful conclusions from these data.

\begin{figure}
    \centering
    \includegraphics[width=\columnwidth]{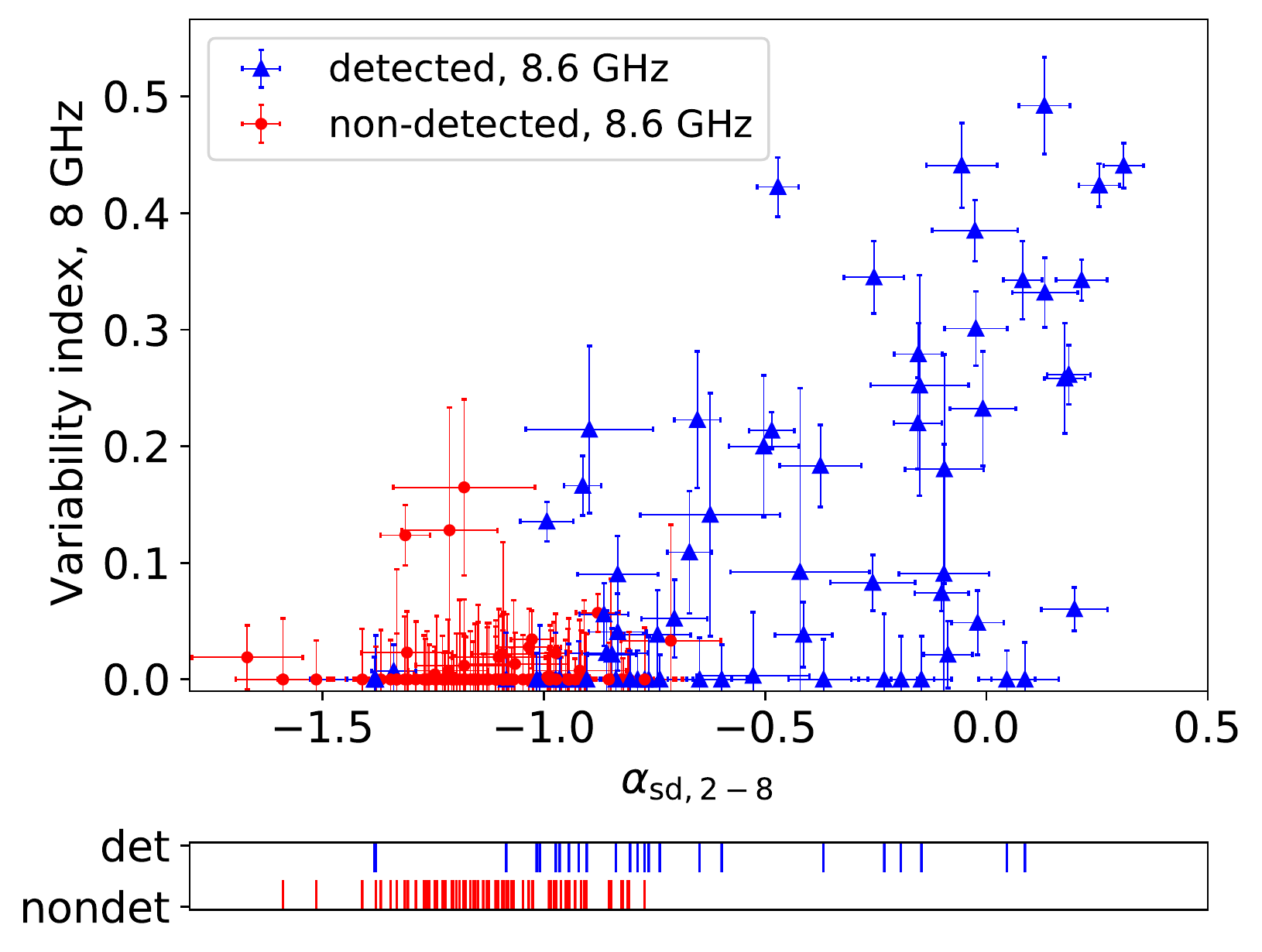}
    \caption{
    Single-dish flux density variability index at 8~GHz, $V_{8}$, vs.\ instant single-dish 2-8~GHz spectral index for VLBA detected and non-detected sources. For many of them the variability index is smaller than the measurement errors and is shown as $V_{8}=0$. Spectral index values for the sources with $V_{8}=0$ are marked with vertical lines in the lower subplot, for detected and non-detected sources separately.
    \label{fig:var_vs_atot}
    }
\end{figure}

The variability index at 8~GHz, $V_\mathrm{8}$ (eq.~(\ref{eq:varind})), is shown for the VLBA detected and non-detected sources in \autoref{fig:var_vs_atot} as a function of the 2-8~GHz single-dish spectral index. As expected from causality arguments, practically all the sources with a significant variability are compact enough to be detected by the VLBA. The exceptions are three non-detected steep-spectrum sources with $V_\mathrm{8}$ between $0.1$ and $0.2$ (namely, J0424+7653, J0920+8628, and J1944+7816). They are rather weak at 8~GHz for RATAN-600 (less than 100~mJy), so their measurement total errors can be underestimated, which causes overestimated variability index.

Most sources in \autoref{fig:var_vs_atot} have a zero variability index, which means that the flux density difference between epochs is less than the measurement errors. To display these sources more clearly, we mark them separately in the lower panel as thin vertical lines at positions equal to their $\alpha_\mathrm{sd}$. The detected sources are shown in the upper row, the non-detected ones --- in the lower row. Extragalactic sources non-detected by VLBA prevail among the non-variable objects, which is another sign of the correlation between source variability and the probability of its VLBI detection, i.e., compactness.

\begin{figure}
    \centering
    \includegraphics[width=\columnwidth]{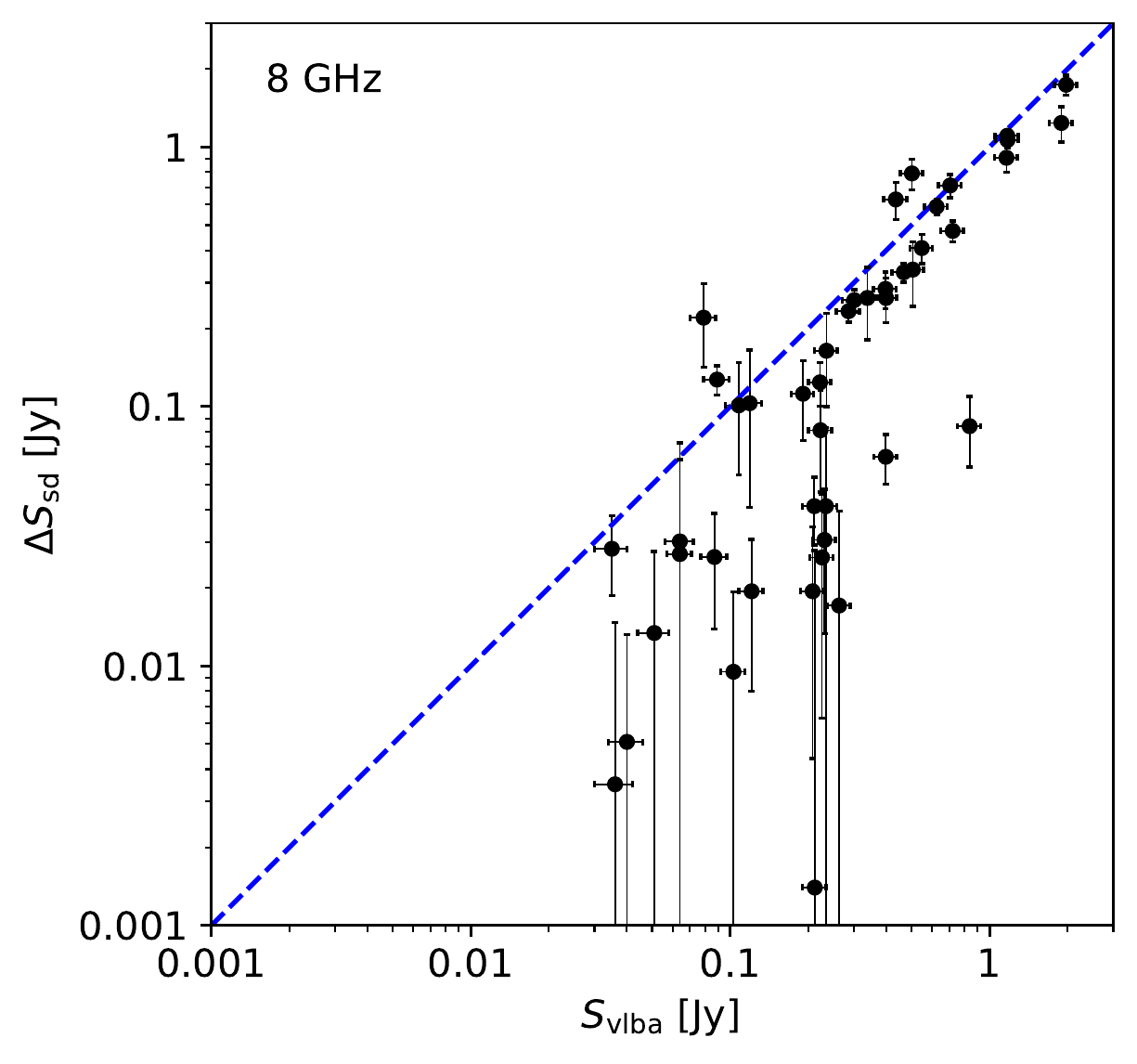}
    \caption{Comparison of the single-dish variability amplitude at 8 GHz and the VLBA flux density at 8.6~GHz for 44 sources with a significant variability.
    The blue dashed line indicates $\Delta S_\mathrm{sd} = S_\mathrm{vlba}$.
    \label{fig:var_vs_vlbaflux}
    }
\end{figure}

\autoref{fig:var_vs_vlbaflux} shows the variability amplitude of the single-dish flux density (eq.~(\ref{eq:varamp})) as a function of the VLBA flux density
for 44 variable sources. Most of the points lie close to the blue dashed line, where $\Delta S_\mathrm{sd} = S_\mathrm{vlba}$, within their error bars. 
This means that indeed the VLBA-detected components dominate the single-dish flux density variability of the objects.
For some sources, the variability amplitude differs significantly from the VLBA flux density. This difference may be due to several following reasons. The sources could have been observed by the VLBA in different states of activity. We may have too few epochs of single-dish flux density measurements, and, thus, the variability amplitude may be underestimated. The VLBA-detected regions could be moderately variable.
There are 26 VLBA-detected sources which show no variability exceeding single-dish flux density errors. 
We conclude that the presence of variability indicates the presence of a compact structure in AGN (as expected) but the reverse is not true: a significant fraction of compact sources might exhibit weak variability.

\begin{figure}
    \centering
    \includegraphics[width=\columnwidth]{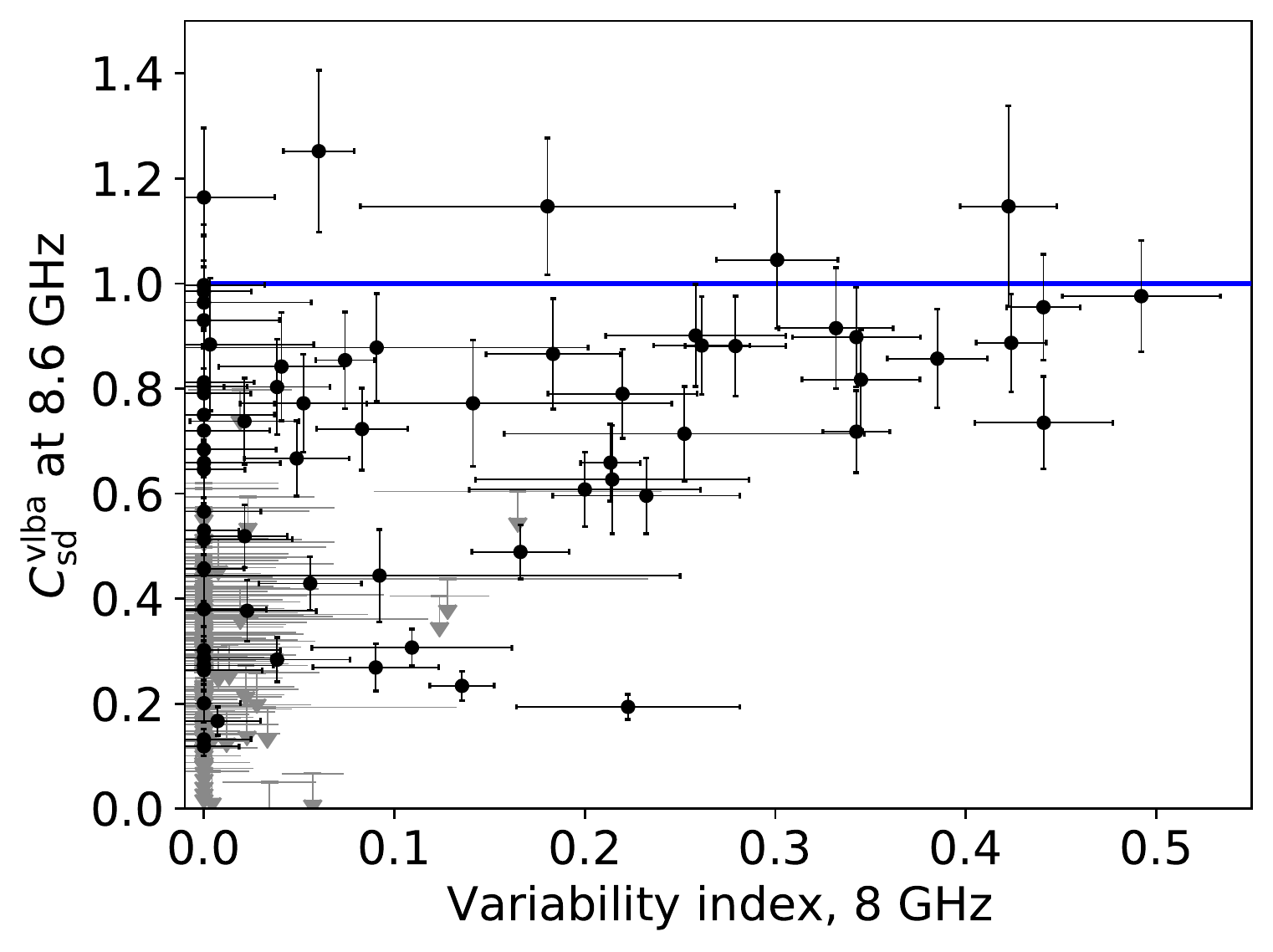}
    \caption{Kiloparsec-scale compactness parameter at 8.6~GHz versus single-dish variability index at 8~GHz. The blue line marks the maximum compactness value of~1.0. A few points moved above the line due to the non-simultaneity of VLBA and single-dish measurements.
    \label{fig:comp_vs_var}}
\end{figure}

A similar trend is shown in \autoref{fig:comp_vs_var} in terms of the kiloparsec-scale compactness parameter $C^\mathrm{vlba}_\mathrm{sd}$. Highly variable sources in our sample have large $C^\mathrm{vlba}_\mathrm{sd}$ since the compact variable component dominates the total emission for them. At the same time, many objects which are very compact at kiloparsec scale show no strong variability.

\begin{figure}
    \centering
    \includegraphics[width=\columnwidth]{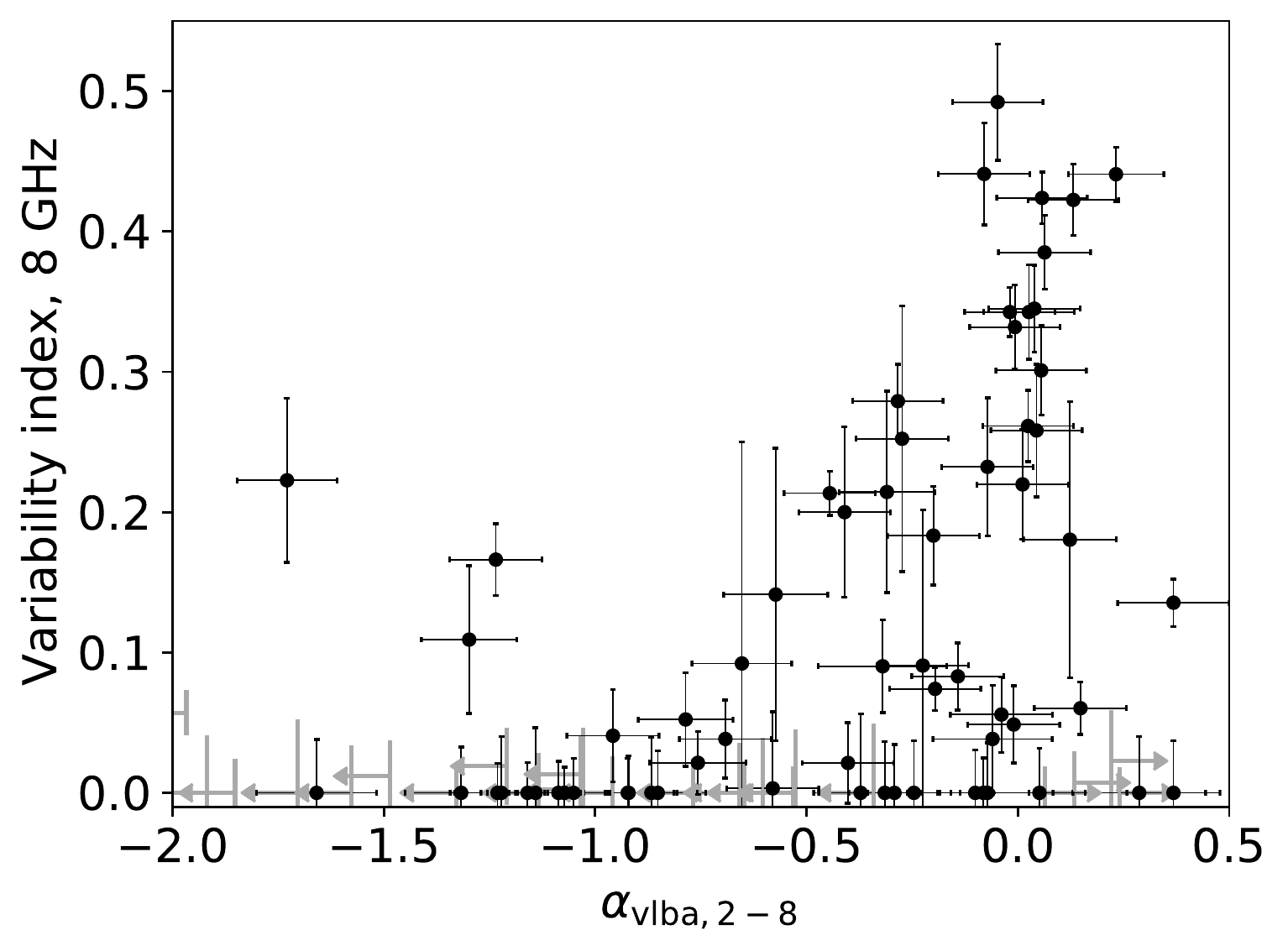}
    \caption{Single-dish variability index at 8~GHz as a function of the VLBA 2-8~GHz spectral index. Note that practically all the sources with significant variability have a flat VLBA spectrum.
    \label{fig:var_vs_avlba}}
\end{figure}

The information about variability sheds light on the nature of the compact sources with a steep VLBA spectrum. In our sample, such sources are numerous.
\autoref{fig:var_vs_avlba} shows the single-dish 8~GHz variability index versus the 2-8~GHz VLBA spectral index. All the sources with high variability have a flat spectrum of a compact structure ($\alpha_\mathrm{vlba} \ge -0.5$). Oppositely, most sources with a steep VLBA spectrum have a variability index close to zero. 
The reason for this dichotomy is that the high variability of emission occurs mostly in the opaque core. If the core dominates the emission, such sources are observed to have flat radio spectra and high variability. 
Vice versa, if the radio emission is dominated by outer parts of the jet or mini-lobes, we see the combination of a steep spectrum and low variability.
Three exceptions from this rule are J1435+7605, J1609+7939, and J2344+8226 (\autoref{fig:var_vs_avlba}). However, their VLBA spectra are measured to be significantly steeper than single-dish spectra. This is unexpected from the general physics picture.
Most probably, the VLBA spectral index is underestimated due to the effect of partial resolution (see the discussion in \autoref{sec:processing:params}).
Another explanation might be related to the VLBA core which spectrum peaks at lower radio frequencies.


\section{Discussion}
\label{sec:discus}

Our results show that there are a significant number of steep-spectrum sources with compact features of structure detectable for the VLBI. We observed with the VLBA the complete sample of sources with the total flux density at 1.4~GHz higher than 200 mJy and we reached the detection
limit of 30~mJy. The number of the steep-spectrum sources detected at 2.3 GHz
is 2.6 times higher than the number of the detected flat-spectrum sources.
At 8.6 GHz this ratio is 1.8.

The probability that at least one source
from the list is falsely detected is 0.01 based on the estimates
of the probability of false detection of a given source.
It is more difficult to evaluate completeness. We re-observed 283
out of 386 NVSS targets not detected at 8.6~GHz in the NPCS in the
framework of the VCS10 observing campaign in 2020. The number of bits
at X-band collected in that campaign is a factor of 2 greater than
in NPCS, and therefore, a detection limit is 40\% lower. We have detected
13 more sources at 8~GHz. We deter a thorough analysis of the VCS10
for the upcoming publication, but these preliminary results already
demonstrate there is no gross miss of sources above the detection limit.

Using our NPCS detection statistics, we can estimate how many compact sources
will be missing in a survey biased towards flat-spectrum sources. The NPCS
sample is complete to the level of the VLBA flux density of 200~mJy at 8.6~GHz, not counting the sources with inverted total spectra with $\alpha>0$.
Among the NPCS sources with the VLBA flux density at 8.6~GHz higher than 200~mJy, 33 sources have flat single-dish spectra and 8 sources have steep single-dish spectra. 
Thus a catalog with the parent sample of flat-spectrum sources will have a completeness at 200~mJy somewhat higher than 80\%. 
Note that among weaker compact sources, steep-spectrum sources dominate, as in our survey with the 30~mJy detection limit; see also \citet{1981AJ.....86..643C,1991SvA....35..563G}.

The statistics of the sources detections, compactness, and spectral shape depends on the sample selection frequency. In our sample, selected from the NVSS at 1.4~GHz, steep-spectrum sources account for 90\%. There were studies of different flux density limited samples, selected at higher frequencies \citep{1999ARep...43..631B, 2000ARep...44..353G, 2003ARep...47..903G, 2006ARep...50..210G}. In these works, simultaneous broadband radio spectra were observed and analyzed. For a flux density limit similar to ours of 200~mJy, the fraction of the steep-spectrum sources decreases to 57\% for the selection frequency 4~GHz, and to 46\% for 5~GHz. In the AT20G catalog at 20~GHz, only 27\% of the sources have steep spectra at GHz frequencies \citep{2013MNRAS.434..956C}.
Concluding, in the samples selected at higher frequencies, the VLBI detection rate is expected to be higher due to a higher fraction of the compact flat-spectrum sources. At the same time, the samples selected at lower frequencies are more suitable for studying compact steep-spectrum sources and they typically go deeper.

Another observational reason for the dependence of the detection statistics on the sample selection frequency and the observing frequency is the following. We selected our sample at 1.4~GHz and observed it at the frequencies several times higher, 2.3 and 8.6~GHz. As a result, we preferably detect, especially at 8.6~GHz, the sources with flatter spectra, since they have a relatively higher flux density at a higher observing frequency. In modern surveys, the detection limit is usually lower, so this limitation is less strict, but it cannot be eliminated completely.

Our results on the relation between the compactness and the spectral index are in a good agreement with earlier works. The percent of compact steep-spectrum sources in our sample is close to that determined by \cite{1988ApJ...328..114P} from a several times smaller sample. Our \autoref{fig:vlba2ratan} is similar to Figure~7 in \cite{2013MNRAS.434..956C}: they show the same smooth transition from compact flat-spectrum sources to steep-spectrum sources with the kiloparsec-scale compactness parameter of the latter in the whole range from zero to one. At the same time, there are some differences: the fraction of the extended steep-spectrum sources is much higher in our sample, than in the sample of \cite{2013MNRAS.434..956C}, and the flat-spectrum sources are more resolved in our observations. The reasons for that are the higher angular resolution of our observations and the lower selection frequency of our sample.

The angular size-frequency dependence for the compact sources in our sample is in agreement with \cite{2015MNRAS.452.4274P}. This dependence is commonly characterized by a power index $k$: $\theta\propto\nu^{-k}$. These authors found that for about 2000 extragalactic sources outside the Milky Way plane (galactic latitude $|b|>10\degree$), the $k$ distribution can be approximated by a Gaussian with a mean of 0.90 and a standard deviation of 0.44. The similar fitting for 60 sources from our sample with sizes measured at both frequencies, 2.3 and 8.6~GHz, yields the peak position at $k=0.82$ and the standard deviation of 0.51, in agreement with that work.

We identified 82 sources with steep spectra of parsec-scale structure. At the same time, many of them are significantly resolved on parsec scales. Morphology of strongly resolved sources cannot be reliably identified with the VLBI. In such cases, there is a possibility that the VLBA detects only the brightest region in a hot spot of an extended radio galaxy. We call these sources ``CSS candidates,'' follow-up observations with e-MERLIN and/or VLA are needed to clarify their morphology.

Our detection statistics is based on the fringe amplitude of individual
pointed observations of the targets. The calibration converts the fringe amplitude at a given scan and a given baseline to the correlated flux density in Jy. An alternative approach is to image a field where a source is supposed to be located. In such a case a decision about the detection is made on the basis of an excessive surface brightness in Jy/beam beyond the image noise level. This approach was used in mJIVE-20 \citep{2014AJ....147...14D}, COSMOS \citep{2017A&A...607A.132H}, and GOODS-N \citep{2018A&A...619A..48R} VLBI surveys, which have utilized phase referencing for a longer coherent integration and higher sensitivity.
For sources that had enough detections to get an image, the total flux density
integrated over the image will be the same in both approaches. The case of a source that is partly resolved and detected only at several short
baselines is more complicated and requires further investigation. Note, mJIVE-20, COSMOS and GOODS-N targeted weaker sources than our survey and reported the detection fraction around 20\,\%. 
The detection fractions of our survey at 2.3 and 8.6~GHz cannot be directly
compared with the detection fractions of the mJIVE-20, COSMOS, or GOODS-N at 1.4~GHz. The surveys have used different observing techniques, different integration times, frequencies, their detections and fractions are defined in a different way, detection limit for extended sources is also not equivalent. All these issues are treatable, homogenization of all these surveys is possible. This work will be done in the future and allow us to extend the statistics from the Jy to the 0.1~mJy level.


\section{Summary}
\label{sec:concl}

We reported the results of the VLBA North Polar Cap Survey ---
VLBA observations at 2.3 and 8.6~GHz of the large complete flux-density limited sample drawn from the NVSS catalog {\it regardless of the spectral index}. Of 482 target sources, 162 were detected. We measured their coordinates as well as the flux density of compact parsec-scale structure and, for most of the detected sources, the angular size and the brightness temperature of their dominant components. For all the target sources, the total (single-dish) continuum radio spectra were published earlier; for most of them, quasi-simultaneous RATAN-600 spectra at 1-22~GHz are available.
This allowed us to analyze the relation between parsec-scale structure and broadband radio spectra.
We also characterised single-dish 8~GHz variability for about one-third of the targets using our RATAN-600 AGN monitoring program and the data from the literature.
    
The VLBA detection statistics shows that a significant fraction of steep-spectrum sources have compact features of a size of several hundreds parsec or less. We detected 116 steep-spectrum sources at least in one band compared to 41 detected flat-spectrum sources and 5 peaked spectrum sources. Despite the detection rate for the latter two spectral types is nearly 100\% and for steep-spectrum sources it is only about 25\%, steep-spectrum sources account for more than 2/3 of the detected sources because they dominate the full sample selected at 1.4~GHz.

The sources detected by the VLBA and having steep single-dish spectra belong to one of two subclasses. The first one consists of flat-spectrum opaque cores of extended steep-spectrum sources. Their parsec-scale properties are similar to those of the sources with flat single-dish spectra: high compactness, high brightness temperature, and high radio variability. This leads to the conclusion that, together with the flat-spectrum sources, they form a subsample with observable relativistic jet cores. Together they comprise 14\% of the full sample.
The second subclass consists of compact steep-spectrum sources. They have steep spectra at the VLBA spatial scales as well. Their lower parsec-scale compactness and brightness temperature, together with practically no variability in most of them, indicate that their emission comes mostly from optically thin outer parts of jets or mini-lobes. They account for 17\% of the full studied sample. 

The compactness parameters, the angular size, and the brightness temperature $T_\mathrm{b}$ of the sources are correlated with their spectral index: compactness and $T_\mathrm{b}$ have a positive correlation, and size --- a negative. 
Our analysis also shows the correlation between the source variability amplitude and its compactness. As expected, the sources variable at 8~GHz are more likely VLBI-detectable. The single-dish variability amplitude of most of variable sources is observed to be very close to the flux density of their components detected by the VLBI. At the same time, we detected with VLBI a significant number of sources showing no variability within the margin of errors.

The selection bias towards flat-spectrum sources was lifted in some
recent VLBI surveys, such as the VLBI Ecliptic Plane Survey
\citep{2017ApJS..230...13S} and the VLBA Calibrator Surveys 7, 8, and 9
\citep{2020arXiv200809243P}. We ran two VLBA follow-up programs
at 4.3 and 7.6~GHz targeting all the AT20G sources with the declination
$>-40\degree$, and all the sources from GB6 \citep{r:gb6} catalog stronger than 70~mJy
within $\pm 7.5\degree$ of the ecliptic band. The programs were
completed in 2020, and the results will be published soon. 


\acknowledgments

We thank Alexander Plavin and Alexander Pushkarev for helpful discussions and comments, the anonymous referee and Eduardo Ros for valuable suggestions which have helped to improve the manuscript.
This study was supported by the Russian Foundation of Basic Research, grant 19-32-90140.
The National Radio Astronomy Observatory is a facility of the National Science Foundation operated under cooperative agreement by Associated Universities, Inc.
This research is based on the observations with RATAN-600 of the Special Astrophysical Observatory, Russian Academy of Sciences (SAO RAS).
The observations with the SAO RAS telescopes are supported by the Ministry of Science and Higher Education of the Russian Federation.
The authors made use of the database CATS \citep{2005BSAO...58..118V} of the Special Astrophysical Observatory.
This research made use of the NASA/IPAC Extragalactic Database (NED),
which is operated by the Jet Propulsion Laboratory, California Institute of Technology, under contract with the National Aeronautics and Space Administration.
\facilities{VLBA, RATAN}
\software{AIPS \citep{2003ASSL..285..109G}, PIMA \citep{2011AJ....142...35P}, Difmap \citep{1997ASPC..125...77S}, astropy \citep{2013A&A...558A..33A, 2018AJ....156..123A}}

\vspace{0 pt}


\bibliography{npcs}
\bibliographystyle{aasjournal}



\figsetstart
\figsetnum{1}
\figsettitle{NVSS maps of the complex sources.}

\figsetgrpstart
\figsetgrpnum{1.1}
\figsetgrptitle{NVSS map of J0123+8056}
\figsetplot{J0123+8056_NVSS.pdf}
\figsetgrpnote{NVSS maps of the complex sources. The source name is given at the top left of each map. The contours are plotted starting at 1~mJy intensity level with an increment of 2.}
\figsetgrpend

\figsetgrpstart
\figsetgrpnum{1.2}
\figsetgrptitle{NVSS map of J0152+7550}
\figsetplot{J0152+7550_NVSS.pdf}
\figsetgrpnote{NVSS maps of the complex sources. The source name is given at the top left of each map. The contours are plotted starting at 1~mJy intensity level with an increment of 2.}
\figsetgrpend

\figsetgrpstart
\figsetgrpnum{1.3}
\figsetgrptitle{NVSS map of J0222+8618}
\figsetplot{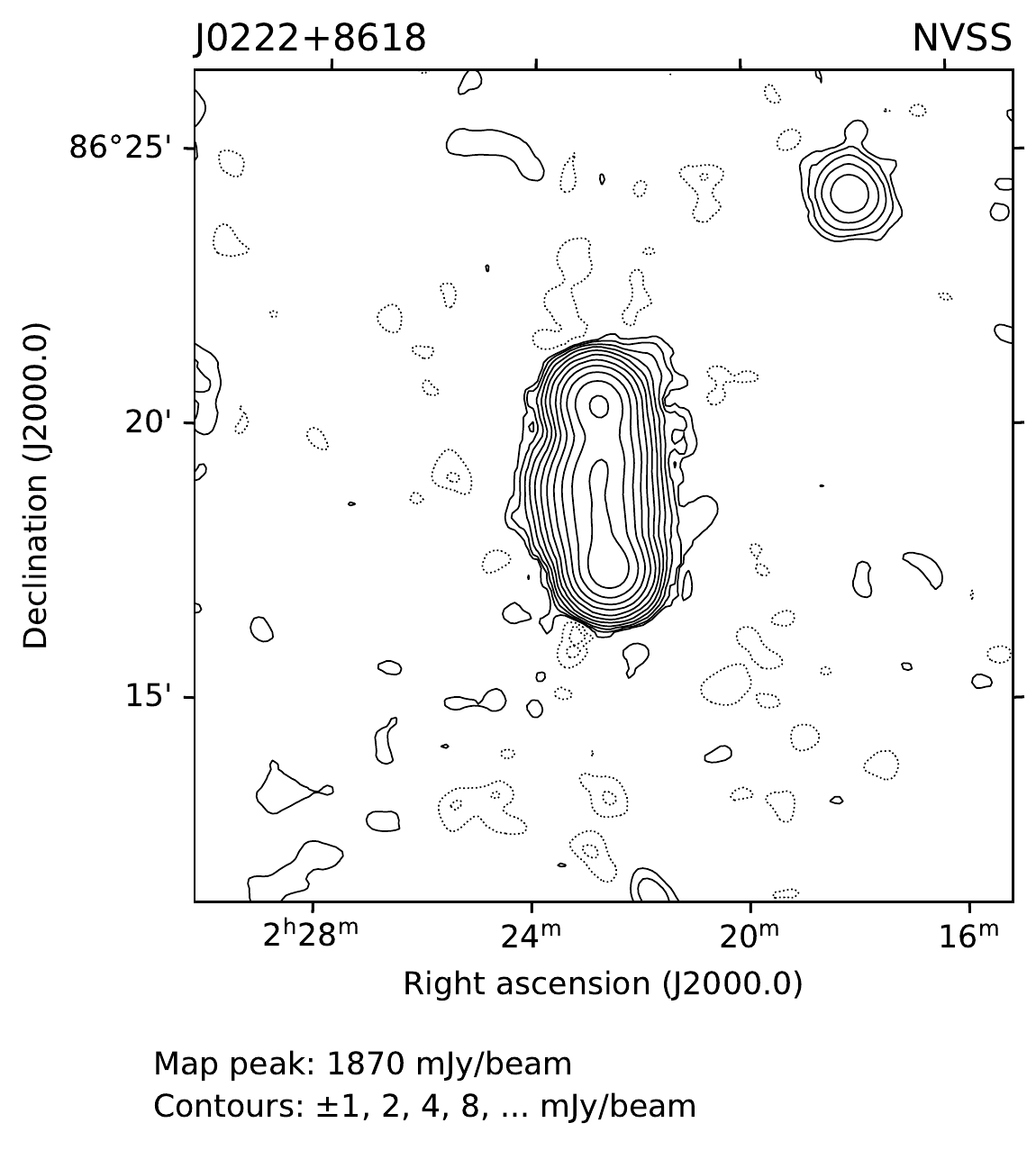}
\figsetgrpnote{NVSS maps of the complex sources. The source name is given at the top left of each map. The contours are plotted starting at 1~mJy intensity level with an increment of 2.}
\figsetgrpend

\figsetgrpstart
\figsetgrpnum{1.4}
\figsetgrptitle{NVSS map of J0403+7616}
\figsetplot{J0403+7616_NVSS.pdf}
\figsetgrpnote{NVSS maps of the complex sources. The source name is given at the top left of each map. The contours are plotted starting at 1~mJy intensity level with an increment of 2.}
\figsetgrpend

\figsetgrpstart
\figsetgrpnum{1.5}
\figsetgrptitle{NVSS map of J0452+7626}
\figsetplot{J0452+7626_NVSS.pdf}
\figsetgrpnote{NVSS maps of the complex sources. The source name is given at the top left of each map. The contours are plotted starting at 1~mJy intensity level with an increment of 2.}
\figsetgrpend

\figsetgrpstart
\figsetgrpnum{1.6}
\figsetgrptitle{NVSS map of J0630+7632}
\figsetplot{J0630+7632_NVSS.pdf}
\figsetgrpnote{NVSS maps of the complex sources. The source name is given at the top left of each map. The contours are plotted starting at 1~mJy intensity level with an increment of 2.}
\figsetgrpend

\figsetgrpstart
\figsetgrpnum{1.7}
\figsetgrptitle{NVSS map of J0645+7755}
\figsetplot{J0645+7755_NVSS.pdf}
\figsetgrpnote{NVSS maps of the complex sources. The source name is given at the top left of each map. The contours are plotted starting at 1~mJy intensity level with an increment of 2.}
\figsetgrpend

\figsetgrpstart
\figsetgrpnum{1.8}
\figsetgrptitle{NVSS map of J0743+8025}
\figsetplot{J0743+8025_NVSS.pdf}
\figsetgrpnote{NVSS maps of the complex sources. The source name is given at the top left of each map. The contours are plotted starting at 1~mJy intensity level with an increment of 2.}
\figsetgrpend

\figsetgrpstart
\figsetgrpnum{1.9}
\figsetgrptitle{NVSS map of J1014+8553}
\figsetplot{J1014+8553_NVSS.pdf}
\figsetgrpnote{NVSS maps of the complex sources. The source name is given at the top left of each map. The contours are plotted starting at 1~mJy intensity level with an increment of 2.}
\figsetgrpend

\figsetgrpstart
\figsetgrpnum{1.10}
\figsetgrptitle{NVSS map of J1234+7730}
\figsetplot{J1234+7730_NVSS.pdf}
\figsetgrpnote{NVSS maps of the complex sources. The source name is given at the top left of each map. The contours are plotted starting at 1~mJy intensity level with an increment of 2.}
\figsetgrpend

\figsetgrpstart
\figsetgrpnum{1.11}
\figsetgrptitle{NVSS map of J1337+8033}
\figsetplot{J1337+8033_NVSS.pdf}
\figsetgrpnote{NVSS maps of the complex sources. The source name is given at the top left of each map. The contours are plotted starting at 1~mJy intensity level with an increment of 2.}
\figsetgrpend

\figsetgrpstart
\figsetgrpnum{1.12}
\figsetgrptitle{NVSS map of J1607+8501}
\figsetplot{J1607+8501_NVSS.pdf}
\figsetgrpnote{NVSS maps of the complex sources. The source name is given at the top left of each map. The contours are plotted starting at 1~mJy intensity level with an increment of 2.}
\figsetgrpend

\figsetgrpstart
\figsetgrpnum{1.13}
\figsetgrptitle{NVSS map of J1632+8232}
\figsetplot{J1632+8232_NVSS.pdf}
\figsetgrpnote{NVSS maps of the complex sources. The source name is given at the top left of each map. The contours are plotted starting at 1~mJy intensity level with an increment of 2.}
\figsetgrpend

\figsetgrpstart
\figsetgrpnum{1.14}
\figsetgrptitle{NVSS map of J1706+7707}
\figsetplot{J1706+7707_NVSS.pdf}
\figsetgrpnote{NVSS maps of the complex sources. The source name is given at the top left of each map. The contours are plotted starting at 1~mJy intensity level with an increment of 2.}
\figsetgrpend

\figsetgrpstart
\figsetgrpnum{1.15}
\figsetgrptitle{NVSS map of J1842+7946}
\figsetplot{J1842+7946_NVSS.pdf}
\figsetgrpnote{NVSS maps of the complex sources. The source name is given at the top left of each map. The contours are plotted starting at 1~mJy intensity level with an increment of 2.}
\figsetgrpend

\figsetgrpstart
\figsetgrpnum{1.16}
\figsetgrptitle{NVSS map of J1845+8150}
\figsetplot{J1845+8150_NVSS.pdf}
\figsetgrpnote{NVSS maps of the complex sources. The source name is given at the top left of each map. The contours are plotted starting at 1~mJy intensity level with an increment of 2.}
\figsetgrpend

\figsetgrpstart
\figsetgrpnum{1.17}
\figsetgrptitle{NVSS map of J1906+8100}
\figsetplot{J1906+8100_NVSS.pdf}
\figsetgrpnote{NVSS maps of the complex sources. The source name is given at the top left of each map. The contours are plotted starting at 1~mJy intensity level with an increment of 2.}
\figsetgrpend

\figsetgrpstart
\figsetgrpnum{1.18}
\figsetgrptitle{NVSS map of J1941+8501}
\figsetplot{J1941+8501_NVSS.pdf}
\figsetgrpnote{NVSS maps of the complex sources. The source name is given at the top left of each map. The contours are plotted starting at 1~mJy intensity level with an increment of 2.}
\figsetgrpend

\figsetgrpstart
\figsetgrpnum{1.19}
\figsetgrptitle{NVSS map of J2042+7508}
\figsetplot{J2042+7508_NVSS.pdf}
\figsetgrpnote{NVSS maps of the complex sources. The source name is given at the top left of each map. The contours are plotted starting at 1~mJy intensity level with an increment of 2.}
\figsetgrpend

\figsetgrpstart
\figsetgrpnum{1.20}
\figsetgrptitle{NVSS map of J2118+7511}
\figsetplot{J2118+7511_NVSS.pdf}
\figsetgrpnote{NVSS maps of the complex sources. The source name is given at the top left of each map. The contours are plotted starting at 1~mJy intensity level with an increment of 2.}
\figsetgrpend

\figsetgrpstart
\figsetgrpnum{1.21}
\figsetgrptitle{NVSS map of J2238+8148}
\figsetplot{J2238+8148_NVSS.pdf}
\figsetgrpnote{NVSS maps of the complex sources. The source name is given at the top left of each map. The contours are plotted starting at 1~mJy intensity level with an increment of 2.}
\figsetgrpend

\figsetgrpstart
\figsetgrpnum{1.22}
\figsetgrptitle{NVSS map of J2355+7954}
\figsetplot{J2355+7954_NVSS.pdf}
\figsetgrpnote{NVSS maps of the complex sources. The source name is given at the top left of each map. The contours are plotted starting at 1~mJy intensity level with an increment of 2.}
\figsetgrpend

\figsetend


\figsetstart
\figsetnum{3}
\figsettitle{Correlated flux density averaged over time and IFs versus the $uv$ radius for all the detected sources.}

\figsetgrpstart
\figsetgrpnum{3.1}
\figsetgrptitle{J0009+7603 at 2.3 GHz}
\figsetplot{J0009+7603_S_2006_02_14_pet_uvt_rad.pdf}
\figsetgrpnote{Correlated flux density averaged over time and IFs versus the $uv$ radius. The source name, the observation date, and the frequency are given at the top of the panel. In the cases when the data for a source were calibrated in AIPS and then underwent hybrid imaging in Difmap with both amplitude and phase self-calibration, they are plotted as filled circles. In the cases when the processing was the same except no amplitude self-calibration was made, the data are plotted as filled triangles. In the cases when no self-calibration was made for the source and the data calibrated in PIMA were used, they are plotted as open circles.}
\figsetgrpend

\figsetgrpstart
\figsetgrpnum{3.2}
\figsetgrptitle{J0009+7603 at 8.6 GHz}
\figsetplot{J0009+7603_X_2006_02_14_avp_uvs_rad.pdf}
\figsetgrpnote{Correlated flux density averaged over time and IFs versus the $uv$ radius. The source name, the observation date, and the frequency are given at the top of the panel. In the cases when the data for a source were calibrated in AIPS and then underwent hybrid imaging in Difmap with both amplitude and phase self-calibration, they are plotted as filled circles. In the cases when the processing was the same except no amplitude self-calibration was made, the data are plotted as filled triangles. In the cases when no self-calibration was made for the source and the data calibrated in PIMA were used, they are plotted as open circles.}
\figsetgrpend

\figsetgrpstart
\figsetgrpnum{3.3}
\figsetgrptitle{J0009+7724 at 2.3 GHz}
\figsetplot{J0009+7724_S_2006_02_23_pet_uvt_rad.pdf}
\figsetgrpnote{Correlated flux density averaged over time and IFs versus the $uv$ radius. The source name, the observation date, and the frequency are given at the top of the panel. In the cases when the data for a source were calibrated in AIPS and then underwent hybrid imaging in Difmap with both amplitude and phase self-calibration, they are plotted as filled circles. In the cases when the processing was the same except no amplitude self-calibration was made, the data are plotted as filled triangles. In the cases when no self-calibration was made for the source and the data calibrated in PIMA were used, they are plotted as open circles.}
\figsetgrpend

\figsetgrpstart
\figsetgrpnum{3.4}
\figsetgrptitle{J0009+7724 at 8.6 GHz}
\figsetplot{J0009+7724_X_2006_02_23_pet_uvt_rad.pdf}
\figsetgrpnote{Correlated flux density averaged over time and IFs versus the $uv$ radius. The source name, the observation date, and the frequency are given at the top of the panel. In the cases when the data for a source were calibrated in AIPS and then underwent hybrid imaging in Difmap with both amplitude and phase self-calibration, they are plotted as filled circles. In the cases when the processing was the same except no amplitude self-calibration was made, the data are plotted as filled triangles. In the cases when no self-calibration was made for the source and the data calibrated in PIMA were used, they are plotted as open circles.}
\figsetgrpend

\figsetgrpstart
\figsetgrpnum{3.5}
\figsetgrptitle{J0013+7748 at 2.3 GHz}
\figsetplot{J0013+7748_S_2006_02_23_pet_uvt_rad.pdf}
\figsetgrpnote{Correlated flux density averaged over time and IFs versus the $uv$ radius. The source name, the observation date, and the frequency are given at the top of the panel. In the cases when the data for a source were calibrated in AIPS and then underwent hybrid imaging in Difmap with both amplitude and phase self-calibration, they are plotted as filled circles. In the cases when the processing was the same except no amplitude self-calibration was made, the data are plotted as filled triangles. In the cases when no self-calibration was made for the source and the data calibrated in PIMA were used, they are plotted as open circles.}
\figsetgrpend

\figsetgrpstart
\figsetgrpnum{3.6}
\figsetgrptitle{J0017+8135 at 2.3 GHz}
\figsetplot{J0017+8135_S_2006_02_14_avp_uvs_rad.pdf}
\figsetgrpnote{Correlated flux density averaged over time and IFs versus the $uv$ radius. The source name, the observation date, and the frequency are given at the top of the panel. In the cases when the data for a source were calibrated in AIPS and then underwent hybrid imaging in Difmap with both amplitude and phase self-calibration, they are plotted as filled circles. In the cases when the processing was the same except no amplitude self-calibration was made, the data are plotted as filled triangles. In the cases when no self-calibration was made for the source and the data calibrated in PIMA were used, they are plotted as open circles.}
\figsetgrpend

\figsetgrpstart
\figsetgrpnum{3.7}
\figsetgrptitle{J0017+8135 at 2.3 GHz}
\figsetplot{J0017+8135_S_2006_02_16_avp_uvs_rad.pdf}
\figsetgrpnote{Correlated flux density averaged over time and IFs versus the $uv$ radius. The source name, the observation date, and the frequency are given at the top of the panel. In the cases when the data for a source were calibrated in AIPS and then underwent hybrid imaging in Difmap with both amplitude and phase self-calibration, they are plotted as filled circles. In the cases when the processing was the same except no amplitude self-calibration was made, the data are plotted as filled triangles. In the cases when no self-calibration was made for the source and the data calibrated in PIMA were used, they are plotted as open circles.}
\figsetgrpend

\figsetgrpstart
\figsetgrpnum{3.8}
\figsetgrptitle{J0017+8135 at 2.3 GHz}
\figsetplot{J0017+8135_S_2006_02_23_avp_uvs_rad.pdf}
\figsetgrpnote{Correlated flux density averaged over time and IFs versus the $uv$ radius. The source name, the observation date, and the frequency are given at the top of the panel. In the cases when the data for a source were calibrated in AIPS and then underwent hybrid imaging in Difmap with both amplitude and phase self-calibration, they are plotted as filled circles. In the cases when the processing was the same except no amplitude self-calibration was made, the data are plotted as filled triangles. In the cases when no self-calibration was made for the source and the data calibrated in PIMA were used, they are plotted as open circles.}
\figsetgrpend

\figsetgrpstart
\figsetgrpnum{3.9}
\figsetgrptitle{J0017+8135 at 8.6 GHz}
\figsetplot{J0017+8135_X_2006_02_14_avp_uvs_rad.pdf}
\figsetgrpnote{Correlated flux density averaged over time and IFs versus the $uv$ radius. The source name, the observation date, and the frequency are given at the top of the panel. In the cases when the data for a source were calibrated in AIPS and then underwent hybrid imaging in Difmap with both amplitude and phase self-calibration, they are plotted as filled circles. In the cases when the processing was the same except no amplitude self-calibration was made, the data are plotted as filled triangles. In the cases when no self-calibration was made for the source and the data calibrated in PIMA were used, they are plotted as open circles.}
\figsetgrpend

\figsetgrpstart
\figsetgrpnum{3.10}
\figsetgrptitle{J0017+8135 at 8.6 GHz}
\figsetplot{J0017+8135_X_2006_02_16_avp_uvs_rad.pdf}
\figsetgrpnote{Correlated flux density averaged over time and IFs versus the $uv$ radius. The source name, the observation date, and the frequency are given at the top of the panel. In the cases when the data for a source were calibrated in AIPS and then underwent hybrid imaging in Difmap with both amplitude and phase self-calibration, they are plotted as filled circles. In the cases when the processing was the same except no amplitude self-calibration was made, the data are plotted as filled triangles. In the cases when no self-calibration was made for the source and the data calibrated in PIMA were used, they are plotted as open circles.}
\figsetgrpend

\figsetgrpstart
\figsetgrpnum{3.11}
\figsetgrptitle{J0017+8135 at 8.6 GHz}
\figsetplot{J0017+8135_X_2006_02_23_avp_uvs_rad.pdf}
\figsetgrpnote{Correlated flux density averaged over time and IFs versus the $uv$ radius. The source name, the observation date, and the frequency are given at the top of the panel. In the cases when the data for a source were calibrated in AIPS and then underwent hybrid imaging in Difmap with both amplitude and phase self-calibration, they are plotted as filled circles. In the cases when the processing was the same except no amplitude self-calibration was made, the data are plotted as filled triangles. In the cases when no self-calibration was made for the source and the data calibrated in PIMA were used, they are plotted as open circles.}
\figsetgrpend

\figsetgrpstart
\figsetgrpnum{3.12}
\figsetgrptitle{J0038+8447 at 2.3 GHz}
\figsetplot{J0038+8447_S_2006_02_16_avp_uvs_rad.pdf}
\figsetgrpnote{Correlated flux density averaged over time and IFs versus the $uv$ radius. The source name, the observation date, and the frequency are given at the top of the panel. In the cases when the data for a source were calibrated in AIPS and then underwent hybrid imaging in Difmap with both amplitude and phase self-calibration, they are plotted as filled circles. In the cases when the processing was the same except no amplitude self-calibration was made, the data are plotted as filled triangles. In the cases when no self-calibration was made for the source and the data calibrated in PIMA were used, they are plotted as open circles.}
\figsetgrpend

\figsetgrpstart
\figsetgrpnum{3.13}
\figsetgrptitle{J0038+8447 at 8.6 GHz}
\figsetplot{J0038+8447_X_2006_02_16_pet_uvt_rad.pdf}
\figsetgrpnote{Correlated flux density averaged over time and IFs versus the $uv$ radius. The source name, the observation date, and the frequency are given at the top of the panel. In the cases when the data for a source were calibrated in AIPS and then underwent hybrid imaging in Difmap with both amplitude and phase self-calibration, they are plotted as filled circles. In the cases when the processing was the same except no amplitude self-calibration was made, the data are plotted as filled triangles. In the cases when no self-calibration was made for the source and the data calibrated in PIMA were used, they are plotted as open circles.}
\figsetgrpend

\figsetgrpstart
\figsetgrpnum{3.14}
\figsetgrptitle{J0041+8114 at 8.6 GHz}
\figsetplot{J0041+8114_X_2006_02_16_pet_uvt_rad.pdf}
\figsetgrpnote{Correlated flux density averaged over time and IFs versus the $uv$ radius. The source name, the observation date, and the frequency are given at the top of the panel. In the cases when the data for a source were calibrated in AIPS and then underwent hybrid imaging in Difmap with both amplitude and phase self-calibration, they are plotted as filled circles. In the cases when the processing was the same except no amplitude self-calibration was made, the data are plotted as filled triangles. In the cases when no self-calibration was made for the source and the data calibrated in PIMA were used, they are plotted as open circles.}
\figsetgrpend

\figsetgrpstart
\figsetgrpnum{3.15}
\figsetgrptitle{J0117+8928 at 2.3 GHz}
\figsetplot{J0117+8928_S_2006_02_16_pet_uvt_rad.pdf}
\figsetgrpnote{Correlated flux density averaged over time and IFs versus the $uv$ radius. The source name, the observation date, and the frequency are given at the top of the panel. In the cases when the data for a source were calibrated in AIPS and then underwent hybrid imaging in Difmap with both amplitude and phase self-calibration, they are plotted as filled circles. In the cases when the processing was the same except no amplitude self-calibration was made, the data are plotted as filled triangles. In the cases when no self-calibration was made for the source and the data calibrated in PIMA were used, they are plotted as open circles.}
\figsetgrpend

\figsetgrpstart
\figsetgrpnum{3.16}
\figsetgrptitle{J0117+8928 at 8.6 GHz}
\figsetplot{J0117+8928_X_2006_02_16_pet_uvt_rad.pdf}
\figsetgrpnote{Correlated flux density averaged over time and IFs versus the $uv$ radius. The source name, the observation date, and the frequency are given at the top of the panel. In the cases when the data for a source were calibrated in AIPS and then underwent hybrid imaging in Difmap with both amplitude and phase self-calibration, they are plotted as filled circles. In the cases when the processing was the same except no amplitude self-calibration was made, the data are plotted as filled triangles. In the cases when no self-calibration was made for the source and the data calibrated in PIMA were used, they are plotted as open circles.}
\figsetgrpend

\figsetgrpstart
\figsetgrpnum{3.17}
\figsetgrptitle{J0125+8424 at 2.3 GHz}
\figsetplot{J0125+8424_S_2006_02_16_pet_uvt_rad.pdf}
\figsetgrpnote{Correlated flux density averaged over time and IFs versus the $uv$ radius. The source name, the observation date, and the frequency are given at the top of the panel. In the cases when the data for a source were calibrated in AIPS and then underwent hybrid imaging in Difmap with both amplitude and phase self-calibration, they are plotted as filled circles. In the cases when the processing was the same except no amplitude self-calibration was made, the data are plotted as filled triangles. In the cases when no self-calibration was made for the source and the data calibrated in PIMA were used, they are plotted as open circles.}
\figsetgrpend

\figsetgrpstart
\figsetgrpnum{3.18}
\figsetgrptitle{J0131+8446 at 2.3 GHz}
\figsetplot{J0131+8446_S_2006_02_23_pet_uvt_rad.pdf}
\figsetgrpnote{Correlated flux density averaged over time and IFs versus the $uv$ radius. The source name, the observation date, and the frequency are given at the top of the panel. In the cases when the data for a source were calibrated in AIPS and then underwent hybrid imaging in Difmap with both amplitude and phase self-calibration, they are plotted as filled circles. In the cases when the processing was the same except no amplitude self-calibration was made, the data are plotted as filled triangles. In the cases when no self-calibration was made for the source and the data calibrated in PIMA were used, they are plotted as open circles.}
\figsetgrpend

\figsetgrpstart
\figsetgrpnum{3.19}
\figsetgrptitle{J0157+7552 at 2.3 GHz}
\figsetplot{J0157+7552_S_2006_02_14_avp_uvs_rad.pdf}
\figsetgrpnote{Correlated flux density averaged over time and IFs versus the $uv$ radius. The source name, the observation date, and the frequency are given at the top of the panel. In the cases when the data for a source were calibrated in AIPS and then underwent hybrid imaging in Difmap with both amplitude and phase self-calibration, they are plotted as filled circles. In the cases when the processing was the same except no amplitude self-calibration was made, the data are plotted as filled triangles. In the cases when no self-calibration was made for the source and the data calibrated in PIMA were used, they are plotted as open circles.}
\figsetgrpend

\figsetgrpstart
\figsetgrpnum{3.20}
\figsetgrptitle{J0202+8115 at 2.3 GHz}
\figsetplot{J0202+8115_S_2006_02_16_pet_uvt_rad.pdf}
\figsetgrpnote{Correlated flux density averaged over time and IFs versus the $uv$ radius. The source name, the observation date, and the frequency are given at the top of the panel. In the cases when the data for a source were calibrated in AIPS and then underwent hybrid imaging in Difmap with both amplitude and phase self-calibration, they are plotted as filled circles. In the cases when the processing was the same except no amplitude self-calibration was made, the data are plotted as filled triangles. In the cases when no self-calibration was made for the source and the data calibrated in PIMA were used, they are plotted as open circles.}
\figsetgrpend

\figsetgrpstart
\figsetgrpnum{3.21}
\figsetgrptitle{J0202+8115 at 8.6 GHz}
\figsetplot{J0202+8115_X_2006_02_16_pet_uvt_rad.pdf}
\figsetgrpnote{Correlated flux density averaged over time and IFs versus the $uv$ radius. The source name, the observation date, and the frequency are given at the top of the panel. In the cases when the data for a source were calibrated in AIPS and then underwent hybrid imaging in Difmap with both amplitude and phase self-calibration, they are plotted as filled circles. In the cases when the processing was the same except no amplitude self-calibration was made, the data are plotted as filled triangles. In the cases when no self-calibration was made for the source and the data calibrated in PIMA were used, they are plotted as open circles.}
\figsetgrpend

\figsetgrpstart
\figsetgrpnum{3.22}
\figsetgrptitle{J0203+8106 at 2.3 GHz}
\figsetplot{J0203+8106_S_2006_02_14_avp_uvs_rad.pdf}
\figsetgrpnote{Correlated flux density averaged over time and IFs versus the $uv$ radius. The source name, the observation date, and the frequency are given at the top of the panel. In the cases when the data for a source were calibrated in AIPS and then underwent hybrid imaging in Difmap with both amplitude and phase self-calibration, they are plotted as filled circles. In the cases when the processing was the same except no amplitude self-calibration was made, the data are plotted as filled triangles. In the cases when no self-calibration was made for the source and the data calibrated in PIMA were used, they are plotted as open circles.}
\figsetgrpend

\figsetgrpstart
\figsetgrpnum{3.23}
\figsetgrptitle{J0203+8106 at 8.6 GHz}
\figsetplot{J0203+8106_X_2006_02_14_avp_uvs_rad.pdf}
\figsetgrpnote{Correlated flux density averaged over time and IFs versus the $uv$ radius. The source name, the observation date, and the frequency are given at the top of the panel. In the cases when the data for a source were calibrated in AIPS and then underwent hybrid imaging in Difmap with both amplitude and phase self-calibration, they are plotted as filled circles. In the cases when the processing was the same except no amplitude self-calibration was made, the data are plotted as filled triangles. In the cases when no self-calibration was made for the source and the data calibrated in PIMA were used, they are plotted as open circles.}
\figsetgrpend

\figsetgrpstart
\figsetgrpnum{3.24}
\figsetgrptitle{J0205+7522 at 2.3 GHz}
\figsetplot{J0205+7522_S_2006_02_16_avp_uvs_rad.pdf}
\figsetgrpnote{Correlated flux density averaged over time and IFs versus the $uv$ radius. The source name, the observation date, and the frequency are given at the top of the panel. In the cases when the data for a source were calibrated in AIPS and then underwent hybrid imaging in Difmap with both amplitude and phase self-calibration, they are plotted as filled circles. In the cases when the processing was the same except no amplitude self-calibration was made, the data are plotted as filled triangles. In the cases when no self-calibration was made for the source and the data calibrated in PIMA were used, they are plotted as open circles.}
\figsetgrpend

\figsetgrpstart
\figsetgrpnum{3.25}
\figsetgrptitle{J0205+7522 at 8.6 GHz}
\figsetplot{J0205+7522_X_2006_02_16_avp_uvs_rad.pdf}
\figsetgrpnote{Correlated flux density averaged over time and IFs versus the $uv$ radius. The source name, the observation date, and the frequency are given at the top of the panel. In the cases when the data for a source were calibrated in AIPS and then underwent hybrid imaging in Difmap with both amplitude and phase self-calibration, they are plotted as filled circles. In the cases when the processing was the same except no amplitude self-calibration was made, the data are plotted as filled triangles. In the cases when no self-calibration was made for the source and the data calibrated in PIMA were used, they are plotted as open circles.}
\figsetgrpend

\figsetgrpstart
\figsetgrpnum{3.26}
\figsetgrptitle{J0207+7956 at 2.3 GHz}
\figsetplot{J0207+7956_S_2006_02_14_pet_uvt_rad.pdf}
\figsetgrpnote{Correlated flux density averaged over time and IFs versus the $uv$ radius. The source name, the observation date, and the frequency are given at the top of the panel. In the cases when the data for a source were calibrated in AIPS and then underwent hybrid imaging in Difmap with both amplitude and phase self-calibration, they are plotted as filled circles. In the cases when the processing was the same except no amplitude self-calibration was made, the data are plotted as filled triangles. In the cases when no self-calibration was made for the source and the data calibrated in PIMA were used, they are plotted as open circles.}
\figsetgrpend

\figsetgrpstart
\figsetgrpnum{3.27}
\figsetgrptitle{J0224+7655 at 2.3 GHz}
\figsetplot{J0224+7655_S_2006_02_16_pet_uvt_rad.pdf}
\figsetgrpnote{Correlated flux density averaged over time and IFs versus the $uv$ radius. The source name, the observation date, and the frequency are given at the top of the panel. In the cases when the data for a source were calibrated in AIPS and then underwent hybrid imaging in Difmap with both amplitude and phase self-calibration, they are plotted as filled circles. In the cases when the processing was the same except no amplitude self-calibration was made, the data are plotted as filled triangles. In the cases when no self-calibration was made for the source and the data calibrated in PIMA were used, they are plotted as open circles.}
\figsetgrpend

\figsetgrpstart
\figsetgrpnum{3.28}
\figsetgrptitle{J0229+7743 at 2.3 GHz}
\figsetplot{J0229+7743_S_2006_02_23_pet_uvt_rad.pdf}
\figsetgrpnote{Correlated flux density averaged over time and IFs versus the $uv$ radius. The source name, the observation date, and the frequency are given at the top of the panel. In the cases when the data for a source were calibrated in AIPS and then underwent hybrid imaging in Difmap with both amplitude and phase self-calibration, they are plotted as filled circles. In the cases when the processing was the same except no amplitude self-calibration was made, the data are plotted as filled triangles. In the cases when no self-calibration was made for the source and the data calibrated in PIMA were used, they are plotted as open circles.}
\figsetgrpend

\figsetgrpstart
\figsetgrpnum{3.29}
\figsetgrptitle{J0230+8141 at 2.3 GHz}
\figsetplot{J0230+8141_S_2006_02_14_pet_uvt_rad.pdf}
\figsetgrpnote{Correlated flux density averaged over time and IFs versus the $uv$ radius. The source name, the observation date, and the frequency are given at the top of the panel. In the cases when the data for a source were calibrated in AIPS and then underwent hybrid imaging in Difmap with both amplitude and phase self-calibration, they are plotted as filled circles. In the cases when the processing was the same except no amplitude self-calibration was made, the data are plotted as filled triangles. In the cases when no self-calibration was made for the source and the data calibrated in PIMA were used, they are plotted as open circles.}
\figsetgrpend

\figsetgrpstart
\figsetgrpnum{3.30}
\figsetgrptitle{J0230+8141 at 8.6 GHz}
\figsetplot{J0230+8141_X_2006_02_14_pet_uvt_rad.pdf}
\figsetgrpnote{Correlated flux density averaged over time and IFs versus the $uv$ radius. The source name, the observation date, and the frequency are given at the top of the panel. In the cases when the data for a source were calibrated in AIPS and then underwent hybrid imaging in Difmap with both amplitude and phase self-calibration, they are plotted as filled circles. In the cases when the processing was the same except no amplitude self-calibration was made, the data are plotted as filled triangles. In the cases when no self-calibration was made for the source and the data calibrated in PIMA were used, they are plotted as open circles.}
\figsetgrpend

\figsetgrpstart
\figsetgrpnum{3.31}
\figsetgrptitle{J0232+7825 at 2.3 GHz}
\figsetplot{J0232+7825_S_2006_02_16_pet_uvt_rad.pdf}
\figsetgrpnote{Correlated flux density averaged over time and IFs versus the $uv$ radius. The source name, the observation date, and the frequency are given at the top of the panel. In the cases when the data for a source were calibrated in AIPS and then underwent hybrid imaging in Difmap with both amplitude and phase self-calibration, they are plotted as filled circles. In the cases when the processing was the same except no amplitude self-calibration was made, the data are plotted as filled triangles. In the cases when no self-calibration was made for the source and the data calibrated in PIMA were used, they are plotted as open circles.}
\figsetgrpend

\figsetgrpstart
\figsetgrpnum{3.32}
\figsetgrptitle{J0257+7843 at 2.3 GHz}
\figsetplot{J0257+7843_S_2006_02_16_avp_uvs_rad.pdf}
\figsetgrpnote{Correlated flux density averaged over time and IFs versus the $uv$ radius. The source name, the observation date, and the frequency are given at the top of the panel. In the cases when the data for a source were calibrated in AIPS and then underwent hybrid imaging in Difmap with both amplitude and phase self-calibration, they are plotted as filled circles. In the cases when the processing was the same except no amplitude self-calibration was made, the data are plotted as filled triangles. In the cases when no self-calibration was made for the source and the data calibrated in PIMA were used, they are plotted as open circles.}
\figsetgrpend

\figsetgrpstart
\figsetgrpnum{3.33}
\figsetgrptitle{J0257+7843 at 8.6 GHz}
\figsetplot{J0257+7843_X_2006_02_16_avp_uvs_rad.pdf}
\figsetgrpnote{Correlated flux density averaged over time and IFs versus the $uv$ radius. The source name, the observation date, and the frequency are given at the top of the panel. In the cases when the data for a source were calibrated in AIPS and then underwent hybrid imaging in Difmap with both amplitude and phase self-calibration, they are plotted as filled circles. In the cases when the processing was the same except no amplitude self-calibration was made, the data are plotted as filled triangles. In the cases when no self-calibration was made for the source and the data calibrated in PIMA were used, they are plotted as open circles.}
\figsetgrpend

\figsetgrpstart
\figsetgrpnum{3.34}
\figsetgrptitle{J0300+8202 at 2.3 GHz}
\figsetplot{J0300+8202_S_2006_02_16_avp_uvs_rad.pdf}
\figsetgrpnote{Correlated flux density averaged over time and IFs versus the $uv$ radius. The source name, the observation date, and the frequency are given at the top of the panel. In the cases when the data for a source were calibrated in AIPS and then underwent hybrid imaging in Difmap with both amplitude and phase self-calibration, they are plotted as filled circles. In the cases when the processing was the same except no amplitude self-calibration was made, the data are plotted as filled triangles. In the cases when no self-calibration was made for the source and the data calibrated in PIMA were used, they are plotted as open circles.}
\figsetgrpend

\figsetgrpstart
\figsetgrpnum{3.35}
\figsetgrptitle{J0300+8202 at 8.6 GHz}
\figsetplot{J0300+8202_X_2006_02_16_pet_uvt_rad.pdf}
\figsetgrpnote{Correlated flux density averaged over time and IFs versus the $uv$ radius. The source name, the observation date, and the frequency are given at the top of the panel. In the cases when the data for a source were calibrated in AIPS and then underwent hybrid imaging in Difmap with both amplitude and phase self-calibration, they are plotted as filled circles. In the cases when the processing was the same except no amplitude self-calibration was made, the data are plotted as filled triangles. In the cases when no self-calibration was made for the source and the data calibrated in PIMA were used, they are plotted as open circles.}
\figsetgrpend

\figsetgrpstart
\figsetgrpnum{3.36}
\figsetgrptitle{J0304+7727 at 2.3 GHz}
\figsetplot{J0304+7727_S_2006_02_14_avp_uvs_rad.pdf}
\figsetgrpnote{Correlated flux density averaged over time and IFs versus the $uv$ radius. The source name, the observation date, and the frequency are given at the top of the panel. In the cases when the data for a source were calibrated in AIPS and then underwent hybrid imaging in Difmap with both amplitude and phase self-calibration, they are plotted as filled circles. In the cases when the processing was the same except no amplitude self-calibration was made, the data are plotted as filled triangles. In the cases when no self-calibration was made for the source and the data calibrated in PIMA were used, they are plotted as open circles.}
\figsetgrpend

\figsetgrpstart
\figsetgrpnum{3.37}
\figsetgrptitle{J0304+7727 at 8.6 GHz}
\figsetplot{J0304+7727_X_2006_02_14_avp_uvs_rad.pdf}
\figsetgrpnote{Correlated flux density averaged over time and IFs versus the $uv$ radius. The source name, the observation date, and the frequency are given at the top of the panel. In the cases when the data for a source were calibrated in AIPS and then underwent hybrid imaging in Difmap with both amplitude and phase self-calibration, they are plotted as filled circles. In the cases when the processing was the same except no amplitude self-calibration was made, the data are plotted as filled triangles. In the cases when no self-calibration was made for the source and the data calibrated in PIMA were used, they are plotted as open circles.}
\figsetgrpend

\figsetgrpstart
\figsetgrpnum{3.38}
\figsetgrptitle{J0306+8200 at 2.3 GHz}
\figsetplot{J0306+8200_S_2006_02_23_pet_uvt_rad.pdf}
\figsetgrpnote{Correlated flux density averaged over time and IFs versus the $uv$ radius. The source name, the observation date, and the frequency are given at the top of the panel. In the cases when the data for a source were calibrated in AIPS and then underwent hybrid imaging in Difmap with both amplitude and phase self-calibration, they are plotted as filled circles. In the cases when the processing was the same except no amplitude self-calibration was made, the data are plotted as filled triangles. In the cases when no self-calibration was made for the source and the data calibrated in PIMA were used, they are plotted as open circles.}
\figsetgrpend

\figsetgrpstart
\figsetgrpnum{3.39}
\figsetgrptitle{J0306+8200 at 8.6 GHz}
\figsetplot{J0306+8200_X_2006_02_23_pet_uvt_rad.pdf}
\figsetgrpnote{Correlated flux density averaged over time and IFs versus the $uv$ radius. The source name, the observation date, and the frequency are given at the top of the panel. In the cases when the data for a source were calibrated in AIPS and then underwent hybrid imaging in Difmap with both amplitude and phase self-calibration, they are plotted as filled circles. In the cases when the processing was the same except no amplitude self-calibration was made, the data are plotted as filled triangles. In the cases when no self-calibration was made for the source and the data calibrated in PIMA were used, they are plotted as open circles.}
\figsetgrpend

\figsetgrpstart
\figsetgrpnum{3.40}
\figsetgrptitle{J0330+7633 at 8.6 GHz}
\figsetplot{J0330+7633_X_2006_02_23_pet_uvt_rad.pdf}
\figsetgrpnote{Correlated flux density averaged over time and IFs versus the $uv$ radius. The source name, the observation date, and the frequency are given at the top of the panel. In the cases when the data for a source were calibrated in AIPS and then underwent hybrid imaging in Difmap with both amplitude and phase self-calibration, they are plotted as filled circles. In the cases when the processing was the same except no amplitude self-calibration was made, the data are plotted as filled triangles. In the cases when no self-calibration was made for the source and the data calibrated in PIMA were used, they are plotted as open circles.}
\figsetgrpend

\figsetgrpstart
\figsetgrpnum{3.41}
\figsetgrptitle{J0354+8009 at 2.3 GHz}
\figsetplot{J0354+8009_S_2006_02_23_avp_uvs_rad.pdf}
\figsetgrpnote{Correlated flux density averaged over time and IFs versus the $uv$ radius. The source name, the observation date, and the frequency are given at the top of the panel. In the cases when the data for a source were calibrated in AIPS and then underwent hybrid imaging in Difmap with both amplitude and phase self-calibration, they are plotted as filled circles. In the cases when the processing was the same except no amplitude self-calibration was made, the data are plotted as filled triangles. In the cases when no self-calibration was made for the source and the data calibrated in PIMA were used, they are plotted as open circles.}
\figsetgrpend

\figsetgrpstart
\figsetgrpnum{3.42}
\figsetgrptitle{J0354+8009 at 8.6 GHz}
\figsetplot{J0354+8009_X_2006_02_23_avp_uvs_rad.pdf}
\figsetgrpnote{Correlated flux density averaged over time and IFs versus the $uv$ radius. The source name, the observation date, and the frequency are given at the top of the panel. In the cases when the data for a source were calibrated in AIPS and then underwent hybrid imaging in Difmap with both amplitude and phase self-calibration, they are plotted as filled circles. In the cases when the processing was the same except no amplitude self-calibration was made, the data are plotted as filled triangles. In the cases when no self-calibration was made for the source and the data calibrated in PIMA were used, they are plotted as open circles.}
\figsetgrpend

\figsetgrpstart
\figsetgrpnum{3.43}
\figsetgrptitle{J0402+8241 at 2.3 GHz}
\figsetplot{J0402+8241_S_2006_02_14_pet_uvt_rad.pdf}
\figsetgrpnote{Correlated flux density averaged over time and IFs versus the $uv$ radius. The source name, the observation date, and the frequency are given at the top of the panel. In the cases when the data for a source were calibrated in AIPS and then underwent hybrid imaging in Difmap with both amplitude and phase self-calibration, they are plotted as filled circles. In the cases when the processing was the same except no amplitude self-calibration was made, the data are plotted as filled triangles. In the cases when no self-calibration was made for the source and the data calibrated in PIMA were used, they are plotted as open circles.}
\figsetgrpend

\figsetgrpstart
\figsetgrpnum{3.44}
\figsetgrptitle{J0402+8241 at 8.6 GHz}
\figsetplot{J0402+8241_X_2006_02_14_pet_uvt_rad.pdf}
\figsetgrpnote{Correlated flux density averaged over time and IFs versus the $uv$ radius. The source name, the observation date, and the frequency are given at the top of the panel. In the cases when the data for a source were calibrated in AIPS and then underwent hybrid imaging in Difmap with both amplitude and phase self-calibration, they are plotted as filled circles. In the cases when the processing was the same except no amplitude self-calibration was made, the data are plotted as filled triangles. In the cases when no self-calibration was made for the source and the data calibrated in PIMA were used, they are plotted as open circles.}
\figsetgrpend

\figsetgrpstart
\figsetgrpnum{3.45}
\figsetgrptitle{J0410+7656 at 2.3 GHz}
\figsetplot{J0410+7656_S_2006_02_23_avp_uvs_rad.pdf}
\figsetgrpnote{Correlated flux density averaged over time and IFs versus the $uv$ radius. The source name, the observation date, and the frequency are given at the top of the panel. In the cases when the data for a source were calibrated in AIPS and then underwent hybrid imaging in Difmap with both amplitude and phase self-calibration, they are plotted as filled circles. In the cases when the processing was the same except no amplitude self-calibration was made, the data are plotted as filled triangles. In the cases when no self-calibration was made for the source and the data calibrated in PIMA were used, they are plotted as open circles.}
\figsetgrpend

\figsetgrpstart
\figsetgrpnum{3.46}
\figsetgrptitle{J0410+7656 at 8.6 GHz}
\figsetplot{J0410+7656_X_2006_02_23_avp_uvs_rad.pdf}
\figsetgrpnote{Correlated flux density averaged over time and IFs versus the $uv$ radius. The source name, the observation date, and the frequency are given at the top of the panel. In the cases when the data for a source were calibrated in AIPS and then underwent hybrid imaging in Difmap with both amplitude and phase self-calibration, they are plotted as filled circles. In the cases when the processing was the same except no amplitude self-calibration was made, the data are plotted as filled triangles. In the cases when no self-calibration was made for the source and the data calibrated in PIMA were used, they are plotted as open circles.}
\figsetgrpend

\figsetgrpstart
\figsetgrpnum{3.47}
\figsetgrptitle{J0410+8208 at 2.3 GHz}
\figsetplot{J0410+8208_S_2006_02_16_avp_uvs_rad.pdf}
\figsetgrpnote{Correlated flux density averaged over time and IFs versus the $uv$ radius. The source name, the observation date, and the frequency are given at the top of the panel. In the cases when the data for a source were calibrated in AIPS and then underwent hybrid imaging in Difmap with both amplitude and phase self-calibration, they are plotted as filled circles. In the cases when the processing was the same except no amplitude self-calibration was made, the data are plotted as filled triangles. In the cases when no self-calibration was made for the source and the data calibrated in PIMA were used, they are plotted as open circles.}
\figsetgrpend

\figsetgrpstart
\figsetgrpnum{3.48}
\figsetgrptitle{J0410+8208 at 8.6 GHz}
\figsetplot{J0410+8208_X_2006_02_16_pet_uvt_rad.pdf}
\figsetgrpnote{Correlated flux density averaged over time and IFs versus the $uv$ radius. The source name, the observation date, and the frequency are given at the top of the panel. In the cases when the data for a source were calibrated in AIPS and then underwent hybrid imaging in Difmap with both amplitude and phase self-calibration, they are plotted as filled circles. In the cases when the processing was the same except no amplitude self-calibration was made, the data are plotted as filled triangles. In the cases when no self-calibration was made for the source and the data calibrated in PIMA were used, they are plotted as open circles.}
\figsetgrpend

\figsetgrpstart
\figsetgrpnum{3.49}
\figsetgrptitle{J0415+7753 at 2.3 GHz}
\figsetplot{J0415+7753_S_2006_02_16_pet_uvt_rad.pdf}
\figsetgrpnote{Correlated flux density averaged over time and IFs versus the $uv$ radius. The source name, the observation date, and the frequency are given at the top of the panel. In the cases when the data for a source were calibrated in AIPS and then underwent hybrid imaging in Difmap with both amplitude and phase self-calibration, they are plotted as filled circles. In the cases when the processing was the same except no amplitude self-calibration was made, the data are plotted as filled triangles. In the cases when no self-calibration was made for the source and the data calibrated in PIMA were used, they are plotted as open circles.}
\figsetgrpend

\figsetgrpstart
\figsetgrpnum{3.50}
\figsetgrptitle{J0415+7753 at 8.6 GHz}
\figsetplot{J0415+7753_X_2006_02_16_pet_uvt_rad.pdf}
\figsetgrpnote{Correlated flux density averaged over time and IFs versus the $uv$ radius. The source name, the observation date, and the frequency are given at the top of the panel. In the cases when the data for a source were calibrated in AIPS and then underwent hybrid imaging in Difmap with both amplitude and phase self-calibration, they are plotted as filled circles. In the cases when the processing was the same except no amplitude self-calibration was made, the data are plotted as filled triangles. In the cases when no self-calibration was made for the source and the data calibrated in PIMA were used, they are plotted as open circles.}
\figsetgrpend

\figsetgrpstart
\figsetgrpnum{3.51}
\figsetgrptitle{J0445+7838 at 2.3 GHz}
\figsetplot{J0445+7838_S_2006_02_16_pet_uvt_rad.pdf}
\figsetgrpnote{Correlated flux density averaged over time and IFs versus the $uv$ radius. The source name, the observation date, and the frequency are given at the top of the panel. In the cases when the data for a source were calibrated in AIPS and then underwent hybrid imaging in Difmap with both amplitude and phase self-calibration, they are plotted as filled circles. In the cases when the processing was the same except no amplitude self-calibration was made, the data are plotted as filled triangles. In the cases when no self-calibration was made for the source and the data calibrated in PIMA were used, they are plotted as open circles.}
\figsetgrpend

\figsetgrpstart
\figsetgrpnum{3.52}
\figsetgrptitle{J0445+7838 at 8.6 GHz}
\figsetplot{J0445+7838_X_2006_02_16_pet_uvt_rad.pdf}
\figsetgrpnote{Correlated flux density averaged over time and IFs versus the $uv$ radius. The source name, the observation date, and the frequency are given at the top of the panel. In the cases when the data for a source were calibrated in AIPS and then underwent hybrid imaging in Difmap with both amplitude and phase self-calibration, they are plotted as filled circles. In the cases when the processing was the same except no amplitude self-calibration was made, the data are plotted as filled triangles. In the cases when no self-calibration was made for the source and the data calibrated in PIMA were used, they are plotted as open circles.}
\figsetgrpend

\figsetgrpstart
\figsetgrpnum{3.53}
\figsetgrptitle{J0449+8233 at 2.3 GHz}
\figsetplot{J0449+8233_S_2006_02_23_avp_uvs_rad.pdf}
\figsetgrpnote{Correlated flux density averaged over time and IFs versus the $uv$ radius. The source name, the observation date, and the frequency are given at the top of the panel. In the cases when the data for a source were calibrated in AIPS and then underwent hybrid imaging in Difmap with both amplitude and phase self-calibration, they are plotted as filled circles. In the cases when the processing was the same except no amplitude self-calibration was made, the data are plotted as filled triangles. In the cases when no self-calibration was made for the source and the data calibrated in PIMA were used, they are plotted as open circles.}
\figsetgrpend

\figsetgrpstart
\figsetgrpnum{3.54}
\figsetgrptitle{J0449+8233 at 8.6 GHz}
\figsetplot{J0449+8233_X_2006_02_23_avp_uvs_rad.pdf}
\figsetgrpnote{Correlated flux density averaged over time and IFs versus the $uv$ radius. The source name, the observation date, and the frequency are given at the top of the panel. In the cases when the data for a source were calibrated in AIPS and then underwent hybrid imaging in Difmap with both amplitude and phase self-calibration, they are plotted as filled circles. In the cases when the processing was the same except no amplitude self-calibration was made, the data are plotted as filled triangles. In the cases when no self-calibration was made for the source and the data calibrated in PIMA were used, they are plotted as open circles.}
\figsetgrpend

\figsetgrpstart
\figsetgrpnum{3.55}
\figsetgrptitle{J0458+7615 at 2.3 GHz}
\figsetplot{J0458+7615_S_2006_02_16_pet_uvt_rad.pdf}
\figsetgrpnote{Correlated flux density averaged over time and IFs versus the $uv$ radius. The source name, the observation date, and the frequency are given at the top of the panel. In the cases when the data for a source were calibrated in AIPS and then underwent hybrid imaging in Difmap with both amplitude and phase self-calibration, they are plotted as filled circles. In the cases when the processing was the same except no amplitude self-calibration was made, the data are plotted as filled triangles. In the cases when no self-calibration was made for the source and the data calibrated in PIMA were used, they are plotted as open circles.}
\figsetgrpend

\figsetgrpstart
\figsetgrpnum{3.56}
\figsetgrptitle{J0507+7912 at 2.3 GHz}
\figsetplot{J0507+7912_S_2006_02_23_pet_uvt_rad.pdf}
\figsetgrpnote{Correlated flux density averaged over time and IFs versus the $uv$ radius. The source name, the observation date, and the frequency are given at the top of the panel. In the cases when the data for a source were calibrated in AIPS and then underwent hybrid imaging in Difmap with both amplitude and phase self-calibration, they are plotted as filled circles. In the cases when the processing was the same except no amplitude self-calibration was made, the data are plotted as filled triangles. In the cases when no self-calibration was made for the source and the data calibrated in PIMA were used, they are plotted as open circles.}
\figsetgrpend

\figsetgrpstart
\figsetgrpnum{3.57}
\figsetgrptitle{J0507+7912 at 8.6 GHz}
\figsetplot{J0507+7912_X_2006_02_23_pet_uvt_rad.pdf}
\figsetgrpnote{Correlated flux density averaged over time and IFs versus the $uv$ radius. The source name, the observation date, and the frequency are given at the top of the panel. In the cases when the data for a source were calibrated in AIPS and then underwent hybrid imaging in Difmap with both amplitude and phase self-calibration, they are plotted as filled circles. In the cases when the processing was the same except no amplitude self-calibration was made, the data are plotted as filled triangles. In the cases when no self-calibration was made for the source and the data calibrated in PIMA were used, they are plotted as open circles.}
\figsetgrpend

\figsetgrpstart
\figsetgrpnum{3.58}
\figsetgrptitle{J0508+8432 at 2.3 GHz}
\figsetplot{J0508+8432_S_2006_02_14_avp_uvs_rad.pdf}
\figsetgrpnote{Correlated flux density averaged over time and IFs versus the $uv$ radius. The source name, the observation date, and the frequency are given at the top of the panel. In the cases when the data for a source were calibrated in AIPS and then underwent hybrid imaging in Difmap with both amplitude and phase self-calibration, they are plotted as filled circles. In the cases when the processing was the same except no amplitude self-calibration was made, the data are plotted as filled triangles. In the cases when no self-calibration was made for the source and the data calibrated in PIMA were used, they are plotted as open circles.}
\figsetgrpend

\figsetgrpstart
\figsetgrpnum{3.59}
\figsetgrptitle{J0508+8432 at 8.6 GHz}
\figsetplot{J0508+8432_X_2006_02_14_avp_uvs_rad.pdf}
\figsetgrpnote{Correlated flux density averaged over time and IFs versus the $uv$ radius. The source name, the observation date, and the frequency are given at the top of the panel. In the cases when the data for a source were calibrated in AIPS and then underwent hybrid imaging in Difmap with both amplitude and phase self-calibration, they are plotted as filled circles. In the cases when the processing was the same except no amplitude self-calibration was made, the data are plotted as filled triangles. In the cases when no self-calibration was made for the source and the data calibrated in PIMA were used, they are plotted as open circles.}
\figsetgrpend

\figsetgrpstart
\figsetgrpnum{3.60}
\figsetgrptitle{J0525+8737 at 2.3 GHz}
\figsetplot{J0525+8737_S_2006_02_16_pet_uvt_rad.pdf}
\figsetgrpnote{Correlated flux density averaged over time and IFs versus the $uv$ radius. The source name, the observation date, and the frequency are given at the top of the panel. In the cases when the data for a source were calibrated in AIPS and then underwent hybrid imaging in Difmap with both amplitude and phase self-calibration, they are plotted as filled circles. In the cases when the processing was the same except no amplitude self-calibration was made, the data are plotted as filled triangles. In the cases when no self-calibration was made for the source and the data calibrated in PIMA were used, they are plotted as open circles.}
\figsetgrpend

\figsetgrpstart
\figsetgrpnum{3.61}
\figsetgrptitle{J0540+7519 at 2.3 GHz}
\figsetplot{J0540+7519_S_2006_02_16_pet_uvt_rad.pdf}
\figsetgrpnote{Correlated flux density averaged over time and IFs versus the $uv$ radius. The source name, the observation date, and the frequency are given at the top of the panel. In the cases when the data for a source were calibrated in AIPS and then underwent hybrid imaging in Difmap with both amplitude and phase self-calibration, they are plotted as filled circles. In the cases when the processing was the same except no amplitude self-calibration was made, the data are plotted as filled triangles. In the cases when no self-calibration was made for the source and the data calibrated in PIMA were used, they are plotted as open circles.}
\figsetgrpend

\figsetgrpstart
\figsetgrpnum{3.62}
\figsetgrptitle{J0543+8118 at 2.3 GHz}
\figsetplot{J0543+8118_S_2006_02_14_pet_uvt_rad.pdf}
\figsetgrpnote{Correlated flux density averaged over time and IFs versus the $uv$ radius. The source name, the observation date, and the frequency are given at the top of the panel. In the cases when the data for a source were calibrated in AIPS and then underwent hybrid imaging in Difmap with both amplitude and phase self-calibration, they are plotted as filled circles. In the cases when the processing was the same except no amplitude self-calibration was made, the data are plotted as filled triangles. In the cases when no self-calibration was made for the source and the data calibrated in PIMA were used, they are plotted as open circles.}
\figsetgrpend

\figsetgrpstart
\figsetgrpnum{3.63}
\figsetgrptitle{J0543+8238 at 2.3 GHz}
\figsetplot{J0543+8238_S_2006_02_23_avp_uvs_rad.pdf}
\figsetgrpnote{Correlated flux density averaged over time and IFs versus the $uv$ radius. The source name, the observation date, and the frequency are given at the top of the panel. In the cases when the data for a source were calibrated in AIPS and then underwent hybrid imaging in Difmap with both amplitude and phase self-calibration, they are plotted as filled circles. In the cases when the processing was the same except no amplitude self-calibration was made, the data are plotted as filled triangles. In the cases when no self-calibration was made for the source and the data calibrated in PIMA were used, they are plotted as open circles.}
\figsetgrpend

\figsetgrpstart
\figsetgrpnum{3.64}
\figsetgrptitle{J0543+8238 at 8.6 GHz}
\figsetplot{J0543+8238_X_2006_02_23_avp_uvs_rad.pdf}
\figsetgrpnote{Correlated flux density averaged over time and IFs versus the $uv$ radius. The source name, the observation date, and the frequency are given at the top of the panel. In the cases when the data for a source were calibrated in AIPS and then underwent hybrid imaging in Difmap with both amplitude and phase self-calibration, they are plotted as filled circles. In the cases when the processing was the same except no amplitude self-calibration was made, the data are plotted as filled triangles. In the cases when no self-calibration was made for the source and the data calibrated in PIMA were used, they are plotted as open circles.}
\figsetgrpend

\figsetgrpstart
\figsetgrpnum{3.65}
\figsetgrptitle{J0621+7605 at 2.3 GHz}
\figsetplot{J0621+7605_S_2006_02_16_avp_uvs_rad.pdf}
\figsetgrpnote{Correlated flux density averaged over time and IFs versus the $uv$ radius. The source name, the observation date, and the frequency are given at the top of the panel. In the cases when the data for a source were calibrated in AIPS and then underwent hybrid imaging in Difmap with both amplitude and phase self-calibration, they are plotted as filled circles. In the cases when the processing was the same except no amplitude self-calibration was made, the data are plotted as filled triangles. In the cases when no self-calibration was made for the source and the data calibrated in PIMA were used, they are plotted as open circles.}
\figsetgrpend

\figsetgrpstart
\figsetgrpnum{3.66}
\figsetgrptitle{J0621+7605 at 8.6 GHz}
\figsetplot{J0621+7605_X_2006_02_16_avp_uvs_rad.pdf}
\figsetgrpnote{Correlated flux density averaged over time and IFs versus the $uv$ radius. The source name, the observation date, and the frequency are given at the top of the panel. In the cases when the data for a source were calibrated in AIPS and then underwent hybrid imaging in Difmap with both amplitude and phase self-calibration, they are plotted as filled circles. In the cases when the processing was the same except no amplitude self-calibration was made, the data are plotted as filled triangles. In the cases when no self-calibration was made for the source and the data calibrated in PIMA were used, they are plotted as open circles.}
\figsetgrpend

\figsetgrpstart
\figsetgrpnum{3.67}
\figsetgrptitle{J0622+8719 at 2.3 GHz}
\figsetplot{J0622+8719_S_2006_02_23_pet_uvt_rad.pdf}
\figsetgrpnote{Correlated flux density averaged over time and IFs versus the $uv$ radius. The source name, the observation date, and the frequency are given at the top of the panel. In the cases when the data for a source were calibrated in AIPS and then underwent hybrid imaging in Difmap with both amplitude and phase self-calibration, they are plotted as filled circles. In the cases when the processing was the same except no amplitude self-calibration was made, the data are plotted as filled triangles. In the cases when no self-calibration was made for the source and the data calibrated in PIMA were used, they are plotted as open circles.}
\figsetgrpend

\figsetgrpstart
\figsetgrpnum{3.68}
\figsetgrptitle{J0626+8202 at 2.3 GHz}
\figsetplot{J0626+8202_S_2006_02_14_avp_uvs_rad.pdf}
\figsetgrpnote{Correlated flux density averaged over time and IFs versus the $uv$ radius. The source name, the observation date, and the frequency are given at the top of the panel. In the cases when the data for a source were calibrated in AIPS and then underwent hybrid imaging in Difmap with both amplitude and phase self-calibration, they are plotted as filled circles. In the cases when the processing was the same except no amplitude self-calibration was made, the data are plotted as filled triangles. In the cases when no self-calibration was made for the source and the data calibrated in PIMA were used, they are plotted as open circles.}
\figsetgrpend

\figsetgrpstart
\figsetgrpnum{3.69}
\figsetgrptitle{J0626+8202 at 8.6 GHz}
\figsetplot{J0626+8202_X_2006_02_14_avp_uvs_rad.pdf}
\figsetgrpnote{Correlated flux density averaged over time and IFs versus the $uv$ radius. The source name, the observation date, and the frequency are given at the top of the panel. In the cases when the data for a source were calibrated in AIPS and then underwent hybrid imaging in Difmap with both amplitude and phase self-calibration, they are plotted as filled circles. In the cases when the processing was the same except no amplitude self-calibration was made, the data are plotted as filled triangles. In the cases when no self-calibration was made for the source and the data calibrated in PIMA were used, they are plotted as open circles.}
\figsetgrpend

\figsetgrpstart
\figsetgrpnum{3.70}
\figsetgrptitle{J0629+8451 at 2.3 GHz}
\figsetplot{J0629+8451_S_2006_02_23_pet_uvt_rad.pdf}
\figsetgrpnote{Correlated flux density averaged over time and IFs versus the $uv$ radius. The source name, the observation date, and the frequency are given at the top of the panel. In the cases when the data for a source were calibrated in AIPS and then underwent hybrid imaging in Difmap with both amplitude and phase self-calibration, they are plotted as filled circles. In the cases when the processing was the same except no amplitude self-calibration was made, the data are plotted as filled triangles. In the cases when no self-calibration was made for the source and the data calibrated in PIMA were used, they are plotted as open circles.}
\figsetgrpend

\figsetgrpstart
\figsetgrpnum{3.71}
\figsetgrptitle{J0632+8020 at 2.3 GHz}
\figsetplot{J0632+8020_S_2006_02_14_avp_uvs_rad.pdf}
\figsetgrpnote{Correlated flux density averaged over time and IFs versus the $uv$ radius. The source name, the observation date, and the frequency are given at the top of the panel. In the cases when the data for a source were calibrated in AIPS and then underwent hybrid imaging in Difmap with both amplitude and phase self-calibration, they are plotted as filled circles. In the cases when the processing was the same except no amplitude self-calibration was made, the data are plotted as filled triangles. In the cases when no self-calibration was made for the source and the data calibrated in PIMA were used, they are plotted as open circles.}
\figsetgrpend

\figsetgrpstart
\figsetgrpnum{3.72}
\figsetgrptitle{J0632+8020 at 8.6 GHz}
\figsetplot{J0632+8020_X_2006_02_14_avp_uvs_rad.pdf}
\figsetgrpnote{Correlated flux density averaged over time and IFs versus the $uv$ radius. The source name, the observation date, and the frequency are given at the top of the panel. In the cases when the data for a source were calibrated in AIPS and then underwent hybrid imaging in Difmap with both amplitude and phase self-calibration, they are plotted as filled circles. In the cases when the processing was the same except no amplitude self-calibration was made, the data are plotted as filled triangles. In the cases when no self-calibration was made for the source and the data calibrated in PIMA were used, they are plotted as open circles.}
\figsetgrpend

\figsetgrpstart
\figsetgrpnum{3.73}
\figsetgrptitle{J0637+8125 at 2.3 GHz}
\figsetplot{J0637+8125_S_2006_02_23_avp_uvs_rad.pdf}
\figsetgrpnote{Correlated flux density averaged over time and IFs versus the $uv$ radius. The source name, the observation date, and the frequency are given at the top of the panel. In the cases when the data for a source were calibrated in AIPS and then underwent hybrid imaging in Difmap with both amplitude and phase self-calibration, they are plotted as filled circles. In the cases when the processing was the same except no amplitude self-calibration was made, the data are plotted as filled triangles. In the cases when no self-calibration was made for the source and the data calibrated in PIMA were used, they are plotted as open circles.}
\figsetgrpend

\figsetgrpstart
\figsetgrpnum{3.74}
\figsetgrptitle{J0637+8125 at 8.6 GHz}
\figsetplot{J0637+8125_X_2006_02_23_avp_uvs_rad.pdf}
\figsetgrpnote{Correlated flux density averaged over time and IFs versus the $uv$ radius. The source name, the observation date, and the frequency are given at the top of the panel. In the cases when the data for a source were calibrated in AIPS and then underwent hybrid imaging in Difmap with both amplitude and phase self-calibration, they are plotted as filled circles. In the cases when the processing was the same except no amplitude self-calibration was made, the data are plotted as filled triangles. In the cases when no self-calibration was made for the source and the data calibrated in PIMA were used, they are plotted as open circles.}
\figsetgrpend

\figsetgrpstart
\figsetgrpnum{3.75}
\figsetgrptitle{J0638+8411 at 2.3 GHz}
\figsetplot{J0638+8411_S_2006_02_14_pet_uvt_rad.pdf}
\figsetgrpnote{Correlated flux density averaged over time and IFs versus the $uv$ radius. The source name, the observation date, and the frequency are given at the top of the panel. In the cases when the data for a source were calibrated in AIPS and then underwent hybrid imaging in Difmap with both amplitude and phase self-calibration, they are plotted as filled circles. In the cases when the processing was the same except no amplitude self-calibration was made, the data are plotted as filled triangles. In the cases when no self-calibration was made for the source and the data calibrated in PIMA were used, they are plotted as open circles.}
\figsetgrpend

\figsetgrpstart
\figsetgrpnum{3.76}
\figsetgrptitle{J0644+8018 at 2.3 GHz}
\figsetplot{J0644+8018_S_2006_02_14_avp_uvs_rad.pdf}
\figsetgrpnote{Correlated flux density averaged over time and IFs versus the $uv$ radius. The source name, the observation date, and the frequency are given at the top of the panel. In the cases when the data for a source were calibrated in AIPS and then underwent hybrid imaging in Difmap with both amplitude and phase self-calibration, they are plotted as filled circles. In the cases when the processing was the same except no amplitude self-calibration was made, the data are plotted as filled triangles. In the cases when no self-calibration was made for the source and the data calibrated in PIMA were used, they are plotted as open circles.}
\figsetgrpend

\figsetgrpstart
\figsetgrpnum{3.77}
\figsetgrptitle{J0644+8018 at 8.6 GHz}
\figsetplot{J0644+8018_X_2006_02_14_pet_uvt_rad.pdf}
\figsetgrpnote{Correlated flux density averaged over time and IFs versus the $uv$ radius. The source name, the observation date, and the frequency are given at the top of the panel. In the cases when the data for a source were calibrated in AIPS and then underwent hybrid imaging in Difmap with both amplitude and phase self-calibration, they are plotted as filled circles. In the cases when the processing was the same except no amplitude self-calibration was made, the data are plotted as filled triangles. In the cases when no self-calibration was made for the source and the data calibrated in PIMA were used, they are plotted as open circles.}
\figsetgrpend

\figsetgrpstart
\figsetgrpnum{3.78}
\figsetgrptitle{J0648+7756 at 2.3 GHz}
\figsetplot{J0648+7756_S_2006_02_16_pet_uvt_rad.pdf}
\figsetgrpnote{Correlated flux density averaged over time and IFs versus the $uv$ radius. The source name, the observation date, and the frequency are given at the top of the panel. In the cases when the data for a source were calibrated in AIPS and then underwent hybrid imaging in Difmap with both amplitude and phase self-calibration, they are plotted as filled circles. In the cases when the processing was the same except no amplitude self-calibration was made, the data are plotted as filled triangles. In the cases when no self-calibration was made for the source and the data calibrated in PIMA were used, they are plotted as open circles.}
\figsetgrpend

\figsetgrpstart
\figsetgrpnum{3.79}
\figsetgrptitle{J0702+8549 at 2.3 GHz}
\figsetplot{J0702+8549_S_2006_02_14_avp_uvs_rad.pdf}
\figsetgrpnote{Correlated flux density averaged over time and IFs versus the $uv$ radius. The source name, the observation date, and the frequency are given at the top of the panel. In the cases when the data for a source were calibrated in AIPS and then underwent hybrid imaging in Difmap with both amplitude and phase self-calibration, they are plotted as filled circles. In the cases when the processing was the same except no amplitude self-calibration was made, the data are plotted as filled triangles. In the cases when no self-calibration was made for the source and the data calibrated in PIMA were used, they are plotted as open circles.}
\figsetgrpend

\figsetgrpstart
\figsetgrpnum{3.80}
\figsetgrptitle{J0702+8549 at 8.6 GHz}
\figsetplot{J0702+8549_X_2006_02_14_avp_uvs_rad.pdf}
\figsetgrpnote{Correlated flux density averaged over time and IFs versus the $uv$ radius. The source name, the observation date, and the frequency are given at the top of the panel. In the cases when the data for a source were calibrated in AIPS and then underwent hybrid imaging in Difmap with both amplitude and phase self-calibration, they are plotted as filled circles. In the cases when the processing was the same except no amplitude self-calibration was made, the data are plotted as filled triangles. In the cases when no self-calibration was made for the source and the data calibrated in PIMA were used, they are plotted as open circles.}
\figsetgrpend

\figsetgrpstart
\figsetgrpnum{3.81}
\figsetgrptitle{J0714+8151 at 2.3 GHz}
\figsetplot{J0714+8151_S_2006_02_14_pet_uvt_rad.pdf}
\figsetgrpnote{Correlated flux density averaged over time and IFs versus the $uv$ radius. The source name, the observation date, and the frequency are given at the top of the panel. In the cases when the data for a source were calibrated in AIPS and then underwent hybrid imaging in Difmap with both amplitude and phase self-calibration, they are plotted as filled circles. In the cases when the processing was the same except no amplitude self-calibration was made, the data are plotted as filled triangles. In the cases when no self-calibration was made for the source and the data calibrated in PIMA were used, they are plotted as open circles.}
\figsetgrpend

\figsetgrpstart
\figsetgrpnum{3.82}
\figsetgrptitle{J0714+8151 at 8.6 GHz}
\figsetplot{J0714+8151_X_2006_02_14_pet_uvt_rad.pdf}
\figsetgrpnote{Correlated flux density averaged over time and IFs versus the $uv$ radius. The source name, the observation date, and the frequency are given at the top of the panel. In the cases when the data for a source were calibrated in AIPS and then underwent hybrid imaging in Difmap with both amplitude and phase self-calibration, they are plotted as filled circles. In the cases when the processing was the same except no amplitude self-calibration was made, the data are plotted as filled triangles. In the cases when no self-calibration was made for the source and the data calibrated in PIMA were used, they are plotted as open circles.}
\figsetgrpend

\figsetgrpstart
\figsetgrpnum{3.83}
\figsetgrptitle{J0726+7911 at 2.3 GHz}
\figsetplot{J0726+7911_S_2006_02_16_avp_uvs_rad.pdf}
\figsetgrpnote{Correlated flux density averaged over time and IFs versus the $uv$ radius. The source name, the observation date, and the frequency are given at the top of the panel. In the cases when the data for a source were calibrated in AIPS and then underwent hybrid imaging in Difmap with both amplitude and phase self-calibration, they are plotted as filled circles. In the cases when the processing was the same except no amplitude self-calibration was made, the data are plotted as filled triangles. In the cases when no self-calibration was made for the source and the data calibrated in PIMA were used, they are plotted as open circles.}
\figsetgrpend

\figsetgrpstart
\figsetgrpnum{3.84}
\figsetgrptitle{J0726+7911 at 8.6 GHz}
\figsetplot{J0726+7911_X_2006_02_16_avp_uvs_rad.pdf}
\figsetgrpnote{Correlated flux density averaged over time and IFs versus the $uv$ radius. The source name, the observation date, and the frequency are given at the top of the panel. In the cases when the data for a source were calibrated in AIPS and then underwent hybrid imaging in Difmap with both amplitude and phase self-calibration, they are plotted as filled circles. In the cases when the processing was the same except no amplitude self-calibration was made, the data are plotted as filled triangles. In the cases when no self-calibration was made for the source and the data calibrated in PIMA were used, they are plotted as open circles.}
\figsetgrpend

\figsetgrpstart
\figsetgrpnum{3.85}
\figsetgrptitle{J0750+8241 at 2.3 GHz}
\figsetplot{J0750+8241_S_2006_02_14_avp_uvs_rad.pdf}
\figsetgrpnote{Correlated flux density averaged over time and IFs versus the $uv$ radius. The source name, the observation date, and the frequency are given at the top of the panel. In the cases when the data for a source were calibrated in AIPS and then underwent hybrid imaging in Difmap with both amplitude and phase self-calibration, they are plotted as filled circles. In the cases when the processing was the same except no amplitude self-calibration was made, the data are plotted as filled triangles. In the cases when no self-calibration was made for the source and the data calibrated in PIMA were used, they are plotted as open circles.}
\figsetgrpend

\figsetgrpstart
\figsetgrpnum{3.86}
\figsetgrptitle{J0750+8241 at 8.6 GHz}
\figsetplot{J0750+8241_X_2006_02_14_avp_uvs_rad.pdf}
\figsetgrpnote{Correlated flux density averaged over time and IFs versus the $uv$ radius. The source name, the observation date, and the frequency are given at the top of the panel. In the cases when the data for a source were calibrated in AIPS and then underwent hybrid imaging in Difmap with both amplitude and phase self-calibration, they are plotted as filled circles. In the cases when the processing was the same except no amplitude self-calibration was made, the data are plotted as filled triangles. In the cases when no self-calibration was made for the source and the data calibrated in PIMA were used, they are plotted as open circles.}
\figsetgrpend

\figsetgrpstart
\figsetgrpnum{3.87}
\figsetgrptitle{J0802+7620 at 2.3 GHz}
\figsetplot{J0802+7620_S_2006_02_23_avp_uvs_rad.pdf}
\figsetgrpnote{Correlated flux density averaged over time and IFs versus the $uv$ radius. The source name, the observation date, and the frequency are given at the top of the panel. In the cases when the data for a source were calibrated in AIPS and then underwent hybrid imaging in Difmap with both amplitude and phase self-calibration, they are plotted as filled circles. In the cases when the processing was the same except no amplitude self-calibration was made, the data are plotted as filled triangles. In the cases when no self-calibration was made for the source and the data calibrated in PIMA were used, they are plotted as open circles.}
\figsetgrpend

\figsetgrpstart
\figsetgrpnum{3.88}
\figsetgrptitle{J0806+7746 at 2.3 GHz}
\figsetplot{J0806+7746_S_2006_02_23_avp_uvs_rad.pdf}
\figsetgrpnote{Correlated flux density averaged over time and IFs versus the $uv$ radius. The source name, the observation date, and the frequency are given at the top of the panel. In the cases when the data for a source were calibrated in AIPS and then underwent hybrid imaging in Difmap with both amplitude and phase self-calibration, they are plotted as filled circles. In the cases when the processing was the same except no amplitude self-calibration was made, the data are plotted as filled triangles. In the cases when no self-calibration was made for the source and the data calibrated in PIMA were used, they are plotted as open circles.}
\figsetgrpend

\figsetgrpstart
\figsetgrpnum{3.89}
\figsetgrptitle{J0806+7746 at 8.6 GHz}
\figsetplot{J0806+7746_X_2006_02_23_avp_uvs_rad.pdf}
\figsetgrpnote{Correlated flux density averaged over time and IFs versus the $uv$ radius. The source name, the observation date, and the frequency are given at the top of the panel. In the cases when the data for a source were calibrated in AIPS and then underwent hybrid imaging in Difmap with both amplitude and phase self-calibration, they are plotted as filled circles. In the cases when the processing was the same except no amplitude self-calibration was made, the data are plotted as filled triangles. In the cases when no self-calibration was made for the source and the data calibrated in PIMA were used, they are plotted as open circles.}
\figsetgrpend

\figsetgrpstart
\figsetgrpnum{3.90}
\figsetgrptitle{J0806+8126 at 2.3 GHz}
\figsetplot{J0806+8126_S_2006_02_16_pet_uvt_rad.pdf}
\figsetgrpnote{Correlated flux density averaged over time and IFs versus the $uv$ radius. The source name, the observation date, and the frequency are given at the top of the panel. In the cases when the data for a source were calibrated in AIPS and then underwent hybrid imaging in Difmap with both amplitude and phase self-calibration, they are plotted as filled circles. In the cases when the processing was the same except no amplitude self-calibration was made, the data are plotted as filled triangles. In the cases when no self-calibration was made for the source and the data calibrated in PIMA were used, they are plotted as open circles.}
\figsetgrpend

\figsetgrpstart
\figsetgrpnum{3.91}
\figsetgrptitle{J0819+8105 at 2.3 GHz}
\figsetplot{J0819+8105_S_2006_02_16_avp_uvs_rad.pdf}
\figsetgrpnote{Correlated flux density averaged over time and IFs versus the $uv$ radius. The source name, the observation date, and the frequency are given at the top of the panel. In the cases when the data for a source were calibrated in AIPS and then underwent hybrid imaging in Difmap with both amplitude and phase self-calibration, they are plotted as filled circles. In the cases when the processing was the same except no amplitude self-calibration was made, the data are plotted as filled triangles. In the cases when no self-calibration was made for the source and the data calibrated in PIMA were used, they are plotted as open circles.}
\figsetgrpend

\figsetgrpstart
\figsetgrpnum{3.92}
\figsetgrptitle{J0819+8105 at 8.6 GHz}
\figsetplot{J0819+8105_X_2006_02_16_pet_uvt_rad.pdf}
\figsetgrpnote{Correlated flux density averaged over time and IFs versus the $uv$ radius. The source name, the observation date, and the frequency are given at the top of the panel. In the cases when the data for a source were calibrated in AIPS and then underwent hybrid imaging in Difmap with both amplitude and phase self-calibration, they are plotted as filled circles. In the cases when the processing was the same except no amplitude self-calibration was made, the data are plotted as filled triangles. In the cases when no self-calibration was made for the source and the data calibrated in PIMA were used, they are plotted as open circles.}
\figsetgrpend

\figsetgrpstart
\figsetgrpnum{3.93}
\figsetgrptitle{J0835+8350 at 2.3 GHz}
\figsetplot{J0835+8350_S_2006_02_14_pet_uvt_rad.pdf}
\figsetgrpnote{Correlated flux density averaged over time and IFs versus the $uv$ radius. The source name, the observation date, and the frequency are given at the top of the panel. In the cases when the data for a source were calibrated in AIPS and then underwent hybrid imaging in Difmap with both amplitude and phase self-calibration, they are plotted as filled circles. In the cases when the processing was the same except no amplitude self-calibration was made, the data are plotted as filled triangles. In the cases when no self-calibration was made for the source and the data calibrated in PIMA were used, they are plotted as open circles.}
\figsetgrpend

\figsetgrpstart
\figsetgrpnum{3.94}
\figsetgrptitle{J0858+7501 at 2.3 GHz}
\figsetplot{J0858+7501_S_2006_02_16_avp_uvs_rad.pdf}
\figsetgrpnote{Correlated flux density averaged over time and IFs versus the $uv$ radius. The source name, the observation date, and the frequency are given at the top of the panel. In the cases when the data for a source were calibrated in AIPS and then underwent hybrid imaging in Difmap with both amplitude and phase self-calibration, they are plotted as filled circles. In the cases when the processing was the same except no amplitude self-calibration was made, the data are plotted as filled triangles. In the cases when no self-calibration was made for the source and the data calibrated in PIMA were used, they are plotted as open circles.}
\figsetgrpend

\figsetgrpstart
\figsetgrpnum{3.95}
\figsetgrptitle{J0909+8327 at 2.3 GHz}
\figsetplot{J0909+8327_S_2006_02_16_avp_uvs_rad.pdf}
\figsetgrpnote{Correlated flux density averaged over time and IFs versus the $uv$ radius. The source name, the observation date, and the frequency are given at the top of the panel. In the cases when the data for a source were calibrated in AIPS and then underwent hybrid imaging in Difmap with both amplitude and phase self-calibration, they are plotted as filled circles. In the cases when the processing was the same except no amplitude self-calibration was made, the data are plotted as filled triangles. In the cases when no self-calibration was made for the source and the data calibrated in PIMA were used, they are plotted as open circles.}
\figsetgrpend

\figsetgrpstart
\figsetgrpnum{3.96}
\figsetgrptitle{J0909+8327 at 8.6 GHz}
\figsetplot{J0909+8327_X_2006_02_16_avp_uvs_rad.pdf}
\figsetgrpnote{Correlated flux density averaged over time and IFs versus the $uv$ radius. The source name, the observation date, and the frequency are given at the top of the panel. In the cases when the data for a source were calibrated in AIPS and then underwent hybrid imaging in Difmap with both amplitude and phase self-calibration, they are plotted as filled circles. In the cases when the processing was the same except no amplitude self-calibration was made, the data are plotted as filled triangles. In the cases when no self-calibration was made for the source and the data calibrated in PIMA were used, they are plotted as open circles.}
\figsetgrpend

\figsetgrpstart
\figsetgrpnum{3.97}
\figsetgrptitle{J0911+8607 at 2.3 GHz}
\figsetplot{J0911+8607_S_2006_02_23_pet_uvt_rad.pdf}
\figsetgrpnote{Correlated flux density averaged over time and IFs versus the $uv$ radius. The source name, the observation date, and the frequency are given at the top of the panel. In the cases when the data for a source were calibrated in AIPS and then underwent hybrid imaging in Difmap with both amplitude and phase self-calibration, they are plotted as filled circles. In the cases when the processing was the same except no amplitude self-calibration was made, the data are plotted as filled triangles. In the cases when no self-calibration was made for the source and the data calibrated in PIMA were used, they are plotted as open circles.}
\figsetgrpend

\figsetgrpstart
\figsetgrpnum{3.98}
\figsetgrptitle{J0911+8607 at 8.6 GHz}
\figsetplot{J0911+8607_X_2006_02_23_avp_uvs_rad.pdf}
\figsetgrpnote{Correlated flux density averaged over time and IFs versus the $uv$ radius. The source name, the observation date, and the frequency are given at the top of the panel. In the cases when the data for a source were calibrated in AIPS and then underwent hybrid imaging in Difmap with both amplitude and phase self-calibration, they are plotted as filled circles. In the cases when the processing was the same except no amplitude self-calibration was made, the data are plotted as filled triangles. In the cases when no self-calibration was made for the source and the data calibrated in PIMA were used, they are plotted as open circles.}
\figsetgrpend

\figsetgrpstart
\figsetgrpnum{3.99}
\figsetgrptitle{J0919+7825 at 2.3 GHz}
\figsetplot{J0919+7825_S_2006_02_14_avp_uvs_rad.pdf}
\figsetgrpnote{Correlated flux density averaged over time and IFs versus the $uv$ radius. The source name, the observation date, and the frequency are given at the top of the panel. In the cases when the data for a source were calibrated in AIPS and then underwent hybrid imaging in Difmap with both amplitude and phase self-calibration, they are plotted as filled circles. In the cases when the processing was the same except no amplitude self-calibration was made, the data are plotted as filled triangles. In the cases when no self-calibration was made for the source and the data calibrated in PIMA were used, they are plotted as open circles.}
\figsetgrpend

\figsetgrpstart
\figsetgrpnum{3.100}
\figsetgrptitle{J0919+7825 at 8.6 GHz}
\figsetplot{J0919+7825_X_2006_02_14_avp_uvs_rad.pdf}
\figsetgrpnote{Correlated flux density averaged over time and IFs versus the $uv$ radius. The source name, the observation date, and the frequency are given at the top of the panel. In the cases when the data for a source were calibrated in AIPS and then underwent hybrid imaging in Difmap with both amplitude and phase self-calibration, they are plotted as filled circles. In the cases when the processing was the same except no amplitude self-calibration was made, the data are plotted as filled triangles. In the cases when no self-calibration was made for the source and the data calibrated in PIMA were used, they are plotted as open circles.}
\figsetgrpend

\figsetgrpstart
\figsetgrpnum{3.101}
\figsetgrptitle{J0932+7906 at 8.6 GHz}
\figsetplot{J0932+7906_X_2006_02_16_pet_uvt_rad.pdf}
\figsetgrpnote{Correlated flux density averaged over time and IFs versus the $uv$ radius. The source name, the observation date, and the frequency are given at the top of the panel. In the cases when the data for a source were calibrated in AIPS and then underwent hybrid imaging in Difmap with both amplitude and phase self-calibration, they are plotted as filled circles. In the cases when the processing was the same except no amplitude self-calibration was made, the data are plotted as filled triangles. In the cases when no self-calibration was made for the source and the data calibrated in PIMA were used, they are plotted as open circles.}
\figsetgrpend

\figsetgrpstart
\figsetgrpnum{3.102}
\figsetgrptitle{J0944+8254 at 2.3 GHz}
\figsetplot{J0944+8254_S_2006_02_16_pet_uvt_rad.pdf}
\figsetgrpnote{Correlated flux density averaged over time and IFs versus the $uv$ radius. The source name, the observation date, and the frequency are given at the top of the panel. In the cases when the data for a source were calibrated in AIPS and then underwent hybrid imaging in Difmap with both amplitude and phase self-calibration, they are plotted as filled circles. In the cases when the processing was the same except no amplitude self-calibration was made, the data are plotted as filled triangles. In the cases when no self-calibration was made for the source and the data calibrated in PIMA were used, they are plotted as open circles.}
\figsetgrpend

\figsetgrpstart
\figsetgrpnum{3.103}
\figsetgrptitle{J0956+7911 at 2.3 GHz}
\figsetplot{J0956+7911_S_2006_02_16_pet_uvt_rad.pdf}
\figsetgrpnote{Correlated flux density averaged over time and IFs versus the $uv$ radius. The source name, the observation date, and the frequency are given at the top of the panel. In the cases when the data for a source were calibrated in AIPS and then underwent hybrid imaging in Difmap with both amplitude and phase self-calibration, they are plotted as filled circles. In the cases when the processing was the same except no amplitude self-calibration was made, the data are plotted as filled triangles. In the cases when no self-calibration was made for the source and the data calibrated in PIMA were used, they are plotted as open circles.}
\figsetgrpend

\figsetgrpstart
\figsetgrpnum{3.104}
\figsetgrptitle{J1000+8127 at 2.3 GHz}
\figsetplot{J1000+8127_S_2006_02_23_pet_uvt_rad.pdf}
\figsetgrpnote{Correlated flux density averaged over time and IFs versus the $uv$ radius. The source name, the observation date, and the frequency are given at the top of the panel. In the cases when the data for a source were calibrated in AIPS and then underwent hybrid imaging in Difmap with both amplitude and phase self-calibration, they are plotted as filled circles. In the cases when the processing was the same except no amplitude self-calibration was made, the data are plotted as filled triangles. In the cases when no self-calibration was made for the source and the data calibrated in PIMA were used, they are plotted as open circles.}
\figsetgrpend

\figsetgrpstart
\figsetgrpnum{3.105}
\figsetgrptitle{J1005+7739 at 8.6 GHz}
\figsetplot{J1005+7739_X_2006_02_16_pet_uvt_rad.pdf}
\figsetgrpnote{Correlated flux density averaged over time and IFs versus the $uv$ radius. The source name, the observation date, and the frequency are given at the top of the panel. In the cases when the data for a source were calibrated in AIPS and then underwent hybrid imaging in Difmap with both amplitude and phase self-calibration, they are plotted as filled circles. In the cases when the processing was the same except no amplitude self-calibration was made, the data are plotted as filled triangles. In the cases when no self-calibration was made for the source and the data calibrated in PIMA were used, they are plotted as open circles.}
\figsetgrpend

\figsetgrpstart
\figsetgrpnum{3.106}
\figsetgrptitle{J1010+8250 at 2.3 GHz}
\figsetplot{J1010+8250_S_2006_02_16_avp_uvs_rad.pdf}
\figsetgrpnote{Correlated flux density averaged over time and IFs versus the $uv$ radius. The source name, the observation date, and the frequency are given at the top of the panel. In the cases when the data for a source were calibrated in AIPS and then underwent hybrid imaging in Difmap with both amplitude and phase self-calibration, they are plotted as filled circles. In the cases when the processing was the same except no amplitude self-calibration was made, the data are plotted as filled triangles. In the cases when no self-calibration was made for the source and the data calibrated in PIMA were used, they are plotted as open circles.}
\figsetgrpend

\figsetgrpstart
\figsetgrpnum{3.107}
\figsetgrptitle{J1010+8250 at 8.6 GHz}
\figsetplot{J1010+8250_X_2006_02_16_avp_uvs_rad.pdf}
\figsetgrpnote{Correlated flux density averaged over time and IFs versus the $uv$ radius. The source name, the observation date, and the frequency are given at the top of the panel. In the cases when the data for a source were calibrated in AIPS and then underwent hybrid imaging in Difmap with both amplitude and phase self-calibration, they are plotted as filled circles. In the cases when the processing was the same except no amplitude self-calibration was made, the data are plotted as filled triangles. In the cases when no self-calibration was made for the source and the data calibrated in PIMA were used, they are plotted as open circles.}
\figsetgrpend

\figsetgrpstart
\figsetgrpnum{3.108}
\figsetgrptitle{J1016+7617 at 2.3 GHz}
\figsetplot{J1016+7617_S_2006_02_23_avp_uvs_rad.pdf}
\figsetgrpnote{Correlated flux density averaged over time and IFs versus the $uv$ radius. The source name, the observation date, and the frequency are given at the top of the panel. In the cases when the data for a source were calibrated in AIPS and then underwent hybrid imaging in Difmap with both amplitude and phase self-calibration, they are plotted as filled circles. In the cases when the processing was the same except no amplitude self-calibration was made, the data are plotted as filled triangles. In the cases when no self-calibration was made for the source and the data calibrated in PIMA were used, they are plotted as open circles.}
\figsetgrpend

\figsetgrpstart
\figsetgrpnum{3.109}
\figsetgrptitle{J1016+7617 at 8.6 GHz}
\figsetplot{J1016+7617_X_2006_02_23_pet_uvt_rad.pdf}
\figsetgrpnote{Correlated flux density averaged over time and IFs versus the $uv$ radius. The source name, the observation date, and the frequency are given at the top of the panel. In the cases when the data for a source were calibrated in AIPS and then underwent hybrid imaging in Difmap with both amplitude and phase self-calibration, they are plotted as filled circles. In the cases when the processing was the same except no amplitude self-calibration was made, the data are plotted as filled triangles. In the cases when no self-calibration was made for the source and the data calibrated in PIMA were used, they are plotted as open circles.}
\figsetgrpend

\figsetgrpstart
\figsetgrpnum{3.110}
\figsetgrptitle{J1021+7655 at 2.3 GHz}
\figsetplot{J1021+7655_S_2006_02_23_pet_uvt_rad.pdf}
\figsetgrpnote{Correlated flux density averaged over time and IFs versus the $uv$ radius. The source name, the observation date, and the frequency are given at the top of the panel. In the cases when the data for a source were calibrated in AIPS and then underwent hybrid imaging in Difmap with both amplitude and phase self-calibration, they are plotted as filled circles. In the cases when the processing was the same except no amplitude self-calibration was made, the data are plotted as filled triangles. In the cases when no self-calibration was made for the source and the data calibrated in PIMA were used, they are plotted as open circles.}
\figsetgrpend

\figsetgrpstart
\figsetgrpnum{3.111}
\figsetgrptitle{J1023+8032 at 8.6 GHz}
\figsetplot{J1023+8032_X_2006_02_14_pet_uvt_rad.pdf}
\figsetgrpnote{Correlated flux density averaged over time and IFs versus the $uv$ radius. The source name, the observation date, and the frequency are given at the top of the panel. In the cases when the data for a source were calibrated in AIPS and then underwent hybrid imaging in Difmap with both amplitude and phase self-calibration, they are plotted as filled circles. In the cases when the processing was the same except no amplitude self-calibration was made, the data are plotted as filled triangles. In the cases when no self-calibration was made for the source and the data calibrated in PIMA were used, they are plotted as open circles.}
\figsetgrpend

\figsetgrpstart
\figsetgrpnum{3.112}
\figsetgrptitle{J1025+8144 at 2.3 GHz}
\figsetplot{J1025+8144_S_2006_02_16_pet_uvt_rad.pdf}
\figsetgrpnote{Correlated flux density averaged over time and IFs versus the $uv$ radius. The source name, the observation date, and the frequency are given at the top of the panel. In the cases when the data for a source were calibrated in AIPS and then underwent hybrid imaging in Difmap with both amplitude and phase self-calibration, they are plotted as filled circles. In the cases when the processing was the same except no amplitude self-calibration was made, the data are plotted as filled triangles. In the cases when no self-calibration was made for the source and the data calibrated in PIMA were used, they are plotted as open circles.}
\figsetgrpend

\figsetgrpstart
\figsetgrpnum{3.113}
\figsetgrptitle{J1044+8054 at 2.3 GHz}
\figsetplot{J1044+8054_S_2006_02_14_avp_uvs_rad.pdf}
\figsetgrpnote{Correlated flux density averaged over time and IFs versus the $uv$ radius. The source name, the observation date, and the frequency are given at the top of the panel. In the cases when the data for a source were calibrated in AIPS and then underwent hybrid imaging in Difmap with both amplitude and phase self-calibration, they are plotted as filled circles. In the cases when the processing was the same except no amplitude self-calibration was made, the data are plotted as filled triangles. In the cases when no self-calibration was made for the source and the data calibrated in PIMA were used, they are plotted as open circles.}
\figsetgrpend

\figsetgrpstart
\figsetgrpnum{3.114}
\figsetgrptitle{J1044+8054 at 8.6 GHz}
\figsetplot{J1044+8054_X_2006_02_14_avp_uvs_rad.pdf}
\figsetgrpnote{Correlated flux density averaged over time and IFs versus the $uv$ radius. The source name, the observation date, and the frequency are given at the top of the panel. In the cases when the data for a source were calibrated in AIPS and then underwent hybrid imaging in Difmap with both amplitude and phase self-calibration, they are plotted as filled circles. In the cases when the processing was the same except no amplitude self-calibration was made, the data are plotted as filled triangles. In the cases when no self-calibration was made for the source and the data calibrated in PIMA were used, they are plotted as open circles.}
\figsetgrpend

\figsetgrpstart
\figsetgrpnum{3.115}
\figsetgrptitle{J1052+8317 at 2.3 GHz}
\figsetplot{J1052+8317_S_2006_02_14_avp_uvs_rad.pdf}
\figsetgrpnote{Correlated flux density averaged over time and IFs versus the $uv$ radius. The source name, the observation date, and the frequency are given at the top of the panel. In the cases when the data for a source were calibrated in AIPS and then underwent hybrid imaging in Difmap with both amplitude and phase self-calibration, they are plotted as filled circles. In the cases when the processing was the same except no amplitude self-calibration was made, the data are plotted as filled triangles. In the cases when no self-calibration was made for the source and the data calibrated in PIMA were used, they are plotted as open circles.}
\figsetgrpend

\figsetgrpstart
\figsetgrpnum{3.116}
\figsetgrptitle{J1052+8317 at 8.6 GHz}
\figsetplot{J1052+8317_X_2006_02_14_pet_uvt_rad.pdf}
\figsetgrpnote{Correlated flux density averaged over time and IFs versus the $uv$ radius. The source name, the observation date, and the frequency are given at the top of the panel. In the cases when the data for a source were calibrated in AIPS and then underwent hybrid imaging in Difmap with both amplitude and phase self-calibration, they are plotted as filled circles. In the cases when the processing was the same except no amplitude self-calibration was made, the data are plotted as filled triangles. In the cases when no self-calibration was made for the source and the data calibrated in PIMA were used, they are plotted as open circles.}
\figsetgrpend

\figsetgrpstart
\figsetgrpnum{3.117}
\figsetgrptitle{J1054+8629 at 2.3 GHz}
\figsetplot{J1054+8629_S_2006_02_16_avp_uvs_rad.pdf}
\figsetgrpnote{Correlated flux density averaged over time and IFs versus the $uv$ radius. The source name, the observation date, and the frequency are given at the top of the panel. In the cases when the data for a source were calibrated in AIPS and then underwent hybrid imaging in Difmap with both amplitude and phase self-calibration, they are plotted as filled circles. In the cases when the processing was the same except no amplitude self-calibration was made, the data are plotted as filled triangles. In the cases when no self-calibration was made for the source and the data calibrated in PIMA were used, they are plotted as open circles.}
\figsetgrpend

\figsetgrpstart
\figsetgrpnum{3.118}
\figsetgrptitle{J1054+8629 at 8.6 GHz}
\figsetplot{J1054+8629_X_2006_02_16_avp_uvs_rad.pdf}
\figsetgrpnote{Correlated flux density averaged over time and IFs versus the $uv$ radius. The source name, the observation date, and the frequency are given at the top of the panel. In the cases when the data for a source were calibrated in AIPS and then underwent hybrid imaging in Difmap with both amplitude and phase self-calibration, they are plotted as filled circles. In the cases when the processing was the same except no amplitude self-calibration was made, the data are plotted as filled triangles. In the cases when no self-calibration was made for the source and the data calibrated in PIMA were used, they are plotted as open circles.}
\figsetgrpend

\figsetgrpstart
\figsetgrpnum{3.119}
\figsetgrptitle{J1057+8858 at 2.3 GHz}
\figsetplot{J1057+8858_S_2006_02_23_pet_uvt_rad.pdf}
\figsetgrpnote{Correlated flux density averaged over time and IFs versus the $uv$ radius. The source name, the observation date, and the frequency are given at the top of the panel. In the cases when the data for a source were calibrated in AIPS and then underwent hybrid imaging in Difmap with both amplitude and phase self-calibration, they are plotted as filled circles. In the cases when the processing was the same except no amplitude self-calibration was made, the data are plotted as filled triangles. In the cases when no self-calibration was made for the source and the data calibrated in PIMA were used, they are plotted as open circles.}
\figsetgrpend

\figsetgrpstart
\figsetgrpnum{3.120}
\figsetgrptitle{J1058+8114 at 2.3 GHz}
\figsetplot{J1058+8114_S_2006_02_14_avp_uvs_rad.pdf}
\figsetgrpnote{Correlated flux density averaged over time and IFs versus the $uv$ radius. The source name, the observation date, and the frequency are given at the top of the panel. In the cases when the data for a source were calibrated in AIPS and then underwent hybrid imaging in Difmap with both amplitude and phase self-calibration, they are plotted as filled circles. In the cases when the processing was the same except no amplitude self-calibration was made, the data are plotted as filled triangles. In the cases when no self-calibration was made for the source and the data calibrated in PIMA were used, they are plotted as open circles.}
\figsetgrpend

\figsetgrpstart
\figsetgrpnum{3.121}
\figsetgrptitle{J1058+8114 at 2.3 GHz}
\figsetplot{J1058+8114_S_2006_02_16_avp_uvs_rad.pdf}
\figsetgrpnote{Correlated flux density averaged over time and IFs versus the $uv$ radius. The source name, the observation date, and the frequency are given at the top of the panel. In the cases when the data for a source were calibrated in AIPS and then underwent hybrid imaging in Difmap with both amplitude and phase self-calibration, they are plotted as filled circles. In the cases when the processing was the same except no amplitude self-calibration was made, the data are plotted as filled triangles. In the cases when no self-calibration was made for the source and the data calibrated in PIMA were used, they are plotted as open circles.}
\figsetgrpend

\figsetgrpstart
\figsetgrpnum{3.122}
\figsetgrptitle{J1058+8114 at 2.3 GHz}
\figsetplot{J1058+8114_S_2006_02_23_avp_uvs_rad.pdf}
\figsetgrpnote{Correlated flux density averaged over time and IFs versus the $uv$ radius. The source name, the observation date, and the frequency are given at the top of the panel. In the cases when the data for a source were calibrated in AIPS and then underwent hybrid imaging in Difmap with both amplitude and phase self-calibration, they are plotted as filled circles. In the cases when the processing was the same except no amplitude self-calibration was made, the data are plotted as filled triangles. In the cases when no self-calibration was made for the source and the data calibrated in PIMA were used, they are plotted as open circles.}
\figsetgrpend

\figsetgrpstart
\figsetgrpnum{3.123}
\figsetgrptitle{J1058+8114 at 8.6 GHz}
\figsetplot{J1058+8114_X_2006_02_14_avp_uvs_rad.pdf}
\figsetgrpnote{Correlated flux density averaged over time and IFs versus the $uv$ radius. The source name, the observation date, and the frequency are given at the top of the panel. In the cases when the data for a source were calibrated in AIPS and then underwent hybrid imaging in Difmap with both amplitude and phase self-calibration, they are plotted as filled circles. In the cases when the processing was the same except no amplitude self-calibration was made, the data are plotted as filled triangles. In the cases when no self-calibration was made for the source and the data calibrated in PIMA were used, they are plotted as open circles.}
\figsetgrpend

\figsetgrpstart
\figsetgrpnum{3.124}
\figsetgrptitle{J1058+8114 at 8.6 GHz}
\figsetplot{J1058+8114_X_2006_02_16_avp_uvs_rad.pdf}
\figsetgrpnote{Correlated flux density averaged over time and IFs versus the $uv$ radius. The source name, the observation date, and the frequency are given at the top of the panel. In the cases when the data for a source were calibrated in AIPS and then underwent hybrid imaging in Difmap with both amplitude and phase self-calibration, they are plotted as filled circles. In the cases when the processing was the same except no amplitude self-calibration was made, the data are plotted as filled triangles. In the cases when no self-calibration was made for the source and the data calibrated in PIMA were used, they are plotted as open circles.}
\figsetgrpend

\figsetgrpstart
\figsetgrpnum{3.125}
\figsetgrptitle{J1058+8114 at 8.6 GHz}
\figsetplot{J1058+8114_X_2006_02_23_avp_uvs_rad.pdf}
\figsetgrpnote{Correlated flux density averaged over time and IFs versus the $uv$ radius. The source name, the observation date, and the frequency are given at the top of the panel. In the cases when the data for a source were calibrated in AIPS and then underwent hybrid imaging in Difmap with both amplitude and phase self-calibration, they are plotted as filled circles. In the cases when the processing was the same except no amplitude self-calibration was made, the data are plotted as filled triangles. In the cases when no self-calibration was made for the source and the data calibrated in PIMA were used, they are plotted as open circles.}
\figsetgrpend

\figsetgrpstart
\figsetgrpnum{3.126}
\figsetgrptitle{J1102+7905 at 2.3 GHz}
\figsetplot{J1102+7905_S_2006_02_16_pet_uvt_rad.pdf}
\figsetgrpnote{Correlated flux density averaged over time and IFs versus the $uv$ radius. The source name, the observation date, and the frequency are given at the top of the panel. In the cases when the data for a source were calibrated in AIPS and then underwent hybrid imaging in Difmap with both amplitude and phase self-calibration, they are plotted as filled circles. In the cases when the processing was the same except no amplitude self-calibration was made, the data are plotted as filled triangles. In the cases when no self-calibration was made for the source and the data calibrated in PIMA were used, they are plotted as open circles.}
\figsetgrpend

\figsetgrpstart
\figsetgrpnum{3.127}
\figsetgrptitle{J1104+7658 at 2.3 GHz}
\figsetplot{J1104+7658_S_2006_02_14_pet_uvt_rad.pdf}
\figsetgrpnote{Correlated flux density averaged over time and IFs versus the $uv$ radius. The source name, the observation date, and the frequency are given at the top of the panel. In the cases when the data for a source were calibrated in AIPS and then underwent hybrid imaging in Difmap with both amplitude and phase self-calibration, they are plotted as filled circles. In the cases when the processing was the same except no amplitude self-calibration was made, the data are plotted as filled triangles. In the cases when no self-calibration was made for the source and the data calibrated in PIMA were used, they are plotted as open circles.}
\figsetgrpend

\figsetgrpstart
\figsetgrpnum{3.128}
\figsetgrptitle{J1104+7658 at 8.6 GHz}
\figsetplot{J1104+7658_X_2006_02_14_avp_uvs_rad.pdf}
\figsetgrpnote{Correlated flux density averaged over time and IFs versus the $uv$ radius. The source name, the observation date, and the frequency are given at the top of the panel. In the cases when the data for a source were calibrated in AIPS and then underwent hybrid imaging in Difmap with both amplitude and phase self-calibration, they are plotted as filled circles. In the cases when the processing was the same except no amplitude self-calibration was made, the data are plotted as filled triangles. In the cases when no self-calibration was made for the source and the data calibrated in PIMA were used, they are plotted as open circles.}
\figsetgrpend

\figsetgrpstart
\figsetgrpnum{3.129}
\figsetgrptitle{J1104+7932 at 2.3 GHz}
\figsetplot{J1104+7932_S_2006_02_23_avp_uvs_rad.pdf}
\figsetgrpnote{Correlated flux density averaged over time and IFs versus the $uv$ radius. The source name, the observation date, and the frequency are given at the top of the panel. In the cases when the data for a source were calibrated in AIPS and then underwent hybrid imaging in Difmap with both amplitude and phase self-calibration, they are plotted as filled circles. In the cases when the processing was the same except no amplitude self-calibration was made, the data are plotted as filled triangles. In the cases when no self-calibration was made for the source and the data calibrated in PIMA were used, they are plotted as open circles.}
\figsetgrpend

\figsetgrpstart
\figsetgrpnum{3.130}
\figsetgrptitle{J1104+7932 at 8.6 GHz}
\figsetplot{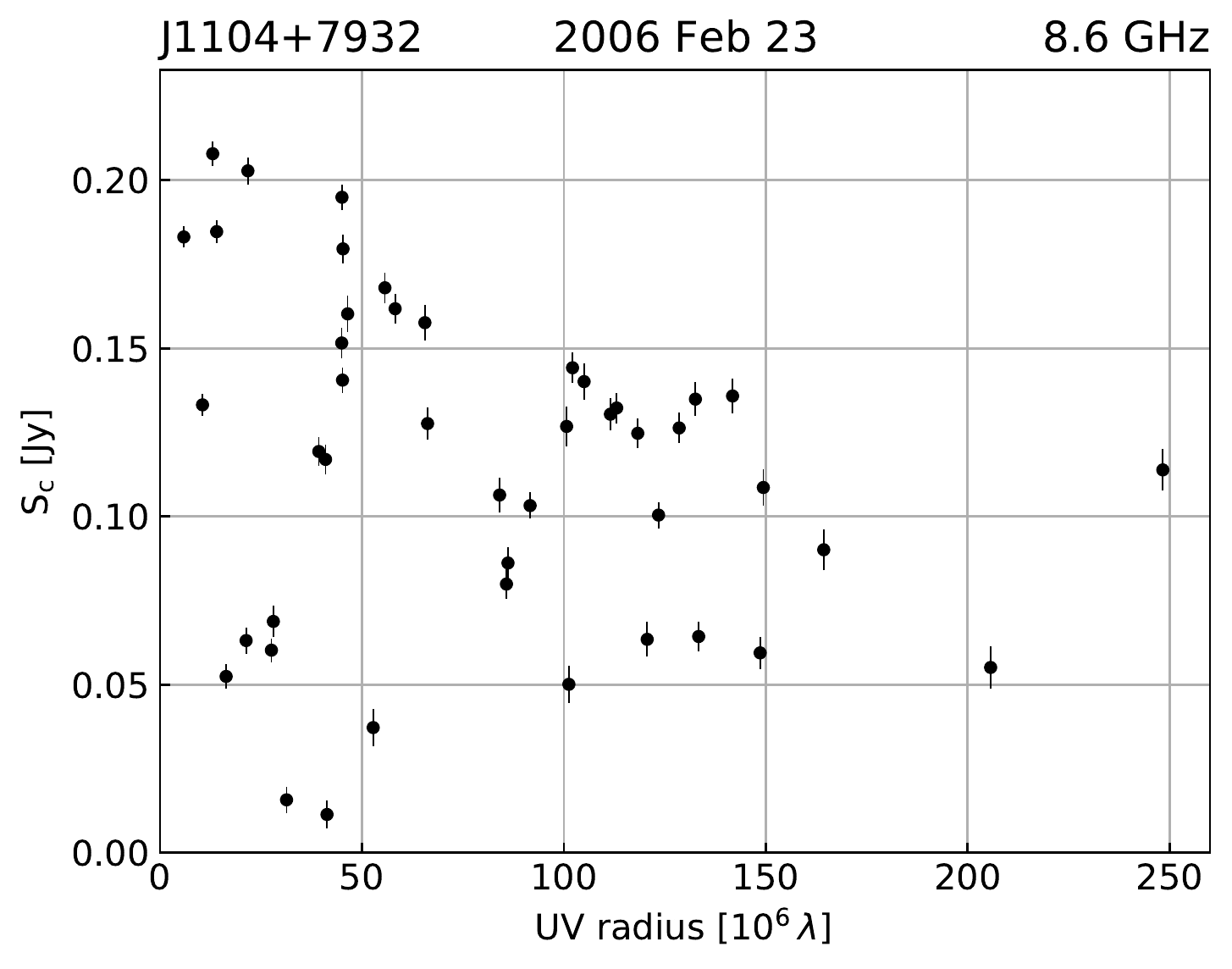}
\figsetgrpnote{Correlated flux density averaged over time and IFs versus the $uv$ radius. The source name, the observation date, and the frequency are given at the top of the panel. In the cases when the data for a source were calibrated in AIPS and then underwent hybrid imaging in Difmap with both amplitude and phase self-calibration, they are plotted as filled circles. In the cases when the processing was the same except no amplitude self-calibration was made, the data are plotted as filled triangles. In the cases when no self-calibration was made for the source and the data calibrated in PIMA were used, they are plotted as open circles.}
\figsetgrpend

\figsetgrpstart
\figsetgrpnum{3.131}
\figsetgrptitle{J1119+8048 at 2.3 GHz}
\figsetplot{J1119+8048_S_2006_02_14_avp_uvs_rad.pdf}
\figsetgrpnote{Correlated flux density averaged over time and IFs versus the $uv$ radius. The source name, the observation date, and the frequency are given at the top of the panel. In the cases when the data for a source were calibrated in AIPS and then underwent hybrid imaging in Difmap with both amplitude and phase self-calibration, they are plotted as filled circles. In the cases when the processing was the same except no amplitude self-calibration was made, the data are plotted as filled triangles. In the cases when no self-calibration was made for the source and the data calibrated in PIMA were used, they are plotted as open circles.}
\figsetgrpend

\figsetgrpstart
\figsetgrpnum{3.132}
\figsetgrptitle{J1119+8048 at 8.6 GHz}
\figsetplot{J1119+8048_X_2006_02_14_pet_uvt_rad.pdf}
\figsetgrpnote{Correlated flux density averaged over time and IFs versus the $uv$ radius. The source name, the observation date, and the frequency are given at the top of the panel. In the cases when the data for a source were calibrated in AIPS and then underwent hybrid imaging in Difmap with both amplitude and phase self-calibration, they are plotted as filled circles. In the cases when the processing was the same except no amplitude self-calibration was made, the data are plotted as filled triangles. In the cases when no self-calibration was made for the source and the data calibrated in PIMA were used, they are plotted as open circles.}
\figsetgrpend

\figsetgrpstart
\figsetgrpnum{3.133}
\figsetgrptitle{J1133+7831 at 2.3 GHz}
\figsetplot{J1133+7831_S_2006_02_23_avp_uvs_rad.pdf}
\figsetgrpnote{Correlated flux density averaged over time and IFs versus the $uv$ radius. The source name, the observation date, and the frequency are given at the top of the panel. In the cases when the data for a source were calibrated in AIPS and then underwent hybrid imaging in Difmap with both amplitude and phase self-calibration, they are plotted as filled circles. In the cases when the processing was the same except no amplitude self-calibration was made, the data are plotted as filled triangles. In the cases when no self-calibration was made for the source and the data calibrated in PIMA were used, they are plotted as open circles.}
\figsetgrpend

\figsetgrpstart
\figsetgrpnum{3.134}
\figsetgrptitle{J1153+8058 at 2.3 GHz}
\figsetplot{J1153+8058_S_2006_02_14_avp_uvs_rad.pdf}
\figsetgrpnote{Correlated flux density averaged over time and IFs versus the $uv$ radius. The source name, the observation date, and the frequency are given at the top of the panel. In the cases when the data for a source were calibrated in AIPS and then underwent hybrid imaging in Difmap with both amplitude and phase self-calibration, they are plotted as filled circles. In the cases when the processing was the same except no amplitude self-calibration was made, the data are plotted as filled triangles. In the cases when no self-calibration was made for the source and the data calibrated in PIMA were used, they are plotted as open circles.}
\figsetgrpend

\figsetgrpstart
\figsetgrpnum{3.135}
\figsetgrptitle{J1153+8058 at 2.3 GHz}
\figsetplot{J1153+8058_S_2006_02_16_avp_uvs_rad.pdf}
\figsetgrpnote{Correlated flux density averaged over time and IFs versus the $uv$ radius. The source name, the observation date, and the frequency are given at the top of the panel. In the cases when the data for a source were calibrated in AIPS and then underwent hybrid imaging in Difmap with both amplitude and phase self-calibration, they are plotted as filled circles. In the cases when the processing was the same except no amplitude self-calibration was made, the data are plotted as filled triangles. In the cases when no self-calibration was made for the source and the data calibrated in PIMA were used, they are plotted as open circles.}
\figsetgrpend

\figsetgrpstart
\figsetgrpnum{3.136}
\figsetgrptitle{J1153+8058 at 2.3 GHz}
\figsetplot{J1153+8058_S_2006_02_23_avp_uvs_rad.pdf}
\figsetgrpnote{Correlated flux density averaged over time and IFs versus the $uv$ radius. The source name, the observation date, and the frequency are given at the top of the panel. In the cases when the data for a source were calibrated in AIPS and then underwent hybrid imaging in Difmap with both amplitude and phase self-calibration, they are plotted as filled circles. In the cases when the processing was the same except no amplitude self-calibration was made, the data are plotted as filled triangles. In the cases when no self-calibration was made for the source and the data calibrated in PIMA were used, they are plotted as open circles.}
\figsetgrpend

\figsetgrpstart
\figsetgrpnum{3.137}
\figsetgrptitle{J1153+8058 at 8.6 GHz}
\figsetplot{J1153+8058_X_2006_02_14_avp_uvs_rad.pdf}
\figsetgrpnote{Correlated flux density averaged over time and IFs versus the $uv$ radius. The source name, the observation date, and the frequency are given at the top of the panel. In the cases when the data for a source were calibrated in AIPS and then underwent hybrid imaging in Difmap with both amplitude and phase self-calibration, they are plotted as filled circles. In the cases when the processing was the same except no amplitude self-calibration was made, the data are plotted as filled triangles. In the cases when no self-calibration was made for the source and the data calibrated in PIMA were used, they are plotted as open circles.}
\figsetgrpend

\figsetgrpstart
\figsetgrpnum{3.138}
\figsetgrptitle{J1153+8058 at 8.6 GHz}
\figsetplot{J1153+8058_X_2006_02_16_avp_uvs_rad.pdf}
\figsetgrpnote{Correlated flux density averaged over time and IFs versus the $uv$ radius. The source name, the observation date, and the frequency are given at the top of the panel. In the cases when the data for a source were calibrated in AIPS and then underwent hybrid imaging in Difmap with both amplitude and phase self-calibration, they are plotted as filled circles. In the cases when the processing was the same except no amplitude self-calibration was made, the data are plotted as filled triangles. In the cases when no self-calibration was made for the source and the data calibrated in PIMA were used, they are plotted as open circles.}
\figsetgrpend

\figsetgrpstart
\figsetgrpnum{3.139}
\figsetgrptitle{J1153+8058 at 8.6 GHz}
\figsetplot{J1153+8058_X_2006_02_23_avp_uvs_rad.pdf}
\figsetgrpnote{Correlated flux density averaged over time and IFs versus the $uv$ radius. The source name, the observation date, and the frequency are given at the top of the panel. In the cases when the data for a source were calibrated in AIPS and then underwent hybrid imaging in Difmap with both amplitude and phase self-calibration, they are plotted as filled circles. In the cases when the processing was the same except no amplitude self-calibration was made, the data are plotted as filled triangles. In the cases when no self-calibration was made for the source and the data calibrated in PIMA were used, they are plotted as open circles.}
\figsetgrpend

\figsetgrpstart
\figsetgrpnum{3.140}
\figsetgrptitle{J1155+7534 at 2.3 GHz}
\figsetplot{J1155+7534_S_2006_02_14_pet_uvt_rad.pdf}
\figsetgrpnote{Correlated flux density averaged over time and IFs versus the $uv$ radius. The source name, the observation date, and the frequency are given at the top of the panel. In the cases when the data for a source were calibrated in AIPS and then underwent hybrid imaging in Difmap with both amplitude and phase self-calibration, they are plotted as filled circles. In the cases when the processing was the same except no amplitude self-calibration was made, the data are plotted as filled triangles. In the cases when no self-calibration was made for the source and the data calibrated in PIMA were used, they are plotted as open circles.}
\figsetgrpend

\figsetgrpstart
\figsetgrpnum{3.141}
\figsetgrptitle{J1155+8157 at 2.3 GHz}
\figsetplot{J1155+8157_S_2006_02_16_pet_uvt_rad.pdf}
\figsetgrpnote{Correlated flux density averaged over time and IFs versus the $uv$ radius. The source name, the observation date, and the frequency are given at the top of the panel. In the cases when the data for a source were calibrated in AIPS and then underwent hybrid imaging in Difmap with both amplitude and phase self-calibration, they are plotted as filled circles. In the cases when the processing was the same except no amplitude self-calibration was made, the data are plotted as filled triangles. In the cases when no self-calibration was made for the source and the data calibrated in PIMA were used, they are plotted as open circles.}
\figsetgrpend

\figsetgrpstart
\figsetgrpnum{3.142}
\figsetgrptitle{J1157+8118 at 2.3 GHz}
\figsetplot{J1157+8118_S_2006_02_14_avp_uvs_rad.pdf}
\figsetgrpnote{Correlated flux density averaged over time and IFs versus the $uv$ radius. The source name, the observation date, and the frequency are given at the top of the panel. In the cases when the data for a source were calibrated in AIPS and then underwent hybrid imaging in Difmap with both amplitude and phase self-calibration, they are plotted as filled circles. In the cases when the processing was the same except no amplitude self-calibration was made, the data are plotted as filled triangles. In the cases when no self-calibration was made for the source and the data calibrated in PIMA were used, they are plotted as open circles.}
\figsetgrpend

\figsetgrpstart
\figsetgrpnum{3.143}
\figsetgrptitle{J1157+8118 at 8.6 GHz}
\figsetplot{J1157+8118_X_2006_02_14_pet_uvt_rad.pdf}
\figsetgrpnote{Correlated flux density averaged over time and IFs versus the $uv$ radius. The source name, the observation date, and the frequency are given at the top of the panel. In the cases when the data for a source were calibrated in AIPS and then underwent hybrid imaging in Difmap with both amplitude and phase self-calibration, they are plotted as filled circles. In the cases when the processing was the same except no amplitude self-calibration was made, the data are plotted as filled triangles. In the cases when no self-calibration was made for the source and the data calibrated in PIMA were used, they are plotted as open circles.}
\figsetgrpend

\figsetgrpstart
\figsetgrpnum{3.144}
\figsetgrptitle{J1218+7934 at 2.3 GHz}
\figsetplot{J1218+7934_S_2006_02_23_pet_uvt_rad.pdf}
\figsetgrpnote{Correlated flux density averaged over time and IFs versus the $uv$ radius. The source name, the observation date, and the frequency are given at the top of the panel. In the cases when the data for a source were calibrated in AIPS and then underwent hybrid imaging in Difmap with both amplitude and phase self-calibration, they are plotted as filled circles. In the cases when the processing was the same except no amplitude self-calibration was made, the data are plotted as filled triangles. In the cases when no self-calibration was made for the source and the data calibrated in PIMA were used, they are plotted as open circles.}
\figsetgrpend

\figsetgrpstart
\figsetgrpnum{3.145}
\figsetgrptitle{J1223+8040 at 2.3 GHz}
\figsetplot{J1223+8040_S_2006_02_23_avp_uvs_rad.pdf}
\figsetgrpnote{Correlated flux density averaged over time and IFs versus the $uv$ radius. The source name, the observation date, and the frequency are given at the top of the panel. In the cases when the data for a source were calibrated in AIPS and then underwent hybrid imaging in Difmap with both amplitude and phase self-calibration, they are plotted as filled circles. In the cases when the processing was the same except no amplitude self-calibration was made, the data are plotted as filled triangles. In the cases when no self-calibration was made for the source and the data calibrated in PIMA were used, they are plotted as open circles.}
\figsetgrpend

\figsetgrpstart
\figsetgrpnum{3.146}
\figsetgrptitle{J1223+8040 at 8.6 GHz}
\figsetplot{J1223+8040_X_2006_02_23_avp_uvs_rad.pdf}
\figsetgrpnote{Correlated flux density averaged over time and IFs versus the $uv$ radius. The source name, the observation date, and the frequency are given at the top of the panel. In the cases when the data for a source were calibrated in AIPS and then underwent hybrid imaging in Difmap with both amplitude and phase self-calibration, they are plotted as filled circles. In the cases when the processing was the same except no amplitude self-calibration was made, the data are plotted as filled triangles. In the cases when no self-calibration was made for the source and the data calibrated in PIMA were used, they are plotted as open circles.}
\figsetgrpend

\figsetgrpstart
\figsetgrpnum{3.147}
\figsetgrptitle{J1223+8436 at 2.3 GHz}
\figsetplot{J1223+8436_S_2006_02_16_avp_uvs_rad.pdf}
\figsetgrpnote{Correlated flux density averaged over time and IFs versus the $uv$ radius. The source name, the observation date, and the frequency are given at the top of the panel. In the cases when the data for a source were calibrated in AIPS and then underwent hybrid imaging in Difmap with both amplitude and phase self-calibration, they are plotted as filled circles. In the cases when the processing was the same except no amplitude self-calibration was made, the data are plotted as filled triangles. In the cases when no self-calibration was made for the source and the data calibrated in PIMA were used, they are plotted as open circles.}
\figsetgrpend

\figsetgrpstart
\figsetgrpnum{3.148}
\figsetgrptitle{J1223+8436 at 8.6 GHz}
\figsetplot{J1223+8436_X_2006_02_16_avp_uvs_rad.pdf}
\figsetgrpnote{Correlated flux density averaged over time and IFs versus the $uv$ radius. The source name, the observation date, and the frequency are given at the top of the panel. In the cases when the data for a source were calibrated in AIPS and then underwent hybrid imaging in Difmap with both amplitude and phase self-calibration, they are plotted as filled circles. In the cases when the processing was the same except no amplitude self-calibration was made, the data are plotted as filled triangles. In the cases when no self-calibration was made for the source and the data calibrated in PIMA were used, they are plotted as open circles.}
\figsetgrpend

\figsetgrpstart
\figsetgrpnum{3.149}
\figsetgrptitle{J1233+8054 at 2.3 GHz}
\figsetplot{J1233+8054_S_2006_02_16_avp_uvs_rad.pdf}
\figsetgrpnote{Correlated flux density averaged over time and IFs versus the $uv$ radius. The source name, the observation date, and the frequency are given at the top of the panel. In the cases when the data for a source were calibrated in AIPS and then underwent hybrid imaging in Difmap with both amplitude and phase self-calibration, they are plotted as filled circles. In the cases when the processing was the same except no amplitude self-calibration was made, the data are plotted as filled triangles. In the cases when no self-calibration was made for the source and the data calibrated in PIMA were used, they are plotted as open circles.}
\figsetgrpend

\figsetgrpstart
\figsetgrpnum{3.150}
\figsetgrptitle{J1233+8054 at 8.6 GHz}
\figsetplot{J1233+8054_X_2006_02_16_avp_uvs_rad.pdf}
\figsetgrpnote{Correlated flux density averaged over time and IFs versus the $uv$ radius. The source name, the observation date, and the frequency are given at the top of the panel. In the cases when the data for a source were calibrated in AIPS and then underwent hybrid imaging in Difmap with both amplitude and phase self-calibration, they are plotted as filled circles. In the cases when the processing was the same except no amplitude self-calibration was made, the data are plotted as filled triangles. In the cases when no self-calibration was made for the source and the data calibrated in PIMA were used, they are plotted as open circles.}
\figsetgrpend

\figsetgrpstart
\figsetgrpnum{3.151}
\figsetgrptitle{J1244+8000 at 2.3 GHz}
\figsetplot{J1244+8000_S_2006_02_16_pet_uvt_rad.pdf}
\figsetgrpnote{Correlated flux density averaged over time and IFs versus the $uv$ radius. The source name, the observation date, and the frequency are given at the top of the panel. In the cases when the data for a source were calibrated in AIPS and then underwent hybrid imaging in Difmap with both amplitude and phase self-calibration, they are plotted as filled circles. In the cases when the processing was the same except no amplitude self-calibration was made, the data are plotted as filled triangles. In the cases when no self-calibration was made for the source and the data calibrated in PIMA were used, they are plotted as open circles.}
\figsetgrpend

\figsetgrpstart
\figsetgrpnum{3.152}
\figsetgrptitle{J1244+8755 at 2.3 GHz}
\figsetplot{J1244+8755_S_2006_02_23_pet_uvt_rad.pdf}
\figsetgrpnote{Correlated flux density averaged over time and IFs versus the $uv$ radius. The source name, the observation date, and the frequency are given at the top of the panel. In the cases when the data for a source were calibrated in AIPS and then underwent hybrid imaging in Difmap with both amplitude and phase self-calibration, they are plotted as filled circles. In the cases when the processing was the same except no amplitude self-calibration was made, the data are plotted as filled triangles. In the cases when no self-calibration was made for the source and the data calibrated in PIMA were used, they are plotted as open circles.}
\figsetgrpend

\figsetgrpstart
\figsetgrpnum{3.153}
\figsetgrptitle{J1257+7958 at 2.3 GHz}
\figsetplot{J1257+7958_S_2006_02_23_avp_uvs_rad.pdf}
\figsetgrpnote{Correlated flux density averaged over time and IFs versus the $uv$ radius. The source name, the observation date, and the frequency are given at the top of the panel. In the cases when the data for a source were calibrated in AIPS and then underwent hybrid imaging in Difmap with both amplitude and phase self-calibration, they are plotted as filled circles. In the cases when the processing was the same except no amplitude self-calibration was made, the data are plotted as filled triangles. In the cases when no self-calibration was made for the source and the data calibrated in PIMA were used, they are plotted as open circles.}
\figsetgrpend

\figsetgrpstart
\figsetgrpnum{3.154}
\figsetgrptitle{J1257+7958 at 8.6 GHz}
\figsetplot{J1257+7958_X_2006_02_23_pet_uvt_rad.pdf}
\figsetgrpnote{Correlated flux density averaged over time and IFs versus the $uv$ radius. The source name, the observation date, and the frequency are given at the top of the panel. In the cases when the data for a source were calibrated in AIPS and then underwent hybrid imaging in Difmap with both amplitude and phase self-calibration, they are plotted as filled circles. In the cases when the processing was the same except no amplitude self-calibration was made, the data are plotted as filled triangles. In the cases when no self-calibration was made for the source and the data calibrated in PIMA were used, they are plotted as open circles.}
\figsetgrpend

\figsetgrpstart
\figsetgrpnum{3.155}
\figsetgrptitle{J1257+8342 at 2.3 GHz}
\figsetplot{J1257+8342_S_2006_02_14_pet_uvt_rad.pdf}
\figsetgrpnote{Correlated flux density averaged over time and IFs versus the $uv$ radius. The source name, the observation date, and the frequency are given at the top of the panel. In the cases when the data for a source were calibrated in AIPS and then underwent hybrid imaging in Difmap with both amplitude and phase self-calibration, they are plotted as filled circles. In the cases when the processing was the same except no amplitude self-calibration was made, the data are plotted as filled triangles. In the cases when no self-calibration was made for the source and the data calibrated in PIMA were used, they are plotted as open circles.}
\figsetgrpend

\figsetgrpstart
\figsetgrpnum{3.156}
\figsetgrptitle{J1257+8342 at 8.6 GHz}
\figsetplot{J1257+8342_X_2006_02_14_pet_uvt_rad.pdf}
\figsetgrpnote{Correlated flux density averaged over time and IFs versus the $uv$ radius. The source name, the observation date, and the frequency are given at the top of the panel. In the cases when the data for a source were calibrated in AIPS and then underwent hybrid imaging in Difmap with both amplitude and phase self-calibration, they are plotted as filled circles. In the cases when the processing was the same except no amplitude self-calibration was made, the data are plotted as filled triangles. In the cases when no self-calibration was made for the source and the data calibrated in PIMA were used, they are plotted as open circles.}
\figsetgrpend

\figsetgrpstart
\figsetgrpnum{3.157}
\figsetgrptitle{J1300+8054 at 2.3 GHz}
\figsetplot{J1300+8054_S_2006_02_16_pet_uvt_rad.pdf}
\figsetgrpnote{Correlated flux density averaged over time and IFs versus the $uv$ radius. The source name, the observation date, and the frequency are given at the top of the panel. In the cases when the data for a source were calibrated in AIPS and then underwent hybrid imaging in Difmap with both amplitude and phase self-calibration, they are plotted as filled circles. In the cases when the processing was the same except no amplitude self-calibration was made, the data are plotted as filled triangles. In the cases when no self-calibration was made for the source and the data calibrated in PIMA were used, they are plotted as open circles.}
\figsetgrpend

\figsetgrpstart
\figsetgrpnum{3.158}
\figsetgrptitle{J1305+7854 at 2.3 GHz}
\figsetplot{J1305+7854_S_2006_02_23_avp_uvs_rad.pdf}
\figsetgrpnote{Correlated flux density averaged over time and IFs versus the $uv$ radius. The source name, the observation date, and the frequency are given at the top of the panel. In the cases when the data for a source were calibrated in AIPS and then underwent hybrid imaging in Difmap with both amplitude and phase self-calibration, they are plotted as filled circles. In the cases when the processing was the same except no amplitude self-calibration was made, the data are plotted as filled triangles. In the cases when no self-calibration was made for the source and the data calibrated in PIMA were used, they are plotted as open circles.}
\figsetgrpend

\figsetgrpstart
\figsetgrpnum{3.159}
\figsetgrptitle{J1305+7854 at 8.6 GHz}
\figsetplot{J1305+7854_X_2006_02_23_avp_uvs_rad.pdf}
\figsetgrpnote{Correlated flux density averaged over time and IFs versus the $uv$ radius. The source name, the observation date, and the frequency are given at the top of the panel. In the cases when the data for a source were calibrated in AIPS and then underwent hybrid imaging in Difmap with both amplitude and phase self-calibration, they are plotted as filled circles. In the cases when the processing was the same except no amplitude self-calibration was made, the data are plotted as filled triangles. In the cases when no self-calibration was made for the source and the data calibrated in PIMA were used, they are plotted as open circles.}
\figsetgrpend

\figsetgrpstart
\figsetgrpnum{3.160}
\figsetgrptitle{J1305+8216 at 2.3 GHz}
\figsetplot{J1305+8216_S_2006_02_14_avp_uvs_rad.pdf}
\figsetgrpnote{Correlated flux density averaged over time and IFs versus the $uv$ radius. The source name, the observation date, and the frequency are given at the top of the panel. In the cases when the data for a source were calibrated in AIPS and then underwent hybrid imaging in Difmap with both amplitude and phase self-calibration, they are plotted as filled circles. In the cases when the processing was the same except no amplitude self-calibration was made, the data are plotted as filled triangles. In the cases when no self-calibration was made for the source and the data calibrated in PIMA were used, they are plotted as open circles.}
\figsetgrpend

\figsetgrpstart
\figsetgrpnum{3.161}
\figsetgrptitle{J1305+8216 at 8.6 GHz}
\figsetplot{J1305+8216_X_2006_02_14_pet_uvt_rad.pdf}
\figsetgrpnote{Correlated flux density averaged over time and IFs versus the $uv$ radius. The source name, the observation date, and the frequency are given at the top of the panel. In the cases when the data for a source were calibrated in AIPS and then underwent hybrid imaging in Difmap with both amplitude and phase self-calibration, they are plotted as filled circles. In the cases when the processing was the same except no amplitude self-calibration was made, the data are plotted as filled triangles. In the cases when no self-calibration was made for the source and the data calibrated in PIMA were used, they are plotted as open circles.}
\figsetgrpend

\figsetgrpstart
\figsetgrpnum{3.162}
\figsetgrptitle{J1306+8008 at 2.3 GHz}
\figsetplot{J1306+8008_S_2006_02_23_avp_uvs_rad.pdf}
\figsetgrpnote{Correlated flux density averaged over time and IFs versus the $uv$ radius. The source name, the observation date, and the frequency are given at the top of the panel. In the cases when the data for a source were calibrated in AIPS and then underwent hybrid imaging in Difmap with both amplitude and phase self-calibration, they are plotted as filled circles. In the cases when the processing was the same except no amplitude self-calibration was made, the data are plotted as filled triangles. In the cases when no self-calibration was made for the source and the data calibrated in PIMA were used, they are plotted as open circles.}
\figsetgrpend

\figsetgrpstart
\figsetgrpnum{3.163}
\figsetgrptitle{J1306+8008 at 8.6 GHz}
\figsetplot{J1306+8008_X_2006_02_23_avp_uvs_rad.pdf}
\figsetgrpnote{Correlated flux density averaged over time and IFs versus the $uv$ radius. The source name, the observation date, and the frequency are given at the top of the panel. In the cases when the data for a source were calibrated in AIPS and then underwent hybrid imaging in Difmap with both amplitude and phase self-calibration, they are plotted as filled circles. In the cases when the processing was the same except no amplitude self-calibration was made, the data are plotted as filled triangles. In the cases when no self-calibration was made for the source and the data calibrated in PIMA were used, they are plotted as open circles.}
\figsetgrpend

\figsetgrpstart
\figsetgrpnum{3.164}
\figsetgrptitle{J1306+8019 at 2.3 GHz}
\figsetplot{J1306+8019_S_2006_02_14_avp_uvs_rad.pdf}
\figsetgrpnote{Correlated flux density averaged over time and IFs versus the $uv$ radius. The source name, the observation date, and the frequency are given at the top of the panel. In the cases when the data for a source were calibrated in AIPS and then underwent hybrid imaging in Difmap with both amplitude and phase self-calibration, they are plotted as filled circles. In the cases when the processing was the same except no amplitude self-calibration was made, the data are plotted as filled triangles. In the cases when no self-calibration was made for the source and the data calibrated in PIMA were used, they are plotted as open circles.}
\figsetgrpend

\figsetgrpstart
\figsetgrpnum{3.165}
\figsetgrptitle{J1306+8019 at 8.6 GHz}
\figsetplot{J1306+8019_X_2006_02_14_pet_uvt_rad.pdf}
\figsetgrpnote{Correlated flux density averaged over time and IFs versus the $uv$ radius. The source name, the observation date, and the frequency are given at the top of the panel. In the cases when the data for a source were calibrated in AIPS and then underwent hybrid imaging in Difmap with both amplitude and phase self-calibration, they are plotted as filled circles. In the cases when the processing was the same except no amplitude self-calibration was made, the data are plotted as filled triangles. In the cases when no self-calibration was made for the source and the data calibrated in PIMA were used, they are plotted as open circles.}
\figsetgrpend

\figsetgrpstart
\figsetgrpnum{3.166}
\figsetgrptitle{J1307+7649 at 2.3 GHz}
\figsetplot{J1307+7649_S_2006_02_16_avp_uvs_rad.pdf}
\figsetgrpnote{Correlated flux density averaged over time and IFs versus the $uv$ radius. The source name, the observation date, and the frequency are given at the top of the panel. In the cases when the data for a source were calibrated in AIPS and then underwent hybrid imaging in Difmap with both amplitude and phase self-calibration, they are plotted as filled circles. In the cases when the processing was the same except no amplitude self-calibration was made, the data are plotted as filled triangles. In the cases when no self-calibration was made for the source and the data calibrated in PIMA were used, they are plotted as open circles.}
\figsetgrpend

\figsetgrpstart
\figsetgrpnum{3.167}
\figsetgrptitle{J1307+7649 at 8.6 GHz}
\figsetplot{J1307+7649_X_2006_02_16_pet_uvt_rad.pdf}
\figsetgrpnote{Correlated flux density averaged over time and IFs versus the $uv$ radius. The source name, the observation date, and the frequency are given at the top of the panel. In the cases when the data for a source were calibrated in AIPS and then underwent hybrid imaging in Difmap with both amplitude and phase self-calibration, they are plotted as filled circles. In the cases when the processing was the same except no amplitude self-calibration was made, the data are plotted as filled triangles. In the cases when no self-calibration was made for the source and the data calibrated in PIMA were used, they are plotted as open circles.}
\figsetgrpend

\figsetgrpstart
\figsetgrpnum{3.168}
\figsetgrptitle{J1320+8450 at 2.3 GHz}
\figsetplot{J1320+8450_S_2006_02_23_avp_uvs_rad.pdf}
\figsetgrpnote{Correlated flux density averaged over time and IFs versus the $uv$ radius. The source name, the observation date, and the frequency are given at the top of the panel. In the cases when the data for a source were calibrated in AIPS and then underwent hybrid imaging in Difmap with both amplitude and phase self-calibration, they are plotted as filled circles. In the cases when the processing was the same except no amplitude self-calibration was made, the data are plotted as filled triangles. In the cases when no self-calibration was made for the source and the data calibrated in PIMA were used, they are plotted as open circles.}
\figsetgrpend

\figsetgrpstart
\figsetgrpnum{3.169}
\figsetgrptitle{J1320+8450 at 8.6 GHz}
\figsetplot{J1320+8450_X_2006_02_23_avp_uvs_rad.pdf}
\figsetgrpnote{Correlated flux density averaged over time and IFs versus the $uv$ radius. The source name, the observation date, and the frequency are given at the top of the panel. In the cases when the data for a source were calibrated in AIPS and then underwent hybrid imaging in Difmap with both amplitude and phase self-calibration, they are plotted as filled circles. In the cases when the processing was the same except no amplitude self-calibration was made, the data are plotted as filled triangles. In the cases when no self-calibration was made for the source and the data calibrated in PIMA were used, they are plotted as open circles.}
\figsetgrpend

\figsetgrpstart
\figsetgrpnum{3.170}
\figsetgrptitle{J1321+8316 at 2.3 GHz}
\figsetplot{J1321+8316_S_2006_02_14_avp_uvs_rad.pdf}
\figsetgrpnote{Correlated flux density averaged over time and IFs versus the $uv$ radius. The source name, the observation date, and the frequency are given at the top of the panel. In the cases when the data for a source were calibrated in AIPS and then underwent hybrid imaging in Difmap with both amplitude and phase self-calibration, they are plotted as filled circles. In the cases when the processing was the same except no amplitude self-calibration was made, the data are plotted as filled triangles. In the cases when no self-calibration was made for the source and the data calibrated in PIMA were used, they are plotted as open circles.}
\figsetgrpend

\figsetgrpstart
\figsetgrpnum{3.171}
\figsetgrptitle{J1321+8316 at 8.6 GHz}
\figsetplot{J1321+8316_X_2006_02_14_avp_uvs_rad.pdf}
\figsetgrpnote{Correlated flux density averaged over time and IFs versus the $uv$ radius. The source name, the observation date, and the frequency are given at the top of the panel. In the cases when the data for a source were calibrated in AIPS and then underwent hybrid imaging in Difmap with both amplitude and phase self-calibration, they are plotted as filled circles. In the cases when the processing was the same except no amplitude self-calibration was made, the data are plotted as filled triangles. In the cases when no self-calibration was made for the source and the data calibrated in PIMA were used, they are plotted as open circles.}
\figsetgrpend

\figsetgrpstart
\figsetgrpnum{3.172}
\figsetgrptitle{J1323+7942 at 2.3 GHz}
\figsetplot{J1323+7942_S_2006_02_14_avp_uvs_rad.pdf}
\figsetgrpnote{Correlated flux density averaged over time and IFs versus the $uv$ radius. The source name, the observation date, and the frequency are given at the top of the panel. In the cases when the data for a source were calibrated in AIPS and then underwent hybrid imaging in Difmap with both amplitude and phase self-calibration, they are plotted as filled circles. In the cases when the processing was the same except no amplitude self-calibration was made, the data are plotted as filled triangles. In the cases when no self-calibration was made for the source and the data calibrated in PIMA were used, they are plotted as open circles.}
\figsetgrpend

\figsetgrpstart
\figsetgrpnum{3.173}
\figsetgrptitle{J1323+7942 at 8.6 GHz}
\figsetplot{J1323+7942_X_2006_02_14_avp_uvs_rad.pdf}
\figsetgrpnote{Correlated flux density averaged over time and IFs versus the $uv$ radius. The source name, the observation date, and the frequency are given at the top of the panel. In the cases when the data for a source were calibrated in AIPS and then underwent hybrid imaging in Difmap with both amplitude and phase self-calibration, they are plotted as filled circles. In the cases when the processing was the same except no amplitude self-calibration was made, the data are plotted as filled triangles. In the cases when no self-calibration was made for the source and the data calibrated in PIMA were used, they are plotted as open circles.}
\figsetgrpend

\figsetgrpstart
\figsetgrpnum{3.174}
\figsetgrptitle{J1325+7535 at 2.3 GHz}
\figsetplot{J1325+7535_S_2006_02_14_avp_uvs_rad.pdf}
\figsetgrpnote{Correlated flux density averaged over time and IFs versus the $uv$ radius. The source name, the observation date, and the frequency are given at the top of the panel. In the cases when the data for a source were calibrated in AIPS and then underwent hybrid imaging in Difmap with both amplitude and phase self-calibration, they are plotted as filled circles. In the cases when the processing was the same except no amplitude self-calibration was made, the data are plotted as filled triangles. In the cases when no self-calibration was made for the source and the data calibrated in PIMA were used, they are plotted as open circles.}
\figsetgrpend

\figsetgrpstart
\figsetgrpnum{3.175}
\figsetgrptitle{J1325+7535 at 8.6 GHz}
\figsetplot{J1325+7535_X_2006_02_14_pet_uvt_rad.pdf}
\figsetgrpnote{Correlated flux density averaged over time and IFs versus the $uv$ radius. The source name, the observation date, and the frequency are given at the top of the panel. In the cases when the data for a source were calibrated in AIPS and then underwent hybrid imaging in Difmap with both amplitude and phase self-calibration, they are plotted as filled circles. In the cases when the processing was the same except no amplitude self-calibration was made, the data are plotted as filled triangles. In the cases when no self-calibration was made for the source and the data calibrated in PIMA were used, they are plotted as open circles.}
\figsetgrpend

\figsetgrpstart
\figsetgrpnum{3.176}
\figsetgrptitle{J1344+7907 at 2.3 GHz}
\figsetplot{J1344+7907_S_2006_02_14_pet_uvt_rad.pdf}
\figsetgrpnote{Correlated flux density averaged over time and IFs versus the $uv$ radius. The source name, the observation date, and the frequency are given at the top of the panel. In the cases when the data for a source were calibrated in AIPS and then underwent hybrid imaging in Difmap with both amplitude and phase self-calibration, they are plotted as filled circles. In the cases when the processing was the same except no amplitude self-calibration was made, the data are plotted as filled triangles. In the cases when no self-calibration was made for the source and the data calibrated in PIMA were used, they are plotted as open circles.}
\figsetgrpend

\figsetgrpstart
\figsetgrpnum{3.177}
\figsetgrptitle{J1344+7907 at 8.6 GHz}
\figsetplot{J1344+7907_X_2006_02_14_pet_uvt_rad.pdf}
\figsetgrpnote{Correlated flux density averaged over time and IFs versus the $uv$ radius. The source name, the observation date, and the frequency are given at the top of the panel. In the cases when the data for a source were calibrated in AIPS and then underwent hybrid imaging in Difmap with both amplitude and phase self-calibration, they are plotted as filled circles. In the cases when the processing was the same except no amplitude self-calibration was made, the data are plotted as filled triangles. In the cases when no self-calibration was made for the source and the data calibrated in PIMA were used, they are plotted as open circles.}
\figsetgrpend

\figsetgrpstart
\figsetgrpnum{3.178}
\figsetgrptitle{J1357+7643 at 2.3 GHz}
\figsetplot{J1357+7643_S_2006_02_14_avp_uvs_rad.pdf}
\figsetgrpnote{Correlated flux density averaged over time and IFs versus the $uv$ radius. The source name, the observation date, and the frequency are given at the top of the panel. In the cases when the data for a source were calibrated in AIPS and then underwent hybrid imaging in Difmap with both amplitude and phase self-calibration, they are plotted as filled circles. In the cases when the processing was the same except no amplitude self-calibration was made, the data are plotted as filled triangles. In the cases when no self-calibration was made for the source and the data calibrated in PIMA were used, they are plotted as open circles.}
\figsetgrpend

\figsetgrpstart
\figsetgrpnum{3.179}
\figsetgrptitle{J1357+7643 at 8.6 GHz}
\figsetplot{J1357+7643_X_2006_02_14_avp_uvs_rad.pdf}
\figsetgrpnote{Correlated flux density averaged over time and IFs versus the $uv$ radius. The source name, the observation date, and the frequency are given at the top of the panel. In the cases when the data for a source were calibrated in AIPS and then underwent hybrid imaging in Difmap with both amplitude and phase self-calibration, they are plotted as filled circles. In the cases when the processing was the same except no amplitude self-calibration was made, the data are plotted as filled triangles. In the cases when no self-calibration was made for the source and the data calibrated in PIMA were used, they are plotted as open circles.}
\figsetgrpend

\figsetgrpstart
\figsetgrpnum{3.180}
\figsetgrptitle{J1358+8340 at 2.3 GHz}
\figsetplot{J1358+8340_S_2006_02_16_avp_uvs_rad.pdf}
\figsetgrpnote{Correlated flux density averaged over time and IFs versus the $uv$ radius. The source name, the observation date, and the frequency are given at the top of the panel. In the cases when the data for a source were calibrated in AIPS and then underwent hybrid imaging in Difmap with both amplitude and phase self-calibration, they are plotted as filled circles. In the cases when the processing was the same except no amplitude self-calibration was made, the data are plotted as filled triangles. In the cases when no self-calibration was made for the source and the data calibrated in PIMA were used, they are plotted as open circles.}
\figsetgrpend

\figsetgrpstart
\figsetgrpnum{3.181}
\figsetgrptitle{J1358+8340 at 8.6 GHz}
\figsetplot{J1358+8340_X_2006_02_16_avp_uvs_rad.pdf}
\figsetgrpnote{Correlated flux density averaged over time and IFs versus the $uv$ radius. The source name, the observation date, and the frequency are given at the top of the panel. In the cases when the data for a source were calibrated in AIPS and then underwent hybrid imaging in Difmap with both amplitude and phase self-calibration, they are plotted as filled circles. In the cases when the processing was the same except no amplitude self-calibration was made, the data are plotted as filled triangles. In the cases when no self-calibration was made for the source and the data calibrated in PIMA were used, they are plotted as open circles.}
\figsetgrpend

\figsetgrpstart
\figsetgrpnum{3.182}
\figsetgrptitle{J1406+7828 at 2.3 GHz}
\figsetplot{J1406+7828_S_2006_02_23_avp_uvs_rad.pdf}
\figsetgrpnote{Correlated flux density averaged over time and IFs versus the $uv$ radius. The source name, the observation date, and the frequency are given at the top of the panel. In the cases when the data for a source were calibrated in AIPS and then underwent hybrid imaging in Difmap with both amplitude and phase self-calibration, they are plotted as filled circles. In the cases when the processing was the same except no amplitude self-calibration was made, the data are plotted as filled triangles. In the cases when no self-calibration was made for the source and the data calibrated in PIMA were used, they are plotted as open circles.}
\figsetgrpend

\figsetgrpstart
\figsetgrpnum{3.183}
\figsetgrptitle{J1406+7828 at 8.6 GHz}
\figsetplot{J1406+7828_X_2006_02_23_avp_uvs_rad.pdf}
\figsetgrpnote{Correlated flux density averaged over time and IFs versus the $uv$ radius. The source name, the observation date, and the frequency are given at the top of the panel. In the cases when the data for a source were calibrated in AIPS and then underwent hybrid imaging in Difmap with both amplitude and phase self-calibration, they are plotted as filled circles. In the cases when the processing was the same except no amplitude self-calibration was made, the data are plotted as filled triangles. In the cases when no self-calibration was made for the source and the data calibrated in PIMA were used, they are plotted as open circles.}
\figsetgrpend

\figsetgrpstart
\figsetgrpnum{3.184}
\figsetgrptitle{J1414+7905 at 2.3 GHz}
\figsetplot{J1414+7905_S_2006_02_14_pet_uvt_rad.pdf}
\figsetgrpnote{Correlated flux density averaged over time and IFs versus the $uv$ radius. The source name, the observation date, and the frequency are given at the top of the panel. In the cases when the data for a source were calibrated in AIPS and then underwent hybrid imaging in Difmap with both amplitude and phase self-calibration, they are plotted as filled circles. In the cases when the processing was the same except no amplitude self-calibration was made, the data are plotted as filled triangles. In the cases when no self-calibration was made for the source and the data calibrated in PIMA were used, they are plotted as open circles.}
\figsetgrpend

\figsetgrpstart
\figsetgrpnum{3.185}
\figsetgrptitle{J1421+7513 at 2.3 GHz}
\figsetplot{J1421+7513_S_2006_02_23_avp_uvs_rad.pdf}
\figsetgrpnote{Correlated flux density averaged over time and IFs versus the $uv$ radius. The source name, the observation date, and the frequency are given at the top of the panel. In the cases when the data for a source were calibrated in AIPS and then underwent hybrid imaging in Difmap with both amplitude and phase self-calibration, they are plotted as filled circles. In the cases when the processing was the same except no amplitude self-calibration was made, the data are plotted as filled triangles. In the cases when no self-calibration was made for the source and the data calibrated in PIMA were used, they are plotted as open circles.}
\figsetgrpend

\figsetgrpstart
\figsetgrpnum{3.186}
\figsetgrptitle{J1421+7513 at 8.6 GHz}
\figsetplot{J1421+7513_X_2006_02_23_pet_uvt_rad.pdf}
\figsetgrpnote{Correlated flux density averaged over time and IFs versus the $uv$ radius. The source name, the observation date, and the frequency are given at the top of the panel. In the cases when the data for a source were calibrated in AIPS and then underwent hybrid imaging in Difmap with both amplitude and phase self-calibration, they are plotted as filled circles. In the cases when the processing was the same except no amplitude self-calibration was made, the data are plotted as filled triangles. In the cases when no self-calibration was made for the source and the data calibrated in PIMA were used, they are plotted as open circles.}
\figsetgrpend

\figsetgrpstart
\figsetgrpnum{3.187}
\figsetgrptitle{J1422+7704 at 2.3 GHz}
\figsetplot{J1422+7704_S_2006_02_16_pet_uvt_rad.pdf}
\figsetgrpnote{Correlated flux density averaged over time and IFs versus the $uv$ radius. The source name, the observation date, and the frequency are given at the top of the panel. In the cases when the data for a source were calibrated in AIPS and then underwent hybrid imaging in Difmap with both amplitude and phase self-calibration, they are plotted as filled circles. In the cases when the processing was the same except no amplitude self-calibration was made, the data are plotted as filled triangles. In the cases when no self-calibration was made for the source and the data calibrated in PIMA were used, they are plotted as open circles.}
\figsetgrpend

\figsetgrpstart
\figsetgrpnum{3.188}
\figsetgrptitle{J1435+7605 at 2.3 GHz}
\figsetplot{J1435+7605_S_2006_02_14_avp_uvs_rad.pdf}
\figsetgrpnote{Correlated flux density averaged over time and IFs versus the $uv$ radius. The source name, the observation date, and the frequency are given at the top of the panel. In the cases when the data for a source were calibrated in AIPS and then underwent hybrid imaging in Difmap with both amplitude and phase self-calibration, they are plotted as filled circles. In the cases when the processing was the same except no amplitude self-calibration was made, the data are plotted as filled triangles. In the cases when no self-calibration was made for the source and the data calibrated in PIMA were used, they are plotted as open circles.}
\figsetgrpend

\figsetgrpstart
\figsetgrpnum{3.189}
\figsetgrptitle{J1435+7605 at 8.6 GHz}
\figsetplot{J1435+7605_X_2006_02_14_pet_uvt_rad.pdf}
\figsetgrpnote{Correlated flux density averaged over time and IFs versus the $uv$ radius. The source name, the observation date, and the frequency are given at the top of the panel. In the cases when the data for a source were calibrated in AIPS and then underwent hybrid imaging in Difmap with both amplitude and phase self-calibration, they are plotted as filled circles. In the cases when the processing was the same except no amplitude self-calibration was made, the data are plotted as filled triangles. In the cases when no self-calibration was made for the source and the data calibrated in PIMA were used, they are plotted as open circles.}
\figsetgrpend

\figsetgrpstart
\figsetgrpnum{3.190}
\figsetgrptitle{J1443+7707 at 2.3 GHz}
\figsetplot{J1443+7707_S_2006_02_23_pet_uvt_rad.pdf}
\figsetgrpnote{Correlated flux density averaged over time and IFs versus the $uv$ radius. The source name, the observation date, and the frequency are given at the top of the panel. In the cases when the data for a source were calibrated in AIPS and then underwent hybrid imaging in Difmap with both amplitude and phase self-calibration, they are plotted as filled circles. In the cases when the processing was the same except no amplitude self-calibration was made, the data are plotted as filled triangles. In the cases when no self-calibration was made for the source and the data calibrated in PIMA were used, they are plotted as open circles.}
\figsetgrpend

\figsetgrpstart
\figsetgrpnum{3.191}
\figsetgrptitle{J1447+7656 at 2.3 GHz}
\figsetplot{J1447+7656_S_2006_02_16_pet_uvt_rad.pdf}
\figsetgrpnote{Correlated flux density averaged over time and IFs versus the $uv$ radius. The source name, the observation date, and the frequency are given at the top of the panel. In the cases when the data for a source were calibrated in AIPS and then underwent hybrid imaging in Difmap with both amplitude and phase self-calibration, they are plotted as filled circles. In the cases when the processing was the same except no amplitude self-calibration was made, the data are plotted as filled triangles. In the cases when no self-calibration was made for the source and the data calibrated in PIMA were used, they are plotted as open circles.}
\figsetgrpend

\figsetgrpstart
\figsetgrpnum{3.192}
\figsetgrptitle{J1506+8319 at 2.3 GHz}
\figsetplot{J1506+8319_S_2006_02_16_avp_uvs_rad.pdf}
\figsetgrpnote{Correlated flux density averaged over time and IFs versus the $uv$ radius. The source name, the observation date, and the frequency are given at the top of the panel. In the cases when the data for a source were calibrated in AIPS and then underwent hybrid imaging in Difmap with both amplitude and phase self-calibration, they are plotted as filled circles. In the cases when the processing was the same except no amplitude self-calibration was made, the data are plotted as filled triangles. In the cases when no self-calibration was made for the source and the data calibrated in PIMA were used, they are plotted as open circles.}
\figsetgrpend

\figsetgrpstart
\figsetgrpnum{3.193}
\figsetgrptitle{J1506+8319 at 8.6 GHz}
\figsetplot{J1506+8319_X_2006_02_16_avp_uvs_rad.pdf}
\figsetgrpnote{Correlated flux density averaged over time and IFs versus the $uv$ radius. The source name, the observation date, and the frequency are given at the top of the panel. In the cases when the data for a source were calibrated in AIPS and then underwent hybrid imaging in Difmap with both amplitude and phase self-calibration, they are plotted as filled circles. In the cases when the processing was the same except no amplitude self-calibration was made, the data are plotted as filled triangles. In the cases when no self-calibration was made for the source and the data calibrated in PIMA were used, they are plotted as open circles.}
\figsetgrpend

\figsetgrpstart
\figsetgrpnum{3.194}
\figsetgrptitle{J1522+7645 at 2.3 GHz}
\figsetplot{J1522+7645_S_2006_02_16_pet_uvt_rad.pdf}
\figsetgrpnote{Correlated flux density averaged over time and IFs versus the $uv$ radius. The source name, the observation date, and the frequency are given at the top of the panel. In the cases when the data for a source were calibrated in AIPS and then underwent hybrid imaging in Difmap with both amplitude and phase self-calibration, they are plotted as filled circles. In the cases when the processing was the same except no amplitude self-calibration was made, the data are plotted as filled triangles. In the cases when no self-calibration was made for the source and the data calibrated in PIMA were used, they are plotted as open circles.}
\figsetgrpend

\figsetgrpstart
\figsetgrpnum{3.195}
\figsetgrptitle{J1522+7645 at 8.6 GHz}
\figsetplot{J1522+7645_X_2006_02_16_pet_uvt_rad.pdf}
\figsetgrpnote{Correlated flux density averaged over time and IFs versus the $uv$ radius. The source name, the observation date, and the frequency are given at the top of the panel. In the cases when the data for a source were calibrated in AIPS and then underwent hybrid imaging in Difmap with both amplitude and phase self-calibration, they are plotted as filled circles. In the cases when the processing was the same except no amplitude self-calibration was made, the data are plotted as filled triangles. In the cases when no self-calibration was made for the source and the data calibrated in PIMA were used, they are plotted as open circles.}
\figsetgrpend

\figsetgrpstart
\figsetgrpnum{3.196}
\figsetgrptitle{J1523+7645 at 2.3 GHz}
\figsetplot{J1523+7645_S_2006_02_16_pet_uvt_rad.pdf}
\figsetgrpnote{Correlated flux density averaged over time and IFs versus the $uv$ radius. The source name, the observation date, and the frequency are given at the top of the panel. In the cases when the data for a source were calibrated in AIPS and then underwent hybrid imaging in Difmap with both amplitude and phase self-calibration, they are plotted as filled circles. In the cases when the processing was the same except no amplitude self-calibration was made, the data are plotted as filled triangles. In the cases when no self-calibration was made for the source and the data calibrated in PIMA were used, they are plotted as open circles.}
\figsetgrpend

\figsetgrpstart
\figsetgrpnum{3.197}
\figsetgrptitle{J1528+8217 at 2.3 GHz}
\figsetplot{J1528+8217_S_2006_02_14_avp_uvs_rad.pdf}
\figsetgrpnote{Correlated flux density averaged over time and IFs versus the $uv$ radius. The source name, the observation date, and the frequency are given at the top of the panel. In the cases when the data for a source were calibrated in AIPS and then underwent hybrid imaging in Difmap with both amplitude and phase self-calibration, they are plotted as filled circles. In the cases when the processing was the same except no amplitude self-calibration was made, the data are plotted as filled triangles. In the cases when no self-calibration was made for the source and the data calibrated in PIMA were used, they are plotted as open circles.}
\figsetgrpend

\figsetgrpstart
\figsetgrpnum{3.198}
\figsetgrptitle{J1528+8217 at 8.6 GHz}
\figsetplot{J1528+8217_X_2006_02_14_pet_uvt_rad.pdf}
\figsetgrpnote{Correlated flux density averaged over time and IFs versus the $uv$ radius. The source name, the observation date, and the frequency are given at the top of the panel. In the cases when the data for a source were calibrated in AIPS and then underwent hybrid imaging in Difmap with both amplitude and phase self-calibration, they are plotted as filled circles. In the cases when the processing was the same except no amplitude self-calibration was made, the data are plotted as filled triangles. In the cases when no self-calibration was made for the source and the data calibrated in PIMA were used, they are plotted as open circles.}
\figsetgrpend

\figsetgrpstart
\figsetgrpnum{3.199}
\figsetgrptitle{J1531+7706 at 2.3 GHz}
\figsetplot{J1531+7706_S_2006_02_14_avp_uvs_rad.pdf}
\figsetgrpnote{Correlated flux density averaged over time and IFs versus the $uv$ radius. The source name, the observation date, and the frequency are given at the top of the panel. In the cases when the data for a source were calibrated in AIPS and then underwent hybrid imaging in Difmap with both amplitude and phase self-calibration, they are plotted as filled circles. In the cases when the processing was the same except no amplitude self-calibration was made, the data are plotted as filled triangles. In the cases when no self-calibration was made for the source and the data calibrated in PIMA were used, they are plotted as open circles.}
\figsetgrpend

\figsetgrpstart
\figsetgrpnum{3.200}
\figsetgrptitle{J1531+7706 at 8.6 GHz}
\figsetplot{J1531+7706_X_2006_02_14_pet_uvt_rad.pdf}
\figsetgrpnote{Correlated flux density averaged over time and IFs versus the $uv$ radius. The source name, the observation date, and the frequency are given at the top of the panel. In the cases when the data for a source were calibrated in AIPS and then underwent hybrid imaging in Difmap with both amplitude and phase self-calibration, they are plotted as filled circles. In the cases when the processing was the same except no amplitude self-calibration was made, the data are plotted as filled triangles. In the cases when no self-calibration was made for the source and the data calibrated in PIMA were used, they are plotted as open circles.}
\figsetgrpend

\figsetgrpstart
\figsetgrpnum{3.201}
\figsetgrptitle{J1537+8154 at 2.3 GHz}
\figsetplot{J1537+8154_S_2006_02_16_avp_uvs_rad.pdf}
\figsetgrpnote{Correlated flux density averaged over time and IFs versus the $uv$ radius. The source name, the observation date, and the frequency are given at the top of the panel. In the cases when the data for a source were calibrated in AIPS and then underwent hybrid imaging in Difmap with both amplitude and phase self-calibration, they are plotted as filled circles. In the cases when the processing was the same except no amplitude self-calibration was made, the data are plotted as filled triangles. In the cases when no self-calibration was made for the source and the data calibrated in PIMA were used, they are plotted as open circles.}
\figsetgrpend

\figsetgrpstart
\figsetgrpnum{3.202}
\figsetgrptitle{J1537+8154 at 8.6 GHz}
\figsetplot{J1537+8154_X_2006_02_16_avp_uvs_rad.pdf}
\figsetgrpnote{Correlated flux density averaged over time and IFs versus the $uv$ radius. The source name, the observation date, and the frequency are given at the top of the panel. In the cases when the data for a source were calibrated in AIPS and then underwent hybrid imaging in Difmap with both amplitude and phase self-calibration, they are plotted as filled circles. In the cases when the processing was the same except no amplitude self-calibration was made, the data are plotted as filled triangles. In the cases when no self-calibration was made for the source and the data calibrated in PIMA were used, they are plotted as open circles.}
\figsetgrpend

\figsetgrpstart
\figsetgrpnum{3.203}
\figsetgrptitle{J1539+7814 at 2.3 GHz}
\figsetplot{J1539+7814_S_2006_02_23_pet_uvt_rad.pdf}
\figsetgrpnote{Correlated flux density averaged over time and IFs versus the $uv$ radius. The source name, the observation date, and the frequency are given at the top of the panel. In the cases when the data for a source were calibrated in AIPS and then underwent hybrid imaging in Difmap with both amplitude and phase self-calibration, they are plotted as filled circles. In the cases when the processing was the same except no amplitude self-calibration was made, the data are plotted as filled triangles. In the cases when no self-calibration was made for the source and the data calibrated in PIMA were used, they are plotted as open circles.}
\figsetgrpend

\figsetgrpstart
\figsetgrpnum{3.204}
\figsetgrptitle{J1606+7658 at 8.6 GHz}
\figsetplot{J1606+7658_X_2006_02_23_pet_uvt_rad.pdf}
\figsetgrpnote{Correlated flux density averaged over time and IFs versus the $uv$ radius. The source name, the observation date, and the frequency are given at the top of the panel. In the cases when the data for a source were calibrated in AIPS and then underwent hybrid imaging in Difmap with both amplitude and phase self-calibration, they are plotted as filled circles. In the cases when the processing was the same except no amplitude self-calibration was made, the data are plotted as filled triangles. In the cases when no self-calibration was made for the source and the data calibrated in PIMA were used, they are plotted as open circles.}
\figsetgrpend

\figsetgrpstart
\figsetgrpnum{3.205}
\figsetgrptitle{J1607+8501 at 2.3 GHz}
\figsetplot{J1607+8501_S_2006_02_23_avp_uvs_rad.pdf}
\figsetgrpnote{Correlated flux density averaged over time and IFs versus the $uv$ radius. The source name, the observation date, and the frequency are given at the top of the panel. In the cases when the data for a source were calibrated in AIPS and then underwent hybrid imaging in Difmap with both amplitude and phase self-calibration, they are plotted as filled circles. In the cases when the processing was the same except no amplitude self-calibration was made, the data are plotted as filled triangles. In the cases when no self-calibration was made for the source and the data calibrated in PIMA were used, they are plotted as open circles.}
\figsetgrpend

\figsetgrpstart
\figsetgrpnum{3.206}
\figsetgrptitle{J1607+8501 at 8.6 GHz}
\figsetplot{J1607+8501_X_2006_02_23_pet_uvt_rad.pdf}
\figsetgrpnote{Correlated flux density averaged over time and IFs versus the $uv$ radius. The source name, the observation date, and the frequency are given at the top of the panel. In the cases when the data for a source were calibrated in AIPS and then underwent hybrid imaging in Difmap with both amplitude and phase self-calibration, they are plotted as filled circles. In the cases when the processing was the same except no amplitude self-calibration was made, the data are plotted as filled triangles. In the cases when no self-calibration was made for the source and the data calibrated in PIMA were used, they are plotted as open circles.}
\figsetgrpend

\figsetgrpstart
\figsetgrpnum{3.207}
\figsetgrptitle{J1609+7939 at 2.3 GHz}
\figsetplot{J1609+7939_S_2006_02_14_avp_uvs_rad.pdf}
\figsetgrpnote{Correlated flux density averaged over time and IFs versus the $uv$ radius. The source name, the observation date, and the frequency are given at the top of the panel. In the cases when the data for a source were calibrated in AIPS and then underwent hybrid imaging in Difmap with both amplitude and phase self-calibration, they are plotted as filled circles. In the cases when the processing was the same except no amplitude self-calibration was made, the data are plotted as filled triangles. In the cases when no self-calibration was made for the source and the data calibrated in PIMA were used, they are plotted as open circles.}
\figsetgrpend

\figsetgrpstart
\figsetgrpnum{3.208}
\figsetgrptitle{J1609+7939 at 8.6 GHz}
\figsetplot{J1609+7939_X_2006_02_14_pet_uvt_rad.pdf}
\figsetgrpnote{Correlated flux density averaged over time and IFs versus the $uv$ radius. The source name, the observation date, and the frequency are given at the top of the panel. In the cases when the data for a source were calibrated in AIPS and then underwent hybrid imaging in Difmap with both amplitude and phase self-calibration, they are plotted as filled circles. In the cases when the processing was the same except no amplitude self-calibration was made, the data are plotted as filled triangles. In the cases when no self-calibration was made for the source and the data calibrated in PIMA were used, they are plotted as open circles.}
\figsetgrpend

\figsetgrpstart
\figsetgrpnum{3.209}
\figsetgrptitle{J1619+8549 at 2.3 GHz}
\figsetplot{J1619+8549_S_2006_02_14_pet_uvt_rad.pdf}
\figsetgrpnote{Correlated flux density averaged over time and IFs versus the $uv$ radius. The source name, the observation date, and the frequency are given at the top of the panel. In the cases when the data for a source were calibrated in AIPS and then underwent hybrid imaging in Difmap with both amplitude and phase self-calibration, they are plotted as filled circles. In the cases when the processing was the same except no amplitude self-calibration was made, the data are plotted as filled triangles. In the cases when no self-calibration was made for the source and the data calibrated in PIMA were used, they are plotted as open circles.}
\figsetgrpend

\figsetgrpstart
\figsetgrpnum{3.210}
\figsetgrptitle{J1632+8232 at 2.3 GHz}
\figsetplot{J1632+8232_S_2006_02_23_avp_uvs_rad.pdf}
\figsetgrpnote{Correlated flux density averaged over time and IFs versus the $uv$ radius. The source name, the observation date, and the frequency are given at the top of the panel. In the cases when the data for a source were calibrated in AIPS and then underwent hybrid imaging in Difmap with both amplitude and phase self-calibration, they are plotted as filled circles. In the cases when the processing was the same except no amplitude self-calibration was made, the data are plotted as filled triangles. In the cases when no self-calibration was made for the source and the data calibrated in PIMA were used, they are plotted as open circles.}
\figsetgrpend

\figsetgrpstart
\figsetgrpnum{3.211}
\figsetgrptitle{J1632+8232 at 8.6 GHz}
\figsetplot{J1632+8232_X_2006_02_23_avp_uvs_rad.pdf}
\figsetgrpnote{Correlated flux density averaged over time and IFs versus the $uv$ radius. The source name, the observation date, and the frequency are given at the top of the panel. In the cases when the data for a source were calibrated in AIPS and then underwent hybrid imaging in Difmap with both amplitude and phase self-calibration, they are plotted as filled circles. In the cases when the processing was the same except no amplitude self-calibration was made, the data are plotted as filled triangles. In the cases when no self-calibration was made for the source and the data calibrated in PIMA were used, they are plotted as open circles.}
\figsetgrpend

\figsetgrpstart
\figsetgrpnum{3.212}
\figsetgrptitle{J1639+8631 at 2.3 GHz}
\figsetplot{J1639+8631_S_2006_02_14_avp_uvs_rad.pdf}
\figsetgrpnote{Correlated flux density averaged over time and IFs versus the $uv$ radius. The source name, the observation date, and the frequency are given at the top of the panel. In the cases when the data for a source were calibrated in AIPS and then underwent hybrid imaging in Difmap with both amplitude and phase self-calibration, they are plotted as filled circles. In the cases when the processing was the same except no amplitude self-calibration was made, the data are plotted as filled triangles. In the cases when no self-calibration was made for the source and the data calibrated in PIMA were used, they are plotted as open circles.}
\figsetgrpend

\figsetgrpstart
\figsetgrpnum{3.213}
\figsetgrptitle{J1639+8631 at 8.6 GHz}
\figsetplot{J1639+8631_X_2006_02_14_avp_uvs_rad.pdf}
\figsetgrpnote{Correlated flux density averaged over time and IFs versus the $uv$ radius. The source name, the observation date, and the frequency are given at the top of the panel. In the cases when the data for a source were calibrated in AIPS and then underwent hybrid imaging in Difmap with both amplitude and phase self-calibration, they are plotted as filled circles. In the cases when the processing was the same except no amplitude self-calibration was made, the data are plotted as filled triangles. In the cases when no self-calibration was made for the source and the data calibrated in PIMA were used, they are plotted as open circles.}
\figsetgrpend

\figsetgrpstart
\figsetgrpnum{3.214}
\figsetgrptitle{J1648+7546 at 2.3 GHz}
\figsetplot{J1648+7546_S_2006_02_14_pet_uvt_rad.pdf}
\figsetgrpnote{Correlated flux density averaged over time and IFs versus the $uv$ radius. The source name, the observation date, and the frequency are given at the top of the panel. In the cases when the data for a source were calibrated in AIPS and then underwent hybrid imaging in Difmap with both amplitude and phase self-calibration, they are plotted as filled circles. In the cases when the processing was the same except no amplitude self-calibration was made, the data are plotted as filled triangles. In the cases when no self-calibration was made for the source and the data calibrated in PIMA were used, they are plotted as open circles.}
\figsetgrpend

\figsetgrpstart
\figsetgrpnum{3.215}
\figsetgrptitle{J1705+7756 at 2.3 GHz}
\figsetplot{J1705+7756_S_2006_02_16_avp_uvs_rad.pdf}
\figsetgrpnote{Correlated flux density averaged over time and IFs versus the $uv$ radius. The source name, the observation date, and the frequency are given at the top of the panel. In the cases when the data for a source were calibrated in AIPS and then underwent hybrid imaging in Difmap with both amplitude and phase self-calibration, they are plotted as filled circles. In the cases when the processing was the same except no amplitude self-calibration was made, the data are plotted as filled triangles. In the cases when no self-calibration was made for the source and the data calibrated in PIMA were used, they are plotted as open circles.}
\figsetgrpend

\figsetgrpstart
\figsetgrpnum{3.216}
\figsetgrptitle{J1705+7756 at 8.6 GHz}
\figsetplot{J1705+7756_X_2006_02_16_avp_uvs_rad.pdf}
\figsetgrpnote{Correlated flux density averaged over time and IFs versus the $uv$ radius. The source name, the observation date, and the frequency are given at the top of the panel. In the cases when the data for a source were calibrated in AIPS and then underwent hybrid imaging in Difmap with both amplitude and phase self-calibration, they are plotted as filled circles. In the cases when the processing was the same except no amplitude self-calibration was made, the data are plotted as filled triangles. In the cases when no self-calibration was made for the source and the data calibrated in PIMA were used, they are plotted as open circles.}
\figsetgrpend

\figsetgrpstart
\figsetgrpnum{3.217}
\figsetgrptitle{J1723+7653 at 2.3 GHz}
\figsetplot{J1723+7653_S_2006_02_23_avp_uvs_rad.pdf}
\figsetgrpnote{Correlated flux density averaged over time and IFs versus the $uv$ radius. The source name, the observation date, and the frequency are given at the top of the panel. In the cases when the data for a source were calibrated in AIPS and then underwent hybrid imaging in Difmap with both amplitude and phase self-calibration, they are plotted as filled circles. In the cases when the processing was the same except no amplitude self-calibration was made, the data are plotted as filled triangles. In the cases when no self-calibration was made for the source and the data calibrated in PIMA were used, they are plotted as open circles.}
\figsetgrpend

\figsetgrpstart
\figsetgrpnum{3.218}
\figsetgrptitle{J1723+7653 at 8.6 GHz}
\figsetplot{J1723+7653_X_2006_02_23_avp_uvs_rad.pdf}
\figsetgrpnote{Correlated flux density averaged over time and IFs versus the $uv$ radius. The source name, the observation date, and the frequency are given at the top of the panel. In the cases when the data for a source were calibrated in AIPS and then underwent hybrid imaging in Difmap with both amplitude and phase self-calibration, they are plotted as filled circles. In the cases when the processing was the same except no amplitude self-calibration was made, the data are plotted as filled triangles. In the cases when no self-calibration was made for the source and the data calibrated in PIMA were used, they are plotted as open circles.}
\figsetgrpend

\figsetgrpstart
\figsetgrpnum{3.219}
\figsetgrptitle{J1725+7708 at 2.3 GHz}
\figsetplot{J1725+7708_S_2006_02_14_avp_uvs_rad.pdf}
\figsetgrpnote{Correlated flux density averaged over time and IFs versus the $uv$ radius. The source name, the observation date, and the frequency are given at the top of the panel. In the cases when the data for a source were calibrated in AIPS and then underwent hybrid imaging in Difmap with both amplitude and phase self-calibration, they are plotted as filled circles. In the cases when the processing was the same except no amplitude self-calibration was made, the data are plotted as filled triangles. In the cases when no self-calibration was made for the source and the data calibrated in PIMA were used, they are plotted as open circles.}
\figsetgrpend

\figsetgrpstart
\figsetgrpnum{3.220}
\figsetgrptitle{J1725+7708 at 8.6 GHz}
\figsetplot{J1725+7708_X_2006_02_14_avp_uvs_rad.pdf}
\figsetgrpnote{Correlated flux density averaged over time and IFs versus the $uv$ radius. The source name, the observation date, and the frequency are given at the top of the panel. In the cases when the data for a source were calibrated in AIPS and then underwent hybrid imaging in Difmap with both amplitude and phase self-calibration, they are plotted as filled circles. In the cases when the processing was the same except no amplitude self-calibration was made, the data are plotted as filled triangles. In the cases when no self-calibration was made for the source and the data calibrated in PIMA were used, they are plotted as open circles.}
\figsetgrpend

\figsetgrpstart
\figsetgrpnum{3.221}
\figsetgrptitle{J1725+7726 at 2.3 GHz}
\figsetplot{J1725+7726_S_2006_02_16_avp_uvs_rad.pdf}
\figsetgrpnote{Correlated flux density averaged over time and IFs versus the $uv$ radius. The source name, the observation date, and the frequency are given at the top of the panel. In the cases when the data for a source were calibrated in AIPS and then underwent hybrid imaging in Difmap with both amplitude and phase self-calibration, they are plotted as filled circles. In the cases when the processing was the same except no amplitude self-calibration was made, the data are plotted as filled triangles. In the cases when no self-calibration was made for the source and the data calibrated in PIMA were used, they are plotted as open circles.}
\figsetgrpend

\figsetgrpstart
\figsetgrpnum{3.222}
\figsetgrptitle{J1725+7726 at 8.6 GHz}
\figsetplot{J1725+7726_X_2006_02_16_pet_uvt_rad.pdf}
\figsetgrpnote{Correlated flux density averaged over time and IFs versus the $uv$ radius. The source name, the observation date, and the frequency are given at the top of the panel. In the cases when the data for a source were calibrated in AIPS and then underwent hybrid imaging in Difmap with both amplitude and phase self-calibration, they are plotted as filled circles. In the cases when the processing was the same except no amplitude self-calibration was made, the data are plotted as filled triangles. In the cases when no self-calibration was made for the source and the data calibrated in PIMA were used, they are plotted as open circles.}
\figsetgrpend

\figsetgrpstart
\figsetgrpnum{3.223}
\figsetgrptitle{J1759+7507 at 2.3 GHz}
\figsetplot{J1759+7507_S_2006_02_14_pet_uvt_rad.pdf}
\figsetgrpnote{Correlated flux density averaged over time and IFs versus the $uv$ radius. The source name, the observation date, and the frequency are given at the top of the panel. In the cases when the data for a source were calibrated in AIPS and then underwent hybrid imaging in Difmap with both amplitude and phase self-calibration, they are plotted as filled circles. In the cases when the processing was the same except no amplitude self-calibration was made, the data are plotted as filled triangles. In the cases when no self-calibration was made for the source and the data calibrated in PIMA were used, they are plotted as open circles.}
\figsetgrpend

\figsetgrpstart
\figsetgrpnum{3.224}
\figsetgrptitle{J1800+7828 at 2.3 GHz}
\figsetplot{J1800+7828_S_2006_02_16_avp_uvs_rad.pdf}
\figsetgrpnote{Correlated flux density averaged over time and IFs versus the $uv$ radius. The source name, the observation date, and the frequency are given at the top of the panel. In the cases when the data for a source were calibrated in AIPS and then underwent hybrid imaging in Difmap with both amplitude and phase self-calibration, they are plotted as filled circles. In the cases when the processing was the same except no amplitude self-calibration was made, the data are plotted as filled triangles. In the cases when no self-calibration was made for the source and the data calibrated in PIMA were used, they are plotted as open circles.}
\figsetgrpend

\figsetgrpstart
\figsetgrpnum{3.225}
\figsetgrptitle{J1800+7828 at 8.6 GHz}
\figsetplot{J1800+7828_X_2006_02_16_avp_uvs_rad.pdf}
\figsetgrpnote{Correlated flux density averaged over time and IFs versus the $uv$ radius. The source name, the observation date, and the frequency are given at the top of the panel. In the cases when the data for a source were calibrated in AIPS and then underwent hybrid imaging in Difmap with both amplitude and phase self-calibration, they are plotted as filled circles. In the cases when the processing was the same except no amplitude self-calibration was made, the data are plotted as filled triangles. In the cases when no self-calibration was made for the source and the data calibrated in PIMA were used, they are plotted as open circles.}
\figsetgrpend

\figsetgrpstart
\figsetgrpnum{3.226}
\figsetgrptitle{J1803+7601 at 2.3 GHz}
\figsetplot{J1803+7601_S_2006_02_23_avp_uvs_rad.pdf}
\figsetgrpnote{Correlated flux density averaged over time and IFs versus the $uv$ radius. The source name, the observation date, and the frequency are given at the top of the panel. In the cases when the data for a source were calibrated in AIPS and then underwent hybrid imaging in Difmap with both amplitude and phase self-calibration, they are plotted as filled circles. In the cases when the processing was the same except no amplitude self-calibration was made, the data are plotted as filled triangles. In the cases when no self-calibration was made for the source and the data calibrated in PIMA were used, they are plotted as open circles.}
\figsetgrpend

\figsetgrpstart
\figsetgrpnum{3.227}
\figsetgrptitle{J1803+7601 at 8.6 GHz}
\figsetplot{J1803+7601_X_2006_02_23_pet_uvt_rad.pdf}
\figsetgrpnote{Correlated flux density averaged over time and IFs versus the $uv$ radius. The source name, the observation date, and the frequency are given at the top of the panel. In the cases when the data for a source were calibrated in AIPS and then underwent hybrid imaging in Difmap with both amplitude and phase self-calibration, they are plotted as filled circles. In the cases when the processing was the same except no amplitude self-calibration was made, the data are plotted as filled triangles. In the cases when no self-calibration was made for the source and the data calibrated in PIMA were used, they are plotted as open circles.}
\figsetgrpend

\figsetgrpstart
\figsetgrpnum{3.228}
\figsetgrptitle{J1822+8257 at 2.3 GHz}
\figsetplot{J1822+8257_S_2006_02_23_avp_uvs_rad.pdf}
\figsetgrpnote{Correlated flux density averaged over time and IFs versus the $uv$ radius. The source name, the observation date, and the frequency are given at the top of the panel. In the cases when the data for a source were calibrated in AIPS and then underwent hybrid imaging in Difmap with both amplitude and phase self-calibration, they are plotted as filled circles. In the cases when the processing was the same except no amplitude self-calibration was made, the data are plotted as filled triangles. In the cases when no self-calibration was made for the source and the data calibrated in PIMA were used, they are plotted as open circles.}
\figsetgrpend

\figsetgrpstart
\figsetgrpnum{3.229}
\figsetgrptitle{J1822+8257 at 8.6 GHz}
\figsetplot{J1822+8257_X_2006_02_23_avp_uvs_rad.pdf}
\figsetgrpnote{Correlated flux density averaged over time and IFs versus the $uv$ radius. The source name, the observation date, and the frequency are given at the top of the panel. In the cases when the data for a source were calibrated in AIPS and then underwent hybrid imaging in Difmap with both amplitude and phase self-calibration, they are plotted as filled circles. In the cases when the processing was the same except no amplitude self-calibration was made, the data are plotted as filled triangles. In the cases when no self-calibration was made for the source and the data calibrated in PIMA were used, they are plotted as open circles.}
\figsetgrpend

\figsetgrpstart
\figsetgrpnum{3.230}
\figsetgrptitle{J1823+7938 at 2.3 GHz}
\figsetplot{J1823+7938_S_2006_02_23_avp_uvs_rad.pdf}
\figsetgrpnote{Correlated flux density averaged over time and IFs versus the $uv$ radius. The source name, the observation date, and the frequency are given at the top of the panel. In the cases when the data for a source were calibrated in AIPS and then underwent hybrid imaging in Difmap with both amplitude and phase self-calibration, they are plotted as filled circles. In the cases when the processing was the same except no amplitude self-calibration was made, the data are plotted as filled triangles. In the cases when no self-calibration was made for the source and the data calibrated in PIMA were used, they are plotted as open circles.}
\figsetgrpend

\figsetgrpstart
\figsetgrpnum{3.231}
\figsetgrptitle{J1823+7938 at 8.6 GHz}
\figsetplot{J1823+7938_X_2006_02_23_avp_uvs_rad.pdf}
\figsetgrpnote{Correlated flux density averaged over time and IFs versus the $uv$ radius. The source name, the observation date, and the frequency are given at the top of the panel. In the cases when the data for a source were calibrated in AIPS and then underwent hybrid imaging in Difmap with both amplitude and phase self-calibration, they are plotted as filled circles. In the cases when the processing was the same except no amplitude self-calibration was made, the data are plotted as filled triangles. In the cases when no self-calibration was made for the source and the data calibrated in PIMA were used, they are plotted as open circles.}
\figsetgrpend

\figsetgrpstart
\figsetgrpnum{3.232}
\figsetgrptitle{J1832+8049 at 2.3 GHz}
\figsetplot{J1832+8049_S_2006_02_14_avp_uvs_rad.pdf}
\figsetgrpnote{Correlated flux density averaged over time and IFs versus the $uv$ radius. The source name, the observation date, and the frequency are given at the top of the panel. In the cases when the data for a source were calibrated in AIPS and then underwent hybrid imaging in Difmap with both amplitude and phase self-calibration, they are plotted as filled circles. In the cases when the processing was the same except no amplitude self-calibration was made, the data are plotted as filled triangles. In the cases when no self-calibration was made for the source and the data calibrated in PIMA were used, they are plotted as open circles.}
\figsetgrpend

\figsetgrpstart
\figsetgrpnum{3.233}
\figsetgrptitle{J1832+8049 at 8.6 GHz}
\figsetplot{J1832+8049_X_2006_02_14_pet_uvt_rad.pdf}
\figsetgrpnote{Correlated flux density averaged over time and IFs versus the $uv$ radius. The source name, the observation date, and the frequency are given at the top of the panel. In the cases when the data for a source were calibrated in AIPS and then underwent hybrid imaging in Difmap with both amplitude and phase self-calibration, they are plotted as filled circles. In the cases when the processing was the same except no amplitude self-calibration was made, the data are plotted as filled triangles. In the cases when no self-calibration was made for the source and the data calibrated in PIMA were used, they are plotted as open circles.}
\figsetgrpend

\figsetgrpstart
\figsetgrpnum{3.234}
\figsetgrptitle{J1842+7946 at 2.3 GHz}
\figsetplot{J1842+7946_S_2006_02_23_avp_uvs_rad.pdf}
\figsetgrpnote{Correlated flux density averaged over time and IFs versus the $uv$ radius. The source name, the observation date, and the frequency are given at the top of the panel. In the cases when the data for a source were calibrated in AIPS and then underwent hybrid imaging in Difmap with both amplitude and phase self-calibration, they are plotted as filled circles. In the cases when the processing was the same except no amplitude self-calibration was made, the data are plotted as filled triangles. In the cases when no self-calibration was made for the source and the data calibrated in PIMA were used, they are plotted as open circles.}
\figsetgrpend

\figsetgrpstart
\figsetgrpnum{3.235}
\figsetgrptitle{J1842+7946 at 8.6 GHz}
\figsetplot{J1842+7946_X_2006_02_23_pet_uvt_rad.pdf}
\figsetgrpnote{Correlated flux density averaged over time and IFs versus the $uv$ radius. The source name, the observation date, and the frequency are given at the top of the panel. In the cases when the data for a source were calibrated in AIPS and then underwent hybrid imaging in Difmap with both amplitude and phase self-calibration, they are plotted as filled circles. In the cases when the processing was the same except no amplitude self-calibration was made, the data are plotted as filled triangles. In the cases when no self-calibration was made for the source and the data calibrated in PIMA were used, they are plotted as open circles.}
\figsetgrpend

\figsetgrpstart
\figsetgrpnum{3.236}
\figsetgrptitle{J1901+8623 at 2.3 GHz}
\figsetplot{J1901+8623_S_2006_02_16_avp_uvs_rad.pdf}
\figsetgrpnote{Correlated flux density averaged over time and IFs versus the $uv$ radius. The source name, the observation date, and the frequency are given at the top of the panel. In the cases when the data for a source were calibrated in AIPS and then underwent hybrid imaging in Difmap with both amplitude and phase self-calibration, they are plotted as filled circles. In the cases when the processing was the same except no amplitude self-calibration was made, the data are plotted as filled triangles. In the cases when no self-calibration was made for the source and the data calibrated in PIMA were used, they are plotted as open circles.}
\figsetgrpend

\figsetgrpstart
\figsetgrpnum{3.237}
\figsetgrptitle{J1901+8623 at 8.6 GHz}
\figsetplot{J1901+8623_X_2006_02_16_pet_uvt_rad.pdf}
\figsetgrpnote{Correlated flux density averaged over time and IFs versus the $uv$ radius. The source name, the observation date, and the frequency are given at the top of the panel. In the cases when the data for a source were calibrated in AIPS and then underwent hybrid imaging in Difmap with both amplitude and phase self-calibration, they are plotted as filled circles. In the cases when the processing was the same except no amplitude self-calibration was made, the data are plotted as filled triangles. In the cases when no self-calibration was made for the source and the data calibrated in PIMA were used, they are plotted as open circles.}
\figsetgrpend

\figsetgrpstart
\figsetgrpnum{3.238}
\figsetgrptitle{J1904+7648 at 2.3 GHz}
\figsetplot{J1904+7648_S_2006_02_16_pet_uvt_rad.pdf}
\figsetgrpnote{Correlated flux density averaged over time and IFs versus the $uv$ radius. The source name, the observation date, and the frequency are given at the top of the panel. In the cases when the data for a source were calibrated in AIPS and then underwent hybrid imaging in Difmap with both amplitude and phase self-calibration, they are plotted as filled circles. In the cases when the processing was the same except no amplitude self-calibration was made, the data are plotted as filled triangles. In the cases when no self-calibration was made for the source and the data calibrated in PIMA were used, they are plotted as open circles.}
\figsetgrpend

\figsetgrpstart
\figsetgrpnum{3.239}
\figsetgrptitle{J1904+7648 at 8.6 GHz}
\figsetplot{J1904+7648_X_2006_02_16_pet_uvt_rad.pdf}
\figsetgrpnote{Correlated flux density averaged over time and IFs versus the $uv$ radius. The source name, the observation date, and the frequency are given at the top of the panel. In the cases when the data for a source were calibrated in AIPS and then underwent hybrid imaging in Difmap with both amplitude and phase self-calibration, they are plotted as filled circles. In the cases when the processing was the same except no amplitude self-calibration was made, the data are plotted as filled triangles. In the cases when no self-calibration was made for the source and the data calibrated in PIMA were used, they are plotted as open circles.}
\figsetgrpend

\figsetgrpstart
\figsetgrpnum{3.240}
\figsetgrptitle{J1909+7813 at 2.3 GHz}
\figsetplot{J1909+7813_S_2006_02_16_avp_uvs_rad.pdf}
\figsetgrpnote{Correlated flux density averaged over time and IFs versus the $uv$ radius. The source name, the observation date, and the frequency are given at the top of the panel. In the cases when the data for a source were calibrated in AIPS and then underwent hybrid imaging in Difmap with both amplitude and phase self-calibration, they are plotted as filled circles. In the cases when the processing was the same except no amplitude self-calibration was made, the data are plotted as filled triangles. In the cases when no self-calibration was made for the source and the data calibrated in PIMA were used, they are plotted as open circles.}
\figsetgrpend

\figsetgrpstart
\figsetgrpnum{3.241}
\figsetgrptitle{J1909+7813 at 8.6 GHz}
\figsetplot{J1909+7813_X_2006_02_16_pet_uvt_rad.pdf}
\figsetgrpnote{Correlated flux density averaged over time and IFs versus the $uv$ radius. The source name, the observation date, and the frequency are given at the top of the panel. In the cases when the data for a source were calibrated in AIPS and then underwent hybrid imaging in Difmap with both amplitude and phase self-calibration, they are plotted as filled circles. In the cases when the processing was the same except no amplitude self-calibration was made, the data are plotted as filled triangles. In the cases when no self-calibration was made for the source and the data calibrated in PIMA were used, they are plotted as open circles.}
\figsetgrpend

\figsetgrpstart
\figsetgrpnum{3.242}
\figsetgrptitle{J1935+8130 at 2.3 GHz}
\figsetplot{J1935+8130_S_2006_02_14_avp_uvs_rad.pdf}
\figsetgrpnote{Correlated flux density averaged over time and IFs versus the $uv$ radius. The source name, the observation date, and the frequency are given at the top of the panel. In the cases when the data for a source were calibrated in AIPS and then underwent hybrid imaging in Difmap with both amplitude and phase self-calibration, they are plotted as filled circles. In the cases when the processing was the same except no amplitude self-calibration was made, the data are plotted as filled triangles. In the cases when no self-calibration was made for the source and the data calibrated in PIMA were used, they are plotted as open circles.}
\figsetgrpend

\figsetgrpstart
\figsetgrpnum{3.243}
\figsetgrptitle{J1935+8130 at 8.6 GHz}
\figsetplot{J1935+8130_X_2006_02_14_avp_uvs_rad.pdf}
\figsetgrpnote{Correlated flux density averaged over time and IFs versus the $uv$ radius. The source name, the observation date, and the frequency are given at the top of the panel. In the cases when the data for a source were calibrated in AIPS and then underwent hybrid imaging in Difmap with both amplitude and phase self-calibration, they are plotted as filled circles. In the cases when the processing was the same except no amplitude self-calibration was made, the data are plotted as filled triangles. In the cases when no self-calibration was made for the source and the data calibrated in PIMA were used, they are plotted as open circles.}
\figsetgrpend

\figsetgrpstart
\figsetgrpnum{3.244}
\figsetgrptitle{J1937+8356 at 2.3 GHz}
\figsetplot{J1937+8356_S_2006_02_23_avp_uvs_rad.pdf}
\figsetgrpnote{Correlated flux density averaged over time and IFs versus the $uv$ radius. The source name, the observation date, and the frequency are given at the top of the panel. In the cases when the data for a source were calibrated in AIPS and then underwent hybrid imaging in Difmap with both amplitude and phase self-calibration, they are plotted as filled circles. In the cases when the processing was the same except no amplitude self-calibration was made, the data are plotted as filled triangles. In the cases when no self-calibration was made for the source and the data calibrated in PIMA were used, they are plotted as open circles.}
\figsetgrpend

\figsetgrpstart
\figsetgrpnum{3.245}
\figsetgrptitle{J1937+8356 at 8.6 GHz}
\figsetplot{J1937+8356_X_2006_02_23_avp_uvs_rad.pdf}
\figsetgrpnote{Correlated flux density averaged over time and IFs versus the $uv$ radius. The source name, the observation date, and the frequency are given at the top of the panel. In the cases when the data for a source were calibrated in AIPS and then underwent hybrid imaging in Difmap with both amplitude and phase self-calibration, they are plotted as filled circles. In the cases when the processing was the same except no amplitude self-calibration was made, the data are plotted as filled triangles. In the cases when no self-calibration was made for the source and the data calibrated in PIMA were used, they are plotted as open circles.}
\figsetgrpend

\figsetgrpstart
\figsetgrpnum{3.246}
\figsetgrptitle{J2005+7752 at 2.3 GHz}
\figsetplot{J2005+7752_S_2006_02_14_avp_uvs_rad.pdf}
\figsetgrpnote{Correlated flux density averaged over time and IFs versus the $uv$ radius. The source name, the observation date, and the frequency are given at the top of the panel. In the cases when the data for a source were calibrated in AIPS and then underwent hybrid imaging in Difmap with both amplitude and phase self-calibration, they are plotted as filled circles. In the cases when the processing was the same except no amplitude self-calibration was made, the data are plotted as filled triangles. In the cases when no self-calibration was made for the source and the data calibrated in PIMA were used, they are plotted as open circles.}
\figsetgrpend

\figsetgrpstart
\figsetgrpnum{3.247}
\figsetgrptitle{J2005+7752 at 8.6 GHz}
\figsetplot{J2005+7752_X_2006_02_14_avp_uvs_rad.pdf}
\figsetgrpnote{Correlated flux density averaged over time and IFs versus the $uv$ radius. The source name, the observation date, and the frequency are given at the top of the panel. In the cases when the data for a source were calibrated in AIPS and then underwent hybrid imaging in Difmap with both amplitude and phase self-calibration, they are plotted as filled circles. In the cases when the processing was the same except no amplitude self-calibration was made, the data are plotted as filled triangles. In the cases when no self-calibration was made for the source and the data calibrated in PIMA were used, they are plotted as open circles.}
\figsetgrpend

\figsetgrpstart
\figsetgrpnum{3.248}
\figsetgrptitle{J2007+7942 at 2.3 GHz}
\figsetplot{J2007+7942_S_2006_02_16_pet_uvt_rad.pdf}
\figsetgrpnote{Correlated flux density averaged over time and IFs versus the $uv$ radius. The source name, the observation date, and the frequency are given at the top of the panel. In the cases when the data for a source were calibrated in AIPS and then underwent hybrid imaging in Difmap with both amplitude and phase self-calibration, they are plotted as filled circles. In the cases when the processing was the same except no amplitude self-calibration was made, the data are plotted as filled triangles. In the cases when no self-calibration was made for the source and the data calibrated in PIMA were used, they are plotted as open circles.}
\figsetgrpend

\figsetgrpstart
\figsetgrpnum{3.249}
\figsetgrptitle{J2022+7611 at 2.3 GHz}
\figsetplot{J2022+7611_S_2006_02_23_avp_uvs_rad.pdf}
\figsetgrpnote{Correlated flux density averaged over time and IFs versus the $uv$ radius. The source name, the observation date, and the frequency are given at the top of the panel. In the cases when the data for a source were calibrated in AIPS and then underwent hybrid imaging in Difmap with both amplitude and phase self-calibration, they are plotted as filled circles. In the cases when the processing was the same except no amplitude self-calibration was made, the data are plotted as filled triangles. In the cases when no self-calibration was made for the source and the data calibrated in PIMA were used, they are plotted as open circles.}
\figsetgrpend

\figsetgrpstart
\figsetgrpnum{3.250}
\figsetgrptitle{J2022+7611 at 8.6 GHz}
\figsetplot{J2022+7611_X_2006_02_23_avp_uvs_rad.pdf}
\figsetgrpnote{Correlated flux density averaged over time and IFs versus the $uv$ radius. The source name, the observation date, and the frequency are given at the top of the panel. In the cases when the data for a source were calibrated in AIPS and then underwent hybrid imaging in Difmap with both amplitude and phase self-calibration, they are plotted as filled circles. In the cases when the processing was the same except no amplitude self-calibration was made, the data are plotted as filled triangles. In the cases when no self-calibration was made for the source and the data calibrated in PIMA were used, they are plotted as open circles.}
\figsetgrpend

\figsetgrpstart
\figsetgrpnum{3.251}
\figsetgrptitle{J2042+7508 at 2.3 GHz}
\figsetplot{J2042+7508_S_2006_02_16_avp_uvs_rad.pdf}
\figsetgrpnote{Correlated flux density averaged over time and IFs versus the $uv$ radius. The source name, the observation date, and the frequency are given at the top of the panel. In the cases when the data for a source were calibrated in AIPS and then underwent hybrid imaging in Difmap with both amplitude and phase self-calibration, they are plotted as filled circles. In the cases when the processing was the same except no amplitude self-calibration was made, the data are plotted as filled triangles. In the cases when no self-calibration was made for the source and the data calibrated in PIMA were used, they are plotted as open circles.}
\figsetgrpend

\figsetgrpstart
\figsetgrpnum{3.252}
\figsetgrptitle{J2042+7508 at 8.6 GHz}
\figsetplot{J2042+7508_X_2006_02_16_avp_uvs_rad.pdf}
\figsetgrpnote{Correlated flux density averaged over time and IFs versus the $uv$ radius. The source name, the observation date, and the frequency are given at the top of the panel. In the cases when the data for a source were calibrated in AIPS and then underwent hybrid imaging in Difmap with both amplitude and phase self-calibration, they are plotted as filled circles. In the cases when the processing was the same except no amplitude self-calibration was made, the data are plotted as filled triangles. In the cases when no self-calibration was made for the source and the data calibrated in PIMA were used, they are plotted as open circles.}
\figsetgrpend

\figsetgrpstart
\figsetgrpnum{3.253}
\figsetgrptitle{J2045+7625 at 2.3 GHz}
\figsetplot{J2045+7625_S_2006_02_23_avp_uvs_rad.pdf}
\figsetgrpnote{Correlated flux density averaged over time and IFs versus the $uv$ radius. The source name, the observation date, and the frequency are given at the top of the panel. In the cases when the data for a source were calibrated in AIPS and then underwent hybrid imaging in Difmap with both amplitude and phase self-calibration, they are plotted as filled circles. In the cases when the processing was the same except no amplitude self-calibration was made, the data are plotted as filled triangles. In the cases when no self-calibration was made for the source and the data calibrated in PIMA were used, they are plotted as open circles.}
\figsetgrpend

\figsetgrpstart
\figsetgrpnum{3.254}
\figsetgrptitle{J2045+7625 at 8.6 GHz}
\figsetplot{J2045+7625_X_2006_02_23_avp_uvs_rad.pdf}
\figsetgrpnote{Correlated flux density averaged over time and IFs versus the $uv$ radius. The source name, the observation date, and the frequency are given at the top of the panel. In the cases when the data for a source were calibrated in AIPS and then underwent hybrid imaging in Difmap with both amplitude and phase self-calibration, they are plotted as filled circles. In the cases when the processing was the same except no amplitude self-calibration was made, the data are plotted as filled triangles. In the cases when no self-calibration was made for the source and the data calibrated in PIMA were used, they are plotted as open circles.}
\figsetgrpend

\figsetgrpstart
\figsetgrpnum{3.255}
\figsetgrptitle{J2109+8021 at 2.3 GHz}
\figsetplot{J2109+8021_S_2006_02_23_pet_uvt_rad.pdf}
\figsetgrpnote{Correlated flux density averaged over time and IFs versus the $uv$ radius. The source name, the observation date, and the frequency are given at the top of the panel. In the cases when the data for a source were calibrated in AIPS and then underwent hybrid imaging in Difmap with both amplitude and phase self-calibration, they are plotted as filled circles. In the cases when the processing was the same except no amplitude self-calibration was made, the data are plotted as filled triangles. In the cases when no self-calibration was made for the source and the data calibrated in PIMA were used, they are plotted as open circles.}
\figsetgrpend

\figsetgrpstart
\figsetgrpnum{3.256}
\figsetgrptitle{J2109+8021 at 8.6 GHz}
\figsetplot{J2109+8021_X_2006_02_23_pet_uvt_rad.pdf}
\figsetgrpnote{Correlated flux density averaged over time and IFs versus the $uv$ radius. The source name, the observation date, and the frequency are given at the top of the panel. In the cases when the data for a source were calibrated in AIPS and then underwent hybrid imaging in Difmap with both amplitude and phase self-calibration, they are plotted as filled circles. In the cases when the processing was the same except no amplitude self-calibration was made, the data are plotted as filled triangles. In the cases when no self-calibration was made for the source and the data calibrated in PIMA were used, they are plotted as open circles.}
\figsetgrpend

\figsetgrpstart
\figsetgrpnum{3.257}
\figsetgrptitle{J2114+8204 at 2.3 GHz}
\figsetplot{J2114+8204_S_2006_02_23_avp_uvs_rad.pdf}
\figsetgrpnote{Correlated flux density averaged over time and IFs versus the $uv$ radius. The source name, the observation date, and the frequency are given at the top of the panel. In the cases when the data for a source were calibrated in AIPS and then underwent hybrid imaging in Difmap with both amplitude and phase self-calibration, they are plotted as filled circles. In the cases when the processing was the same except no amplitude self-calibration was made, the data are plotted as filled triangles. In the cases when no self-calibration was made for the source and the data calibrated in PIMA were used, they are plotted as open circles.}
\figsetgrpend

\figsetgrpstart
\figsetgrpnum{3.258}
\figsetgrptitle{J2114+8204 at 8.6 GHz}
\figsetplot{J2114+8204_X_2006_02_23_avp_uvs_rad.pdf}
\figsetgrpnote{Correlated flux density averaged over time and IFs versus the $uv$ radius. The source name, the observation date, and the frequency are given at the top of the panel. In the cases when the data for a source were calibrated in AIPS and then underwent hybrid imaging in Difmap with both amplitude and phase self-calibration, they are plotted as filled circles. In the cases when the processing was the same except no amplitude self-calibration was made, the data are plotted as filled triangles. In the cases when no self-calibration was made for the source and the data calibrated in PIMA were used, they are plotted as open circles.}
\figsetgrpend

\figsetgrpstart
\figsetgrpnum{3.259}
\figsetgrptitle{J2130+8357 at 8.6 GHz}
\figsetplot{J2130+8357_X_2006_02_16_pet_uvt_rad.pdf}
\figsetgrpnote{Correlated flux density averaged over time and IFs versus the $uv$ radius. The source name, the observation date, and the frequency are given at the top of the panel. In the cases when the data for a source were calibrated in AIPS and then underwent hybrid imaging in Difmap with both amplitude and phase self-calibration, they are plotted as filled circles. In the cases when the processing was the same except no amplitude self-calibration was made, the data are plotted as filled triangles. In the cases when no self-calibration was made for the source and the data calibrated in PIMA were used, they are plotted as open circles.}
\figsetgrpend

\figsetgrpstart
\figsetgrpnum{3.260}
\figsetgrptitle{J2131+8430 at 2.3 GHz}
\figsetplot{J2131+8430_S_2006_02_23_avp_uvs_rad.pdf}
\figsetgrpnote{Correlated flux density averaged over time and IFs versus the $uv$ radius. The source name, the observation date, and the frequency are given at the top of the panel. In the cases when the data for a source were calibrated in AIPS and then underwent hybrid imaging in Difmap with both amplitude and phase self-calibration, they are plotted as filled circles. In the cases when the processing was the same except no amplitude self-calibration was made, the data are plotted as filled triangles. In the cases when no self-calibration was made for the source and the data calibrated in PIMA were used, they are plotted as open circles.}
\figsetgrpend

\figsetgrpstart
\figsetgrpnum{3.261}
\figsetgrptitle{J2131+8430 at 8.6 GHz}
\figsetplot{J2131+8430_X_2006_02_23_avp_uvs_rad.pdf}
\figsetgrpnote{Correlated flux density averaged over time and IFs versus the $uv$ radius. The source name, the observation date, and the frequency are given at the top of the panel. In the cases when the data for a source were calibrated in AIPS and then underwent hybrid imaging in Difmap with both amplitude and phase self-calibration, they are plotted as filled circles. In the cases when the processing was the same except no amplitude self-calibration was made, the data are plotted as filled triangles. In the cases when no self-calibration was made for the source and the data calibrated in PIMA were used, they are plotted as open circles.}
\figsetgrpend

\figsetgrpstart
\figsetgrpnum{3.262}
\figsetgrptitle{J2133+8239 at 2.3 GHz}
\figsetplot{J2133+8239_S_2006_02_23_avp_uvs_rad.pdf}
\figsetgrpnote{Correlated flux density averaged over time and IFs versus the $uv$ radius. The source name, the observation date, and the frequency are given at the top of the panel. In the cases when the data for a source were calibrated in AIPS and then underwent hybrid imaging in Difmap with both amplitude and phase self-calibration, they are plotted as filled circles. In the cases when the processing was the same except no amplitude self-calibration was made, the data are plotted as filled triangles. In the cases when no self-calibration was made for the source and the data calibrated in PIMA were used, they are plotted as open circles.}
\figsetgrpend

\figsetgrpstart
\figsetgrpnum{3.263}
\figsetgrptitle{J2133+8239 at 8.6 GHz}
\figsetplot{J2133+8239_X_2006_02_23_avp_uvs_rad.pdf}
\figsetgrpnote{Correlated flux density averaged over time and IFs versus the $uv$ radius. The source name, the observation date, and the frequency are given at the top of the panel. In the cases when the data for a source were calibrated in AIPS and then underwent hybrid imaging in Difmap with both amplitude and phase self-calibration, they are plotted as filled circles. In the cases when the processing was the same except no amplitude self-calibration was made, the data are plotted as filled triangles. In the cases when no self-calibration was made for the source and the data calibrated in PIMA were used, they are plotted as open circles.}
\figsetgrpend

\figsetgrpstart
\figsetgrpnum{3.264}
\figsetgrptitle{J2140+7505 at 2.3 GHz}
\figsetplot{J2140+7505_S_2006_02_14_pet_uvt_rad.pdf}
\figsetgrpnote{Correlated flux density averaged over time and IFs versus the $uv$ radius. The source name, the observation date, and the frequency are given at the top of the panel. In the cases when the data for a source were calibrated in AIPS and then underwent hybrid imaging in Difmap with both amplitude and phase self-calibration, they are plotted as filled circles. In the cases when the processing was the same except no amplitude self-calibration was made, the data are plotted as filled triangles. In the cases when no self-calibration was made for the source and the data calibrated in PIMA were used, they are plotted as open circles.}
\figsetgrpend

\figsetgrpstart
\figsetgrpnum{3.265}
\figsetgrptitle{J2140+7505 at 8.6 GHz}
\figsetplot{J2140+7505_X_2006_02_14_pet_uvt_rad.pdf}
\figsetgrpnote{Correlated flux density averaged over time and IFs versus the $uv$ radius. The source name, the observation date, and the frequency are given at the top of the panel. In the cases when the data for a source were calibrated in AIPS and then underwent hybrid imaging in Difmap with both amplitude and phase self-calibration, they are plotted as filled circles. In the cases when the processing was the same except no amplitude self-calibration was made, the data are plotted as filled triangles. In the cases when no self-calibration was made for the source and the data calibrated in PIMA were used, they are plotted as open circles.}
\figsetgrpend

\figsetgrpstart
\figsetgrpnum{3.266}
\figsetgrptitle{J2149+7540 at 8.6 GHz}
\figsetplot{J2149+7540_X_2006_02_14_pet_uvt_rad.pdf}
\figsetgrpnote{Correlated flux density averaged over time and IFs versus the $uv$ radius. The source name, the observation date, and the frequency are given at the top of the panel. In the cases when the data for a source were calibrated in AIPS and then underwent hybrid imaging in Difmap with both amplitude and phase self-calibration, they are plotted as filled circles. In the cases when the processing was the same except no amplitude self-calibration was made, the data are plotted as filled triangles. In the cases when no self-calibration was made for the source and the data calibrated in PIMA were used, they are plotted as open circles.}
\figsetgrpend

\figsetgrpstart
\figsetgrpnum{3.267}
\figsetgrptitle{J2156+8337 at 2.3 GHz}
\figsetplot{J2156+8337_S_2006_02_16_avp_uvs_rad.pdf}
\figsetgrpnote{Correlated flux density averaged over time and IFs versus the $uv$ radius. The source name, the observation date, and the frequency are given at the top of the panel. In the cases when the data for a source were calibrated in AIPS and then underwent hybrid imaging in Difmap with both amplitude and phase self-calibration, they are plotted as filled circles. In the cases when the processing was the same except no amplitude self-calibration was made, the data are plotted as filled triangles. In the cases when no self-calibration was made for the source and the data calibrated in PIMA were used, they are plotted as open circles.}
\figsetgrpend

\figsetgrpstart
\figsetgrpnum{3.268}
\figsetgrptitle{J2156+8337 at 8.6 GHz}
\figsetplot{J2156+8337_X_2006_02_16_avp_uvs_rad.pdf}
\figsetgrpnote{Correlated flux density averaged over time and IFs versus the $uv$ radius. The source name, the observation date, and the frequency are given at the top of the panel. In the cases when the data for a source were calibrated in AIPS and then underwent hybrid imaging in Difmap with both amplitude and phase self-calibration, they are plotted as filled circles. In the cases when the processing was the same except no amplitude self-calibration was made, the data are plotted as filled triangles. In the cases when no self-calibration was made for the source and the data calibrated in PIMA were used, they are plotted as open circles.}
\figsetgrpend

\figsetgrpstart
\figsetgrpnum{3.269}
\figsetgrptitle{J2242+8224 at 2.3 GHz}
\figsetplot{J2242+8224_S_2006_02_23_avp_uvs_rad.pdf}
\figsetgrpnote{Correlated flux density averaged over time and IFs versus the $uv$ radius. The source name, the observation date, and the frequency are given at the top of the panel. In the cases when the data for a source were calibrated in AIPS and then underwent hybrid imaging in Difmap with both amplitude and phase self-calibration, they are plotted as filled circles. In the cases when the processing was the same except no amplitude self-calibration was made, the data are plotted as filled triangles. In the cases when no self-calibration was made for the source and the data calibrated in PIMA were used, they are plotted as open circles.}
\figsetgrpend

\figsetgrpstart
\figsetgrpnum{3.270}
\figsetgrptitle{J2242+8224 at 8.6 GHz}
\figsetplot{J2242+8224_X_2006_02_23_pet_uvt_rad.pdf}
\figsetgrpnote{Correlated flux density averaged over time and IFs versus the $uv$ radius. The source name, the observation date, and the frequency are given at the top of the panel. In the cases when the data for a source were calibrated in AIPS and then underwent hybrid imaging in Difmap with both amplitude and phase self-calibration, they are plotted as filled circles. In the cases when the processing was the same except no amplitude self-calibration was made, the data are plotted as filled triangles. In the cases when no self-calibration was made for the source and the data calibrated in PIMA were used, they are plotted as open circles.}
\figsetgrpend

\figsetgrpstart
\figsetgrpnum{3.271}
\figsetgrptitle{J2248+7718 at 2.3 GHz}
\figsetplot{J2248+7718_S_2006_02_16_avp_uvs_rad.pdf}
\figsetgrpnote{Correlated flux density averaged over time and IFs versus the $uv$ radius. The source name, the observation date, and the frequency are given at the top of the panel. In the cases when the data for a source were calibrated in AIPS and then underwent hybrid imaging in Difmap with both amplitude and phase self-calibration, they are plotted as filled circles. In the cases when the processing was the same except no amplitude self-calibration was made, the data are plotted as filled triangles. In the cases when no self-calibration was made for the source and the data calibrated in PIMA were used, they are plotted as open circles.}
\figsetgrpend

\figsetgrpstart
\figsetgrpnum{3.272}
\figsetgrptitle{J2301+8200 at 2.3 GHz}
\figsetplot{J2301+8200_S_2006_02_14_avp_uvs_rad.pdf}
\figsetgrpnote{Correlated flux density averaged over time and IFs versus the $uv$ radius. The source name, the observation date, and the frequency are given at the top of the panel. In the cases when the data for a source were calibrated in AIPS and then underwent hybrid imaging in Difmap with both amplitude and phase self-calibration, they are plotted as filled circles. In the cases when the processing was the same except no amplitude self-calibration was made, the data are plotted as filled triangles. In the cases when no self-calibration was made for the source and the data calibrated in PIMA were used, they are plotted as open circles.}
\figsetgrpend

\figsetgrpstart
\figsetgrpnum{3.273}
\figsetgrptitle{J2310+8857 at 2.3 GHz}
\figsetplot{J2310+8857_S_2006_02_16_avp_uvs_rad.pdf}
\figsetgrpnote{Correlated flux density averaged over time and IFs versus the $uv$ radius. The source name, the observation date, and the frequency are given at the top of the panel. In the cases when the data for a source were calibrated in AIPS and then underwent hybrid imaging in Difmap with both amplitude and phase self-calibration, they are plotted as filled circles. In the cases when the processing was the same except no amplitude self-calibration was made, the data are plotted as filled triangles. In the cases when no self-calibration was made for the source and the data calibrated in PIMA were used, they are plotted as open circles.}
\figsetgrpend

\figsetgrpstart
\figsetgrpnum{3.274}
\figsetgrptitle{J2310+8857 at 8.6 GHz}
\figsetplot{J2310+8857_X_2006_02_16_avp_uvs_rad.pdf}
\figsetgrpnote{Correlated flux density averaged over time and IFs versus the $uv$ radius. The source name, the observation date, and the frequency are given at the top of the panel. In the cases when the data for a source were calibrated in AIPS and then underwent hybrid imaging in Difmap with both amplitude and phase self-calibration, they are plotted as filled circles. In the cases when the processing was the same except no amplitude self-calibration was made, the data are plotted as filled triangles. In the cases when no self-calibration was made for the source and the data calibrated in PIMA were used, they are plotted as open circles.}
\figsetgrpend

\figsetgrpstart
\figsetgrpnum{3.275}
\figsetgrptitle{J2325+7917 at 2.3 GHz}
\figsetplot{J2325+7917_S_2006_02_16_avp_uvs_rad.pdf}
\figsetgrpnote{Correlated flux density averaged over time and IFs versus the $uv$ radius. The source name, the observation date, and the frequency are given at the top of the panel. In the cases when the data for a source were calibrated in AIPS and then underwent hybrid imaging in Difmap with both amplitude and phase self-calibration, they are plotted as filled circles. In the cases when the processing was the same except no amplitude self-calibration was made, the data are plotted as filled triangles. In the cases when no self-calibration was made for the source and the data calibrated in PIMA were used, they are plotted as open circles.}
\figsetgrpend

\figsetgrpstart
\figsetgrpnum{3.276}
\figsetgrptitle{J2325+7917 at 8.6 GHz}
\figsetplot{J2325+7917_X_2006_02_16_pet_uvt_rad.pdf}
\figsetgrpnote{Correlated flux density averaged over time and IFs versus the $uv$ radius. The source name, the observation date, and the frequency are given at the top of the panel. In the cases when the data for a source were calibrated in AIPS and then underwent hybrid imaging in Difmap with both amplitude and phase self-calibration, they are plotted as filled circles. In the cases when the processing was the same except no amplitude self-calibration was made, the data are plotted as filled triangles. In the cases when no self-calibration was made for the source and the data calibrated in PIMA were used, they are plotted as open circles.}
\figsetgrpend

\figsetgrpstart
\figsetgrpnum{3.277}
\figsetgrptitle{J2329+8131 at 2.3 GHz}
\figsetplot{J2329+8131_S_2006_02_23_pet_uvt_rad.pdf}
\figsetgrpnote{Correlated flux density averaged over time and IFs versus the $uv$ radius. The source name, the observation date, and the frequency are given at the top of the panel. In the cases when the data for a source were calibrated in AIPS and then underwent hybrid imaging in Difmap with both amplitude and phase self-calibration, they are plotted as filled circles. In the cases when the processing was the same except no amplitude self-calibration was made, the data are plotted as filled triangles. In the cases when no self-calibration was made for the source and the data calibrated in PIMA were used, they are plotted as open circles.}
\figsetgrpend

\figsetgrpstart
\figsetgrpnum{3.278}
\figsetgrptitle{J2330+7742 at 8.6 GHz}
\figsetplot{J2330+7742_X_2006_02_14_pet_uvt_rad.pdf}
\figsetgrpnote{Correlated flux density averaged over time and IFs versus the $uv$ radius. The source name, the observation date, and the frequency are given at the top of the panel. In the cases when the data for a source were calibrated in AIPS and then underwent hybrid imaging in Difmap with both amplitude and phase self-calibration, they are plotted as filled circles. In the cases when the processing was the same except no amplitude self-calibration was made, the data are plotted as filled triangles. In the cases when no self-calibration was made for the source and the data calibrated in PIMA were used, they are plotted as open circles.}
\figsetgrpend

\figsetgrpstart
\figsetgrpnum{3.279}
\figsetgrptitle{J2344+8226 at 2.3 GHz}
\figsetplot{J2344+8226_S_2006_02_14_avp_uvs_rad.pdf}
\figsetgrpnote{Correlated flux density averaged over time and IFs versus the $uv$ radius. The source name, the observation date, and the frequency are given at the top of the panel. In the cases when the data for a source were calibrated in AIPS and then underwent hybrid imaging in Difmap with both amplitude and phase self-calibration, they are plotted as filled circles. In the cases when the processing was the same except no amplitude self-calibration was made, the data are plotted as filled triangles. In the cases when no self-calibration was made for the source and the data calibrated in PIMA were used, they are plotted as open circles.}
\figsetgrpend

\figsetgrpstart
\figsetgrpnum{3.280}
\figsetgrptitle{J2344+8226 at 8.6 GHz}
\figsetplot{J2344+8226_X_2006_02_14_pet_uvt_rad.pdf}
\figsetgrpnote{Correlated flux density averaged over time and IFs versus the $uv$ radius. The source name, the observation date, and the frequency are given at the top of the panel. In the cases when the data for a source were calibrated in AIPS and then underwent hybrid imaging in Difmap with both amplitude and phase self-calibration, they are plotted as filled circles. In the cases when the processing was the same except no amplitude self-calibration was made, the data are plotted as filled triangles. In the cases when no self-calibration was made for the source and the data calibrated in PIMA were used, they are plotted as open circles.}
\figsetgrpend

\figsetgrpstart
\figsetgrpnum{3.281}
\figsetgrptitle{J2349+7517 at 2.3 GHz}
\figsetplot{J2349+7517_S_2006_02_23_pet_uvt_rad.pdf}
\figsetgrpnote{Correlated flux density averaged over time and IFs versus the $uv$ radius. The source name, the observation date, and the frequency are given at the top of the panel. In the cases when the data for a source were calibrated in AIPS and then underwent hybrid imaging in Difmap with both amplitude and phase self-calibration, they are plotted as filled circles. In the cases when the processing was the same except no amplitude self-calibration was made, the data are plotted as filled triangles. In the cases when no self-calibration was made for the source and the data calibrated in PIMA were used, they are plotted as open circles.}
\figsetgrpend

\figsetgrpstart
\figsetgrpnum{3.282}
\figsetgrptitle{J2349+7517 at 8.6 GHz}
\figsetplot{J2349+7517_X_2006_02_23_pet_uvt_rad.pdf}
\figsetgrpnote{Correlated flux density averaged over time and IFs versus the $uv$ radius. The source name, the observation date, and the frequency are given at the top of the panel. In the cases when the data for a source were calibrated in AIPS and then underwent hybrid imaging in Difmap with both amplitude and phase self-calibration, they are plotted as filled circles. In the cases when the processing was the same except no amplitude self-calibration was made, the data are plotted as filled triangles. In the cases when no self-calibration was made for the source and the data calibrated in PIMA were used, they are plotted as open circles.}
\figsetgrpend

\figsetgrpstart
\figsetgrpnum{3.283}
\figsetgrptitle{J2356+8152 at 2.3 GHz}
\figsetplot{J2356+8152_S_2006_02_16_avp_uvs_rad.pdf}
\figsetgrpnote{Correlated flux density averaged over time and IFs versus the $uv$ radius. The source name, the observation date, and the frequency are given at the top of the panel. In the cases when the data for a source were calibrated in AIPS and then underwent hybrid imaging in Difmap with both amplitude and phase self-calibration, they are plotted as filled circles. In the cases when the processing was the same except no amplitude self-calibration was made, the data are plotted as filled triangles. In the cases when no self-calibration was made for the source and the data calibrated in PIMA were used, they are plotted as open circles.}
\figsetgrpend

\figsetgrpstart
\figsetgrpnum{3.284}
\figsetgrptitle{J2356+8152 at 8.6 GHz}
\figsetplot{J2356+8152_X_2006_02_16_avp_uvs_rad.pdf}
\figsetgrpnote{Correlated flux density averaged over time and IFs versus the $uv$ radius. The source name, the observation date, and the frequency are given at the top of the panel. In the cases when the data for a source were calibrated in AIPS and then underwent hybrid imaging in Difmap with both amplitude and phase self-calibration, they are plotted as filled circles. In the cases when the processing was the same except no amplitude self-calibration was made, the data are plotted as filled triangles. In the cases when no self-calibration was made for the source and the data calibrated in PIMA were used, they are plotted as open circles.}
\figsetgrpend

\figsetgrpstart
\figsetgrpnum{3.285}
\figsetgrptitle{J2358+8142 at 2.3 GHz}
\figsetplot{J2358+8142_S_2006_02_23_pet_uvt_rad.pdf}
\figsetgrpnote{Correlated flux density averaged over time and IFs versus the $uv$ radius. The source name, the observation date, and the frequency are given at the top of the panel. In the cases when the data for a source were calibrated in AIPS and then underwent hybrid imaging in Difmap with both amplitude and phase self-calibration, they are plotted as filled circles. In the cases when the processing was the same except no amplitude self-calibration was made, the data are plotted as filled triangles. In the cases when no self-calibration was made for the source and the data calibrated in PIMA were used, they are plotted as open circles.}
\figsetgrpend

\figsetend

\figsetstart
\figsetnum{4}
\figsettitle{CLEAN maps of the sources.}

\figsetgrpstart
\figsetgrpnum{4.1}
\figsetgrptitle{J0009+7603 at 8.6 GHz}
\figsetplot{J0009+7603_X_2006_02_14_avp_map.pdf}
\figsetgrpnote{The source name, the observation date, and the frequency are given at the top of the panel. The intensity is shown by contours: solid lines are the positive contours, dotted lines are the negative contours. The contour levels in percents of the map peak are specified below the map, as well as the total flux density of the CLEAN model of the source and the intensity of the map peak. First contours correspond to the map noise level~$\times\,3$; each subsequent contour marks intensity increase by the factor of two. The CLEAN beam at the half-maximum level is shown in the map lower left corner as a black ellipse; its major and minor axes and position angle are specified below the map.}
\figsetgrpend

\figsetgrpstart
\figsetgrpnum{4.2}
\figsetgrptitle{J0017+8135 at 2.3 GHz}
\figsetplot{J0017+8135_S_2006_02_14_avp_map.pdf}
\figsetgrpnote{The source name, the observation date, and the frequency are given at the top of the panel. The intensity is shown by contours: solid lines are the positive contours, dotted lines are the negative contours. The contour levels in percents of the map peak are specified below the map, as well as the total flux density of the CLEAN model of the source and the intensity of the map peak. First contours correspond to the map noise level~$\times\,3$; each subsequent contour marks intensity increase by the factor of two. The CLEAN beam at the half-maximum level is shown in the map lower left corner as a black ellipse; its major and minor axes and position angle are specified below the map.}
\figsetgrpend

\figsetgrpstart
\figsetgrpnum{4.3}
\figsetgrptitle{J0017+8135 at 2.3 GHz}
\figsetplot{J0017+8135_S_2006_02_16_avp_map.pdf}
\figsetgrpnote{The source name, the observation date, and the frequency are given at the top of the panel. The intensity is shown by contours: solid lines are the positive contours, dotted lines are the negative contours. The contour levels in percents of the map peak are specified below the map, as well as the total flux density of the CLEAN model of the source and the intensity of the map peak. First contours correspond to the map noise level~$\times\,3$; each subsequent contour marks intensity increase by the factor of two. The CLEAN beam at the half-maximum level is shown in the map lower left corner as a black ellipse; its major and minor axes and position angle are specified below the map.}
\figsetgrpend

\figsetgrpstart
\figsetgrpnum{4.4}
\figsetgrptitle{J0017+8135 at 2.3 GHz}
\figsetplot{J0017+8135_S_2006_02_23_avp_map.pdf}
\figsetgrpnote{The source name, the observation date, and the frequency are given at the top of the panel. The intensity is shown by contours: solid lines are the positive contours, dotted lines are the negative contours. The contour levels in percents of the map peak are specified below the map, as well as the total flux density of the CLEAN model of the source and the intensity of the map peak. First contours correspond to the map noise level~$\times\,3$; each subsequent contour marks intensity increase by the factor of two. The CLEAN beam at the half-maximum level is shown in the map lower left corner as a black ellipse; its major and minor axes and position angle are specified below the map.}
\figsetgrpend

\figsetgrpstart
\figsetgrpnum{4.5}
\figsetgrptitle{J0017+8135 at 8.6 GHz}
\figsetplot{J0017+8135_X_2006_02_14_avp_map.pdf}
\figsetgrpnote{The source name, the observation date, and the frequency are given at the top of the panel. The intensity is shown by contours: solid lines are the positive contours, dotted lines are the negative contours. The contour levels in percents of the map peak are specified below the map, as well as the total flux density of the CLEAN model of the source and the intensity of the map peak. First contours correspond to the map noise level~$\times\,3$; each subsequent contour marks intensity increase by the factor of two. The CLEAN beam at the half-maximum level is shown in the map lower left corner as a black ellipse; its major and minor axes and position angle are specified below the map.}
\figsetgrpend

\figsetgrpstart
\figsetgrpnum{4.6}
\figsetgrptitle{J0017+8135 at 8.6 GHz}
\figsetplot{J0017+8135_X_2006_02_16_avp_map.pdf}
\figsetgrpnote{The source name, the observation date, and the frequency are given at the top of the panel. The intensity is shown by contours: solid lines are the positive contours, dotted lines are the negative contours. The contour levels in percents of the map peak are specified below the map, as well as the total flux density of the CLEAN model of the source and the intensity of the map peak. First contours correspond to the map noise level~$\times\,3$; each subsequent contour marks intensity increase by the factor of two. The CLEAN beam at the half-maximum level is shown in the map lower left corner as a black ellipse; its major and minor axes and position angle are specified below the map.}
\figsetgrpend

\figsetgrpstart
\figsetgrpnum{4.7}
\figsetgrptitle{J0017+8135 at 8.6 GHz}
\figsetplot{J0017+8135_X_2006_02_23_avp_map.pdf}
\figsetgrpnote{The source name, the observation date, and the frequency are given at the top of the panel. The intensity is shown by contours: solid lines are the positive contours, dotted lines are the negative contours. The contour levels in percents of the map peak are specified below the map, as well as the total flux density of the CLEAN model of the source and the intensity of the map peak. First contours correspond to the map noise level~$\times\,3$; each subsequent contour marks intensity increase by the factor of two. The CLEAN beam at the half-maximum level is shown in the map lower left corner as a black ellipse; its major and minor axes and position angle are specified below the map.}
\figsetgrpend

\figsetgrpstart
\figsetgrpnum{4.8}
\figsetgrptitle{J0038+8447 at 2.3 GHz}
\figsetplot{J0038+8447_S_2006_02_16_avp_map.pdf}
\figsetgrpnote{The source name, the observation date, and the frequency are given at the top of the panel. The intensity is shown by contours: solid lines are the positive contours, dotted lines are the negative contours. The contour levels in percents of the map peak are specified below the map, as well as the total flux density of the CLEAN model of the source and the intensity of the map peak. First contours correspond to the map noise level~$\times\,3$; each subsequent contour marks intensity increase by the factor of two. The CLEAN beam at the half-maximum level is shown in the map lower left corner as a black ellipse; its major and minor axes and position angle are specified below the map.}
\figsetgrpend

\figsetgrpstart
\figsetgrpnum{4.9}
\figsetgrptitle{J0157+7552 at 2.3 GHz}
\figsetplot{J0157+7552_S_2006_02_14_avp_map.pdf}
\figsetgrpnote{The source name, the observation date, and the frequency are given at the top of the panel. The intensity is shown by contours: solid lines are the positive contours, dotted lines are the negative contours. The contour levels in percents of the map peak are specified below the map, as well as the total flux density of the CLEAN model of the source and the intensity of the map peak. First contours correspond to the map noise level~$\times\,3$; each subsequent contour marks intensity increase by the factor of two. The CLEAN beam at the half-maximum level is shown in the map lower left corner as a black ellipse; its major and minor axes and position angle are specified below the map.}
\figsetgrpend

\figsetgrpstart
\figsetgrpnum{4.10}
\figsetgrptitle{J0203+8106 at 2.3 GHz}
\figsetplot{J0203+8106_S_2006_02_14_avp_map.pdf}
\figsetgrpnote{The source name, the observation date, and the frequency are given at the top of the panel. The intensity is shown by contours: solid lines are the positive contours, dotted lines are the negative contours. The contour levels in percents of the map peak are specified below the map, as well as the total flux density of the CLEAN model of the source and the intensity of the map peak. First contours correspond to the map noise level~$\times\,3$; each subsequent contour marks intensity increase by the factor of two. The CLEAN beam at the half-maximum level is shown in the map lower left corner as a black ellipse; its major and minor axes and position angle are specified below the map.}
\figsetgrpend

\figsetgrpstart
\figsetgrpnum{4.11}
\figsetgrptitle{J0203+8106 at 8.6 GHz}
\figsetplot{J0203+8106_X_2006_02_14_avp_map.pdf}
\figsetgrpnote{The source name, the observation date, and the frequency are given at the top of the panel. The intensity is shown by contours: solid lines are the positive contours, dotted lines are the negative contours. The contour levels in percents of the map peak are specified below the map, as well as the total flux density of the CLEAN model of the source and the intensity of the map peak. First contours correspond to the map noise level~$\times\,3$; each subsequent contour marks intensity increase by the factor of two. The CLEAN beam at the half-maximum level is shown in the map lower left corner as a black ellipse; its major and minor axes and position angle are specified below the map.}
\figsetgrpend

\figsetgrpstart
\figsetgrpnum{4.12}
\figsetgrptitle{J0205+7522 at 2.3 GHz}
\figsetplot{J0205+7522_S_2006_02_16_avp_map.pdf}
\figsetgrpnote{The source name, the observation date, and the frequency are given at the top of the panel. The intensity is shown by contours: solid lines are the positive contours, dotted lines are the negative contours. The contour levels in percents of the map peak are specified below the map, as well as the total flux density of the CLEAN model of the source and the intensity of the map peak. First contours correspond to the map noise level~$\times\,3$; each subsequent contour marks intensity increase by the factor of two. The CLEAN beam at the half-maximum level is shown in the map lower left corner as a black ellipse; its major and minor axes and position angle are specified below the map.}
\figsetgrpend

\figsetgrpstart
\figsetgrpnum{4.13}
\figsetgrptitle{J0205+7522 at 8.6 GHz}
\figsetplot{J0205+7522_X_2006_02_16_avp_map.pdf}
\figsetgrpnote{The source name, the observation date, and the frequency are given at the top of the panel. The intensity is shown by contours: solid lines are the positive contours, dotted lines are the negative contours. The contour levels in percents of the map peak are specified below the map, as well as the total flux density of the CLEAN model of the source and the intensity of the map peak. First contours correspond to the map noise level~$\times\,3$; each subsequent contour marks intensity increase by the factor of two. The CLEAN beam at the half-maximum level is shown in the map lower left corner as a black ellipse; its major and minor axes and position angle are specified below the map.}
\figsetgrpend

\figsetgrpstart
\figsetgrpnum{4.14}
\figsetgrptitle{J0257+7843 at 2.3 GHz}
\figsetplot{J0257+7843_S_2006_02_16_avp_map.pdf}
\figsetgrpnote{The source name, the observation date, and the frequency are given at the top of the panel. The intensity is shown by contours: solid lines are the positive contours, dotted lines are the negative contours. The contour levels in percents of the map peak are specified below the map, as well as the total flux density of the CLEAN model of the source and the intensity of the map peak. First contours correspond to the map noise level~$\times\,3$; each subsequent contour marks intensity increase by the factor of two. The CLEAN beam at the half-maximum level is shown in the map lower left corner as a black ellipse; its major and minor axes and position angle are specified below the map.}
\figsetgrpend

\figsetgrpstart
\figsetgrpnum{4.15}
\figsetgrptitle{J0257+7843 at 8.6 GHz}
\figsetplot{J0257+7843_X_2006_02_16_avp_map.pdf}
\figsetgrpnote{The source name, the observation date, and the frequency are given at the top of the panel. The intensity is shown by contours: solid lines are the positive contours, dotted lines are the negative contours. The contour levels in percents of the map peak are specified below the map, as well as the total flux density of the CLEAN model of the source and the intensity of the map peak. First contours correspond to the map noise level~$\times\,3$; each subsequent contour marks intensity increase by the factor of two. The CLEAN beam at the half-maximum level is shown in the map lower left corner as a black ellipse; its major and minor axes and position angle are specified below the map.}
\figsetgrpend

\figsetgrpstart
\figsetgrpnum{4.16}
\figsetgrptitle{J0300+8202 at 2.3 GHz}
\figsetplot{J0300+8202_S_2006_02_16_avp_map.pdf}
\figsetgrpnote{The source name, the observation date, and the frequency are given at the top of the panel. The intensity is shown by contours: solid lines are the positive contours, dotted lines are the negative contours. The contour levels in percents of the map peak are specified below the map, as well as the total flux density of the CLEAN model of the source and the intensity of the map peak. First contours correspond to the map noise level~$\times\,3$; each subsequent contour marks intensity increase by the factor of two. The CLEAN beam at the half-maximum level is shown in the map lower left corner as a black ellipse; its major and minor axes and position angle are specified below the map.}
\figsetgrpend

\figsetgrpstart
\figsetgrpnum{4.17}
\figsetgrptitle{J0304+7727 at 2.3 GHz}
\figsetplot{J0304+7727_S_2006_02_14_avp_map.pdf}
\figsetgrpnote{The source name, the observation date, and the frequency are given at the top of the panel. The intensity is shown by contours: solid lines are the positive contours, dotted lines are the negative contours. The contour levels in percents of the map peak are specified below the map, as well as the total flux density of the CLEAN model of the source and the intensity of the map peak. First contours correspond to the map noise level~$\times\,3$; each subsequent contour marks intensity increase by the factor of two. The CLEAN beam at the half-maximum level is shown in the map lower left corner as a black ellipse; its major and minor axes and position angle are specified below the map.}
\figsetgrpend

\figsetgrpstart
\figsetgrpnum{4.18}
\figsetgrptitle{J0304+7727 at 8.6 GHz}
\figsetplot{J0304+7727_X_2006_02_14_avp_map.pdf}
\figsetgrpnote{The source name, the observation date, and the frequency are given at the top of the panel. The intensity is shown by contours: solid lines are the positive contours, dotted lines are the negative contours. The contour levels in percents of the map peak are specified below the map, as well as the total flux density of the CLEAN model of the source and the intensity of the map peak. First contours correspond to the map noise level~$\times\,3$; each subsequent contour marks intensity increase by the factor of two. The CLEAN beam at the half-maximum level is shown in the map lower left corner as a black ellipse; its major and minor axes and position angle are specified below the map.}
\figsetgrpend

\figsetgrpstart
\figsetgrpnum{4.19}
\figsetgrptitle{J0354+8009 at 2.3 GHz}
\figsetplot{J0354+8009_S_2006_02_23_avp_map.pdf}
\figsetgrpnote{The source name, the observation date, and the frequency are given at the top of the panel. The intensity is shown by contours: solid lines are the positive contours, dotted lines are the negative contours. The contour levels in percents of the map peak are specified below the map, as well as the total flux density of the CLEAN model of the source and the intensity of the map peak. First contours correspond to the map noise level~$\times\,3$; each subsequent contour marks intensity increase by the factor of two. The CLEAN beam at the half-maximum level is shown in the map lower left corner as a black ellipse; its major and minor axes and position angle are specified below the map.}
\figsetgrpend

\figsetgrpstart
\figsetgrpnum{4.20}
\figsetgrptitle{J0354+8009 at 8.6 GHz}
\figsetplot{J0354+8009_X_2006_02_23_avp_map.pdf}
\figsetgrpnote{The source name, the observation date, and the frequency are given at the top of the panel. The intensity is shown by contours: solid lines are the positive contours, dotted lines are the negative contours. The contour levels in percents of the map peak are specified below the map, as well as the total flux density of the CLEAN model of the source and the intensity of the map peak. First contours correspond to the map noise level~$\times\,3$; each subsequent contour marks intensity increase by the factor of two. The CLEAN beam at the half-maximum level is shown in the map lower left corner as a black ellipse; its major and minor axes and position angle are specified below the map.}
\figsetgrpend

\figsetgrpstart
\figsetgrpnum{4.21}
\figsetgrptitle{J0410+7656 at 2.3 GHz}
\figsetplot{J0410+7656_S_2006_02_23_avp_map.pdf}
\figsetgrpnote{The source name, the observation date, and the frequency are given at the top of the panel. The intensity is shown by contours: solid lines are the positive contours, dotted lines are the negative contours. The contour levels in percents of the map peak are specified below the map, as well as the total flux density of the CLEAN model of the source and the intensity of the map peak. First contours correspond to the map noise level~$\times\,3$; each subsequent contour marks intensity increase by the factor of two. The CLEAN beam at the half-maximum level is shown in the map lower left corner as a black ellipse; its major and minor axes and position angle are specified below the map.}
\figsetgrpend

\figsetgrpstart
\figsetgrpnum{4.22}
\figsetgrptitle{J0410+7656 at 8.6 GHz}
\figsetplot{J0410+7656_X_2006_02_23_avp_map.pdf}
\figsetgrpnote{The source name, the observation date, and the frequency are given at the top of the panel. The intensity is shown by contours: solid lines are the positive contours, dotted lines are the negative contours. The contour levels in percents of the map peak are specified below the map, as well as the total flux density of the CLEAN model of the source and the intensity of the map peak. First contours correspond to the map noise level~$\times\,3$; each subsequent contour marks intensity increase by the factor of two. The CLEAN beam at the half-maximum level is shown in the map lower left corner as a black ellipse; its major and minor axes and position angle are specified below the map.}
\figsetgrpend

\figsetgrpstart
\figsetgrpnum{4.23}
\figsetgrptitle{J0410+8208 at 2.3 GHz}
\figsetplot{J0410+8208_S_2006_02_16_avp_map.pdf}
\figsetgrpnote{The source name, the observation date, and the frequency are given at the top of the panel. The intensity is shown by contours: solid lines are the positive contours, dotted lines are the negative contours. The contour levels in percents of the map peak are specified below the map, as well as the total flux density of the CLEAN model of the source and the intensity of the map peak. First contours correspond to the map noise level~$\times\,3$; each subsequent contour marks intensity increase by the factor of two. The CLEAN beam at the half-maximum level is shown in the map lower left corner as a black ellipse; its major and minor axes and position angle are specified below the map.}
\figsetgrpend

\figsetgrpstart
\figsetgrpnum{4.24}
\figsetgrptitle{J0449+8233 at 2.3 GHz}
\figsetplot{J0449+8233_S_2006_02_23_avp_map.pdf}
\figsetgrpnote{The source name, the observation date, and the frequency are given at the top of the panel. The intensity is shown by contours: solid lines are the positive contours, dotted lines are the negative contours. The contour levels in percents of the map peak are specified below the map, as well as the total flux density of the CLEAN model of the source and the intensity of the map peak. First contours correspond to the map noise level~$\times\,3$; each subsequent contour marks intensity increase by the factor of two. The CLEAN beam at the half-maximum level is shown in the map lower left corner as a black ellipse; its major and minor axes and position angle are specified below the map.}
\figsetgrpend

\figsetgrpstart
\figsetgrpnum{4.25}
\figsetgrptitle{J0449+8233 at 8.6 GHz}
\figsetplot{J0449+8233_X_2006_02_23_avp_map.pdf}
\figsetgrpnote{The source name, the observation date, and the frequency are given at the top of the panel. The intensity is shown by contours: solid lines are the positive contours, dotted lines are the negative contours. The contour levels in percents of the map peak are specified below the map, as well as the total flux density of the CLEAN model of the source and the intensity of the map peak. First contours correspond to the map noise level~$\times\,3$; each subsequent contour marks intensity increase by the factor of two. The CLEAN beam at the half-maximum level is shown in the map lower left corner as a black ellipse; its major and minor axes and position angle are specified below the map.}
\figsetgrpend

\figsetgrpstart
\figsetgrpnum{4.26}
\figsetgrptitle{J0508+8432 at 2.3 GHz}
\figsetplot{J0508+8432_S_2006_02_14_avp_map.pdf}
\figsetgrpnote{The source name, the observation date, and the frequency are given at the top of the panel. The intensity is shown by contours: solid lines are the positive contours, dotted lines are the negative contours. The contour levels in percents of the map peak are specified below the map, as well as the total flux density of the CLEAN model of the source and the intensity of the map peak. First contours correspond to the map noise level~$\times\,3$; each subsequent contour marks intensity increase by the factor of two. The CLEAN beam at the half-maximum level is shown in the map lower left corner as a black ellipse; its major and minor axes and position angle are specified below the map.}
\figsetgrpend

\figsetgrpstart
\figsetgrpnum{4.27}
\figsetgrptitle{J0508+8432 at 8.6 GHz}
\figsetplot{J0508+8432_X_2006_02_14_avp_map.pdf}
\figsetgrpnote{The source name, the observation date, and the frequency are given at the top of the panel. The intensity is shown by contours: solid lines are the positive contours, dotted lines are the negative contours. The contour levels in percents of the map peak are specified below the map, as well as the total flux density of the CLEAN model of the source and the intensity of the map peak. First contours correspond to the map noise level~$\times\,3$; each subsequent contour marks intensity increase by the factor of two. The CLEAN beam at the half-maximum level is shown in the map lower left corner as a black ellipse; its major and minor axes and position angle are specified below the map.}
\figsetgrpend

\figsetgrpstart
\figsetgrpnum{4.28}
\figsetgrptitle{J0543+8238 at 2.3 GHz}
\figsetplot{J0543+8238_S_2006_02_23_avp_map.pdf}
\figsetgrpnote{The source name, the observation date, and the frequency are given at the top of the panel. The intensity is shown by contours: solid lines are the positive contours, dotted lines are the negative contours. The contour levels in percents of the map peak are specified below the map, as well as the total flux density of the CLEAN model of the source and the intensity of the map peak. First contours correspond to the map noise level~$\times\,3$; each subsequent contour marks intensity increase by the factor of two. The CLEAN beam at the half-maximum level is shown in the map lower left corner as a black ellipse; its major and minor axes and position angle are specified below the map.}
\figsetgrpend

\figsetgrpstart
\figsetgrpnum{4.29}
\figsetgrptitle{J0543+8238 at 8.6 GHz}
\figsetplot{J0543+8238_X_2006_02_23_avp_map.pdf}
\figsetgrpnote{The source name, the observation date, and the frequency are given at the top of the panel. The intensity is shown by contours: solid lines are the positive contours, dotted lines are the negative contours. The contour levels in percents of the map peak are specified below the map, as well as the total flux density of the CLEAN model of the source and the intensity of the map peak. First contours correspond to the map noise level~$\times\,3$; each subsequent contour marks intensity increase by the factor of two. The CLEAN beam at the half-maximum level is shown in the map lower left corner as a black ellipse; its major and minor axes and position angle are specified below the map.}
\figsetgrpend

\figsetgrpstart
\figsetgrpnum{4.30}
\figsetgrptitle{J0621+7605 at 2.3 GHz}
\figsetplot{J0621+7605_S_2006_02_16_avp_map.pdf}
\figsetgrpnote{The source name, the observation date, and the frequency are given at the top of the panel. The intensity is shown by contours: solid lines are the positive contours, dotted lines are the negative contours. The contour levels in percents of the map peak are specified below the map, as well as the total flux density of the CLEAN model of the source and the intensity of the map peak. First contours correspond to the map noise level~$\times\,3$; each subsequent contour marks intensity increase by the factor of two. The CLEAN beam at the half-maximum level is shown in the map lower left corner as a black ellipse; its major and minor axes and position angle are specified below the map.}
\figsetgrpend

\figsetgrpstart
\figsetgrpnum{4.31}
\figsetgrptitle{J0621+7605 at 8.6 GHz}
\figsetplot{J0621+7605_X_2006_02_16_avp_map.pdf}
\figsetgrpnote{The source name, the observation date, and the frequency are given at the top of the panel. The intensity is shown by contours: solid lines are the positive contours, dotted lines are the negative contours. The contour levels in percents of the map peak are specified below the map, as well as the total flux density of the CLEAN model of the source and the intensity of the map peak. First contours correspond to the map noise level~$\times\,3$; each subsequent contour marks intensity increase by the factor of two. The CLEAN beam at the half-maximum level is shown in the map lower left corner as a black ellipse; its major and minor axes and position angle are specified below the map.}
\figsetgrpend

\figsetgrpstart
\figsetgrpnum{4.32}
\figsetgrptitle{J0626+8202 at 2.3 GHz}
\figsetplot{J0626+8202_S_2006_02_14_avp_map.pdf}
\figsetgrpnote{The source name, the observation date, and the frequency are given at the top of the panel. The intensity is shown by contours: solid lines are the positive contours, dotted lines are the negative contours. The contour levels in percents of the map peak are specified below the map, as well as the total flux density of the CLEAN model of the source and the intensity of the map peak. First contours correspond to the map noise level~$\times\,3$; each subsequent contour marks intensity increase by the factor of two. The CLEAN beam at the half-maximum level is shown in the map lower left corner as a black ellipse; its major and minor axes and position angle are specified below the map.}
\figsetgrpend

\figsetgrpstart
\figsetgrpnum{4.33}
\figsetgrptitle{J0626+8202 at 8.6 GHz}
\figsetplot{J0626+8202_X_2006_02_14_avp_map.pdf}
\figsetgrpnote{The source name, the observation date, and the frequency are given at the top of the panel. The intensity is shown by contours: solid lines are the positive contours, dotted lines are the negative contours. The contour levels in percents of the map peak are specified below the map, as well as the total flux density of the CLEAN model of the source and the intensity of the map peak. First contours correspond to the map noise level~$\times\,3$; each subsequent contour marks intensity increase by the factor of two. The CLEAN beam at the half-maximum level is shown in the map lower left corner as a black ellipse; its major and minor axes and position angle are specified below the map.}
\figsetgrpend

\figsetgrpstart
\figsetgrpnum{4.34}
\figsetgrptitle{J0632+8020 at 2.3 GHz}
\figsetplot{J0632+8020_S_2006_02_14_avp_map.pdf}
\figsetgrpnote{The source name, the observation date, and the frequency are given at the top of the panel. The intensity is shown by contours: solid lines are the positive contours, dotted lines are the negative contours. The contour levels in percents of the map peak are specified below the map, as well as the total flux density of the CLEAN model of the source and the intensity of the map peak. First contours correspond to the map noise level~$\times\,3$; each subsequent contour marks intensity increase by the factor of two. The CLEAN beam at the half-maximum level is shown in the map lower left corner as a black ellipse; its major and minor axes and position angle are specified below the map.}
\figsetgrpend

\figsetgrpstart
\figsetgrpnum{4.35}
\figsetgrptitle{J0632+8020 at 8.6 GHz}
\figsetplot{J0632+8020_X_2006_02_14_avp_map.pdf}
\figsetgrpnote{The source name, the observation date, and the frequency are given at the top of the panel. The intensity is shown by contours: solid lines are the positive contours, dotted lines are the negative contours. The contour levels in percents of the map peak are specified below the map, as well as the total flux density of the CLEAN model of the source and the intensity of the map peak. First contours correspond to the map noise level~$\times\,3$; each subsequent contour marks intensity increase by the factor of two. The CLEAN beam at the half-maximum level is shown in the map lower left corner as a black ellipse; its major and minor axes and position angle are specified below the map.}
\figsetgrpend

\figsetgrpstart
\figsetgrpnum{4.36}
\figsetgrptitle{J0637+8125 at 2.3 GHz}
\figsetplot{J0637+8125_S_2006_02_23_avp_map.pdf}
\figsetgrpnote{The source name, the observation date, and the frequency are given at the top of the panel. The intensity is shown by contours: solid lines are the positive contours, dotted lines are the negative contours. The contour levels in percents of the map peak are specified below the map, as well as the total flux density of the CLEAN model of the source and the intensity of the map peak. First contours correspond to the map noise level~$\times\,3$; each subsequent contour marks intensity increase by the factor of two. The CLEAN beam at the half-maximum level is shown in the map lower left corner as a black ellipse; its major and minor axes and position angle are specified below the map.}
\figsetgrpend

\figsetgrpstart
\figsetgrpnum{4.37}
\figsetgrptitle{J0637+8125 at 8.6 GHz}
\figsetplot{J0637+8125_X_2006_02_23_avp_map.pdf}
\figsetgrpnote{The source name, the observation date, and the frequency are given at the top of the panel. The intensity is shown by contours: solid lines are the positive contours, dotted lines are the negative contours. The contour levels in percents of the map peak are specified below the map, as well as the total flux density of the CLEAN model of the source and the intensity of the map peak. First contours correspond to the map noise level~$\times\,3$; each subsequent contour marks intensity increase by the factor of two. The CLEAN beam at the half-maximum level is shown in the map lower left corner as a black ellipse; its major and minor axes and position angle are specified below the map.}
\figsetgrpend

\figsetgrpstart
\figsetgrpnum{4.38}
\figsetgrptitle{J0644+8018 at 2.3 GHz}
\figsetplot{J0644+8018_S_2006_02_14_avp_map.pdf}
\figsetgrpnote{The source name, the observation date, and the frequency are given at the top of the panel. The intensity is shown by contours: solid lines are the positive contours, dotted lines are the negative contours. The contour levels in percents of the map peak are specified below the map, as well as the total flux density of the CLEAN model of the source and the intensity of the map peak. First contours correspond to the map noise level~$\times\,3$; each subsequent contour marks intensity increase by the factor of two. The CLEAN beam at the half-maximum level is shown in the map lower left corner as a black ellipse; its major and minor axes and position angle are specified below the map.}
\figsetgrpend

\figsetgrpstart
\figsetgrpnum{4.39}
\figsetgrptitle{J0702+8549 at 2.3 GHz}
\figsetplot{J0702+8549_S_2006_02_14_avp_map.pdf}
\figsetgrpnote{The source name, the observation date, and the frequency are given at the top of the panel. The intensity is shown by contours: solid lines are the positive contours, dotted lines are the negative contours. The contour levels in percents of the map peak are specified below the map, as well as the total flux density of the CLEAN model of the source and the intensity of the map peak. First contours correspond to the map noise level~$\times\,3$; each subsequent contour marks intensity increase by the factor of two. The CLEAN beam at the half-maximum level is shown in the map lower left corner as a black ellipse; its major and minor axes and position angle are specified below the map.}
\figsetgrpend

\figsetgrpstart
\figsetgrpnum{4.40}
\figsetgrptitle{J0702+8549 at 8.6 GHz}
\figsetplot{J0702+8549_X_2006_02_14_avp_map.pdf}
\figsetgrpnote{The source name, the observation date, and the frequency are given at the top of the panel. The intensity is shown by contours: solid lines are the positive contours, dotted lines are the negative contours. The contour levels in percents of the map peak are specified below the map, as well as the total flux density of the CLEAN model of the source and the intensity of the map peak. First contours correspond to the map noise level~$\times\,3$; each subsequent contour marks intensity increase by the factor of two. The CLEAN beam at the half-maximum level is shown in the map lower left corner as a black ellipse; its major and minor axes and position angle are specified below the map.}
\figsetgrpend

\figsetgrpstart
\figsetgrpnum{4.41}
\figsetgrptitle{J0726+7911 at 2.3 GHz}
\figsetplot{J0726+7911_S_2006_02_16_avp_map.pdf}
\figsetgrpnote{The source name, the observation date, and the frequency are given at the top of the panel. The intensity is shown by contours: solid lines are the positive contours, dotted lines are the negative contours. The contour levels in percents of the map peak are specified below the map, as well as the total flux density of the CLEAN model of the source and the intensity of the map peak. First contours correspond to the map noise level~$\times\,3$; each subsequent contour marks intensity increase by the factor of two. The CLEAN beam at the half-maximum level is shown in the map lower left corner as a black ellipse; its major and minor axes and position angle are specified below the map.}
\figsetgrpend

\figsetgrpstart
\figsetgrpnum{4.42}
\figsetgrptitle{J0726+7911 at 8.6 GHz}
\figsetplot{J0726+7911_X_2006_02_16_avp_map.pdf}
\figsetgrpnote{The source name, the observation date, and the frequency are given at the top of the panel. The intensity is shown by contours: solid lines are the positive contours, dotted lines are the negative contours. The contour levels in percents of the map peak are specified below the map, as well as the total flux density of the CLEAN model of the source and the intensity of the map peak. First contours correspond to the map noise level~$\times\,3$; each subsequent contour marks intensity increase by the factor of two. The CLEAN beam at the half-maximum level is shown in the map lower left corner as a black ellipse; its major and minor axes and position angle are specified below the map.}
\figsetgrpend

\figsetgrpstart
\figsetgrpnum{4.43}
\figsetgrptitle{J0750+8241 at 2.3 GHz}
\figsetplot{J0750+8241_S_2006_02_14_avp_map.pdf}
\figsetgrpnote{The source name, the observation date, and the frequency are given at the top of the panel. The intensity is shown by contours: solid lines are the positive contours, dotted lines are the negative contours. The contour levels in percents of the map peak are specified below the map, as well as the total flux density of the CLEAN model of the source and the intensity of the map peak. First contours correspond to the map noise level~$\times\,3$; each subsequent contour marks intensity increase by the factor of two. The CLEAN beam at the half-maximum level is shown in the map lower left corner as a black ellipse; its major and minor axes and position angle are specified below the map.}
\figsetgrpend

\figsetgrpstart
\figsetgrpnum{4.44}
\figsetgrptitle{J0750+8241 at 8.6 GHz}
\figsetplot{J0750+8241_X_2006_02_14_avp_map.pdf}
\figsetgrpnote{The source name, the observation date, and the frequency are given at the top of the panel. The intensity is shown by contours: solid lines are the positive contours, dotted lines are the negative contours. The contour levels in percents of the map peak are specified below the map, as well as the total flux density of the CLEAN model of the source and the intensity of the map peak. First contours correspond to the map noise level~$\times\,3$; each subsequent contour marks intensity increase by the factor of two. The CLEAN beam at the half-maximum level is shown in the map lower left corner as a black ellipse; its major and minor axes and position angle are specified below the map.}
\figsetgrpend

\figsetgrpstart
\figsetgrpnum{4.45}
\figsetgrptitle{J0802+7620 at 2.3 GHz}
\figsetplot{J0802+7620_S_2006_02_23_avp_map.pdf}
\figsetgrpnote{The source name, the observation date, and the frequency are given at the top of the panel. The intensity is shown by contours: solid lines are the positive contours, dotted lines are the negative contours. The contour levels in percents of the map peak are specified below the map, as well as the total flux density of the CLEAN model of the source and the intensity of the map peak. First contours correspond to the map noise level~$\times\,3$; each subsequent contour marks intensity increase by the factor of two. The CLEAN beam at the half-maximum level is shown in the map lower left corner as a black ellipse; its major and minor axes and position angle are specified below the map.}
\figsetgrpend

\figsetgrpstart
\figsetgrpnum{4.46}
\figsetgrptitle{J0806+7746 at 2.3 GHz}
\figsetplot{J0806+7746_S_2006_02_23_avp_map.pdf}
\figsetgrpnote{The source name, the observation date, and the frequency are given at the top of the panel. The intensity is shown by contours: solid lines are the positive contours, dotted lines are the negative contours. The contour levels in percents of the map peak are specified below the map, as well as the total flux density of the CLEAN model of the source and the intensity of the map peak. First contours correspond to the map noise level~$\times\,3$; each subsequent contour marks intensity increase by the factor of two. The CLEAN beam at the half-maximum level is shown in the map lower left corner as a black ellipse; its major and minor axes and position angle are specified below the map.}
\figsetgrpend

\figsetgrpstart
\figsetgrpnum{4.47}
\figsetgrptitle{J0806+7746 at 8.6 GHz}
\figsetplot{J0806+7746_X_2006_02_23_avp_map.pdf}
\figsetgrpnote{The source name, the observation date, and the frequency are given at the top of the panel. The intensity is shown by contours: solid lines are the positive contours, dotted lines are the negative contours. The contour levels in percents of the map peak are specified below the map, as well as the total flux density of the CLEAN model of the source and the intensity of the map peak. First contours correspond to the map noise level~$\times\,3$; each subsequent contour marks intensity increase by the factor of two. The CLEAN beam at the half-maximum level is shown in the map lower left corner as a black ellipse; its major and minor axes and position angle are specified below the map.}
\figsetgrpend

\figsetgrpstart
\figsetgrpnum{4.48}
\figsetgrptitle{J0819+8105 at 2.3 GHz}
\figsetplot{J0819+8105_S_2006_02_16_avp_map.pdf}
\figsetgrpnote{The source name, the observation date, and the frequency are given at the top of the panel. The intensity is shown by contours: solid lines are the positive contours, dotted lines are the negative contours. The contour levels in percents of the map peak are specified below the map, as well as the total flux density of the CLEAN model of the source and the intensity of the map peak. First contours correspond to the map noise level~$\times\,3$; each subsequent contour marks intensity increase by the factor of two. The CLEAN beam at the half-maximum level is shown in the map lower left corner as a black ellipse; its major and minor axes and position angle are specified below the map.}
\figsetgrpend

\figsetgrpstart
\figsetgrpnum{4.49}
\figsetgrptitle{J0858+7501 at 2.3 GHz}
\figsetplot{J0858+7501_S_2006_02_16_avp_map.pdf}
\figsetgrpnote{The source name, the observation date, and the frequency are given at the top of the panel. The intensity is shown by contours: solid lines are the positive contours, dotted lines are the negative contours. The contour levels in percents of the map peak are specified below the map, as well as the total flux density of the CLEAN model of the source and the intensity of the map peak. First contours correspond to the map noise level~$\times\,3$; each subsequent contour marks intensity increase by the factor of two. The CLEAN beam at the half-maximum level is shown in the map lower left corner as a black ellipse; its major and minor axes and position angle are specified below the map.}
\figsetgrpend

\figsetgrpstart
\figsetgrpnum{4.50}
\figsetgrptitle{J0909+8327 at 2.3 GHz}
\figsetplot{J0909+8327_S_2006_02_16_avp_map.pdf}
\figsetgrpnote{The source name, the observation date, and the frequency are given at the top of the panel. The intensity is shown by contours: solid lines are the positive contours, dotted lines are the negative contours. The contour levels in percents of the map peak are specified below the map, as well as the total flux density of the CLEAN model of the source and the intensity of the map peak. First contours correspond to the map noise level~$\times\,3$; each subsequent contour marks intensity increase by the factor of two. The CLEAN beam at the half-maximum level is shown in the map lower left corner as a black ellipse; its major and minor axes and position angle are specified below the map.}
\figsetgrpend

\figsetgrpstart
\figsetgrpnum{4.51}
\figsetgrptitle{J0909+8327 at 8.6 GHz}
\figsetplot{J0909+8327_X_2006_02_16_avp_map.pdf}
\figsetgrpnote{The source name, the observation date, and the frequency are given at the top of the panel. The intensity is shown by contours: solid lines are the positive contours, dotted lines are the negative contours. The contour levels in percents of the map peak are specified below the map, as well as the total flux density of the CLEAN model of the source and the intensity of the map peak. First contours correspond to the map noise level~$\times\,3$; each subsequent contour marks intensity increase by the factor of two. The CLEAN beam at the half-maximum level is shown in the map lower left corner as a black ellipse; its major and minor axes and position angle are specified below the map.}
\figsetgrpend

\figsetgrpstart
\figsetgrpnum{4.52}
\figsetgrptitle{J0911+8607 at 8.6 GHz}
\figsetplot{J0911+8607_X_2006_02_23_avp_map.pdf}
\figsetgrpnote{The source name, the observation date, and the frequency are given at the top of the panel. The intensity is shown by contours: solid lines are the positive contours, dotted lines are the negative contours. The contour levels in percents of the map peak are specified below the map, as well as the total flux density of the CLEAN model of the source and the intensity of the map peak. First contours correspond to the map noise level~$\times\,3$; each subsequent contour marks intensity increase by the factor of two. The CLEAN beam at the half-maximum level is shown in the map lower left corner as a black ellipse; its major and minor axes and position angle are specified below the map.}
\figsetgrpend

\figsetgrpstart
\figsetgrpnum{4.53}
\figsetgrptitle{J0919+7825 at 2.3 GHz}
\figsetplot{J0919+7825_S_2006_02_14_avp_map.pdf}
\figsetgrpnote{The source name, the observation date, and the frequency are given at the top of the panel. The intensity is shown by contours: solid lines are the positive contours, dotted lines are the negative contours. The contour levels in percents of the map peak are specified below the map, as well as the total flux density of the CLEAN model of the source and the intensity of the map peak. First contours correspond to the map noise level~$\times\,3$; each subsequent contour marks intensity increase by the factor of two. The CLEAN beam at the half-maximum level is shown in the map lower left corner as a black ellipse; its major and minor axes and position angle are specified below the map.}
\figsetgrpend

\figsetgrpstart
\figsetgrpnum{4.54}
\figsetgrptitle{J0919+7825 at 8.6 GHz}
\figsetplot{J0919+7825_X_2006_02_14_avp_map.pdf}
\figsetgrpnote{The source name, the observation date, and the frequency are given at the top of the panel. The intensity is shown by contours: solid lines are the positive contours, dotted lines are the negative contours. The contour levels in percents of the map peak are specified below the map, as well as the total flux density of the CLEAN model of the source and the intensity of the map peak. First contours correspond to the map noise level~$\times\,3$; each subsequent contour marks intensity increase by the factor of two. The CLEAN beam at the half-maximum level is shown in the map lower left corner as a black ellipse; its major and minor axes and position angle are specified below the map.}
\figsetgrpend

\figsetgrpstart
\figsetgrpnum{4.55}
\figsetgrptitle{J1010+8250 at 2.3 GHz}
\figsetplot{J1010+8250_S_2006_02_16_avp_map.pdf}
\figsetgrpnote{The source name, the observation date, and the frequency are given at the top of the panel. The intensity is shown by contours: solid lines are the positive contours, dotted lines are the negative contours. The contour levels in percents of the map peak are specified below the map, as well as the total flux density of the CLEAN model of the source and the intensity of the map peak. First contours correspond to the map noise level~$\times\,3$; each subsequent contour marks intensity increase by the factor of two. The CLEAN beam at the half-maximum level is shown in the map lower left corner as a black ellipse; its major and minor axes and position angle are specified below the map.}
\figsetgrpend

\figsetgrpstart
\figsetgrpnum{4.56}
\figsetgrptitle{J1010+8250 at 8.6 GHz}
\figsetplot{J1010+8250_X_2006_02_16_avp_map.pdf}
\figsetgrpnote{The source name, the observation date, and the frequency are given at the top of the panel. The intensity is shown by contours: solid lines are the positive contours, dotted lines are the negative contours. The contour levels in percents of the map peak are specified below the map, as well as the total flux density of the CLEAN model of the source and the intensity of the map peak. First contours correspond to the map noise level~$\times\,3$; each subsequent contour marks intensity increase by the factor of two. The CLEAN beam at the half-maximum level is shown in the map lower left corner as a black ellipse; its major and minor axes and position angle are specified below the map.}
\figsetgrpend

\figsetgrpstart
\figsetgrpnum{4.57}
\figsetgrptitle{J1016+7617 at 2.3 GHz}
\figsetplot{J1016+7617_S_2006_02_23_avp_map.pdf}
\figsetgrpnote{The source name, the observation date, and the frequency are given at the top of the panel. The intensity is shown by contours: solid lines are the positive contours, dotted lines are the negative contours. The contour levels in percents of the map peak are specified below the map, as well as the total flux density of the CLEAN model of the source and the intensity of the map peak. First contours correspond to the map noise level~$\times\,3$; each subsequent contour marks intensity increase by the factor of two. The CLEAN beam at the half-maximum level is shown in the map lower left corner as a black ellipse; its major and minor axes and position angle are specified below the map.}
\figsetgrpend

\figsetgrpstart
\figsetgrpnum{4.58}
\figsetgrptitle{J1044+8054 at 2.3 GHz}
\figsetplot{J1044+8054_S_2006_02_14_avp_map.pdf}
\figsetgrpnote{The source name, the observation date, and the frequency are given at the top of the panel. The intensity is shown by contours: solid lines are the positive contours, dotted lines are the negative contours. The contour levels in percents of the map peak are specified below the map, as well as the total flux density of the CLEAN model of the source and the intensity of the map peak. First contours correspond to the map noise level~$\times\,3$; each subsequent contour marks intensity increase by the factor of two. The CLEAN beam at the half-maximum level is shown in the map lower left corner as a black ellipse; its major and minor axes and position angle are specified below the map.}
\figsetgrpend

\figsetgrpstart
\figsetgrpnum{4.59}
\figsetgrptitle{J1044+8054 at 8.6 GHz}
\figsetplot{J1044+8054_X_2006_02_14_avp_map.pdf}
\figsetgrpnote{The source name, the observation date, and the frequency are given at the top of the panel. The intensity is shown by contours: solid lines are the positive contours, dotted lines are the negative contours. The contour levels in percents of the map peak are specified below the map, as well as the total flux density of the CLEAN model of the source and the intensity of the map peak. First contours correspond to the map noise level~$\times\,3$; each subsequent contour marks intensity increase by the factor of two. The CLEAN beam at the half-maximum level is shown in the map lower left corner as a black ellipse; its major and minor axes and position angle are specified below the map.}
\figsetgrpend

\figsetgrpstart
\figsetgrpnum{4.60}
\figsetgrptitle{J1052+8317 at 2.3 GHz}
\figsetplot{J1052+8317_S_2006_02_14_avp_map.pdf}
\figsetgrpnote{The source name, the observation date, and the frequency are given at the top of the panel. The intensity is shown by contours: solid lines are the positive contours, dotted lines are the negative contours. The contour levels in percents of the map peak are specified below the map, as well as the total flux density of the CLEAN model of the source and the intensity of the map peak. First contours correspond to the map noise level~$\times\,3$; each subsequent contour marks intensity increase by the factor of two. The CLEAN beam at the half-maximum level is shown in the map lower left corner as a black ellipse; its major and minor axes and position angle are specified below the map.}
\figsetgrpend

\figsetgrpstart
\figsetgrpnum{4.61}
\figsetgrptitle{J1054+8629 at 2.3 GHz}
\figsetplot{J1054+8629_S_2006_02_16_avp_map.pdf}
\figsetgrpnote{The source name, the observation date, and the frequency are given at the top of the panel. The intensity is shown by contours: solid lines are the positive contours, dotted lines are the negative contours. The contour levels in percents of the map peak are specified below the map, as well as the total flux density of the CLEAN model of the source and the intensity of the map peak. First contours correspond to the map noise level~$\times\,3$; each subsequent contour marks intensity increase by the factor of two. The CLEAN beam at the half-maximum level is shown in the map lower left corner as a black ellipse; its major and minor axes and position angle are specified below the map.}
\figsetgrpend

\figsetgrpstart
\figsetgrpnum{4.62}
\figsetgrptitle{J1054+8629 at 8.6 GHz}
\figsetplot{J1054+8629_X_2006_02_16_avp_map.pdf}
\figsetgrpnote{The source name, the observation date, and the frequency are given at the top of the panel. The intensity is shown by contours: solid lines are the positive contours, dotted lines are the negative contours. The contour levels in percents of the map peak are specified below the map, as well as the total flux density of the CLEAN model of the source and the intensity of the map peak. First contours correspond to the map noise level~$\times\,3$; each subsequent contour marks intensity increase by the factor of two. The CLEAN beam at the half-maximum level is shown in the map lower left corner as a black ellipse; its major and minor axes and position angle are specified below the map.}
\figsetgrpend

\figsetgrpstart
\figsetgrpnum{4.63}
\figsetgrptitle{J1058+8114 at 2.3 GHz}
\figsetplot{J1058+8114_S_2006_02_14_avp_map.pdf}
\figsetgrpnote{The source name, the observation date, and the frequency are given at the top of the panel. The intensity is shown by contours: solid lines are the positive contours, dotted lines are the negative contours. The contour levels in percents of the map peak are specified below the map, as well as the total flux density of the CLEAN model of the source and the intensity of the map peak. First contours correspond to the map noise level~$\times\,3$; each subsequent contour marks intensity increase by the factor of two. The CLEAN beam at the half-maximum level is shown in the map lower left corner as a black ellipse; its major and minor axes and position angle are specified below the map.}
\figsetgrpend

\figsetgrpstart
\figsetgrpnum{4.64}
\figsetgrptitle{J1058+8114 at 2.3 GHz}
\figsetplot{J1058+8114_S_2006_02_16_avp_map.pdf}
\figsetgrpnote{The source name, the observation date, and the frequency are given at the top of the panel. The intensity is shown by contours: solid lines are the positive contours, dotted lines are the negative contours. The contour levels in percents of the map peak are specified below the map, as well as the total flux density of the CLEAN model of the source and the intensity of the map peak. First contours correspond to the map noise level~$\times\,3$; each subsequent contour marks intensity increase by the factor of two. The CLEAN beam at the half-maximum level is shown in the map lower left corner as a black ellipse; its major and minor axes and position angle are specified below the map.}
\figsetgrpend

\figsetgrpstart
\figsetgrpnum{4.65}
\figsetgrptitle{J1058+8114 at 2.3 GHz}
\figsetplot{J1058+8114_S_2006_02_23_avp_map.pdf}
\figsetgrpnote{The source name, the observation date, and the frequency are given at the top of the panel. The intensity is shown by contours: solid lines are the positive contours, dotted lines are the negative contours. The contour levels in percents of the map peak are specified below the map, as well as the total flux density of the CLEAN model of the source and the intensity of the map peak. First contours correspond to the map noise level~$\times\,3$; each subsequent contour marks intensity increase by the factor of two. The CLEAN beam at the half-maximum level is shown in the map lower left corner as a black ellipse; its major and minor axes and position angle are specified below the map.}
\figsetgrpend

\figsetgrpstart
\figsetgrpnum{4.66}
\figsetgrptitle{J1058+8114 at 8.6 GHz}
\figsetplot{J1058+8114_X_2006_02_14_avp_map.pdf}
\figsetgrpnote{The source name, the observation date, and the frequency are given at the top of the panel. The intensity is shown by contours: solid lines are the positive contours, dotted lines are the negative contours. The contour levels in percents of the map peak are specified below the map, as well as the total flux density of the CLEAN model of the source and the intensity of the map peak. First contours correspond to the map noise level~$\times\,3$; each subsequent contour marks intensity increase by the factor of two. The CLEAN beam at the half-maximum level is shown in the map lower left corner as a black ellipse; its major and minor axes and position angle are specified below the map.}
\figsetgrpend

\figsetgrpstart
\figsetgrpnum{4.67}
\figsetgrptitle{J1058+8114 at 8.6 GHz}
\figsetplot{J1058+8114_X_2006_02_16_avp_map.pdf}
\figsetgrpnote{The source name, the observation date, and the frequency are given at the top of the panel. The intensity is shown by contours: solid lines are the positive contours, dotted lines are the negative contours. The contour levels in percents of the map peak are specified below the map, as well as the total flux density of the CLEAN model of the source and the intensity of the map peak. First contours correspond to the map noise level~$\times\,3$; each subsequent contour marks intensity increase by the factor of two. The CLEAN beam at the half-maximum level is shown in the map lower left corner as a black ellipse; its major and minor axes and position angle are specified below the map.}
\figsetgrpend

\figsetgrpstart
\figsetgrpnum{4.68}
\figsetgrptitle{J1058+8114 at 8.6 GHz}
\figsetplot{J1058+8114_X_2006_02_23_avp_map.pdf}
\figsetgrpnote{The source name, the observation date, and the frequency are given at the top of the panel. The intensity is shown by contours: solid lines are the positive contours, dotted lines are the negative contours. The contour levels in percents of the map peak are specified below the map, as well as the total flux density of the CLEAN model of the source and the intensity of the map peak. First contours correspond to the map noise level~$\times\,3$; each subsequent contour marks intensity increase by the factor of two. The CLEAN beam at the half-maximum level is shown in the map lower left corner as a black ellipse; its major and minor axes and position angle are specified below the map.}
\figsetgrpend

\figsetgrpstart
\figsetgrpnum{4.69}
\figsetgrptitle{J1104+7658 at 8.6 GHz}
\figsetplot{J1104+7658_X_2006_02_14_avp_map.pdf}
\figsetgrpnote{The source name, the observation date, and the frequency are given at the top of the panel. The intensity is shown by contours: solid lines are the positive contours, dotted lines are the negative contours. The contour levels in percents of the map peak are specified below the map, as well as the total flux density of the CLEAN model of the source and the intensity of the map peak. First contours correspond to the map noise level~$\times\,3$; each subsequent contour marks intensity increase by the factor of two. The CLEAN beam at the half-maximum level is shown in the map lower left corner as a black ellipse; its major and minor axes and position angle are specified below the map.}
\figsetgrpend

\figsetgrpstart
\figsetgrpnum{4.70}
\figsetgrptitle{J1104+7932 at 2.3 GHz}
\figsetplot{J1104+7932_S_2006_02_23_avp_map.pdf}
\figsetgrpnote{The source name, the observation date, and the frequency are given at the top of the panel. The intensity is shown by contours: solid lines are the positive contours, dotted lines are the negative contours. The contour levels in percents of the map peak are specified below the map, as well as the total flux density of the CLEAN model of the source and the intensity of the map peak. First contours correspond to the map noise level~$\times\,3$; each subsequent contour marks intensity increase by the factor of two. The CLEAN beam at the half-maximum level is shown in the map lower left corner as a black ellipse; its major and minor axes and position angle are specified below the map.}
\figsetgrpend

\figsetgrpstart
\figsetgrpnum{4.71}
\figsetgrptitle{J1104+7932 at 8.6 GHz}
\figsetplot{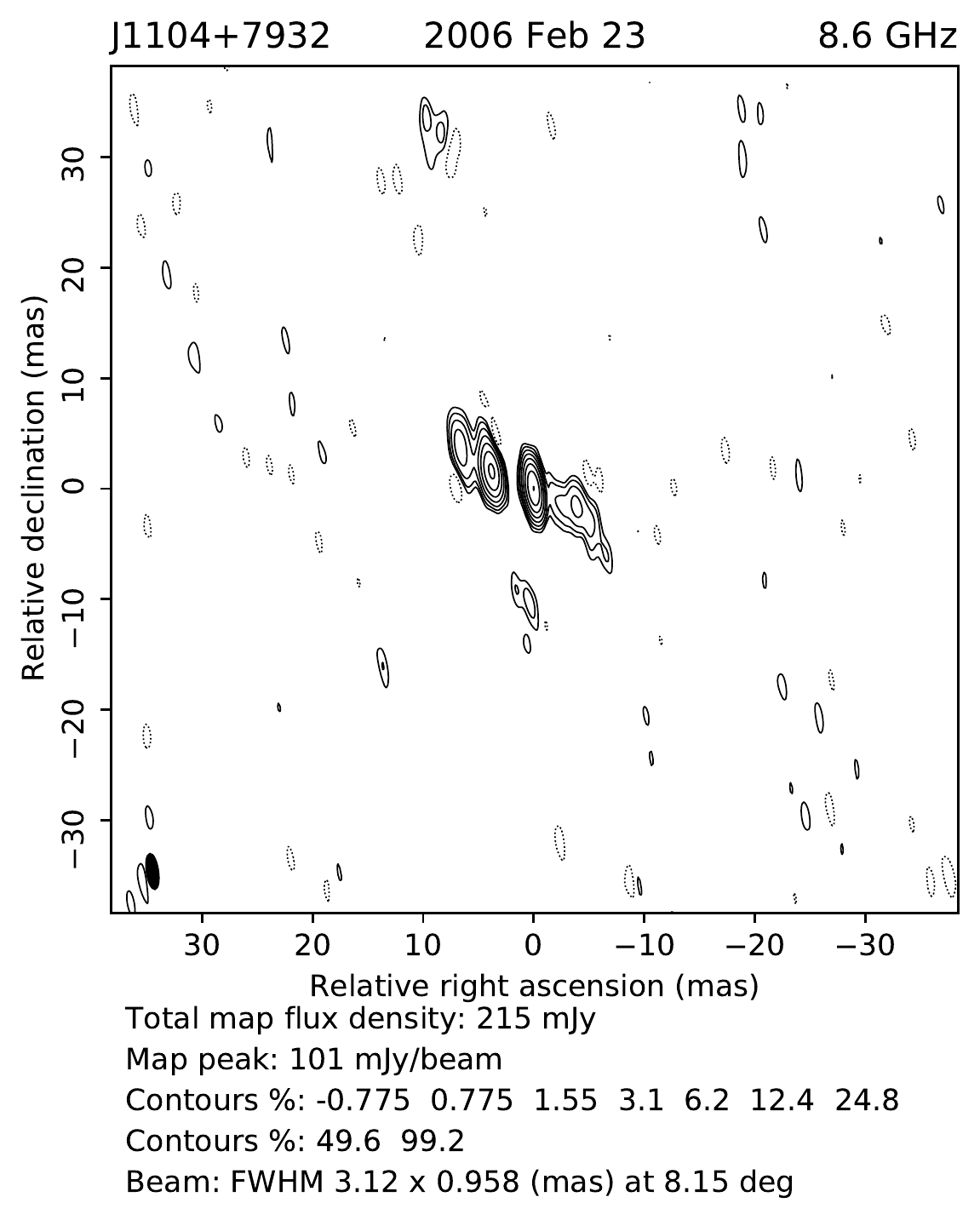}
\figsetgrpnote{The source name, the observation date, and the frequency are given at the top of the panel. The intensity is shown by contours: solid lines are the positive contours, dotted lines are the negative contours. The contour levels in percents of the map peak are specified below the map, as well as the total flux density of the CLEAN model of the source and the intensity of the map peak. First contours correspond to the map noise level~$\times\,3$; each subsequent contour marks intensity increase by the factor of two. The CLEAN beam at the half-maximum level is shown in the map lower left corner as a black ellipse; its major and minor axes and position angle are specified below the map.}
\figsetgrpend

\figsetgrpstart
\figsetgrpnum{4.72}
\figsetgrptitle{J1119+8048 at 2.3 GHz}
\figsetplot{J1119+8048_S_2006_02_14_avp_map.pdf}
\figsetgrpnote{The source name, the observation date, and the frequency are given at the top of the panel. The intensity is shown by contours: solid lines are the positive contours, dotted lines are the negative contours. The contour levels in percents of the map peak are specified below the map, as well as the total flux density of the CLEAN model of the source and the intensity of the map peak. First contours correspond to the map noise level~$\times\,3$; each subsequent contour marks intensity increase by the factor of two. The CLEAN beam at the half-maximum level is shown in the map lower left corner as a black ellipse; its major and minor axes and position angle are specified below the map.}
\figsetgrpend

\figsetgrpstart
\figsetgrpnum{4.73}
\figsetgrptitle{J1133+7831 at 2.3 GHz}
\figsetplot{J1133+7831_S_2006_02_23_avp_map.pdf}
\figsetgrpnote{The source name, the observation date, and the frequency are given at the top of the panel. The intensity is shown by contours: solid lines are the positive contours, dotted lines are the negative contours. The contour levels in percents of the map peak are specified below the map, as well as the total flux density of the CLEAN model of the source and the intensity of the map peak. First contours correspond to the map noise level~$\times\,3$; each subsequent contour marks intensity increase by the factor of two. The CLEAN beam at the half-maximum level is shown in the map lower left corner as a black ellipse; its major and minor axes and position angle are specified below the map.}
\figsetgrpend

\figsetgrpstart
\figsetgrpnum{4.74}
\figsetgrptitle{J1153+8058 at 2.3 GHz}
\figsetplot{J1153+8058_S_2006_02_14_avp_map.pdf}
\figsetgrpnote{The source name, the observation date, and the frequency are given at the top of the panel. The intensity is shown by contours: solid lines are the positive contours, dotted lines are the negative contours. The contour levels in percents of the map peak are specified below the map, as well as the total flux density of the CLEAN model of the source and the intensity of the map peak. First contours correspond to the map noise level~$\times\,3$; each subsequent contour marks intensity increase by the factor of two. The CLEAN beam at the half-maximum level is shown in the map lower left corner as a black ellipse; its major and minor axes and position angle are specified below the map.}
\figsetgrpend

\figsetgrpstart
\figsetgrpnum{4.75}
\figsetgrptitle{J1153+8058 at 2.3 GHz}
\figsetplot{J1153+8058_S_2006_02_16_avp_map.pdf}
\figsetgrpnote{The source name, the observation date, and the frequency are given at the top of the panel. The intensity is shown by contours: solid lines are the positive contours, dotted lines are the negative contours. The contour levels in percents of the map peak are specified below the map, as well as the total flux density of the CLEAN model of the source and the intensity of the map peak. First contours correspond to the map noise level~$\times\,3$; each subsequent contour marks intensity increase by the factor of two. The CLEAN beam at the half-maximum level is shown in the map lower left corner as a black ellipse; its major and minor axes and position angle are specified below the map.}
\figsetgrpend

\figsetgrpstart
\figsetgrpnum{4.76}
\figsetgrptitle{J1153+8058 at 2.3 GHz}
\figsetplot{J1153+8058_S_2006_02_23_avp_map.pdf}
\figsetgrpnote{The source name, the observation date, and the frequency are given at the top of the panel. The intensity is shown by contours: solid lines are the positive contours, dotted lines are the negative contours. The contour levels in percents of the map peak are specified below the map, as well as the total flux density of the CLEAN model of the source and the intensity of the map peak. First contours correspond to the map noise level~$\times\,3$; each subsequent contour marks intensity increase by the factor of two. The CLEAN beam at the half-maximum level is shown in the map lower left corner as a black ellipse; its major and minor axes and position angle are specified below the map.}
\figsetgrpend

\figsetgrpstart
\figsetgrpnum{4.77}
\figsetgrptitle{J1153+8058 at 8.6 GHz}
\figsetplot{J1153+8058_X_2006_02_14_avp_map.pdf}
\figsetgrpnote{The source name, the observation date, and the frequency are given at the top of the panel. The intensity is shown by contours: solid lines are the positive contours, dotted lines are the negative contours. The contour levels in percents of the map peak are specified below the map, as well as the total flux density of the CLEAN model of the source and the intensity of the map peak. First contours correspond to the map noise level~$\times\,3$; each subsequent contour marks intensity increase by the factor of two. The CLEAN beam at the half-maximum level is shown in the map lower left corner as a black ellipse; its major and minor axes and position angle are specified below the map.}
\figsetgrpend

\figsetgrpstart
\figsetgrpnum{4.78}
\figsetgrptitle{J1153+8058 at 8.6 GHz}
\figsetplot{J1153+8058_X_2006_02_16_avp_map.pdf}
\figsetgrpnote{The source name, the observation date, and the frequency are given at the top of the panel. The intensity is shown by contours: solid lines are the positive contours, dotted lines are the negative contours. The contour levels in percents of the map peak are specified below the map, as well as the total flux density of the CLEAN model of the source and the intensity of the map peak. First contours correspond to the map noise level~$\times\,3$; each subsequent contour marks intensity increase by the factor of two. The CLEAN beam at the half-maximum level is shown in the map lower left corner as a black ellipse; its major and minor axes and position angle are specified below the map.}
\figsetgrpend

\figsetgrpstart
\figsetgrpnum{4.79}
\figsetgrptitle{J1153+8058 at 8.6 GHz}
\figsetplot{J1153+8058_X_2006_02_23_avp_map.pdf}
\figsetgrpnote{The source name, the observation date, and the frequency are given at the top of the panel. The intensity is shown by contours: solid lines are the positive contours, dotted lines are the negative contours. The contour levels in percents of the map peak are specified below the map, as well as the total flux density of the CLEAN model of the source and the intensity of the map peak. First contours correspond to the map noise level~$\times\,3$; each subsequent contour marks intensity increase by the factor of two. The CLEAN beam at the half-maximum level is shown in the map lower left corner as a black ellipse; its major and minor axes and position angle are specified below the map.}
\figsetgrpend

\figsetgrpstart
\figsetgrpnum{4.80}
\figsetgrptitle{J1157+8118 at 2.3 GHz}
\figsetplot{J1157+8118_S_2006_02_14_avp_map.pdf}
\figsetgrpnote{The source name, the observation date, and the frequency are given at the top of the panel. The intensity is shown by contours: solid lines are the positive contours, dotted lines are the negative contours. The contour levels in percents of the map peak are specified below the map, as well as the total flux density of the CLEAN model of the source and the intensity of the map peak. First contours correspond to the map noise level~$\times\,3$; each subsequent contour marks intensity increase by the factor of two. The CLEAN beam at the half-maximum level is shown in the map lower left corner as a black ellipse; its major and minor axes and position angle are specified below the map.}
\figsetgrpend

\figsetgrpstart
\figsetgrpnum{4.81}
\figsetgrptitle{J1223+8040 at 2.3 GHz}
\figsetplot{J1223+8040_S_2006_02_23_avp_map.pdf}
\figsetgrpnote{The source name, the observation date, and the frequency are given at the top of the panel. The intensity is shown by contours: solid lines are the positive contours, dotted lines are the negative contours. The contour levels in percents of the map peak are specified below the map, as well as the total flux density of the CLEAN model of the source and the intensity of the map peak. First contours correspond to the map noise level~$\times\,3$; each subsequent contour marks intensity increase by the factor of two. The CLEAN beam at the half-maximum level is shown in the map lower left corner as a black ellipse; its major and minor axes and position angle are specified below the map.}
\figsetgrpend

\figsetgrpstart
\figsetgrpnum{4.82}
\figsetgrptitle{J1223+8040 at 8.6 GHz}
\figsetplot{J1223+8040_X_2006_02_23_avp_map.pdf}
\figsetgrpnote{The source name, the observation date, and the frequency are given at the top of the panel. The intensity is shown by contours: solid lines are the positive contours, dotted lines are the negative contours. The contour levels in percents of the map peak are specified below the map, as well as the total flux density of the CLEAN model of the source and the intensity of the map peak. First contours correspond to the map noise level~$\times\,3$; each subsequent contour marks intensity increase by the factor of two. The CLEAN beam at the half-maximum level is shown in the map lower left corner as a black ellipse; its major and minor axes and position angle are specified below the map.}
\figsetgrpend

\figsetgrpstart
\figsetgrpnum{4.83}
\figsetgrptitle{J1223+8436 at 2.3 GHz}
\figsetplot{J1223+8436_S_2006_02_16_avp_map.pdf}
\figsetgrpnote{The source name, the observation date, and the frequency are given at the top of the panel. The intensity is shown by contours: solid lines are the positive contours, dotted lines are the negative contours. The contour levels in percents of the map peak are specified below the map, as well as the total flux density of the CLEAN model of the source and the intensity of the map peak. First contours correspond to the map noise level~$\times\,3$; each subsequent contour marks intensity increase by the factor of two. The CLEAN beam at the half-maximum level is shown in the map lower left corner as a black ellipse; its major and minor axes and position angle are specified below the map.}
\figsetgrpend

\figsetgrpstart
\figsetgrpnum{4.84}
\figsetgrptitle{J1223+8436 at 8.6 GHz}
\figsetplot{J1223+8436_X_2006_02_16_avp_map.pdf}
\figsetgrpnote{The source name, the observation date, and the frequency are given at the top of the panel. The intensity is shown by contours: solid lines are the positive contours, dotted lines are the negative contours. The contour levels in percents of the map peak are specified below the map, as well as the total flux density of the CLEAN model of the source and the intensity of the map peak. First contours correspond to the map noise level~$\times\,3$; each subsequent contour marks intensity increase by the factor of two. The CLEAN beam at the half-maximum level is shown in the map lower left corner as a black ellipse; its major and minor axes and position angle are specified below the map.}
\figsetgrpend

\figsetgrpstart
\figsetgrpnum{4.85}
\figsetgrptitle{J1233+8054 at 2.3 GHz}
\figsetplot{J1233+8054_S_2006_02_16_avp_map.pdf}
\figsetgrpnote{The source name, the observation date, and the frequency are given at the top of the panel. The intensity is shown by contours: solid lines are the positive contours, dotted lines are the negative contours. The contour levels in percents of the map peak are specified below the map, as well as the total flux density of the CLEAN model of the source and the intensity of the map peak. First contours correspond to the map noise level~$\times\,3$; each subsequent contour marks intensity increase by the factor of two. The CLEAN beam at the half-maximum level is shown in the map lower left corner as a black ellipse; its major and minor axes and position angle are specified below the map.}
\figsetgrpend

\figsetgrpstart
\figsetgrpnum{4.86}
\figsetgrptitle{J1233+8054 at 8.6 GHz}
\figsetplot{J1233+8054_X_2006_02_16_avp_map.pdf}
\figsetgrpnote{The source name, the observation date, and the frequency are given at the top of the panel. The intensity is shown by contours: solid lines are the positive contours, dotted lines are the negative contours. The contour levels in percents of the map peak are specified below the map, as well as the total flux density of the CLEAN model of the source and the intensity of the map peak. First contours correspond to the map noise level~$\times\,3$; each subsequent contour marks intensity increase by the factor of two. The CLEAN beam at the half-maximum level is shown in the map lower left corner as a black ellipse; its major and minor axes and position angle are specified below the map.}
\figsetgrpend

\figsetgrpstart
\figsetgrpnum{4.87}
\figsetgrptitle{J1257+7958 at 2.3 GHz}
\figsetplot{J1257+7958_S_2006_02_23_avp_map.pdf}
\figsetgrpnote{The source name, the observation date, and the frequency are given at the top of the panel. The intensity is shown by contours: solid lines are the positive contours, dotted lines are the negative contours. The contour levels in percents of the map peak are specified below the map, as well as the total flux density of the CLEAN model of the source and the intensity of the map peak. First contours correspond to the map noise level~$\times\,3$; each subsequent contour marks intensity increase by the factor of two. The CLEAN beam at the half-maximum level is shown in the map lower left corner as a black ellipse; its major and minor axes and position angle are specified below the map.}
\figsetgrpend

\figsetgrpstart
\figsetgrpnum{4.88}
\figsetgrptitle{J1305+7854 at 2.3 GHz}
\figsetplot{J1305+7854_S_2006_02_23_avp_map.pdf}
\figsetgrpnote{The source name, the observation date, and the frequency are given at the top of the panel. The intensity is shown by contours: solid lines are the positive contours, dotted lines are the negative contours. The contour levels in percents of the map peak are specified below the map, as well as the total flux density of the CLEAN model of the source and the intensity of the map peak. First contours correspond to the map noise level~$\times\,3$; each subsequent contour marks intensity increase by the factor of two. The CLEAN beam at the half-maximum level is shown in the map lower left corner as a black ellipse; its major and minor axes and position angle are specified below the map.}
\figsetgrpend

\figsetgrpstart
\figsetgrpnum{4.89}
\figsetgrptitle{J1305+7854 at 8.6 GHz}
\figsetplot{J1305+7854_X_2006_02_23_avp_map.pdf}
\figsetgrpnote{The source name, the observation date, and the frequency are given at the top of the panel. The intensity is shown by contours: solid lines are the positive contours, dotted lines are the negative contours. The contour levels in percents of the map peak are specified below the map, as well as the total flux density of the CLEAN model of the source and the intensity of the map peak. First contours correspond to the map noise level~$\times\,3$; each subsequent contour marks intensity increase by the factor of two. The CLEAN beam at the half-maximum level is shown in the map lower left corner as a black ellipse; its major and minor axes and position angle are specified below the map.}
\figsetgrpend

\figsetgrpstart
\figsetgrpnum{4.90}
\figsetgrptitle{J1305+8216 at 2.3 GHz}
\figsetplot{J1305+8216_S_2006_02_14_avp_map.pdf}
\figsetgrpnote{The source name, the observation date, and the frequency are given at the top of the panel. The intensity is shown by contours: solid lines are the positive contours, dotted lines are the negative contours. The contour levels in percents of the map peak are specified below the map, as well as the total flux density of the CLEAN model of the source and the intensity of the map peak. First contours correspond to the map noise level~$\times\,3$; each subsequent contour marks intensity increase by the factor of two. The CLEAN beam at the half-maximum level is shown in the map lower left corner as a black ellipse; its major and minor axes and position angle are specified below the map.}
\figsetgrpend

\figsetgrpstart
\figsetgrpnum{4.91}
\figsetgrptitle{J1306+8008 at 2.3 GHz}
\figsetplot{J1306+8008_S_2006_02_23_avp_map.pdf}
\figsetgrpnote{The source name, the observation date, and the frequency are given at the top of the panel. The intensity is shown by contours: solid lines are the positive contours, dotted lines are the negative contours. The contour levels in percents of the map peak are specified below the map, as well as the total flux density of the CLEAN model of the source and the intensity of the map peak. First contours correspond to the map noise level~$\times\,3$; each subsequent contour marks intensity increase by the factor of two. The CLEAN beam at the half-maximum level is shown in the map lower left corner as a black ellipse; its major and minor axes and position angle are specified below the map.}
\figsetgrpend

\figsetgrpstart
\figsetgrpnum{4.92}
\figsetgrptitle{J1306+8008 at 8.6 GHz}
\figsetplot{J1306+8008_X_2006_02_23_avp_map.pdf}
\figsetgrpnote{The source name, the observation date, and the frequency are given at the top of the panel. The intensity is shown by contours: solid lines are the positive contours, dotted lines are the negative contours. The contour levels in percents of the map peak are specified below the map, as well as the total flux density of the CLEAN model of the source and the intensity of the map peak. First contours correspond to the map noise level~$\times\,3$; each subsequent contour marks intensity increase by the factor of two. The CLEAN beam at the half-maximum level is shown in the map lower left corner as a black ellipse; its major and minor axes and position angle are specified below the map.}
\figsetgrpend

\figsetgrpstart
\figsetgrpnum{4.93}
\figsetgrptitle{J1306+8019 at 2.3 GHz}
\figsetplot{J1306+8019_S_2006_02_14_avp_map.pdf}
\figsetgrpnote{The source name, the observation date, and the frequency are given at the top of the panel. The intensity is shown by contours: solid lines are the positive contours, dotted lines are the negative contours. The contour levels in percents of the map peak are specified below the map, as well as the total flux density of the CLEAN model of the source and the intensity of the map peak. First contours correspond to the map noise level~$\times\,3$; each subsequent contour marks intensity increase by the factor of two. The CLEAN beam at the half-maximum level is shown in the map lower left corner as a black ellipse; its major and minor axes and position angle are specified below the map.}
\figsetgrpend

\figsetgrpstart
\figsetgrpnum{4.94}
\figsetgrptitle{J1307+7649 at 2.3 GHz}
\figsetplot{J1307+7649_S_2006_02_16_avp_map.pdf}
\figsetgrpnote{The source name, the observation date, and the frequency are given at the top of the panel. The intensity is shown by contours: solid lines are the positive contours, dotted lines are the negative contours. The contour levels in percents of the map peak are specified below the map, as well as the total flux density of the CLEAN model of the source and the intensity of the map peak. First contours correspond to the map noise level~$\times\,3$; each subsequent contour marks intensity increase by the factor of two. The CLEAN beam at the half-maximum level is shown in the map lower left corner as a black ellipse; its major and minor axes and position angle are specified below the map.}
\figsetgrpend

\figsetgrpstart
\figsetgrpnum{4.95}
\figsetgrptitle{J1320+8450 at 2.3 GHz}
\figsetplot{J1320+8450_S_2006_02_23_avp_map.pdf}
\figsetgrpnote{The source name, the observation date, and the frequency are given at the top of the panel. The intensity is shown by contours: solid lines are the positive contours, dotted lines are the negative contours. The contour levels in percents of the map peak are specified below the map, as well as the total flux density of the CLEAN model of the source and the intensity of the map peak. First contours correspond to the map noise level~$\times\,3$; each subsequent contour marks intensity increase by the factor of two. The CLEAN beam at the half-maximum level is shown in the map lower left corner as a black ellipse; its major and minor axes and position angle are specified below the map.}
\figsetgrpend

\figsetgrpstart
\figsetgrpnum{4.96}
\figsetgrptitle{J1320+8450 at 8.6 GHz}
\figsetplot{J1320+8450_X_2006_02_23_avp_map.pdf}
\figsetgrpnote{The source name, the observation date, and the frequency are given at the top of the panel. The intensity is shown by contours: solid lines are the positive contours, dotted lines are the negative contours. The contour levels in percents of the map peak are specified below the map, as well as the total flux density of the CLEAN model of the source and the intensity of the map peak. First contours correspond to the map noise level~$\times\,3$; each subsequent contour marks intensity increase by the factor of two. The CLEAN beam at the half-maximum level is shown in the map lower left corner as a black ellipse; its major and minor axes and position angle are specified below the map.}
\figsetgrpend

\figsetgrpstart
\figsetgrpnum{4.97}
\figsetgrptitle{J1321+8316 at 2.3 GHz}
\figsetplot{J1321+8316_S_2006_02_14_avp_map.pdf}
\figsetgrpnote{The source name, the observation date, and the frequency are given at the top of the panel. The intensity is shown by contours: solid lines are the positive contours, dotted lines are the negative contours. The contour levels in percents of the map peak are specified below the map, as well as the total flux density of the CLEAN model of the source and the intensity of the map peak. First contours correspond to the map noise level~$\times\,3$; each subsequent contour marks intensity increase by the factor of two. The CLEAN beam at the half-maximum level is shown in the map lower left corner as a black ellipse; its major and minor axes and position angle are specified below the map.}
\figsetgrpend

\figsetgrpstart
\figsetgrpnum{4.98}
\figsetgrptitle{J1321+8316 at 8.6 GHz}
\figsetplot{J1321+8316_X_2006_02_14_avp_map.pdf}
\figsetgrpnote{The source name, the observation date, and the frequency are given at the top of the panel. The intensity is shown by contours: solid lines are the positive contours, dotted lines are the negative contours. The contour levels in percents of the map peak are specified below the map, as well as the total flux density of the CLEAN model of the source and the intensity of the map peak. First contours correspond to the map noise level~$\times\,3$; each subsequent contour marks intensity increase by the factor of two. The CLEAN beam at the half-maximum level is shown in the map lower left corner as a black ellipse; its major and minor axes and position angle are specified below the map.}
\figsetgrpend

\figsetgrpstart
\figsetgrpnum{4.99}
\figsetgrptitle{J1323+7942 at 2.3 GHz}
\figsetplot{J1323+7942_S_2006_02_14_avp_map.pdf}
\figsetgrpnote{The source name, the observation date, and the frequency are given at the top of the panel. The intensity is shown by contours: solid lines are the positive contours, dotted lines are the negative contours. The contour levels in percents of the map peak are specified below the map, as well as the total flux density of the CLEAN model of the source and the intensity of the map peak. First contours correspond to the map noise level~$\times\,3$; each subsequent contour marks intensity increase by the factor of two. The CLEAN beam at the half-maximum level is shown in the map lower left corner as a black ellipse; its major and minor axes and position angle are specified below the map.}
\figsetgrpend

\figsetgrpstart
\figsetgrpnum{4.100}
\figsetgrptitle{J1323+7942 at 8.6 GHz}
\figsetplot{J1323+7942_X_2006_02_14_avp_map.pdf}
\figsetgrpnote{The source name, the observation date, and the frequency are given at the top of the panel. The intensity is shown by contours: solid lines are the positive contours, dotted lines are the negative contours. The contour levels in percents of the map peak are specified below the map, as well as the total flux density of the CLEAN model of the source and the intensity of the map peak. First contours correspond to the map noise level~$\times\,3$; each subsequent contour marks intensity increase by the factor of two. The CLEAN beam at the half-maximum level is shown in the map lower left corner as a black ellipse; its major and minor axes and position angle are specified below the map.}
\figsetgrpend

\figsetgrpstart
\figsetgrpnum{4.101}
\figsetgrptitle{J1325+7535 at 2.3 GHz}
\figsetplot{J1325+7535_S_2006_02_14_avp_map.pdf}
\figsetgrpnote{The source name, the observation date, and the frequency are given at the top of the panel. The intensity is shown by contours: solid lines are the positive contours, dotted lines are the negative contours. The contour levels in percents of the map peak are specified below the map, as well as the total flux density of the CLEAN model of the source and the intensity of the map peak. First contours correspond to the map noise level~$\times\,3$; each subsequent contour marks intensity increase by the factor of two. The CLEAN beam at the half-maximum level is shown in the map lower left corner as a black ellipse; its major and minor axes and position angle are specified below the map.}
\figsetgrpend

\figsetgrpstart
\figsetgrpnum{4.102}
\figsetgrptitle{J1357+7643 at 2.3 GHz}
\figsetplot{J1357+7643_S_2006_02_14_avp_map.pdf}
\figsetgrpnote{The source name, the observation date, and the frequency are given at the top of the panel. The intensity is shown by contours: solid lines are the positive contours, dotted lines are the negative contours. The contour levels in percents of the map peak are specified below the map, as well as the total flux density of the CLEAN model of the source and the intensity of the map peak. First contours correspond to the map noise level~$\times\,3$; each subsequent contour marks intensity increase by the factor of two. The CLEAN beam at the half-maximum level is shown in the map lower left corner as a black ellipse; its major and minor axes and position angle are specified below the map.}
\figsetgrpend

\figsetgrpstart
\figsetgrpnum{4.103}
\figsetgrptitle{J1357+7643 at 8.6 GHz}
\figsetplot{J1357+7643_X_2006_02_14_avp_map.pdf}
\figsetgrpnote{The source name, the observation date, and the frequency are given at the top of the panel. The intensity is shown by contours: solid lines are the positive contours, dotted lines are the negative contours. The contour levels in percents of the map peak are specified below the map, as well as the total flux density of the CLEAN model of the source and the intensity of the map peak. First contours correspond to the map noise level~$\times\,3$; each subsequent contour marks intensity increase by the factor of two. The CLEAN beam at the half-maximum level is shown in the map lower left corner as a black ellipse; its major and minor axes and position angle are specified below the map.}
\figsetgrpend

\figsetgrpstart
\figsetgrpnum{4.104}
\figsetgrptitle{J1358+8340 at 2.3 GHz}
\figsetplot{J1358+8340_S_2006_02_16_avp_map.pdf}
\figsetgrpnote{The source name, the observation date, and the frequency are given at the top of the panel. The intensity is shown by contours: solid lines are the positive contours, dotted lines are the negative contours. The contour levels in percents of the map peak are specified below the map, as well as the total flux density of the CLEAN model of the source and the intensity of the map peak. First contours correspond to the map noise level~$\times\,3$; each subsequent contour marks intensity increase by the factor of two. The CLEAN beam at the half-maximum level is shown in the map lower left corner as a black ellipse; its major and minor axes and position angle are specified below the map.}
\figsetgrpend

\figsetgrpstart
\figsetgrpnum{4.105}
\figsetgrptitle{J1358+8340 at 8.6 GHz}
\figsetplot{J1358+8340_X_2006_02_16_avp_map.pdf}
\figsetgrpnote{The source name, the observation date, and the frequency are given at the top of the panel. The intensity is shown by contours: solid lines are the positive contours, dotted lines are the negative contours. The contour levels in percents of the map peak are specified below the map, as well as the total flux density of the CLEAN model of the source and the intensity of the map peak. First contours correspond to the map noise level~$\times\,3$; each subsequent contour marks intensity increase by the factor of two. The CLEAN beam at the half-maximum level is shown in the map lower left corner as a black ellipse; its major and minor axes and position angle are specified below the map.}
\figsetgrpend

\figsetgrpstart
\figsetgrpnum{4.106}
\figsetgrptitle{J1406+7828 at 2.3 GHz}
\figsetplot{J1406+7828_S_2006_02_23_avp_map.pdf}
\figsetgrpnote{The source name, the observation date, and the frequency are given at the top of the panel. The intensity is shown by contours: solid lines are the positive contours, dotted lines are the negative contours. The contour levels in percents of the map peak are specified below the map, as well as the total flux density of the CLEAN model of the source and the intensity of the map peak. First contours correspond to the map noise level~$\times\,3$; each subsequent contour marks intensity increase by the factor of two. The CLEAN beam at the half-maximum level is shown in the map lower left corner as a black ellipse; its major and minor axes and position angle are specified below the map.}
\figsetgrpend

\figsetgrpstart
\figsetgrpnum{4.107}
\figsetgrptitle{J1406+7828 at 8.6 GHz}
\figsetplot{J1406+7828_X_2006_02_23_avp_map.pdf}
\figsetgrpnote{The source name, the observation date, and the frequency are given at the top of the panel. The intensity is shown by contours: solid lines are the positive contours, dotted lines are the negative contours. The contour levels in percents of the map peak are specified below the map, as well as the total flux density of the CLEAN model of the source and the intensity of the map peak. First contours correspond to the map noise level~$\times\,3$; each subsequent contour marks intensity increase by the factor of two. The CLEAN beam at the half-maximum level is shown in the map lower left corner as a black ellipse; its major and minor axes and position angle are specified below the map.}
\figsetgrpend

\figsetgrpstart
\figsetgrpnum{4.108}
\figsetgrptitle{J1421+7513 at 2.3 GHz}
\figsetplot{J1421+7513_S_2006_02_23_avp_map.pdf}
\figsetgrpnote{The source name, the observation date, and the frequency are given at the top of the panel. The intensity is shown by contours: solid lines are the positive contours, dotted lines are the negative contours. The contour levels in percents of the map peak are specified below the map, as well as the total flux density of the CLEAN model of the source and the intensity of the map peak. First contours correspond to the map noise level~$\times\,3$; each subsequent contour marks intensity increase by the factor of two. The CLEAN beam at the half-maximum level is shown in the map lower left corner as a black ellipse; its major and minor axes and position angle are specified below the map.}
\figsetgrpend

\figsetgrpstart
\figsetgrpnum{4.109}
\figsetgrptitle{J1435+7605 at 2.3 GHz}
\figsetplot{J1435+7605_S_2006_02_14_avp_map.pdf}
\figsetgrpnote{The source name, the observation date, and the frequency are given at the top of the panel. The intensity is shown by contours: solid lines are the positive contours, dotted lines are the negative contours. The contour levels in percents of the map peak are specified below the map, as well as the total flux density of the CLEAN model of the source and the intensity of the map peak. First contours correspond to the map noise level~$\times\,3$; each subsequent contour marks intensity increase by the factor of two. The CLEAN beam at the half-maximum level is shown in the map lower left corner as a black ellipse; its major and minor axes and position angle are specified below the map.}
\figsetgrpend

\figsetgrpstart
\figsetgrpnum{4.110}
\figsetgrptitle{J1506+8319 at 2.3 GHz}
\figsetplot{J1506+8319_S_2006_02_16_avp_map.pdf}
\figsetgrpnote{The source name, the observation date, and the frequency are given at the top of the panel. The intensity is shown by contours: solid lines are the positive contours, dotted lines are the negative contours. The contour levels in percents of the map peak are specified below the map, as well as the total flux density of the CLEAN model of the source and the intensity of the map peak. First contours correspond to the map noise level~$\times\,3$; each subsequent contour marks intensity increase by the factor of two. The CLEAN beam at the half-maximum level is shown in the map lower left corner as a black ellipse; its major and minor axes and position angle are specified below the map.}
\figsetgrpend

\figsetgrpstart
\figsetgrpnum{4.111}
\figsetgrptitle{J1506+8319 at 8.6 GHz}
\figsetplot{J1506+8319_X_2006_02_16_avp_map.pdf}
\figsetgrpnote{The source name, the observation date, and the frequency are given at the top of the panel. The intensity is shown by contours: solid lines are the positive contours, dotted lines are the negative contours. The contour levels in percents of the map peak are specified below the map, as well as the total flux density of the CLEAN model of the source and the intensity of the map peak. First contours correspond to the map noise level~$\times\,3$; each subsequent contour marks intensity increase by the factor of two. The CLEAN beam at the half-maximum level is shown in the map lower left corner as a black ellipse; its major and minor axes and position angle are specified below the map.}
\figsetgrpend

\figsetgrpstart
\figsetgrpnum{4.112}
\figsetgrptitle{J1528+8217 at 2.3 GHz}
\figsetplot{J1528+8217_S_2006_02_14_avp_map.pdf}
\figsetgrpnote{The source name, the observation date, and the frequency are given at the top of the panel. The intensity is shown by contours: solid lines are the positive contours, dotted lines are the negative contours. The contour levels in percents of the map peak are specified below the map, as well as the total flux density of the CLEAN model of the source and the intensity of the map peak. First contours correspond to the map noise level~$\times\,3$; each subsequent contour marks intensity increase by the factor of two. The CLEAN beam at the half-maximum level is shown in the map lower left corner as a black ellipse; its major and minor axes and position angle are specified below the map.}
\figsetgrpend

\figsetgrpstart
\figsetgrpnum{4.113}
\figsetgrptitle{J1531+7706 at 2.3 GHz}
\figsetplot{J1531+7706_S_2006_02_14_avp_map.pdf}
\figsetgrpnote{The source name, the observation date, and the frequency are given at the top of the panel. The intensity is shown by contours: solid lines are the positive contours, dotted lines are the negative contours. The contour levels in percents of the map peak are specified below the map, as well as the total flux density of the CLEAN model of the source and the intensity of the map peak. First contours correspond to the map noise level~$\times\,3$; each subsequent contour marks intensity increase by the factor of two. The CLEAN beam at the half-maximum level is shown in the map lower left corner as a black ellipse; its major and minor axes and position angle are specified below the map.}
\figsetgrpend

\figsetgrpstart
\figsetgrpnum{4.114}
\figsetgrptitle{J1537+8154 at 2.3 GHz}
\figsetplot{J1537+8154_S_2006_02_16_avp_map.pdf}
\figsetgrpnote{The source name, the observation date, and the frequency are given at the top of the panel. The intensity is shown by contours: solid lines are the positive contours, dotted lines are the negative contours. The contour levels in percents of the map peak are specified below the map, as well as the total flux density of the CLEAN model of the source and the intensity of the map peak. First contours correspond to the map noise level~$\times\,3$; each subsequent contour marks intensity increase by the factor of two. The CLEAN beam at the half-maximum level is shown in the map lower left corner as a black ellipse; its major and minor axes and position angle are specified below the map.}
\figsetgrpend

\figsetgrpstart
\figsetgrpnum{4.115}
\figsetgrptitle{J1537+8154 at 8.6 GHz}
\figsetplot{J1537+8154_X_2006_02_16_avp_map.pdf}
\figsetgrpnote{The source name, the observation date, and the frequency are given at the top of the panel. The intensity is shown by contours: solid lines are the positive contours, dotted lines are the negative contours. The contour levels in percents of the map peak are specified below the map, as well as the total flux density of the CLEAN model of the source and the intensity of the map peak. First contours correspond to the map noise level~$\times\,3$; each subsequent contour marks intensity increase by the factor of two. The CLEAN beam at the half-maximum level is shown in the map lower left corner as a black ellipse; its major and minor axes and position angle are specified below the map.}
\figsetgrpend

\figsetgrpstart
\figsetgrpnum{4.116}
\figsetgrptitle{J1607+8501 at 2.3 GHz}
\figsetplot{J1607+8501_S_2006_02_23_avp_map.pdf}
\figsetgrpnote{The source name, the observation date, and the frequency are given at the top of the panel. The intensity is shown by contours: solid lines are the positive contours, dotted lines are the negative contours. The contour levels in percents of the map peak are specified below the map, as well as the total flux density of the CLEAN model of the source and the intensity of the map peak. First contours correspond to the map noise level~$\times\,3$; each subsequent contour marks intensity increase by the factor of two. The CLEAN beam at the half-maximum level is shown in the map lower left corner as a black ellipse; its major and minor axes and position angle are specified below the map.}
\figsetgrpend

\figsetgrpstart
\figsetgrpnum{4.117}
\figsetgrptitle{J1609+7939 at 2.3 GHz}
\figsetplot{J1609+7939_S_2006_02_14_avp_map.pdf}
\figsetgrpnote{The source name, the observation date, and the frequency are given at the top of the panel. The intensity is shown by contours: solid lines are the positive contours, dotted lines are the negative contours. The contour levels in percents of the map peak are specified below the map, as well as the total flux density of the CLEAN model of the source and the intensity of the map peak. First contours correspond to the map noise level~$\times\,3$; each subsequent contour marks intensity increase by the factor of two. The CLEAN beam at the half-maximum level is shown in the map lower left corner as a black ellipse; its major and minor axes and position angle are specified below the map.}
\figsetgrpend

\figsetgrpstart
\figsetgrpnum{4.118}
\figsetgrptitle{J1632+8232 at 2.3 GHz}
\figsetplot{J1632+8232_S_2006_02_23_avp_map.pdf}
\figsetgrpnote{The source name, the observation date, and the frequency are given at the top of the panel. The intensity is shown by contours: solid lines are the positive contours, dotted lines are the negative contours. The contour levels in percents of the map peak are specified below the map, as well as the total flux density of the CLEAN model of the source and the intensity of the map peak. First contours correspond to the map noise level~$\times\,3$; each subsequent contour marks intensity increase by the factor of two. The CLEAN beam at the half-maximum level is shown in the map lower left corner as a black ellipse; its major and minor axes and position angle are specified below the map.}
\figsetgrpend

\figsetgrpstart
\figsetgrpnum{4.119}
\figsetgrptitle{J1632+8232 at 8.6 GHz}
\figsetplot{J1632+8232_X_2006_02_23_avp_map.pdf}
\figsetgrpnote{The source name, the observation date, and the frequency are given at the top of the panel. The intensity is shown by contours: solid lines are the positive contours, dotted lines are the negative contours. The contour levels in percents of the map peak are specified below the map, as well as the total flux density of the CLEAN model of the source and the intensity of the map peak. First contours correspond to the map noise level~$\times\,3$; each subsequent contour marks intensity increase by the factor of two. The CLEAN beam at the half-maximum level is shown in the map lower left corner as a black ellipse; its major and minor axes and position angle are specified below the map.}
\figsetgrpend

\figsetgrpstart
\figsetgrpnum{4.120}
\figsetgrptitle{J1639+8631 at 2.3 GHz}
\figsetplot{J1639+8631_S_2006_02_14_avp_map.pdf}
\figsetgrpnote{The source name, the observation date, and the frequency are given at the top of the panel. The intensity is shown by contours: solid lines are the positive contours, dotted lines are the negative contours. The contour levels in percents of the map peak are specified below the map, as well as the total flux density of the CLEAN model of the source and the intensity of the map peak. First contours correspond to the map noise level~$\times\,3$; each subsequent contour marks intensity increase by the factor of two. The CLEAN beam at the half-maximum level is shown in the map lower left corner as a black ellipse; its major and minor axes and position angle are specified below the map.}
\figsetgrpend

\figsetgrpstart
\figsetgrpnum{4.121}
\figsetgrptitle{J1639+8631 at 8.6 GHz}
\figsetplot{J1639+8631_X_2006_02_14_avp_map.pdf}
\figsetgrpnote{The source name, the observation date, and the frequency are given at the top of the panel. The intensity is shown by contours: solid lines are the positive contours, dotted lines are the negative contours. The contour levels in percents of the map peak are specified below the map, as well as the total flux density of the CLEAN model of the source and the intensity of the map peak. First contours correspond to the map noise level~$\times\,3$; each subsequent contour marks intensity increase by the factor of two. The CLEAN beam at the half-maximum level is shown in the map lower left corner as a black ellipse; its major and minor axes and position angle are specified below the map.}
\figsetgrpend

\figsetgrpstart
\figsetgrpnum{4.122}
\figsetgrptitle{J1705+7756 at 2.3 GHz}
\figsetplot{J1705+7756_S_2006_02_16_avp_map.pdf}
\figsetgrpnote{The source name, the observation date, and the frequency are given at the top of the panel. The intensity is shown by contours: solid lines are the positive contours, dotted lines are the negative contours. The contour levels in percents of the map peak are specified below the map, as well as the total flux density of the CLEAN model of the source and the intensity of the map peak. First contours correspond to the map noise level~$\times\,3$; each subsequent contour marks intensity increase by the factor of two. The CLEAN beam at the half-maximum level is shown in the map lower left corner as a black ellipse; its major and minor axes and position angle are specified below the map.}
\figsetgrpend

\figsetgrpstart
\figsetgrpnum{4.123}
\figsetgrptitle{J1705+7756 at 8.6 GHz}
\figsetplot{J1705+7756_X_2006_02_16_avp_map.pdf}
\figsetgrpnote{The source name, the observation date, and the frequency are given at the top of the panel. The intensity is shown by contours: solid lines are the positive contours, dotted lines are the negative contours. The contour levels in percents of the map peak are specified below the map, as well as the total flux density of the CLEAN model of the source and the intensity of the map peak. First contours correspond to the map noise level~$\times\,3$; each subsequent contour marks intensity increase by the factor of two. The CLEAN beam at the half-maximum level is shown in the map lower left corner as a black ellipse; its major and minor axes and position angle are specified below the map.}
\figsetgrpend

\figsetgrpstart
\figsetgrpnum{4.124}
\figsetgrptitle{J1723+7653 at 2.3 GHz}
\figsetplot{J1723+7653_S_2006_02_23_avp_map.pdf}
\figsetgrpnote{The source name, the observation date, and the frequency are given at the top of the panel. The intensity is shown by contours: solid lines are the positive contours, dotted lines are the negative contours. The contour levels in percents of the map peak are specified below the map, as well as the total flux density of the CLEAN model of the source and the intensity of the map peak. First contours correspond to the map noise level~$\times\,3$; each subsequent contour marks intensity increase by the factor of two. The CLEAN beam at the half-maximum level is shown in the map lower left corner as a black ellipse; its major and minor axes and position angle are specified below the map.}
\figsetgrpend

\figsetgrpstart
\figsetgrpnum{4.125}
\figsetgrptitle{J1723+7653 at 8.6 GHz}
\figsetplot{J1723+7653_X_2006_02_23_avp_map.pdf}
\figsetgrpnote{The source name, the observation date, and the frequency are given at the top of the panel. The intensity is shown by contours: solid lines are the positive contours, dotted lines are the negative contours. The contour levels in percents of the map peak are specified below the map, as well as the total flux density of the CLEAN model of the source and the intensity of the map peak. First contours correspond to the map noise level~$\times\,3$; each subsequent contour marks intensity increase by the factor of two. The CLEAN beam at the half-maximum level is shown in the map lower left corner as a black ellipse; its major and minor axes and position angle are specified below the map.}
\figsetgrpend

\figsetgrpstart
\figsetgrpnum{4.126}
\figsetgrptitle{J1725+7708 at 2.3 GHz}
\figsetplot{J1725+7708_S_2006_02_14_avp_map.pdf}
\figsetgrpnote{The source name, the observation date, and the frequency are given at the top of the panel. The intensity is shown by contours: solid lines are the positive contours, dotted lines are the negative contours. The contour levels in percents of the map peak are specified below the map, as well as the total flux density of the CLEAN model of the source and the intensity of the map peak. First contours correspond to the map noise level~$\times\,3$; each subsequent contour marks intensity increase by the factor of two. The CLEAN beam at the half-maximum level is shown in the map lower left corner as a black ellipse; its major and minor axes and position angle are specified below the map.}
\figsetgrpend

\figsetgrpstart
\figsetgrpnum{4.127}
\figsetgrptitle{J1725+7708 at 8.6 GHz}
\figsetplot{J1725+7708_X_2006_02_14_avp_map.pdf}
\figsetgrpnote{The source name, the observation date, and the frequency are given at the top of the panel. The intensity is shown by contours: solid lines are the positive contours, dotted lines are the negative contours. The contour levels in percents of the map peak are specified below the map, as well as the total flux density of the CLEAN model of the source and the intensity of the map peak. First contours correspond to the map noise level~$\times\,3$; each subsequent contour marks intensity increase by the factor of two. The CLEAN beam at the half-maximum level is shown in the map lower left corner as a black ellipse; its major and minor axes and position angle are specified below the map.}
\figsetgrpend

\figsetgrpstart
\figsetgrpnum{4.128}
\figsetgrptitle{J1725+7726 at 2.3 GHz}
\figsetplot{J1725+7726_S_2006_02_16_avp_map.pdf}
\figsetgrpnote{The source name, the observation date, and the frequency are given at the top of the panel. The intensity is shown by contours: solid lines are the positive contours, dotted lines are the negative contours. The contour levels in percents of the map peak are specified below the map, as well as the total flux density of the CLEAN model of the source and the intensity of the map peak. First contours correspond to the map noise level~$\times\,3$; each subsequent contour marks intensity increase by the factor of two. The CLEAN beam at the half-maximum level is shown in the map lower left corner as a black ellipse; its major and minor axes and position angle are specified below the map.}
\figsetgrpend

\figsetgrpstart
\figsetgrpnum{4.129}
\figsetgrptitle{J1800+7828 at 2.3 GHz}
\figsetplot{J1800+7828_S_2006_02_16_avp_map.pdf}
\figsetgrpnote{The source name, the observation date, and the frequency are given at the top of the panel. The intensity is shown by contours: solid lines are the positive contours, dotted lines are the negative contours. The contour levels in percents of the map peak are specified below the map, as well as the total flux density of the CLEAN model of the source and the intensity of the map peak. First contours correspond to the map noise level~$\times\,3$; each subsequent contour marks intensity increase by the factor of two. The CLEAN beam at the half-maximum level is shown in the map lower left corner as a black ellipse; its major and minor axes and position angle are specified below the map.}
\figsetgrpend

\figsetgrpstart
\figsetgrpnum{4.130}
\figsetgrptitle{J1800+7828 at 8.6 GHz}
\figsetplot{J1800+7828_X_2006_02_16_avp_map.pdf}
\figsetgrpnote{The source name, the observation date, and the frequency are given at the top of the panel. The intensity is shown by contours: solid lines are the positive contours, dotted lines are the negative contours. The contour levels in percents of the map peak are specified below the map, as well as the total flux density of the CLEAN model of the source and the intensity of the map peak. First contours correspond to the map noise level~$\times\,3$; each subsequent contour marks intensity increase by the factor of two. The CLEAN beam at the half-maximum level is shown in the map lower left corner as a black ellipse; its major and minor axes and position angle are specified below the map.}
\figsetgrpend

\figsetgrpstart
\figsetgrpnum{4.131}
\figsetgrptitle{J1803+7601 at 2.3 GHz}
\figsetplot{J1803+7601_S_2006_02_23_avp_map.pdf}
\figsetgrpnote{The source name, the observation date, and the frequency are given at the top of the panel. The intensity is shown by contours: solid lines are the positive contours, dotted lines are the negative contours. The contour levels in percents of the map peak are specified below the map, as well as the total flux density of the CLEAN model of the source and the intensity of the map peak. First contours correspond to the map noise level~$\times\,3$; each subsequent contour marks intensity increase by the factor of two. The CLEAN beam at the half-maximum level is shown in the map lower left corner as a black ellipse; its major and minor axes and position angle are specified below the map.}
\figsetgrpend

\figsetgrpstart
\figsetgrpnum{4.132}
\figsetgrptitle{J1822+8257 at 2.3 GHz}
\figsetplot{J1822+8257_S_2006_02_23_avp_map.pdf}
\figsetgrpnote{The source name, the observation date, and the frequency are given at the top of the panel. The intensity is shown by contours: solid lines are the positive contours, dotted lines are the negative contours. The contour levels in percents of the map peak are specified below the map, as well as the total flux density of the CLEAN model of the source and the intensity of the map peak. First contours correspond to the map noise level~$\times\,3$; each subsequent contour marks intensity increase by the factor of two. The CLEAN beam at the half-maximum level is shown in the map lower left corner as a black ellipse; its major and minor axes and position angle are specified below the map.}
\figsetgrpend

\figsetgrpstart
\figsetgrpnum{4.133}
\figsetgrptitle{J1822+8257 at 8.6 GHz}
\figsetplot{J1822+8257_X_2006_02_23_avp_map.pdf}
\figsetgrpnote{The source name, the observation date, and the frequency are given at the top of the panel. The intensity is shown by contours: solid lines are the positive contours, dotted lines are the negative contours. The contour levels in percents of the map peak are specified below the map, as well as the total flux density of the CLEAN model of the source and the intensity of the map peak. First contours correspond to the map noise level~$\times\,3$; each subsequent contour marks intensity increase by the factor of two. The CLEAN beam at the half-maximum level is shown in the map lower left corner as a black ellipse; its major and minor axes and position angle are specified below the map.}
\figsetgrpend

\figsetgrpstart
\figsetgrpnum{4.134}
\figsetgrptitle{J1823+7938 at 2.3 GHz}
\figsetplot{J1823+7938_S_2006_02_23_avp_map.pdf}
\figsetgrpnote{The source name, the observation date, and the frequency are given at the top of the panel. The intensity is shown by contours: solid lines are the positive contours, dotted lines are the negative contours. The contour levels in percents of the map peak are specified below the map, as well as the total flux density of the CLEAN model of the source and the intensity of the map peak. First contours correspond to the map noise level~$\times\,3$; each subsequent contour marks intensity increase by the factor of two. The CLEAN beam at the half-maximum level is shown in the map lower left corner as a black ellipse; its major and minor axes and position angle are specified below the map.}
\figsetgrpend

\figsetgrpstart
\figsetgrpnum{4.135}
\figsetgrptitle{J1823+7938 at 8.6 GHz}
\figsetplot{J1823+7938_X_2006_02_23_avp_map.pdf}
\figsetgrpnote{The source name, the observation date, and the frequency are given at the top of the panel. The intensity is shown by contours: solid lines are the positive contours, dotted lines are the negative contours. The contour levels in percents of the map peak are specified below the map, as well as the total flux density of the CLEAN model of the source and the intensity of the map peak. First contours correspond to the map noise level~$\times\,3$; each subsequent contour marks intensity increase by the factor of two. The CLEAN beam at the half-maximum level is shown in the map lower left corner as a black ellipse; its major and minor axes and position angle are specified below the map.}
\figsetgrpend

\figsetgrpstart
\figsetgrpnum{4.136}
\figsetgrptitle{J1832+8049 at 2.3 GHz}
\figsetplot{J1832+8049_S_2006_02_14_avp_map.pdf}
\figsetgrpnote{The source name, the observation date, and the frequency are given at the top of the panel. The intensity is shown by contours: solid lines are the positive contours, dotted lines are the negative contours. The contour levels in percents of the map peak are specified below the map, as well as the total flux density of the CLEAN model of the source and the intensity of the map peak. First contours correspond to the map noise level~$\times\,3$; each subsequent contour marks intensity increase by the factor of two. The CLEAN beam at the half-maximum level is shown in the map lower left corner as a black ellipse; its major and minor axes and position angle are specified below the map.}
\figsetgrpend

\figsetgrpstart
\figsetgrpnum{4.137}
\figsetgrptitle{J1842+7946 at 2.3 GHz}
\figsetplot{J1842+7946_S_2006_02_23_avp_map.pdf}
\figsetgrpnote{The source name, the observation date, and the frequency are given at the top of the panel. The intensity is shown by contours: solid lines are the positive contours, dotted lines are the negative contours. The contour levels in percents of the map peak are specified below the map, as well as the total flux density of the CLEAN model of the source and the intensity of the map peak. First contours correspond to the map noise level~$\times\,3$; each subsequent contour marks intensity increase by the factor of two. The CLEAN beam at the half-maximum level is shown in the map lower left corner as a black ellipse; its major and minor axes and position angle are specified below the map.}
\figsetgrpend

\figsetgrpstart
\figsetgrpnum{4.138}
\figsetgrptitle{J1901+8623 at 2.3 GHz}
\figsetplot{J1901+8623_S_2006_02_16_avp_map.pdf}
\figsetgrpnote{The source name, the observation date, and the frequency are given at the top of the panel. The intensity is shown by contours: solid lines are the positive contours, dotted lines are the negative contours. The contour levels in percents of the map peak are specified below the map, as well as the total flux density of the CLEAN model of the source and the intensity of the map peak. First contours correspond to the map noise level~$\times\,3$; each subsequent contour marks intensity increase by the factor of two. The CLEAN beam at the half-maximum level is shown in the map lower left corner as a black ellipse; its major and minor axes and position angle are specified below the map.}
\figsetgrpend

\figsetgrpstart
\figsetgrpnum{4.139}
\figsetgrptitle{J1909+7813 at 2.3 GHz}
\figsetplot{J1909+7813_S_2006_02_16_avp_map.pdf}
\figsetgrpnote{The source name, the observation date, and the frequency are given at the top of the panel. The intensity is shown by contours: solid lines are the positive contours, dotted lines are the negative contours. The contour levels in percents of the map peak are specified below the map, as well as the total flux density of the CLEAN model of the source and the intensity of the map peak. First contours correspond to the map noise level~$\times\,3$; each subsequent contour marks intensity increase by the factor of two. The CLEAN beam at the half-maximum level is shown in the map lower left corner as a black ellipse; its major and minor axes and position angle are specified below the map.}
\figsetgrpend

\figsetgrpstart
\figsetgrpnum{4.140}
\figsetgrptitle{J1935+8130 at 2.3 GHz}
\figsetplot{J1935+8130_S_2006_02_14_avp_map.pdf}
\figsetgrpnote{The source name, the observation date, and the frequency are given at the top of the panel. The intensity is shown by contours: solid lines are the positive contours, dotted lines are the negative contours. The contour levels in percents of the map peak are specified below the map, as well as the total flux density of the CLEAN model of the source and the intensity of the map peak. First contours correspond to the map noise level~$\times\,3$; each subsequent contour marks intensity increase by the factor of two. The CLEAN beam at the half-maximum level is shown in the map lower left corner as a black ellipse; its major and minor axes and position angle are specified below the map.}
\figsetgrpend

\figsetgrpstart
\figsetgrpnum{4.141}
\figsetgrptitle{J1935+8130 at 8.6 GHz}
\figsetplot{J1935+8130_X_2006_02_14_avp_map.pdf}
\figsetgrpnote{The source name, the observation date, and the frequency are given at the top of the panel. The intensity is shown by contours: solid lines are the positive contours, dotted lines are the negative contours. The contour levels in percents of the map peak are specified below the map, as well as the total flux density of the CLEAN model of the source and the intensity of the map peak. First contours correspond to the map noise level~$\times\,3$; each subsequent contour marks intensity increase by the factor of two. The CLEAN beam at the half-maximum level is shown in the map lower left corner as a black ellipse; its major and minor axes and position angle are specified below the map.}
\figsetgrpend

\figsetgrpstart
\figsetgrpnum{4.142}
\figsetgrptitle{J1937+8356 at 2.3 GHz}
\figsetplot{J1937+8356_S_2006_02_23_avp_map.pdf}
\figsetgrpnote{The source name, the observation date, and the frequency are given at the top of the panel. The intensity is shown by contours: solid lines are the positive contours, dotted lines are the negative contours. The contour levels in percents of the map peak are specified below the map, as well as the total flux density of the CLEAN model of the source and the intensity of the map peak. First contours correspond to the map noise level~$\times\,3$; each subsequent contour marks intensity increase by the factor of two. The CLEAN beam at the half-maximum level is shown in the map lower left corner as a black ellipse; its major and minor axes and position angle are specified below the map.}
\figsetgrpend

\figsetgrpstart
\figsetgrpnum{4.143}
\figsetgrptitle{J1937+8356 at 8.6 GHz}
\figsetplot{J1937+8356_X_2006_02_23_avp_map.pdf}
\figsetgrpnote{The source name, the observation date, and the frequency are given at the top of the panel. The intensity is shown by contours: solid lines are the positive contours, dotted lines are the negative contours. The contour levels in percents of the map peak are specified below the map, as well as the total flux density of the CLEAN model of the source and the intensity of the map peak. First contours correspond to the map noise level~$\times\,3$; each subsequent contour marks intensity increase by the factor of two. The CLEAN beam at the half-maximum level is shown in the map lower left corner as a black ellipse; its major and minor axes and position angle are specified below the map.}
\figsetgrpend

\figsetgrpstart
\figsetgrpnum{4.144}
\figsetgrptitle{J2005+7752 at 2.3 GHz}
\figsetplot{J2005+7752_S_2006_02_14_avp_map.pdf}
\figsetgrpnote{The source name, the observation date, and the frequency are given at the top of the panel. The intensity is shown by contours: solid lines are the positive contours, dotted lines are the negative contours. The contour levels in percents of the map peak are specified below the map, as well as the total flux density of the CLEAN model of the source and the intensity of the map peak. First contours correspond to the map noise level~$\times\,3$; each subsequent contour marks intensity increase by the factor of two. The CLEAN beam at the half-maximum level is shown in the map lower left corner as a black ellipse; its major and minor axes and position angle are specified below the map.}
\figsetgrpend

\figsetgrpstart
\figsetgrpnum{4.145}
\figsetgrptitle{J2005+7752 at 8.6 GHz}
\figsetplot{J2005+7752_X_2006_02_14_avp_map.pdf}
\figsetgrpnote{The source name, the observation date, and the frequency are given at the top of the panel. The intensity is shown by contours: solid lines are the positive contours, dotted lines are the negative contours. The contour levels in percents of the map peak are specified below the map, as well as the total flux density of the CLEAN model of the source and the intensity of the map peak. First contours correspond to the map noise level~$\times\,3$; each subsequent contour marks intensity increase by the factor of two. The CLEAN beam at the half-maximum level is shown in the map lower left corner as a black ellipse; its major and minor axes and position angle are specified below the map.}
\figsetgrpend

\figsetgrpstart
\figsetgrpnum{4.146}
\figsetgrptitle{J2022+7611 at 2.3 GHz}
\figsetplot{J2022+7611_S_2006_02_23_avp_map.pdf}
\figsetgrpnote{The source name, the observation date, and the frequency are given at the top of the panel. The intensity is shown by contours: solid lines are the positive contours, dotted lines are the negative contours. The contour levels in percents of the map peak are specified below the map, as well as the total flux density of the CLEAN model of the source and the intensity of the map peak. First contours correspond to the map noise level~$\times\,3$; each subsequent contour marks intensity increase by the factor of two. The CLEAN beam at the half-maximum level is shown in the map lower left corner as a black ellipse; its major and minor axes and position angle are specified below the map.}
\figsetgrpend

\figsetgrpstart
\figsetgrpnum{4.147}
\figsetgrptitle{J2022+7611 at 8.6 GHz}
\figsetplot{J2022+7611_X_2006_02_23_avp_map.pdf}
\figsetgrpnote{The source name, the observation date, and the frequency are given at the top of the panel. The intensity is shown by contours: solid lines are the positive contours, dotted lines are the negative contours. The contour levels in percents of the map peak are specified below the map, as well as the total flux density of the CLEAN model of the source and the intensity of the map peak. First contours correspond to the map noise level~$\times\,3$; each subsequent contour marks intensity increase by the factor of two. The CLEAN beam at the half-maximum level is shown in the map lower left corner as a black ellipse; its major and minor axes and position angle are specified below the map.}
\figsetgrpend

\figsetgrpstart
\figsetgrpnum{4.148}
\figsetgrptitle{J2042+7508 at 2.3 GHz}
\figsetplot{J2042+7508_S_2006_02_16_avp_map.pdf}
\figsetgrpnote{The source name, the observation date, and the frequency are given at the top of the panel. The intensity is shown by contours: solid lines are the positive contours, dotted lines are the negative contours. The contour levels in percents of the map peak are specified below the map, as well as the total flux density of the CLEAN model of the source and the intensity of the map peak. First contours correspond to the map noise level~$\times\,3$; each subsequent contour marks intensity increase by the factor of two. The CLEAN beam at the half-maximum level is shown in the map lower left corner as a black ellipse; its major and minor axes and position angle are specified below the map.}
\figsetgrpend

\figsetgrpstart
\figsetgrpnum{4.149}
\figsetgrptitle{J2042+7508 at 8.6 GHz}
\figsetplot{J2042+7508_X_2006_02_16_avp_map.pdf}
\figsetgrpnote{The source name, the observation date, and the frequency are given at the top of the panel. The intensity is shown by contours: solid lines are the positive contours, dotted lines are the negative contours. The contour levels in percents of the map peak are specified below the map, as well as the total flux density of the CLEAN model of the source and the intensity of the map peak. First contours correspond to the map noise level~$\times\,3$; each subsequent contour marks intensity increase by the factor of two. The CLEAN beam at the half-maximum level is shown in the map lower left corner as a black ellipse; its major and minor axes and position angle are specified below the map.}
\figsetgrpend

\figsetgrpstart
\figsetgrpnum{4.150}
\figsetgrptitle{J2045+7625 at 2.3 GHz}
\figsetplot{J2045+7625_S_2006_02_23_avp_map.pdf}
\figsetgrpnote{The source name, the observation date, and the frequency are given at the top of the panel. The intensity is shown by contours: solid lines are the positive contours, dotted lines are the negative contours. The contour levels in percents of the map peak are specified below the map, as well as the total flux density of the CLEAN model of the source and the intensity of the map peak. First contours correspond to the map noise level~$\times\,3$; each subsequent contour marks intensity increase by the factor of two. The CLEAN beam at the half-maximum level is shown in the map lower left corner as a black ellipse; its major and minor axes and position angle are specified below the map.}
\figsetgrpend

\figsetgrpstart
\figsetgrpnum{4.151}
\figsetgrptitle{J2045+7625 at 8.6 GHz}
\figsetplot{J2045+7625_X_2006_02_23_avp_map.pdf}
\figsetgrpnote{The source name, the observation date, and the frequency are given at the top of the panel. The intensity is shown by contours: solid lines are the positive contours, dotted lines are the negative contours. The contour levels in percents of the map peak are specified below the map, as well as the total flux density of the CLEAN model of the source and the intensity of the map peak. First contours correspond to the map noise level~$\times\,3$; each subsequent contour marks intensity increase by the factor of two. The CLEAN beam at the half-maximum level is shown in the map lower left corner as a black ellipse; its major and minor axes and position angle are specified below the map.}
\figsetgrpend

\figsetgrpstart
\figsetgrpnum{4.152}
\figsetgrptitle{J2114+8204 at 2.3 GHz}
\figsetplot{J2114+8204_S_2006_02_23_avp_map.pdf}
\figsetgrpnote{The source name, the observation date, and the frequency are given at the top of the panel. The intensity is shown by contours: solid lines are the positive contours, dotted lines are the negative contours. The contour levels in percents of the map peak are specified below the map, as well as the total flux density of the CLEAN model of the source and the intensity of the map peak. First contours correspond to the map noise level~$\times\,3$; each subsequent contour marks intensity increase by the factor of two. The CLEAN beam at the half-maximum level is shown in the map lower left corner as a black ellipse; its major and minor axes and position angle are specified below the map.}
\figsetgrpend

\figsetgrpstart
\figsetgrpnum{4.153}
\figsetgrptitle{J2114+8204 at 8.6 GHz}
\figsetplot{J2114+8204_X_2006_02_23_avp_map.pdf}
\figsetgrpnote{The source name, the observation date, and the frequency are given at the top of the panel. The intensity is shown by contours: solid lines are the positive contours, dotted lines are the negative contours. The contour levels in percents of the map peak are specified below the map, as well as the total flux density of the CLEAN model of the source and the intensity of the map peak. First contours correspond to the map noise level~$\times\,3$; each subsequent contour marks intensity increase by the factor of two. The CLEAN beam at the half-maximum level is shown in the map lower left corner as a black ellipse; its major and minor axes and position angle are specified below the map.}
\figsetgrpend

\figsetgrpstart
\figsetgrpnum{4.154}
\figsetgrptitle{J2131+8430 at 2.3 GHz}
\figsetplot{J2131+8430_S_2006_02_23_avp_map.pdf}
\figsetgrpnote{The source name, the observation date, and the frequency are given at the top of the panel. The intensity is shown by contours: solid lines are the positive contours, dotted lines are the negative contours. The contour levels in percents of the map peak are specified below the map, as well as the total flux density of the CLEAN model of the source and the intensity of the map peak. First contours correspond to the map noise level~$\times\,3$; each subsequent contour marks intensity increase by the factor of two. The CLEAN beam at the half-maximum level is shown in the map lower left corner as a black ellipse; its major and minor axes and position angle are specified below the map.}
\figsetgrpend

\figsetgrpstart
\figsetgrpnum{4.155}
\figsetgrptitle{J2131+8430 at 8.6 GHz}
\figsetplot{J2131+8430_X_2006_02_23_avp_map.pdf}
\figsetgrpnote{The source name, the observation date, and the frequency are given at the top of the panel. The intensity is shown by contours: solid lines are the positive contours, dotted lines are the negative contours. The contour levels in percents of the map peak are specified below the map, as well as the total flux density of the CLEAN model of the source and the intensity of the map peak. First contours correspond to the map noise level~$\times\,3$; each subsequent contour marks intensity increase by the factor of two. The CLEAN beam at the half-maximum level is shown in the map lower left corner as a black ellipse; its major and minor axes and position angle are specified below the map.}
\figsetgrpend

\figsetgrpstart
\figsetgrpnum{4.156}
\figsetgrptitle{J2133+8239 at 2.3 GHz}
\figsetplot{J2133+8239_S_2006_02_23_avp_map.pdf}
\figsetgrpnote{The source name, the observation date, and the frequency are given at the top of the panel. The intensity is shown by contours: solid lines are the positive contours, dotted lines are the negative contours. The contour levels in percents of the map peak are specified below the map, as well as the total flux density of the CLEAN model of the source and the intensity of the map peak. First contours correspond to the map noise level~$\times\,3$; each subsequent contour marks intensity increase by the factor of two. The CLEAN beam at the half-maximum level is shown in the map lower left corner as a black ellipse; its major and minor axes and position angle are specified below the map.}
\figsetgrpend

\figsetgrpstart
\figsetgrpnum{4.157}
\figsetgrptitle{J2133+8239 at 8.6 GHz}
\figsetplot{J2133+8239_X_2006_02_23_avp_map.pdf}
\figsetgrpnote{The source name, the observation date, and the frequency are given at the top of the panel. The intensity is shown by contours: solid lines are the positive contours, dotted lines are the negative contours. The contour levels in percents of the map peak are specified below the map, as well as the total flux density of the CLEAN model of the source and the intensity of the map peak. First contours correspond to the map noise level~$\times\,3$; each subsequent contour marks intensity increase by the factor of two. The CLEAN beam at the half-maximum level is shown in the map lower left corner as a black ellipse; its major and minor axes and position angle are specified below the map.}
\figsetgrpend

\figsetgrpstart
\figsetgrpnum{4.158}
\figsetgrptitle{J2156+8337 at 2.3 GHz}
\figsetplot{J2156+8337_S_2006_02_16_avp_map.pdf}
\figsetgrpnote{The source name, the observation date, and the frequency are given at the top of the panel. The intensity is shown by contours: solid lines are the positive contours, dotted lines are the negative contours. The contour levels in percents of the map peak are specified below the map, as well as the total flux density of the CLEAN model of the source and the intensity of the map peak. First contours correspond to the map noise level~$\times\,3$; each subsequent contour marks intensity increase by the factor of two. The CLEAN beam at the half-maximum level is shown in the map lower left corner as a black ellipse; its major and minor axes and position angle are specified below the map.}
\figsetgrpend

\figsetgrpstart
\figsetgrpnum{4.159}
\figsetgrptitle{J2156+8337 at 8.6 GHz}
\figsetplot{J2156+8337_X_2006_02_16_avp_map.pdf}
\figsetgrpnote{The source name, the observation date, and the frequency are given at the top of the panel. The intensity is shown by contours: solid lines are the positive contours, dotted lines are the negative contours. The contour levels in percents of the map peak are specified below the map, as well as the total flux density of the CLEAN model of the source and the intensity of the map peak. First contours correspond to the map noise level~$\times\,3$; each subsequent contour marks intensity increase by the factor of two. The CLEAN beam at the half-maximum level is shown in the map lower left corner as a black ellipse; its major and minor axes and position angle are specified below the map.}
\figsetgrpend

\figsetgrpstart
\figsetgrpnum{4.160}
\figsetgrptitle{J2242+8224 at 2.3 GHz}
\figsetplot{J2242+8224_S_2006_02_23_avp_map.pdf}
\figsetgrpnote{The source name, the observation date, and the frequency are given at the top of the panel. The intensity is shown by contours: solid lines are the positive contours, dotted lines are the negative contours. The contour levels in percents of the map peak are specified below the map, as well as the total flux density of the CLEAN model of the source and the intensity of the map peak. First contours correspond to the map noise level~$\times\,3$; each subsequent contour marks intensity increase by the factor of two. The CLEAN beam at the half-maximum level is shown in the map lower left corner as a black ellipse; its major and minor axes and position angle are specified below the map.}
\figsetgrpend

\figsetgrpstart
\figsetgrpnum{4.161}
\figsetgrptitle{J2248+7718 at 2.3 GHz}
\figsetplot{J2248+7718_S_2006_02_16_avp_map.pdf}
\figsetgrpnote{The source name, the observation date, and the frequency are given at the top of the panel. The intensity is shown by contours: solid lines are the positive contours, dotted lines are the negative contours. The contour levels in percents of the map peak are specified below the map, as well as the total flux density of the CLEAN model of the source and the intensity of the map peak. First contours correspond to the map noise level~$\times\,3$; each subsequent contour marks intensity increase by the factor of two. The CLEAN beam at the half-maximum level is shown in the map lower left corner as a black ellipse; its major and minor axes and position angle are specified below the map.}
\figsetgrpend

\figsetgrpstart
\figsetgrpnum{4.162}
\figsetgrptitle{J2301+8200 at 2.3 GHz}
\figsetplot{J2301+8200_S_2006_02_14_avp_map.pdf}
\figsetgrpnote{The source name, the observation date, and the frequency are given at the top of the panel. The intensity is shown by contours: solid lines are the positive contours, dotted lines are the negative contours. The contour levels in percents of the map peak are specified below the map, as well as the total flux density of the CLEAN model of the source and the intensity of the map peak. First contours correspond to the map noise level~$\times\,3$; each subsequent contour marks intensity increase by the factor of two. The CLEAN beam at the half-maximum level is shown in the map lower left corner as a black ellipse; its major and minor axes and position angle are specified below the map.}
\figsetgrpend

\figsetgrpstart
\figsetgrpnum{4.163}
\figsetgrptitle{J2310+8857 at 2.3 GHz}
\figsetplot{J2310+8857_S_2006_02_16_avp_map.pdf}
\figsetgrpnote{The source name, the observation date, and the frequency are given at the top of the panel. The intensity is shown by contours: solid lines are the positive contours, dotted lines are the negative contours. The contour levels in percents of the map peak are specified below the map, as well as the total flux density of the CLEAN model of the source and the intensity of the map peak. First contours correspond to the map noise level~$\times\,3$; each subsequent contour marks intensity increase by the factor of two. The CLEAN beam at the half-maximum level is shown in the map lower left corner as a black ellipse; its major and minor axes and position angle are specified below the map.}
\figsetgrpend

\figsetgrpstart
\figsetgrpnum{4.164}
\figsetgrptitle{J2310+8857 at 8.6 GHz}
\figsetplot{J2310+8857_X_2006_02_16_avp_map.pdf}
\figsetgrpnote{The source name, the observation date, and the frequency are given at the top of the panel. The intensity is shown by contours: solid lines are the positive contours, dotted lines are the negative contours. The contour levels in percents of the map peak are specified below the map, as well as the total flux density of the CLEAN model of the source and the intensity of the map peak. First contours correspond to the map noise level~$\times\,3$; each subsequent contour marks intensity increase by the factor of two. The CLEAN beam at the half-maximum level is shown in the map lower left corner as a black ellipse; its major and minor axes and position angle are specified below the map.}
\figsetgrpend

\figsetgrpstart
\figsetgrpnum{4.165}
\figsetgrptitle{J2325+7917 at 2.3 GHz}
\figsetplot{J2325+7917_S_2006_02_16_avp_map.pdf}
\figsetgrpnote{The source name, the observation date, and the frequency are given at the top of the panel. The intensity is shown by contours: solid lines are the positive contours, dotted lines are the negative contours. The contour levels in percents of the map peak are specified below the map, as well as the total flux density of the CLEAN model of the source and the intensity of the map peak. First contours correspond to the map noise level~$\times\,3$; each subsequent contour marks intensity increase by the factor of two. The CLEAN beam at the half-maximum level is shown in the map lower left corner as a black ellipse; its major and minor axes and position angle are specified below the map.}
\figsetgrpend

\figsetgrpstart
\figsetgrpnum{4.166}
\figsetgrptitle{J2344+8226 at 2.3 GHz}
\figsetplot{J2344+8226_S_2006_02_14_avp_map.pdf}
\figsetgrpnote{The source name, the observation date, and the frequency are given at the top of the panel. The intensity is shown by contours: solid lines are the positive contours, dotted lines are the negative contours. The contour levels in percents of the map peak are specified below the map, as well as the total flux density of the CLEAN model of the source and the intensity of the map peak. First contours correspond to the map noise level~$\times\,3$; each subsequent contour marks intensity increase by the factor of two. The CLEAN beam at the half-maximum level is shown in the map lower left corner as a black ellipse; its major and minor axes and position angle are specified below the map.}
\figsetgrpend

\figsetgrpstart
\figsetgrpnum{4.167}
\figsetgrptitle{J2356+8152 at 2.3 GHz}
\figsetplot{J2356+8152_S_2006_02_16_avp_map.pdf}
\figsetgrpnote{The source name, the observation date, and the frequency are given at the top of the panel. The intensity is shown by contours: solid lines are the positive contours, dotted lines are the negative contours. The contour levels in percents of the map peak are specified below the map, as well as the total flux density of the CLEAN model of the source and the intensity of the map peak. First contours correspond to the map noise level~$\times\,3$; each subsequent contour marks intensity increase by the factor of two. The CLEAN beam at the half-maximum level is shown in the map lower left corner as a black ellipse; its major and minor axes and position angle are specified below the map.}
\figsetgrpend

\figsetgrpstart
\figsetgrpnum{4.168}
\figsetgrptitle{J2356+8152 at 8.6 GHz}
\figsetplot{J2356+8152_X_2006_02_16_avp_map.pdf}
\figsetgrpnote{The source name, the observation date, and the frequency are given at the top of the panel. The intensity is shown by contours: solid lines are the positive contours, dotted lines are the negative contours. The contour levels in percents of the map peak are specified below the map, as well as the total flux density of the CLEAN model of the source and the intensity of the map peak. First contours correspond to the map noise level~$\times\,3$; each subsequent contour marks intensity increase by the factor of two. The CLEAN beam at the half-maximum level is shown in the map lower left corner as a black ellipse; its major and minor axes and position angle are specified below the map.}
\figsetgrpend

\figsetend


\figsetstart
\figsetnum{7}
\figsettitle{Single-dish and VLBA broadband spectra of all the sources of our sample.}

\figsetgrpstart
\figsetgrpnum{7.1}
\figsetgrptitle{Total and VLBA spectra of J0000+8123}
\figsetplot{J0000+8123_spectra.pdf}
\figsetgrpnote{Single-dish and VLBA broadband spectra of all the sources of our sample. The source name is specified at the top right of each plot. The single-dish spectra are plotted black, the VLBA spectra are plotted by the red color. In cases of the VLBA non-detection, the upper limits on the VLBA flux density are shown by red arrows.}
\figsetgrpend

\figsetgrpstart
\figsetgrpnum{7.2}
\figsetgrptitle{Total and VLBA spectra of J0005+8135}
\figsetplot{J0005+8135_spectra.pdf}
\figsetgrpnote{Single-dish and VLBA broadband spectra of all the sources of our sample. The source name is specified at the top right of each plot. The single-dish spectra are plotted black, the VLBA spectra are plotted by the red color. In cases of the VLBA non-detection, the upper limits on the VLBA flux density are shown by red arrows.}
\figsetgrpend

\figsetgrpstart
\figsetgrpnum{7.3}
\figsetgrptitle{Total and VLBA spectra of J0008+8426}
\figsetplot{J0008+8426_spectra.pdf}
\figsetgrpnote{Single-dish and VLBA broadband spectra of all the sources of our sample. The source name is specified at the top right of each plot. The single-dish spectra are plotted black, the VLBA spectra are plotted by the red color. In cases of the VLBA non-detection, the upper limits on the VLBA flux density are shown by red arrows.}
\figsetgrpend

\figsetgrpstart
\figsetgrpnum{7.4}
\figsetgrptitle{Total and VLBA spectra of J0009+7603}
\figsetplot{J0009+7603_spectra.pdf}
\figsetgrpnote{Single-dish and VLBA broadband spectra of all the sources of our sample. The source name is specified at the top right of each plot. The single-dish spectra are plotted black, the VLBA spectra are plotted by the red color. In cases of the VLBA non-detection, the upper limits on the VLBA flux density are shown by red arrows.}
\figsetgrpend

\figsetgrpstart
\figsetgrpnum{7.5}
\figsetgrptitle{Total and VLBA spectra of J0009+7724}
\figsetplot{J0009+7724_spectra.pdf}
\figsetgrpnote{Single-dish and VLBA broadband spectra of all the sources of our sample. The source name is specified at the top right of each plot. The single-dish spectra are plotted black, the VLBA spectra are plotted by the red color. In cases of the VLBA non-detection, the upper limits on the VLBA flux density are shown by red arrows.}
\figsetgrpend

\figsetgrpstart
\figsetgrpnum{7.6}
\figsetgrptitle{Total and VLBA spectra of J0012+8543}
\figsetplot{J0012+8543_spectra.pdf}
\figsetgrpnote{Single-dish and VLBA broadband spectra of all the sources of our sample. The source name is specified at the top right of each plot. The single-dish spectra are plotted black, the VLBA spectra are plotted by the red color. In cases of the VLBA non-detection, the upper limits on the VLBA flux density are shown by red arrows.}
\figsetgrpend

\figsetgrpstart
\figsetgrpnum{7.7}
\figsetgrptitle{Total and VLBA spectra of J0013+7748}
\figsetplot{J0013+7748_spectra.pdf}
\figsetgrpnote{Single-dish and VLBA broadband spectra of all the sources of our sample. The source name is specified at the top right of each plot. The single-dish spectra are plotted black, the VLBA spectra are plotted by the red color. In cases of the VLBA non-detection, the upper limits on the VLBA flux density are shown by red arrows.}
\figsetgrpend

\figsetgrpstart
\figsetgrpnum{7.8}
\figsetgrptitle{Total and VLBA spectra of J0015+7756}
\figsetplot{J0015+7756_spectra.pdf}
\figsetgrpnote{Single-dish and VLBA broadband spectra of all the sources of our sample. The source name is specified at the top right of each plot. The single-dish spectra are plotted black, the VLBA spectra are plotted by the red color. In cases of the VLBA non-detection, the upper limits on the VLBA flux density are shown by red arrows.}
\figsetgrpend

\figsetgrpstart
\figsetgrpnum{7.9}
\figsetgrptitle{Total and VLBA spectra of J0016+7916}
\figsetplot{J0016+7916_spectra.pdf}
\figsetgrpnote{Single-dish and VLBA broadband spectra of all the sources of our sample. The source name is specified at the top right of each plot. The single-dish spectra are plotted black, the VLBA spectra are plotted by the red color. In cases of the VLBA non-detection, the upper limits on the VLBA flux density are shown by red arrows.}
\figsetgrpend

\figsetgrpstart
\figsetgrpnum{7.10}
\figsetgrptitle{Total and VLBA spectra of J0017+8135}
\figsetplot{J0017+8135_spectra.pdf}
\figsetgrpnote{Single-dish and VLBA broadband spectra of all the sources of our sample. The source name is specified at the top right of each plot. The single-dish spectra are plotted black, the VLBA spectra are plotted by the red color. In cases of the VLBA non-detection, the upper limits on the VLBA flux density are shown by red arrows.}
\figsetgrpend

\figsetgrpstart
\figsetgrpnum{7.11}
\figsetgrptitle{Total and VLBA spectra of J0018+7827}
\figsetplot{J0018+7827_spectra.pdf}
\figsetgrpnote{Single-dish and VLBA broadband spectra of all the sources of our sample. The source name is specified at the top right of each plot. The single-dish spectra are plotted black, the VLBA spectra are plotted by the red color. In cases of the VLBA non-detection, the upper limits on the VLBA flux density are shown by red arrows.}
\figsetgrpend

\figsetgrpstart
\figsetgrpnum{7.12}
\figsetgrptitle{Total and VLBA spectra of J0019+8039}
\figsetplot{J0019+8039_spectra.pdf}
\figsetgrpnote{Single-dish and VLBA broadband spectra of all the sources of our sample. The source name is specified at the top right of each plot. The single-dish spectra are plotted black, the VLBA spectra are plotted by the red color. In cases of the VLBA non-detection, the upper limits on the VLBA flux density are shown by red arrows.}
\figsetgrpend

\figsetgrpstart
\figsetgrpnum{7.13}
\figsetgrptitle{Total and VLBA spectra of J0028+7958}
\figsetplot{J0028+7958_spectra.pdf}
\figsetgrpnote{Single-dish and VLBA broadband spectra of all the sources of our sample. The source name is specified at the top right of each plot. The single-dish spectra are plotted black, the VLBA spectra are plotted by the red color. In cases of the VLBA non-detection, the upper limits on the VLBA flux density are shown by red arrows.}
\figsetgrpend

\figsetgrpstart
\figsetgrpnum{7.14}
\figsetgrptitle{Total and VLBA spectra of J0030+8449}
\figsetplot{J0030+8449_spectra.pdf}
\figsetgrpnote{Single-dish and VLBA broadband spectra of all the sources of our sample. The source name is specified at the top right of each plot. The single-dish spectra are plotted black, the VLBA spectra are plotted by the red color. In cases of the VLBA non-detection, the upper limits on the VLBA flux density are shown by red arrows.}
\figsetgrpend

\figsetgrpstart
\figsetgrpnum{7.15}
\figsetgrptitle{Total and VLBA spectra of J0034+7647}
\figsetplot{J0034+7647_spectra.pdf}
\figsetgrpnote{Single-dish and VLBA broadband spectra of all the sources of our sample. The source name is specified at the top right of each plot. The single-dish spectra are plotted black, the VLBA spectra are plotted by the red color. In cases of the VLBA non-detection, the upper limits on the VLBA flux density are shown by red arrows.}
\figsetgrpend

\figsetgrpstart
\figsetgrpnum{7.16}
\figsetgrptitle{Total and VLBA spectra of J0034+8013}
\figsetplot{J0034+8013_spectra.pdf}
\figsetgrpnote{Single-dish and VLBA broadband spectra of all the sources of our sample. The source name is specified at the top right of each plot. The single-dish spectra are plotted black, the VLBA spectra are plotted by the red color. In cases of the VLBA non-detection, the upper limits on the VLBA flux density are shown by red arrows.}
\figsetgrpend

\figsetgrpstart
\figsetgrpnum{7.17}
\figsetgrptitle{Total and VLBA spectra of J0035+8025}
\figsetplot{J0035+8025_spectra.pdf}
\figsetgrpnote{Single-dish and VLBA broadband spectra of all the sources of our sample. The source name is specified at the top right of each plot. The single-dish spectra are plotted black, the VLBA spectra are plotted by the red color. In cases of the VLBA non-detection, the upper limits on the VLBA flux density are shown by red arrows.}
\figsetgrpend

\figsetgrpstart
\figsetgrpnum{7.18}
\figsetgrptitle{Total and VLBA spectra of J0038+8447}
\figsetplot{J0038+8447_spectra.pdf}
\figsetgrpnote{Single-dish and VLBA broadband spectra of all the sources of our sample. The source name is specified at the top right of each plot. The single-dish spectra are plotted black, the VLBA spectra are plotted by the red color. In cases of the VLBA non-detection, the upper limits on the VLBA flux density are shown by red arrows.}
\figsetgrpend

\figsetgrpstart
\figsetgrpnum{7.19}
\figsetgrptitle{Total and VLBA spectra of J0039+7505}
\figsetplot{J0039+7505_spectra.pdf}
\figsetgrpnote{Single-dish and VLBA broadband spectra of all the sources of our sample. The source name is specified at the top right of each plot. The single-dish spectra are plotted black, the VLBA spectra are plotted by the red color. In cases of the VLBA non-detection, the upper limits on the VLBA flux density are shown by red arrows.}
\figsetgrpend

\figsetgrpstart
\figsetgrpnum{7.20}
\figsetgrptitle{Total and VLBA spectra of J0039+7700}
\figsetplot{J0039+7700_spectra.pdf}
\figsetgrpnote{Single-dish and VLBA broadband spectra of all the sources of our sample. The source name is specified at the top right of each plot. The single-dish spectra are plotted black, the VLBA spectra are plotted by the red color. In cases of the VLBA non-detection, the upper limits on the VLBA flux density are shown by red arrows.}
\figsetgrpend

\figsetgrpstart
\figsetgrpnum{7.21}
\figsetgrptitle{Total and VLBA spectra of J0041+8114}
\figsetplot{J0041+8114_spectra.pdf}
\figsetgrpnote{Single-dish and VLBA broadband spectra of all the sources of our sample. The source name is specified at the top right of each plot. The single-dish spectra are plotted black, the VLBA spectra are plotted by the red color. In cases of the VLBA non-detection, the upper limits on the VLBA flux density are shown by red arrows.}
\figsetgrpend

\figsetgrpstart
\figsetgrpnum{7.22}
\figsetgrptitle{Total and VLBA spectra of J0044+8929}
\figsetplot{J0044+8929_spectra.pdf}
\figsetgrpnote{Single-dish and VLBA broadband spectra of all the sources of our sample. The source name is specified at the top right of each plot. The single-dish spectra are plotted black, the VLBA spectra are plotted by the red color. In cases of the VLBA non-detection, the upper limits on the VLBA flux density are shown by red arrows.}
\figsetgrpend

\figsetgrpstart
\figsetgrpnum{7.23}
\figsetgrptitle{Total and VLBA spectra of J0045+7542}
\figsetplot{J0045+7542_spectra.pdf}
\figsetgrpnote{Single-dish and VLBA broadband spectra of all the sources of our sample. The source name is specified at the top right of each plot. The single-dish spectra are plotted black, the VLBA spectra are plotted by the red color. In cases of the VLBA non-detection, the upper limits on the VLBA flux density are shown by red arrows.}
\figsetgrpend

\figsetgrpstart
\figsetgrpnum{7.24}
\figsetgrptitle{Total and VLBA spectra of J0046+7517}
\figsetplot{J0046+7517_spectra.pdf}
\figsetgrpnote{Single-dish and VLBA broadband spectra of all the sources of our sample. The source name is specified at the top right of each plot. The single-dish spectra are plotted black, the VLBA spectra are plotted by the red color. In cases of the VLBA non-detection, the upper limits on the VLBA flux density are shown by red arrows.}
\figsetgrpend

\figsetgrpstart
\figsetgrpnum{7.25}
\figsetgrptitle{Total and VLBA spectra of J0051+8411}
\figsetplot{J0051+8411_spectra.pdf}
\figsetgrpnote{Single-dish and VLBA broadband spectra of all the sources of our sample. The source name is specified at the top right of each plot. The single-dish spectra are plotted black, the VLBA spectra are plotted by the red color. In cases of the VLBA non-detection, the upper limits on the VLBA flux density are shown by red arrows.}
\figsetgrpend

\figsetgrpstart
\figsetgrpnum{7.26}
\figsetgrptitle{Total and VLBA spectra of J0056+7823}
\figsetplot{J0056+7823_spectra.pdf}
\figsetgrpnote{Single-dish and VLBA broadband spectra of all the sources of our sample. The source name is specified at the top right of each plot. The single-dish spectra are plotted black, the VLBA spectra are plotted by the red color. In cases of the VLBA non-detection, the upper limits on the VLBA flux density are shown by red arrows.}
\figsetgrpend

\figsetgrpstart
\figsetgrpnum{7.27}
\figsetgrptitle{Total and VLBA spectra of J0059+7955}
\figsetplot{J0059+7955_spectra.pdf}
\figsetgrpnote{Single-dish and VLBA broadband spectra of all the sources of our sample. The source name is specified at the top right of each plot. The single-dish spectra are plotted black, the VLBA spectra are plotted by the red color. In cases of the VLBA non-detection, the upper limits on the VLBA flux density are shown by red arrows.}
\figsetgrpend

\figsetgrpstart
\figsetgrpnum{7.28}
\figsetgrptitle{Total and VLBA spectra of J0110+7846}
\figsetplot{J0110+7846_spectra.pdf}
\figsetgrpnote{Single-dish and VLBA broadband spectra of all the sources of our sample. The source name is specified at the top right of each plot. The single-dish spectra are plotted black, the VLBA spectra are plotted by the red color. In cases of the VLBA non-detection, the upper limits on the VLBA flux density are shown by red arrows.}
\figsetgrpend

\figsetgrpstart
\figsetgrpnum{7.29}
\figsetgrptitle{Total and VLBA spectra of J0110+8738}
\figsetplot{J0110+8738_spectra.pdf}
\figsetgrpnote{Single-dish and VLBA broadband spectra of all the sources of our sample. The source name is specified at the top right of each plot. The single-dish spectra are plotted black, the VLBA spectra are plotted by the red color. In cases of the VLBA non-detection, the upper limits on the VLBA flux density are shown by red arrows.}
\figsetgrpend

\figsetgrpstart
\figsetgrpnum{7.30}
\figsetgrptitle{Total and VLBA spectra of J0117+8928}
\figsetplot{J0117+8928_spectra.pdf}
\figsetgrpnote{Single-dish and VLBA broadband spectra of all the sources of our sample. The source name is specified at the top right of each plot. The single-dish spectra are plotted black, the VLBA spectra are plotted by the red color. In cases of the VLBA non-detection, the upper limits on the VLBA flux density are shown by red arrows.}
\figsetgrpend

\figsetgrpstart
\figsetgrpnum{7.31}
\figsetgrptitle{Total and VLBA spectra of J0121+8328}
\figsetplot{J0121+8328_spectra.pdf}
\figsetgrpnote{Single-dish and VLBA broadband spectra of all the sources of our sample. The source name is specified at the top right of each plot. The single-dish spectra are plotted black, the VLBA spectra are plotted by the red color. In cases of the VLBA non-detection, the upper limits on the VLBA flux density are shown by red arrows.}
\figsetgrpend

\figsetgrpstart
\figsetgrpnum{7.32}
\figsetgrptitle{Total and VLBA spectra of J0123+8056}
\figsetplot{J0123+8056_spectra.pdf}
\figsetgrpnote{Single-dish and VLBA broadband spectra of all the sources of our sample. The source name is specified at the top right of each plot. The single-dish spectra are plotted black, the VLBA spectra are plotted by the red color. In cases of the VLBA non-detection, the upper limits on the VLBA flux density are shown by red arrows.}
\figsetgrpend

\figsetgrpstart
\figsetgrpnum{7.33}
\figsetgrptitle{Total and VLBA spectra of J0125+8424}
\figsetplot{J0125+8424_spectra.pdf}
\figsetgrpnote{Single-dish and VLBA broadband spectra of all the sources of our sample. The source name is specified at the top right of each plot. The single-dish spectra are plotted black, the VLBA spectra are plotted by the red color. In cases of the VLBA non-detection, the upper limits on the VLBA flux density are shown by red arrows.}
\figsetgrpend

\figsetgrpstart
\figsetgrpnum{7.34}
\figsetgrptitle{Total and VLBA spectra of J0131+8446}
\figsetplot{J0131+8446_spectra.pdf}
\figsetgrpnote{Single-dish and VLBA broadband spectra of all the sources of our sample. The source name is specified at the top right of each plot. The single-dish spectra are plotted black, the VLBA spectra are plotted by the red color. In cases of the VLBA non-detection, the upper limits on the VLBA flux density are shown by red arrows.}
\figsetgrpend

\figsetgrpstart
\figsetgrpnum{7.35}
\figsetgrptitle{Total and VLBA spectra of J0138+7609}
\figsetplot{J0138+7609_spectra.pdf}
\figsetgrpnote{Single-dish and VLBA broadband spectra of all the sources of our sample. The source name is specified at the top right of each plot. The single-dish spectra are plotted black, the VLBA spectra are plotted by the red color. In cases of the VLBA non-detection, the upper limits on the VLBA flux density are shown by red arrows.}
\figsetgrpend

\figsetgrpstart
\figsetgrpnum{7.36}
\figsetgrptitle{Total and VLBA spectra of J0144+8200}
\figsetplot{J0144+8200_spectra.pdf}
\figsetgrpnote{Single-dish and VLBA broadband spectra of all the sources of our sample. The source name is specified at the top right of each plot. The single-dish spectra are plotted black, the VLBA spectra are plotted by the red color. In cases of the VLBA non-detection, the upper limits on the VLBA flux density are shown by red arrows.}
\figsetgrpend

\figsetgrpstart
\figsetgrpnum{7.37}
\figsetgrptitle{Total and VLBA spectra of J0152+7550}
\figsetplot{J0152+7550_spectra.pdf}
\figsetgrpnote{Single-dish and VLBA broadband spectra of all the sources of our sample. The source name is specified at the top right of each plot. The single-dish spectra are plotted black, the VLBA spectra are plotted by the red color. In cases of the VLBA non-detection, the upper limits on the VLBA flux density are shown by red arrows.}
\figsetgrpend

\figsetgrpstart
\figsetgrpnum{7.38}
\figsetgrptitle{Total and VLBA spectra of J0157+7552}
\figsetplot{J0157+7552_spectra.pdf}
\figsetgrpnote{Single-dish and VLBA broadband spectra of all the sources of our sample. The source name is specified at the top right of each plot. The single-dish spectra are plotted black, the VLBA spectra are plotted by the red color. In cases of the VLBA non-detection, the upper limits on the VLBA flux density are shown by red arrows.}
\figsetgrpend

\figsetgrpstart
\figsetgrpnum{7.39}
\figsetgrptitle{Total and VLBA spectra of J0202+7949}
\figsetplot{J0202+7949_spectra.pdf}
\figsetgrpnote{Single-dish and VLBA broadband spectra of all the sources of our sample. The source name is specified at the top right of each plot. The single-dish spectra are plotted black, the VLBA spectra are plotted by the red color. In cases of the VLBA non-detection, the upper limits on the VLBA flux density are shown by red arrows.}
\figsetgrpend

\figsetgrpstart
\figsetgrpnum{7.40}
\figsetgrptitle{Total and VLBA spectra of J0202+8115}
\figsetplot{J0202+8115_spectra.pdf}
\figsetgrpnote{Single-dish and VLBA broadband spectra of all the sources of our sample. The source name is specified at the top right of each plot. The single-dish spectra are plotted black, the VLBA spectra are plotted by the red color. In cases of the VLBA non-detection, the upper limits on the VLBA flux density are shown by red arrows.}
\figsetgrpend

\figsetgrpstart
\figsetgrpnum{7.41}
\figsetgrptitle{Total and VLBA spectra of J0203+8106}
\figsetplot{J0203+8106_spectra.pdf}
\figsetgrpnote{Single-dish and VLBA broadband spectra of all the sources of our sample. The source name is specified at the top right of each plot. The single-dish spectra are plotted black, the VLBA spectra are plotted by the red color. In cases of the VLBA non-detection, the upper limits on the VLBA flux density are shown by red arrows.}
\figsetgrpend

\figsetgrpstart
\figsetgrpnum{7.42}
\figsetgrptitle{Total and VLBA spectra of J0205+7522}
\figsetplot{J0205+7522_spectra.pdf}
\figsetgrpnote{Single-dish and VLBA broadband spectra of all the sources of our sample. The source name is specified at the top right of each plot. The single-dish spectra are plotted black, the VLBA spectra are plotted by the red color. In cases of the VLBA non-detection, the upper limits on the VLBA flux density are shown by red arrows.}
\figsetgrpend

\figsetgrpstart
\figsetgrpnum{7.43}
\figsetgrptitle{Total and VLBA spectra of J0206+8325}
\figsetplot{J0206+8325_spectra.pdf}
\figsetgrpnote{Single-dish and VLBA broadband spectra of all the sources of our sample. The source name is specified at the top right of each plot. The single-dish spectra are plotted black, the VLBA spectra are plotted by the red color. In cases of the VLBA non-detection, the upper limits on the VLBA flux density are shown by red arrows.}
\figsetgrpend

\figsetgrpstart
\figsetgrpnum{7.44}
\figsetgrptitle{Total and VLBA spectra of J0207+7956}
\figsetplot{J0207+7956_spectra.pdf}
\figsetgrpnote{Single-dish and VLBA broadband spectra of all the sources of our sample. The source name is specified at the top right of each plot. The single-dish spectra are plotted black, the VLBA spectra are plotted by the red color. In cases of the VLBA non-detection, the upper limits on the VLBA flux density are shown by red arrows.}
\figsetgrpend

\figsetgrpstart
\figsetgrpnum{7.45}
\figsetgrptitle{Total and VLBA spectra of J0217+7945}
\figsetplot{J0217+7945_spectra.pdf}
\figsetgrpnote{Single-dish and VLBA broadband spectra of all the sources of our sample. The source name is specified at the top right of each plot. The single-dish spectra are plotted black, the VLBA spectra are plotted by the red color. In cases of the VLBA non-detection, the upper limits on the VLBA flux density are shown by red arrows.}
\figsetgrpend

\figsetgrpstart
\figsetgrpnum{7.46}
\figsetgrptitle{Total and VLBA spectra of J0220+8004}
\figsetplot{J0220+8004_spectra.pdf}
\figsetgrpnote{Single-dish and VLBA broadband spectra of all the sources of our sample. The source name is specified at the top right of each plot. The single-dish spectra are plotted black, the VLBA spectra are plotted by the red color. In cases of the VLBA non-detection, the upper limits on the VLBA flux density are shown by red arrows.}
\figsetgrpend

\figsetgrpstart
\figsetgrpnum{7.47}
\figsetgrptitle{Total and VLBA spectra of J0222+8055}
\figsetplot{J0222+8055_spectra.pdf}
\figsetgrpnote{Single-dish and VLBA broadband spectra of all the sources of our sample. The source name is specified at the top right of each plot. The single-dish spectra are plotted black, the VLBA spectra are plotted by the red color. In cases of the VLBA non-detection, the upper limits on the VLBA flux density are shown by red arrows.}
\figsetgrpend

\figsetgrpstart
\figsetgrpnum{7.48}
\figsetgrptitle{Total and VLBA spectra of J0222+8618}
\figsetplot{J0222+8618_spectra.pdf}
\figsetgrpnote{Single-dish and VLBA broadband spectra of all the sources of our sample. The source name is specified at the top right of each plot. The single-dish spectra are plotted black, the VLBA spectra are plotted by the red color. In cases of the VLBA non-detection, the upper limits on the VLBA flux density are shown by red arrows.}
\figsetgrpend

\figsetgrpstart
\figsetgrpnum{7.49}
\figsetgrptitle{Total and VLBA spectra of J0224+7655}
\figsetplot{J0224+7655_spectra.pdf}
\figsetgrpnote{Single-dish and VLBA broadband spectra of all the sources of our sample. The source name is specified at the top right of each plot. The single-dish spectra are plotted black, the VLBA spectra are plotted by the red color. In cases of the VLBA non-detection, the upper limits on the VLBA flux density are shown by red arrows.}
\figsetgrpend

\figsetgrpstart
\figsetgrpnum{7.50}
\figsetgrptitle{Total and VLBA spectra of J0229+7743}
\figsetplot{J0229+7743_spectra.pdf}
\figsetgrpnote{Single-dish and VLBA broadband spectra of all the sources of our sample. The source name is specified at the top right of each plot. The single-dish spectra are plotted black, the VLBA spectra are plotted by the red color. In cases of the VLBA non-detection, the upper limits on the VLBA flux density are shown by red arrows.}
\figsetgrpend

\figsetgrpstart
\figsetgrpnum{7.51}
\figsetgrptitle{Total and VLBA spectra of J0230+8141}
\figsetplot{J0230+8141_spectra.pdf}
\figsetgrpnote{Single-dish and VLBA broadband spectra of all the sources of our sample. The source name is specified at the top right of each plot. The single-dish spectra are plotted black, the VLBA spectra are plotted by the red color. In cases of the VLBA non-detection, the upper limits on the VLBA flux density are shown by red arrows.}
\figsetgrpend

\figsetgrpstart
\figsetgrpnum{7.52}
\figsetgrptitle{Total and VLBA spectra of J0232+7825}
\figsetplot{J0232+7825_spectra.pdf}
\figsetgrpnote{Single-dish and VLBA broadband spectra of all the sources of our sample. The source name is specified at the top right of each plot. The single-dish spectra are plotted black, the VLBA spectra are plotted by the red color. In cases of the VLBA non-detection, the upper limits on the VLBA flux density are shown by red arrows.}
\figsetgrpend

\figsetgrpstart
\figsetgrpnum{7.53}
\figsetgrptitle{Total and VLBA spectra of J0251+7914}
\figsetplot{J0251+7914_spectra.pdf}
\figsetgrpnote{Single-dish and VLBA broadband spectra of all the sources of our sample. The source name is specified at the top right of each plot. The single-dish spectra are plotted black, the VLBA spectra are plotted by the red color. In cases of the VLBA non-detection, the upper limits on the VLBA flux density are shown by red arrows.}
\figsetgrpend

\figsetgrpstart
\figsetgrpnum{7.54}
\figsetgrptitle{Total and VLBA spectra of J0254+7911}
\figsetplot{J0254+7911_spectra.pdf}
\figsetgrpnote{Single-dish and VLBA broadband spectra of all the sources of our sample. The source name is specified at the top right of each plot. The single-dish spectra are plotted black, the VLBA spectra are plotted by the red color. In cases of the VLBA non-detection, the upper limits on the VLBA flux density are shown by red arrows.}
\figsetgrpend

\figsetgrpstart
\figsetgrpnum{7.55}
\figsetgrptitle{Total and VLBA spectra of J0257+7843}
\figsetplot{J0257+7843_spectra.pdf}
\figsetgrpnote{Single-dish and VLBA broadband spectra of all the sources of our sample. The source name is specified at the top right of each plot. The single-dish spectra are plotted black, the VLBA spectra are plotted by the red color. In cases of the VLBA non-detection, the upper limits on the VLBA flux density are shown by red arrows.}
\figsetgrpend

\figsetgrpstart
\figsetgrpnum{7.56}
\figsetgrptitle{Total and VLBA spectra of J0258+7943}
\figsetplot{J0258+7943_spectra.pdf}
\figsetgrpnote{Single-dish and VLBA broadband spectra of all the sources of our sample. The source name is specified at the top right of each plot. The single-dish spectra are plotted black, the VLBA spectra are plotted by the red color. In cases of the VLBA non-detection, the upper limits on the VLBA flux density are shown by red arrows.}
\figsetgrpend

\figsetgrpstart
\figsetgrpnum{7.57}
\figsetgrptitle{Total and VLBA spectra of J0300+7906}
\figsetplot{J0300+7906_spectra.pdf}
\figsetgrpnote{Single-dish and VLBA broadband spectra of all the sources of our sample. The source name is specified at the top right of each plot. The single-dish spectra are plotted black, the VLBA spectra are plotted by the red color. In cases of the VLBA non-detection, the upper limits on the VLBA flux density are shown by red arrows.}
\figsetgrpend

\figsetgrpstart
\figsetgrpnum{7.58}
\figsetgrptitle{Total and VLBA spectra of J0300+8202}
\figsetplot{J0300+8202_spectra.pdf}
\figsetgrpnote{Single-dish and VLBA broadband spectra of all the sources of our sample. The source name is specified at the top right of each plot. The single-dish spectra are plotted black, the VLBA spectra are plotted by the red color. In cases of the VLBA non-detection, the upper limits on the VLBA flux density are shown by red arrows.}
\figsetgrpend

\figsetgrpstart
\figsetgrpnum{7.59}
\figsetgrptitle{Total and VLBA spectra of J0302+7843}
\figsetplot{J0302+7843_spectra.pdf}
\figsetgrpnote{Single-dish and VLBA broadband spectra of all the sources of our sample. The source name is specified at the top right of each plot. The single-dish spectra are plotted black, the VLBA spectra are plotted by the red color. In cases of the VLBA non-detection, the upper limits on the VLBA flux density are shown by red arrows.}
\figsetgrpend

\figsetgrpstart
\figsetgrpnum{7.60}
\figsetgrptitle{Total and VLBA spectra of J0304+7727}
\figsetplot{J0304+7727_spectra.pdf}
\figsetgrpnote{Single-dish and VLBA broadband spectra of all the sources of our sample. The source name is specified at the top right of each plot. The single-dish spectra are plotted black, the VLBA spectra are plotted by the red color. In cases of the VLBA non-detection, the upper limits on the VLBA flux density are shown by red arrows.}
\figsetgrpend

\figsetgrpstart
\figsetgrpnum{7.61}
\figsetgrptitle{Total and VLBA spectra of J0306+7553}
\figsetplot{J0306+7553_spectra.pdf}
\figsetgrpnote{Single-dish and VLBA broadband spectra of all the sources of our sample. The source name is specified at the top right of each plot. The single-dish spectra are plotted black, the VLBA spectra are plotted by the red color. In cases of the VLBA non-detection, the upper limits on the VLBA flux density are shown by red arrows.}
\figsetgrpend

\figsetgrpstart
\figsetgrpnum{7.62}
\figsetgrptitle{Total and VLBA spectra of J0306+7933}
\figsetplot{J0306+7933_spectra.pdf}
\figsetgrpnote{Single-dish and VLBA broadband spectra of all the sources of our sample. The source name is specified at the top right of each plot. The single-dish spectra are plotted black, the VLBA spectra are plotted by the red color. In cases of the VLBA non-detection, the upper limits on the VLBA flux density are shown by red arrows.}
\figsetgrpend

\figsetgrpstart
\figsetgrpnum{7.63}
\figsetgrptitle{Total and VLBA spectra of J0306+8200}
\figsetplot{J0306+8200_spectra.pdf}
\figsetgrpnote{Single-dish and VLBA broadband spectra of all the sources of our sample. The source name is specified at the top right of each plot. The single-dish spectra are plotted black, the VLBA spectra are plotted by the red color. In cases of the VLBA non-detection, the upper limits on the VLBA flux density are shown by red arrows.}
\figsetgrpend

\figsetgrpstart
\figsetgrpnum{7.64}
\figsetgrptitle{Total and VLBA spectra of J0307+7735}
\figsetplot{J0307+7735_spectra.pdf}
\figsetgrpnote{Single-dish and VLBA broadband spectra of all the sources of our sample. The source name is specified at the top right of each plot. The single-dish spectra are plotted black, the VLBA spectra are plotted by the red color. In cases of the VLBA non-detection, the upper limits on the VLBA flux density are shown by red arrows.}
\figsetgrpend

\figsetgrpstart
\figsetgrpnum{7.65}
\figsetgrptitle{Total and VLBA spectra of J0330+7633}
\figsetplot{J0330+7633_spectra.pdf}
\figsetgrpnote{Single-dish and VLBA broadband spectra of all the sources of our sample. The source name is specified at the top right of each plot. The single-dish spectra are plotted black, the VLBA spectra are plotted by the red color. In cases of the VLBA non-detection, the upper limits on the VLBA flux density are shown by red arrows.}
\figsetgrpend

\figsetgrpstart
\figsetgrpnum{7.66}
\figsetgrptitle{Total and VLBA spectra of J0344+8620}
\figsetplot{J0344+8620_spectra.pdf}
\figsetgrpnote{Single-dish and VLBA broadband spectra of all the sources of our sample. The source name is specified at the top right of each plot. The single-dish spectra are plotted black, the VLBA spectra are plotted by the red color. In cases of the VLBA non-detection, the upper limits on the VLBA flux density are shown by red arrows.}
\figsetgrpend

\figsetgrpstart
\figsetgrpnum{7.67}
\figsetgrptitle{Total and VLBA spectra of J0351+8004}
\figsetplot{J0351+8004_spectra.pdf}
\figsetgrpnote{Single-dish and VLBA broadband spectra of all the sources of our sample. The source name is specified at the top right of each plot. The single-dish spectra are plotted black, the VLBA spectra are plotted by the red color. In cases of the VLBA non-detection, the upper limits on the VLBA flux density are shown by red arrows.}
\figsetgrpend

\figsetgrpstart
\figsetgrpnum{7.68}
\figsetgrptitle{Total and VLBA spectra of J0354+8009}
\figsetplot{J0354+8009_spectra.pdf}
\figsetgrpnote{Single-dish and VLBA broadband spectra of all the sources of our sample. The source name is specified at the top right of each plot. The single-dish spectra are plotted black, the VLBA spectra are plotted by the red color. In cases of the VLBA non-detection, the upper limits on the VLBA flux density are shown by red arrows.}
\figsetgrpend

\figsetgrpstart
\figsetgrpnum{7.69}
\figsetgrptitle{Total and VLBA spectra of J0356+7637}
\figsetplot{J0356+7637_spectra.pdf}
\figsetgrpnote{Single-dish and VLBA broadband spectra of all the sources of our sample. The source name is specified at the top right of each plot. The single-dish spectra are plotted black, the VLBA spectra are plotted by the red color. In cases of the VLBA non-detection, the upper limits on the VLBA flux density are shown by red arrows.}
\figsetgrpend

\figsetgrpstart
\figsetgrpnum{7.70}
\figsetgrptitle{Total and VLBA spectra of J0358+7837}
\figsetplot{J0358+7837_spectra.pdf}
\figsetgrpnote{Single-dish and VLBA broadband spectra of all the sources of our sample. The source name is specified at the top right of each plot. The single-dish spectra are plotted black, the VLBA spectra are plotted by the red color. In cases of the VLBA non-detection, the upper limits on the VLBA flux density are shown by red arrows.}
\figsetgrpend

\figsetgrpstart
\figsetgrpnum{7.71}
\figsetgrptitle{Total and VLBA spectra of J0400+7621}
\figsetplot{J0400+7621_spectra.pdf}
\figsetgrpnote{Single-dish and VLBA broadband spectra of all the sources of our sample. The source name is specified at the top right of each plot. The single-dish spectra are plotted black, the VLBA spectra are plotted by the red color. In cases of the VLBA non-detection, the upper limits on the VLBA flux density are shown by red arrows.}
\figsetgrpend

\figsetgrpstart
\figsetgrpnum{7.72}
\figsetgrptitle{Total and VLBA spectra of J0402+8241}
\figsetplot{J0402+8241_spectra.pdf}
\figsetgrpnote{Single-dish and VLBA broadband spectra of all the sources of our sample. The source name is specified at the top right of each plot. The single-dish spectra are plotted black, the VLBA spectra are plotted by the red color. In cases of the VLBA non-detection, the upper limits on the VLBA flux density are shown by red arrows.}
\figsetgrpend

\figsetgrpstart
\figsetgrpnum{7.73}
\figsetgrptitle{Total and VLBA spectra of J0403+7616}
\figsetplot{J0403+7616_spectra.pdf}
\figsetgrpnote{Single-dish and VLBA broadband spectra of all the sources of our sample. The source name is specified at the top right of each plot. The single-dish spectra are plotted black, the VLBA spectra are plotted by the red color. In cases of the VLBA non-detection, the upper limits on the VLBA flux density are shown by red arrows.}
\figsetgrpend

\figsetgrpstart
\figsetgrpnum{7.74}
\figsetgrptitle{Total and VLBA spectra of J0406+7633}
\figsetplot{J0406+7633_spectra.pdf}
\figsetgrpnote{Single-dish and VLBA broadband spectra of all the sources of our sample. The source name is specified at the top right of each plot. The single-dish spectra are plotted black, the VLBA spectra are plotted by the red color. In cases of the VLBA non-detection, the upper limits on the VLBA flux density are shown by red arrows.}
\figsetgrpend

\figsetgrpstart
\figsetgrpnum{7.75}
\figsetgrptitle{Total and VLBA spectra of J0406+8146}
\figsetplot{J0406+8146_spectra.pdf}
\figsetgrpnote{Single-dish and VLBA broadband spectra of all the sources of our sample. The source name is specified at the top right of each plot. The single-dish spectra are plotted black, the VLBA spectra are plotted by the red color. In cases of the VLBA non-detection, the upper limits on the VLBA flux density are shown by red arrows.}
\figsetgrpend

\figsetgrpstart
\figsetgrpnum{7.76}
\figsetgrptitle{Total and VLBA spectra of J0410+7656}
\figsetplot{J0410+7656_spectra.pdf}
\figsetgrpnote{Single-dish and VLBA broadband spectra of all the sources of our sample. The source name is specified at the top right of each plot. The single-dish spectra are plotted black, the VLBA spectra are plotted by the red color. In cases of the VLBA non-detection, the upper limits on the VLBA flux density are shown by red arrows.}
\figsetgrpend

\figsetgrpstart
\figsetgrpnum{7.77}
\figsetgrptitle{Total and VLBA spectra of J0410+8208}
\figsetplot{J0410+8208_spectra.pdf}
\figsetgrpnote{Single-dish and VLBA broadband spectra of all the sources of our sample. The source name is specified at the top right of each plot. The single-dish spectra are plotted black, the VLBA spectra are plotted by the red color. In cases of the VLBA non-detection, the upper limits on the VLBA flux density are shown by red arrows.}
\figsetgrpend

\figsetgrpstart
\figsetgrpnum{7.78}
\figsetgrptitle{Total and VLBA spectra of J0414+7612}
\figsetplot{J0414+7612_spectra.pdf}
\figsetgrpnote{Single-dish and VLBA broadband spectra of all the sources of our sample. The source name is specified at the top right of each plot. The single-dish spectra are plotted black, the VLBA spectra are plotted by the red color. In cases of the VLBA non-detection, the upper limits on the VLBA flux density are shown by red arrows.}
\figsetgrpend

\figsetgrpstart
\figsetgrpnum{7.79}
\figsetgrptitle{Total and VLBA spectra of J0415+7753}
\figsetplot{J0415+7753_spectra.pdf}
\figsetgrpnote{Single-dish and VLBA broadband spectra of all the sources of our sample. The source name is specified at the top right of each plot. The single-dish spectra are plotted black, the VLBA spectra are plotted by the red color. In cases of the VLBA non-detection, the upper limits on the VLBA flux density are shown by red arrows.}
\figsetgrpend

\figsetgrpstart
\figsetgrpnum{7.80}
\figsetgrptitle{Total and VLBA spectra of J0415+8424}
\figsetplot{J0415+8424_spectra.pdf}
\figsetgrpnote{Single-dish and VLBA broadband spectra of all the sources of our sample. The source name is specified at the top right of each plot. The single-dish spectra are plotted black, the VLBA spectra are plotted by the red color. In cases of the VLBA non-detection, the upper limits on the VLBA flux density are shown by red arrows.}
\figsetgrpend

\figsetgrpstart
\figsetgrpnum{7.81}
\figsetgrptitle{Total and VLBA spectra of J0418+8048}
\figsetplot{J0418+8048_spectra.pdf}
\figsetgrpnote{Single-dish and VLBA broadband spectra of all the sources of our sample. The source name is specified at the top right of each plot. The single-dish spectra are plotted black, the VLBA spectra are plotted by the red color. In cases of the VLBA non-detection, the upper limits on the VLBA flux density are shown by red arrows.}
\figsetgrpend

\figsetgrpstart
\figsetgrpnum{7.82}
\figsetgrptitle{Total and VLBA spectra of J0419+7559}
\figsetplot{J0419+7559_spectra.pdf}
\figsetgrpnote{Single-dish and VLBA broadband spectra of all the sources of our sample. The source name is specified at the top right of each plot. The single-dish spectra are plotted black, the VLBA spectra are plotted by the red color. In cases of the VLBA non-detection, the upper limits on the VLBA flux density are shown by red arrows.}
\figsetgrpend

\figsetgrpstart
\figsetgrpnum{7.83}
\figsetgrptitle{Total and VLBA spectra of J0422+7627}
\figsetplot{J0422+7627_spectra.pdf}
\figsetgrpnote{Single-dish and VLBA broadband spectra of all the sources of our sample. The source name is specified at the top right of each plot. The single-dish spectra are plotted black, the VLBA spectra are plotted by the red color. In cases of the VLBA non-detection, the upper limits on the VLBA flux density are shown by red arrows.}
\figsetgrpend

\figsetgrpstart
\figsetgrpnum{7.84}
\figsetgrptitle{Total and VLBA spectra of J0422+7919}
\figsetplot{J0422+7919_spectra.pdf}
\figsetgrpnote{Single-dish and VLBA broadband spectra of all the sources of our sample. The source name is specified at the top right of each plot. The single-dish spectra are plotted black, the VLBA spectra are plotted by the red color. In cases of the VLBA non-detection, the upper limits on the VLBA flux density are shown by red arrows.}
\figsetgrpend

\figsetgrpstart
\figsetgrpnum{7.85}
\figsetgrptitle{Total and VLBA spectra of J0422+8548}
\figsetplot{J0422+8548_spectra.pdf}
\figsetgrpnote{Single-dish and VLBA broadband spectra of all the sources of our sample. The source name is specified at the top right of each plot. The single-dish spectra are plotted black, the VLBA spectra are plotted by the red color. In cases of the VLBA non-detection, the upper limits on the VLBA flux density are shown by red arrows.}
\figsetgrpend

\figsetgrpstart
\figsetgrpnum{7.86}
\figsetgrptitle{Total and VLBA spectra of J0424+7653}
\figsetplot{J0424+7653_spectra.pdf}
\figsetgrpnote{Single-dish and VLBA broadband spectra of all the sources of our sample. The source name is specified at the top right of each plot. The single-dish spectra are plotted black, the VLBA spectra are plotted by the red color. In cases of the VLBA non-detection, the upper limits on the VLBA flux density are shown by red arrows.}
\figsetgrpend

\figsetgrpstart
\figsetgrpnum{7.87}
\figsetgrptitle{Total and VLBA spectra of J0429+7709}
\figsetplot{J0429+7709_spectra.pdf}
\figsetgrpnote{Single-dish and VLBA broadband spectra of all the sources of our sample. The source name is specified at the top right of each plot. The single-dish spectra are plotted black, the VLBA spectra are plotted by the red color. In cases of the VLBA non-detection, the upper limits on the VLBA flux density are shown by red arrows.}
\figsetgrpend

\figsetgrpstart
\figsetgrpnum{7.88}
\figsetgrptitle{Total and VLBA spectra of J0429+8705}
\figsetplot{J0429+8705_spectra.pdf}
\figsetgrpnote{Single-dish and VLBA broadband spectra of all the sources of our sample. The source name is specified at the top right of each plot. The single-dish spectra are plotted black, the VLBA spectra are plotted by the red color. In cases of the VLBA non-detection, the upper limits on the VLBA flux density are shown by red arrows.}
\figsetgrpend

\figsetgrpstart
\figsetgrpnum{7.89}
\figsetgrptitle{Total and VLBA spectra of J0437+8050}
\figsetplot{J0437+8050_spectra.pdf}
\figsetgrpnote{Single-dish and VLBA broadband spectra of all the sources of our sample. The source name is specified at the top right of each plot. The single-dish spectra are plotted black, the VLBA spectra are plotted by the red color. In cases of the VLBA non-detection, the upper limits on the VLBA flux density are shown by red arrows.}
\figsetgrpend

\figsetgrpstart
\figsetgrpnum{7.90}
\figsetgrptitle{Total and VLBA spectra of J0441+7601}
\figsetplot{J0441+7601_spectra.pdf}
\figsetgrpnote{Single-dish and VLBA broadband spectra of all the sources of our sample. The source name is specified at the top right of each plot. The single-dish spectra are plotted black, the VLBA spectra are plotted by the red color. In cases of the VLBA non-detection, the upper limits on the VLBA flux density are shown by red arrows.}
\figsetgrpend

\figsetgrpstart
\figsetgrpnum{7.91}
\figsetgrptitle{Total and VLBA spectra of J0445+7838}
\figsetplot{J0445+7838_spectra.pdf}
\figsetgrpnote{Single-dish and VLBA broadband spectra of all the sources of our sample. The source name is specified at the top right of each plot. The single-dish spectra are plotted black, the VLBA spectra are plotted by the red color. In cases of the VLBA non-detection, the upper limits on the VLBA flux density are shown by red arrows.}
\figsetgrpend

\figsetgrpstart
\figsetgrpnum{7.92}
\figsetgrptitle{Total and VLBA spectra of J0449+8233}
\figsetplot{J0449+8233_spectra.pdf}
\figsetgrpnote{Single-dish and VLBA broadband spectra of all the sources of our sample. The source name is specified at the top right of each plot. The single-dish spectra are plotted black, the VLBA spectra are plotted by the red color. In cases of the VLBA non-detection, the upper limits on the VLBA flux density are shown by red arrows.}
\figsetgrpend

\figsetgrpstart
\figsetgrpnum{7.93}
\figsetgrptitle{Total and VLBA spectra of J0451+8013}
\figsetplot{J0451+8013_spectra.pdf}
\figsetgrpnote{Single-dish and VLBA broadband spectra of all the sources of our sample. The source name is specified at the top right of each plot. The single-dish spectra are plotted black, the VLBA spectra are plotted by the red color. In cases of the VLBA non-detection, the upper limits on the VLBA flux density are shown by red arrows.}
\figsetgrpend

\figsetgrpstart
\figsetgrpnum{7.94}
\figsetgrptitle{Total and VLBA spectra of J0452+7626}
\figsetplot{J0452+7626_spectra.pdf}
\figsetgrpnote{Single-dish and VLBA broadband spectra of all the sources of our sample. The source name is specified at the top right of each plot. The single-dish spectra are plotted black, the VLBA spectra are plotted by the red color. In cases of the VLBA non-detection, the upper limits on the VLBA flux density are shown by red arrows.}
\figsetgrpend

\figsetgrpstart
\figsetgrpnum{7.95}
\figsetgrptitle{Total and VLBA spectra of J0455+7626}
\figsetplot{J0455+7626_spectra.pdf}
\figsetgrpnote{Single-dish and VLBA broadband spectra of all the sources of our sample. The source name is specified at the top right of each plot. The single-dish spectra are plotted black, the VLBA spectra are plotted by the red color. In cases of the VLBA non-detection, the upper limits on the VLBA flux density are shown by red arrows.}
\figsetgrpend

\figsetgrpstart
\figsetgrpnum{7.96}
\figsetgrptitle{Total and VLBA spectra of J0456+8309}
\figsetplot{J0456+8309_spectra.pdf}
\figsetgrpnote{Single-dish and VLBA broadband spectra of all the sources of our sample. The source name is specified at the top right of each plot. The single-dish spectra are plotted black, the VLBA spectra are plotted by the red color. In cases of the VLBA non-detection, the upper limits on the VLBA flux density are shown by red arrows.}
\figsetgrpend

\figsetgrpstart
\figsetgrpnum{7.97}
\figsetgrptitle{Total and VLBA spectra of J0459+7612}
\figsetplot{J0459+7612_spectra.pdf}
\figsetgrpnote{Single-dish and VLBA broadband spectra of all the sources of our sample. The source name is specified at the top right of each plot. The single-dish spectra are plotted black, the VLBA spectra are plotted by the red color. In cases of the VLBA non-detection, the upper limits on the VLBA flux density are shown by red arrows.}
\figsetgrpend

\figsetgrpstart
\figsetgrpnum{7.98}
\figsetgrptitle{Total and VLBA spectra of J0507+7912}
\figsetplot{J0507+7912_spectra.pdf}
\figsetgrpnote{Single-dish and VLBA broadband spectra of all the sources of our sample. The source name is specified at the top right of each plot. The single-dish spectra are plotted black, the VLBA spectra are plotted by the red color. In cases of the VLBA non-detection, the upper limits on the VLBA flux density are shown by red arrows.}
\figsetgrpend

\figsetgrpstart
\figsetgrpnum{7.99}
\figsetgrptitle{Total and VLBA spectra of J0508+8432}
\figsetplot{J0508+8432_spectra.pdf}
\figsetgrpnote{Single-dish and VLBA broadband spectra of all the sources of our sample. The source name is specified at the top right of each plot. The single-dish spectra are plotted black, the VLBA spectra are plotted by the red color. In cases of the VLBA non-detection, the upper limits on the VLBA flux density are shown by red arrows.}
\figsetgrpend

\figsetgrpstart
\figsetgrpnum{7.100}
\figsetgrptitle{Total and VLBA spectra of J0518+7650}
\figsetplot{J0518+7650_spectra.pdf}
\figsetgrpnote{Single-dish and VLBA broadband spectra of all the sources of our sample. The source name is specified at the top right of each plot. The single-dish spectra are plotted black, the VLBA spectra are plotted by the red color. In cases of the VLBA non-detection, the upper limits on the VLBA flux density are shown by red arrows.}
\figsetgrpend

\figsetgrpstart
\figsetgrpnum{7.101}
\figsetgrptitle{Total and VLBA spectra of J0520+8400}
\figsetplot{J0520+8400_spectra.pdf}
\figsetgrpnote{Single-dish and VLBA broadband spectra of all the sources of our sample. The source name is specified at the top right of each plot. The single-dish spectra are plotted black, the VLBA spectra are plotted by the red color. In cases of the VLBA non-detection, the upper limits on the VLBA flux density are shown by red arrows.}
\figsetgrpend

\figsetgrpstart
\figsetgrpnum{7.102}
\figsetgrptitle{Total and VLBA spectra of J0520+8600}
\figsetplot{J0520+8600_spectra.pdf}
\figsetgrpnote{Single-dish and VLBA broadband spectra of all the sources of our sample. The source name is specified at the top right of each plot. The single-dish spectra are plotted black, the VLBA spectra are plotted by the red color. In cases of the VLBA non-detection, the upper limits on the VLBA flux density are shown by red arrows.}
\figsetgrpend

\figsetgrpstart
\figsetgrpnum{7.103}
\figsetgrptitle{Total and VLBA spectra of J0525+7613}
\figsetplot{J0525+7613_spectra.pdf}
\figsetgrpnote{Single-dish and VLBA broadband spectra of all the sources of our sample. The source name is specified at the top right of each plot. The single-dish spectra are plotted black, the VLBA spectra are plotted by the red color. In cases of the VLBA non-detection, the upper limits on the VLBA flux density are shown by red arrows.}
\figsetgrpend

\figsetgrpstart
\figsetgrpnum{7.104}
\figsetgrptitle{Total and VLBA spectra of J0525+8737}
\figsetplot{J0525+8737_spectra.pdf}
\figsetgrpnote{Single-dish and VLBA broadband spectra of all the sources of our sample. The source name is specified at the top right of each plot. The single-dish spectra are plotted black, the VLBA spectra are plotted by the red color. In cases of the VLBA non-detection, the upper limits on the VLBA flux density are shown by red arrows.}
\figsetgrpend

\figsetgrpstart
\figsetgrpnum{7.105}
\figsetgrptitle{Total and VLBA spectra of J0532+8731}
\figsetplot{J0532+8731_spectra.pdf}
\figsetgrpnote{Single-dish and VLBA broadband spectra of all the sources of our sample. The source name is specified at the top right of each plot. The single-dish spectra are plotted black, the VLBA spectra are plotted by the red color. In cases of the VLBA non-detection, the upper limits on the VLBA flux density are shown by red arrows.}
\figsetgrpend

\figsetgrpstart
\figsetgrpnum{7.106}
\figsetgrptitle{Total and VLBA spectra of J0540+7519}
\figsetplot{J0540+7519_spectra.pdf}
\figsetgrpnote{Single-dish and VLBA broadband spectra of all the sources of our sample. The source name is specified at the top right of each plot. The single-dish spectra are plotted black, the VLBA spectra are plotted by the red color. In cases of the VLBA non-detection, the upper limits on the VLBA flux density are shown by red arrows.}
\figsetgrpend

\figsetgrpstart
\figsetgrpnum{7.107}
\figsetgrptitle{Total and VLBA spectra of J0543+8118}
\figsetplot{J0543+8118_spectra.pdf}
\figsetgrpnote{Single-dish and VLBA broadband spectra of all the sources of our sample. The source name is specified at the top right of each plot. The single-dish spectra are plotted black, the VLBA spectra are plotted by the red color. In cases of the VLBA non-detection, the upper limits on the VLBA flux density are shown by red arrows.}
\figsetgrpend

\figsetgrpstart
\figsetgrpnum{7.108}
\figsetgrptitle{Total and VLBA spectra of J0543+8238}
\figsetplot{J0543+8238_spectra.pdf}
\figsetgrpnote{Single-dish and VLBA broadband spectra of all the sources of our sample. The source name is specified at the top right of each plot. The single-dish spectra are plotted black, the VLBA spectra are plotted by the red color. In cases of the VLBA non-detection, the upper limits on the VLBA flux density are shown by red arrows.}
\figsetgrpend

\figsetgrpstart
\figsetgrpnum{7.109}
\figsetgrptitle{Total and VLBA spectra of J0546+7514}
\figsetplot{J0546+7514_spectra.pdf}
\figsetgrpnote{Single-dish and VLBA broadband spectra of all the sources of our sample. The source name is specified at the top right of each plot. The single-dish spectra are plotted black, the VLBA spectra are plotted by the red color. In cases of the VLBA non-detection, the upper limits on the VLBA flux density are shown by red arrows.}
\figsetgrpend

\figsetgrpstart
\figsetgrpnum{7.110}
\figsetgrptitle{Total and VLBA spectra of J0548+7559}
\figsetplot{J0548+7559_spectra.pdf}
\figsetgrpnote{Single-dish and VLBA broadband spectra of all the sources of our sample. The source name is specified at the top right of each plot. The single-dish spectra are plotted black, the VLBA spectra are plotted by the red color. In cases of the VLBA non-detection, the upper limits on the VLBA flux density are shown by red arrows.}
\figsetgrpend

\figsetgrpstart
\figsetgrpnum{7.111}
\figsetgrptitle{Total and VLBA spectra of J0557+8001}
\figsetplot{J0557+8001_spectra.pdf}
\figsetgrpnote{Single-dish and VLBA broadband spectra of all the sources of our sample. The source name is specified at the top right of each plot. The single-dish spectra are plotted black, the VLBA spectra are plotted by the red color. In cases of the VLBA non-detection, the upper limits on the VLBA flux density are shown by red arrows.}
\figsetgrpend

\figsetgrpstart
\figsetgrpnum{7.112}
\figsetgrptitle{Total and VLBA spectra of J0604+8007}
\figsetplot{J0604+8007_spectra.pdf}
\figsetgrpnote{Single-dish and VLBA broadband spectra of all the sources of our sample. The source name is specified at the top right of each plot. The single-dish spectra are plotted black, the VLBA spectra are plotted by the red color. In cases of the VLBA non-detection, the upper limits on the VLBA flux density are shown by red arrows.}
\figsetgrpend

\figsetgrpstart
\figsetgrpnum{7.113}
\figsetgrptitle{Total and VLBA spectra of J0618+7821}
\figsetplot{J0618+7821_spectra.pdf}
\figsetgrpnote{Single-dish and VLBA broadband spectra of all the sources of our sample. The source name is specified at the top right of each plot. The single-dish spectra are plotted black, the VLBA spectra are plotted by the red color. In cases of the VLBA non-detection, the upper limits on the VLBA flux density are shown by red arrows.}
\figsetgrpend

\figsetgrpstart
\figsetgrpnum{7.114}
\figsetgrptitle{Total and VLBA spectra of J0619+7531}
\figsetplot{J0619+7531_spectra.pdf}
\figsetgrpnote{Single-dish and VLBA broadband spectra of all the sources of our sample. The source name is specified at the top right of each plot. The single-dish spectra are plotted black, the VLBA spectra are plotted by the red color. In cases of the VLBA non-detection, the upper limits on the VLBA flux density are shown by red arrows.}
\figsetgrpend

\figsetgrpstart
\figsetgrpnum{7.115}
\figsetgrptitle{Total and VLBA spectra of J0621+7605}
\figsetplot{J0621+7605_spectra.pdf}
\figsetgrpnote{Single-dish and VLBA broadband spectra of all the sources of our sample. The source name is specified at the top right of each plot. The single-dish spectra are plotted black, the VLBA spectra are plotted by the red color. In cases of the VLBA non-detection, the upper limits on the VLBA flux density are shown by red arrows.}
\figsetgrpend

\figsetgrpstart
\figsetgrpnum{7.116}
\figsetgrptitle{Total and VLBA spectra of J0622+8719}
\figsetplot{J0622+8719_spectra.pdf}
\figsetgrpnote{Single-dish and VLBA broadband spectra of all the sources of our sample. The source name is specified at the top right of each plot. The single-dish spectra are plotted black, the VLBA spectra are plotted by the red color. In cases of the VLBA non-detection, the upper limits on the VLBA flux density are shown by red arrows.}
\figsetgrpend

\figsetgrpstart
\figsetgrpnum{7.117}
\figsetgrptitle{Total and VLBA spectra of J0626+8202}
\figsetplot{J0626+8202_spectra.pdf}
\figsetgrpnote{Single-dish and VLBA broadband spectra of all the sources of our sample. The source name is specified at the top right of each plot. The single-dish spectra are plotted black, the VLBA spectra are plotted by the red color. In cases of the VLBA non-detection, the upper limits on the VLBA flux density are shown by red arrows.}
\figsetgrpend

\figsetgrpstart
\figsetgrpnum{7.118}
\figsetgrptitle{Total and VLBA spectra of J0627+7940}
\figsetplot{J0627+7940_spectra.pdf}
\figsetgrpnote{Single-dish and VLBA broadband spectra of all the sources of our sample. The source name is specified at the top right of each plot. The single-dish spectra are plotted black, the VLBA spectra are plotted by the red color. In cases of the VLBA non-detection, the upper limits on the VLBA flux density are shown by red arrows.}
\figsetgrpend

\figsetgrpstart
\figsetgrpnum{7.119}
\figsetgrptitle{Total and VLBA spectra of J0629+8451}
\figsetplot{J0629+8451_spectra.pdf}
\figsetgrpnote{Single-dish and VLBA broadband spectra of all the sources of our sample. The source name is specified at the top right of each plot. The single-dish spectra are plotted black, the VLBA spectra are plotted by the red color. In cases of the VLBA non-detection, the upper limits on the VLBA flux density are shown by red arrows.}
\figsetgrpend

\figsetgrpstart
\figsetgrpnum{7.120}
\figsetgrptitle{Total and VLBA spectra of J0630+7632}
\figsetplot{J0630+7632_spectra.pdf}
\figsetgrpnote{Single-dish and VLBA broadband spectra of all the sources of our sample. The source name is specified at the top right of each plot. The single-dish spectra are plotted black, the VLBA spectra are plotted by the red color. In cases of the VLBA non-detection, the upper limits on the VLBA flux density are shown by red arrows.}
\figsetgrpend

\figsetgrpstart
\figsetgrpnum{7.121}
\figsetgrptitle{Total and VLBA spectra of J0630+8123}
\figsetplot{J0630+8123_spectra.pdf}
\figsetgrpnote{Single-dish and VLBA broadband spectra of all the sources of our sample. The source name is specified at the top right of each plot. The single-dish spectra are plotted black, the VLBA spectra are plotted by the red color. In cases of the VLBA non-detection, the upper limits on the VLBA flux density are shown by red arrows.}
\figsetgrpend

\figsetgrpstart
\figsetgrpnum{7.122}
\figsetgrptitle{Total and VLBA spectra of J0632+8020}
\figsetplot{J0632+8020_spectra.pdf}
\figsetgrpnote{Single-dish and VLBA broadband spectra of all the sources of our sample. The source name is specified at the top right of each plot. The single-dish spectra are plotted black, the VLBA spectra are plotted by the red color. In cases of the VLBA non-detection, the upper limits on the VLBA flux density are shown by red arrows.}
\figsetgrpend

\figsetgrpstart
\figsetgrpnum{7.123}
\figsetgrptitle{Total and VLBA spectra of J0637+7937}
\figsetplot{J0637+7937_spectra.pdf}
\figsetgrpnote{Single-dish and VLBA broadband spectra of all the sources of our sample. The source name is specified at the top right of each plot. The single-dish spectra are plotted black, the VLBA spectra are plotted by the red color. In cases of the VLBA non-detection, the upper limits on the VLBA flux density are shown by red arrows.}
\figsetgrpend

\figsetgrpstart
\figsetgrpnum{7.124}
\figsetgrptitle{Total and VLBA spectra of J0637+8125}
\figsetplot{J0637+8125_spectra.pdf}
\figsetgrpnote{Single-dish and VLBA broadband spectra of all the sources of our sample. The source name is specified at the top right of each plot. The single-dish spectra are plotted black, the VLBA spectra are plotted by the red color. In cases of the VLBA non-detection, the upper limits on the VLBA flux density are shown by red arrows.}
\figsetgrpend

\figsetgrpstart
\figsetgrpnum{7.125}
\figsetgrptitle{Total and VLBA spectra of J0638+8411}
\figsetplot{J0638+8411_spectra.pdf}
\figsetgrpnote{Single-dish and VLBA broadband spectra of all the sources of our sample. The source name is specified at the top right of each plot. The single-dish spectra are plotted black, the VLBA spectra are plotted by the red color. In cases of the VLBA non-detection, the upper limits on the VLBA flux density are shown by red arrows.}
\figsetgrpend

\figsetgrpstart
\figsetgrpnum{7.126}
\figsetgrptitle{Total and VLBA spectra of J0640+7813}
\figsetplot{J0640+7813_spectra.pdf}
\figsetgrpnote{Single-dish and VLBA broadband spectra of all the sources of our sample. The source name is specified at the top right of each plot. The single-dish spectra are plotted black, the VLBA spectra are plotted by the red color. In cases of the VLBA non-detection, the upper limits on the VLBA flux density are shown by red arrows.}
\figsetgrpend

\figsetgrpstart
\figsetgrpnum{7.127}
\figsetgrptitle{Total and VLBA spectra of J0642+8031}
\figsetplot{J0642+8031_spectra.pdf}
\figsetgrpnote{Single-dish and VLBA broadband spectra of all the sources of our sample. The source name is specified at the top right of each plot. The single-dish spectra are plotted black, the VLBA spectra are plotted by the red color. In cases of the VLBA non-detection, the upper limits on the VLBA flux density are shown by red arrows.}
\figsetgrpend

\figsetgrpstart
\figsetgrpnum{7.128}
\figsetgrptitle{Total and VLBA spectra of J0644+8018}
\figsetplot{J0644+8018_spectra.pdf}
\figsetgrpnote{Single-dish and VLBA broadband spectra of all the sources of our sample. The source name is specified at the top right of each plot. The single-dish spectra are plotted black, the VLBA spectra are plotted by the red color. In cases of the VLBA non-detection, the upper limits on the VLBA flux density are shown by red arrows.}
\figsetgrpend

\figsetgrpstart
\figsetgrpnum{7.129}
\figsetgrptitle{Total and VLBA spectra of J0645+7755}
\figsetplot{J0645+7755_spectra.pdf}
\figsetgrpnote{Single-dish and VLBA broadband spectra of all the sources of our sample. The source name is specified at the top right of each plot. The single-dish spectra are plotted black, the VLBA spectra are plotted by the red color. In cases of the VLBA non-detection, the upper limits on the VLBA flux density are shown by red arrows.}
\figsetgrpend

\figsetgrpstart
\figsetgrpnum{7.130}
\figsetgrptitle{Total and VLBA spectra of J0646+8215}
\figsetplot{J0646+8215_spectra.pdf}
\figsetgrpnote{Single-dish and VLBA broadband spectra of all the sources of our sample. The source name is specified at the top right of each plot. The single-dish spectra are plotted black, the VLBA spectra are plotted by the red color. In cases of the VLBA non-detection, the upper limits on the VLBA flux density are shown by red arrows.}
\figsetgrpend

\figsetgrpstart
\figsetgrpnum{7.131}
\figsetgrptitle{Total and VLBA spectra of J0648+7756}
\figsetplot{J0648+7756_spectra.pdf}
\figsetgrpnote{Single-dish and VLBA broadband spectra of all the sources of our sample. The source name is specified at the top right of each plot. The single-dish spectra are plotted black, the VLBA spectra are plotted by the red color. In cases of the VLBA non-detection, the upper limits on the VLBA flux density are shown by red arrows.}
\figsetgrpend

\figsetgrpstart
\figsetgrpnum{7.132}
\figsetgrptitle{Total and VLBA spectra of J0655+7655}
\figsetplot{J0655+7655_spectra.pdf}
\figsetgrpnote{Single-dish and VLBA broadband spectra of all the sources of our sample. The source name is specified at the top right of each plot. The single-dish spectra are plotted black, the VLBA spectra are plotted by the red color. In cases of the VLBA non-detection, the upper limits on the VLBA flux density are shown by red arrows.}
\figsetgrpend

\figsetgrpstart
\figsetgrpnum{7.133}
\figsetgrptitle{Total and VLBA spectra of J0702+8549}
\figsetplot{J0702+8549_spectra.pdf}
\figsetgrpnote{Single-dish and VLBA broadband spectra of all the sources of our sample. The source name is specified at the top right of each plot. The single-dish spectra are plotted black, the VLBA spectra are plotted by the red color. In cases of the VLBA non-detection, the upper limits on the VLBA flux density are shown by red arrows.}
\figsetgrpend

\figsetgrpstart
\figsetgrpnum{7.134}
\figsetgrptitle{Total and VLBA spectra of J0702+8605}
\figsetplot{J0702+8605_spectra.pdf}
\figsetgrpnote{Single-dish and VLBA broadband spectra of all the sources of our sample. The source name is specified at the top right of each plot. The single-dish spectra are plotted black, the VLBA spectra are plotted by the red color. In cases of the VLBA non-detection, the upper limits on the VLBA flux density are shown by red arrows.}
\figsetgrpend

\figsetgrpstart
\figsetgrpnum{7.135}
\figsetgrptitle{Total and VLBA spectra of J0702+8634}
\figsetplot{J0702+8634_spectra.pdf}
\figsetgrpnote{Single-dish and VLBA broadband spectra of all the sources of our sample. The source name is specified at the top right of each plot. The single-dish spectra are plotted black, the VLBA spectra are plotted by the red color. In cases of the VLBA non-detection, the upper limits on the VLBA flux density are shown by red arrows.}
\figsetgrpend

\figsetgrpstart
\figsetgrpnum{7.136}
\figsetgrptitle{Total and VLBA spectra of J0714+8151}
\figsetplot{J0714+8151_spectra.pdf}
\figsetgrpnote{Single-dish and VLBA broadband spectra of all the sources of our sample. The source name is specified at the top right of each plot. The single-dish spectra are plotted black, the VLBA spectra are plotted by the red color. In cases of the VLBA non-detection, the upper limits on the VLBA flux density are shown by red arrows.}
\figsetgrpend

\figsetgrpstart
\figsetgrpnum{7.137}
\figsetgrptitle{Total and VLBA spectra of J0717+7642}
\figsetplot{J0717+7642_spectra.pdf}
\figsetgrpnote{Single-dish and VLBA broadband spectra of all the sources of our sample. The source name is specified at the top right of each plot. The single-dish spectra are plotted black, the VLBA spectra are plotted by the red color. In cases of the VLBA non-detection, the upper limits on the VLBA flux density are shown by red arrows.}
\figsetgrpend

\figsetgrpstart
\figsetgrpnum{7.138}
\figsetgrptitle{Total and VLBA spectra of J0719+7515}
\figsetplot{J0719+7515_spectra.pdf}
\figsetgrpnote{Single-dish and VLBA broadband spectra of all the sources of our sample. The source name is specified at the top right of each plot. The single-dish spectra are plotted black, the VLBA spectra are plotted by the red color. In cases of the VLBA non-detection, the upper limits on the VLBA flux density are shown by red arrows.}
\figsetgrpend

\figsetgrpstart
\figsetgrpnum{7.139}
\figsetgrptitle{Total and VLBA spectra of J0721+7750}
\figsetplot{J0721+7750_spectra.pdf}
\figsetgrpnote{Single-dish and VLBA broadband spectra of all the sources of our sample. The source name is specified at the top right of each plot. The single-dish spectra are plotted black, the VLBA spectra are plotted by the red color. In cases of the VLBA non-detection, the upper limits on the VLBA flux density are shown by red arrows.}
\figsetgrpend

\figsetgrpstart
\figsetgrpnum{7.140}
\figsetgrptitle{Total and VLBA spectra of J0726+7911}
\figsetplot{J0726+7911_spectra.pdf}
\figsetgrpnote{Single-dish and VLBA broadband spectra of all the sources of our sample. The source name is specified at the top right of each plot. The single-dish spectra are plotted black, the VLBA spectra are plotted by the red color. In cases of the VLBA non-detection, the upper limits on the VLBA flux density are shown by red arrows.}
\figsetgrpend

\figsetgrpstart
\figsetgrpnum{7.141}
\figsetgrptitle{Total and VLBA spectra of J0726+8711}
\figsetplot{J0726+8711_spectra.pdf}
\figsetgrpnote{Single-dish and VLBA broadband spectra of all the sources of our sample. The source name is specified at the top right of each plot. The single-dish spectra are plotted black, the VLBA spectra are plotted by the red color. In cases of the VLBA non-detection, the upper limits on the VLBA flux density are shown by red arrows.}
\figsetgrpend

\figsetgrpstart
\figsetgrpnum{7.142}
\figsetgrptitle{Total and VLBA spectra of J0727+8223}
\figsetplot{J0727+8223_spectra.pdf}
\figsetgrpnote{Single-dish and VLBA broadband spectra of all the sources of our sample. The source name is specified at the top right of each plot. The single-dish spectra are plotted black, the VLBA spectra are plotted by the red color. In cases of the VLBA non-detection, the upper limits on the VLBA flux density are shown by red arrows.}
\figsetgrpend

\figsetgrpstart
\figsetgrpnum{7.143}
\figsetgrptitle{Total and VLBA spectra of J0727+8545}
\figsetplot{J0727+8545_spectra.pdf}
\figsetgrpnote{Single-dish and VLBA broadband spectra of all the sources of our sample. The source name is specified at the top right of each plot. The single-dish spectra are plotted black, the VLBA spectra are plotted by the red color. In cases of the VLBA non-detection, the upper limits on the VLBA flux density are shown by red arrows.}
\figsetgrpend

\figsetgrpstart
\figsetgrpnum{7.144}
\figsetgrptitle{Total and VLBA spectra of J0734+7658}
\figsetplot{J0734+7658_spectra.pdf}
\figsetgrpnote{Single-dish and VLBA broadband spectra of all the sources of our sample. The source name is specified at the top right of each plot. The single-dish spectra are plotted black, the VLBA spectra are plotted by the red color. In cases of the VLBA non-detection, the upper limits on the VLBA flux density are shown by red arrows.}
\figsetgrpend

\figsetgrpstart
\figsetgrpnum{7.145}
\figsetgrptitle{Total and VLBA spectra of J0743+8025}
\figsetplot{J0743+8025_spectra.pdf}
\figsetgrpnote{Single-dish and VLBA broadband spectra of all the sources of our sample. The source name is specified at the top right of each plot. The single-dish spectra are plotted black, the VLBA spectra are plotted by the red color. In cases of the VLBA non-detection, the upper limits on the VLBA flux density are shown by red arrows.}
\figsetgrpend

\figsetgrpstart
\figsetgrpnum{7.146}
\figsetgrptitle{Total and VLBA spectra of J0748+8340}
\figsetplot{J0748+8340_spectra.pdf}
\figsetgrpnote{Single-dish and VLBA broadband spectra of all the sources of our sample. The source name is specified at the top right of each plot. The single-dish spectra are plotted black, the VLBA spectra are plotted by the red color. In cases of the VLBA non-detection, the upper limits on the VLBA flux density are shown by red arrows.}
\figsetgrpend

\figsetgrpstart
\figsetgrpnum{7.147}
\figsetgrptitle{Total and VLBA spectra of J0750+8241}
\figsetplot{J0750+8241_spectra.pdf}
\figsetgrpnote{Single-dish and VLBA broadband spectra of all the sources of our sample. The source name is specified at the top right of each plot. The single-dish spectra are plotted black, the VLBA spectra are plotted by the red color. In cases of the VLBA non-detection, the upper limits on the VLBA flux density are shown by red arrows.}
\figsetgrpend

\figsetgrpstart
\figsetgrpnum{7.148}
\figsetgrptitle{Total and VLBA spectra of J0800+7709}
\figsetplot{J0800+7709_spectra.pdf}
\figsetgrpnote{Single-dish and VLBA broadband spectra of all the sources of our sample. The source name is specified at the top right of each plot. The single-dish spectra are plotted black, the VLBA spectra are plotted by the red color. In cases of the VLBA non-detection, the upper limits on the VLBA flux density are shown by red arrows.}
\figsetgrpend

\figsetgrpstart
\figsetgrpnum{7.149}
\figsetgrptitle{Total and VLBA spectra of J0802+7620}
\figsetplot{J0802+7620_spectra.pdf}
\figsetgrpnote{Single-dish and VLBA broadband spectra of all the sources of our sample. The source name is specified at the top right of each plot. The single-dish spectra are plotted black, the VLBA spectra are plotted by the red color. In cases of the VLBA non-detection, the upper limits on the VLBA flux density are shown by red arrows.}
\figsetgrpend

\figsetgrpstart
\figsetgrpnum{7.150}
\figsetgrptitle{Total and VLBA spectra of J0806+7556}
\figsetplot{J0806+7556_spectra.pdf}
\figsetgrpnote{Single-dish and VLBA broadband spectra of all the sources of our sample. The source name is specified at the top right of each plot. The single-dish spectra are plotted black, the VLBA spectra are plotted by the red color. In cases of the VLBA non-detection, the upper limits on the VLBA flux density are shown by red arrows.}
\figsetgrpend

\figsetgrpstart
\figsetgrpnum{7.151}
\figsetgrptitle{Total and VLBA spectra of J0806+7746}
\figsetplot{J0806+7746_spectra.pdf}
\figsetgrpnote{Single-dish and VLBA broadband spectra of all the sources of our sample. The source name is specified at the top right of each plot. The single-dish spectra are plotted black, the VLBA spectra are plotted by the red color. In cases of the VLBA non-detection, the upper limits on the VLBA flux density are shown by red arrows.}
\figsetgrpend

\figsetgrpstart
\figsetgrpnum{7.152}
\figsetgrptitle{Total and VLBA spectra of J0806+8126}
\figsetplot{J0806+8126_spectra.pdf}
\figsetgrpnote{Single-dish and VLBA broadband spectra of all the sources of our sample. The source name is specified at the top right of each plot. The single-dish spectra are plotted black, the VLBA spectra are plotted by the red color. In cases of the VLBA non-detection, the upper limits on the VLBA flux density are shown by red arrows.}
\figsetgrpend

\figsetgrpstart
\figsetgrpnum{7.153}
\figsetgrptitle{Total and VLBA spectra of J0807+7846}
\figsetplot{J0807+7846_spectra.pdf}
\figsetgrpnote{Single-dish and VLBA broadband spectra of all the sources of our sample. The source name is specified at the top right of each plot. The single-dish spectra are plotted black, the VLBA spectra are plotted by the red color. In cases of the VLBA non-detection, the upper limits on the VLBA flux density are shown by red arrows.}
\figsetgrpend

\figsetgrpstart
\figsetgrpnum{7.154}
\figsetgrptitle{Total and VLBA spectra of J0809+8702}
\figsetplot{J0809+8702_spectra.pdf}
\figsetgrpnote{Single-dish and VLBA broadband spectra of all the sources of our sample. The source name is specified at the top right of each plot. The single-dish spectra are plotted black, the VLBA spectra are plotted by the red color. In cases of the VLBA non-detection, the upper limits on the VLBA flux density are shown by red arrows.}
\figsetgrpend

\figsetgrpstart
\figsetgrpnum{7.155}
\figsetgrptitle{Total and VLBA spectra of J0813+7619}
\figsetplot{J0813+7619_spectra.pdf}
\figsetgrpnote{Single-dish and VLBA broadband spectra of all the sources of our sample. The source name is specified at the top right of each plot. The single-dish spectra are plotted black, the VLBA spectra are plotted by the red color. In cases of the VLBA non-detection, the upper limits on the VLBA flux density are shown by red arrows.}
\figsetgrpend

\figsetgrpstart
\figsetgrpnum{7.156}
\figsetgrptitle{Total and VLBA spectra of J0819+7537}
\figsetplot{J0819+7537_spectra.pdf}
\figsetgrpnote{Single-dish and VLBA broadband spectra of all the sources of our sample. The source name is specified at the top right of each plot. The single-dish spectra are plotted black, the VLBA spectra are plotted by the red color. In cases of the VLBA non-detection, the upper limits on the VLBA flux density are shown by red arrows.}
\figsetgrpend

\figsetgrpstart
\figsetgrpnum{7.157}
\figsetgrptitle{Total and VLBA spectra of J0819+8105}
\figsetplot{J0819+8105_spectra.pdf}
\figsetgrpnote{Single-dish and VLBA broadband spectra of all the sources of our sample. The source name is specified at the top right of each plot. The single-dish spectra are plotted black, the VLBA spectra are plotted by the red color. In cases of the VLBA non-detection, the upper limits on the VLBA flux density are shown by red arrows.}
\figsetgrpend

\figsetgrpstart
\figsetgrpnum{7.158}
\figsetgrptitle{Total and VLBA spectra of J0821+8614}
\figsetplot{J0821+8614_spectra.pdf}
\figsetgrpnote{Single-dish and VLBA broadband spectra of all the sources of our sample. The source name is specified at the top right of each plot. The single-dish spectra are plotted black, the VLBA spectra are plotted by the red color. In cases of the VLBA non-detection, the upper limits on the VLBA flux density are shown by red arrows.}
\figsetgrpend

\figsetgrpstart
\figsetgrpnum{7.159}
\figsetgrptitle{Total and VLBA spectra of J0825+7653}
\figsetplot{J0825+7653_spectra.pdf}
\figsetgrpnote{Single-dish and VLBA broadband spectra of all the sources of our sample. The source name is specified at the top right of each plot. The single-dish spectra are plotted black, the VLBA spectra are plotted by the red color. In cases of the VLBA non-detection, the upper limits on the VLBA flux density are shown by red arrows.}
\figsetgrpend

\figsetgrpstart
\figsetgrpnum{7.160}
\figsetgrptitle{Total and VLBA spectra of J0830+8425}
\figsetplot{J0830+8425_spectra.pdf}
\figsetgrpnote{Single-dish and VLBA broadband spectra of all the sources of our sample. The source name is specified at the top right of each plot. The single-dish spectra are plotted black, the VLBA spectra are plotted by the red color. In cases of the VLBA non-detection, the upper limits on the VLBA flux density are shown by red arrows.}
\figsetgrpend

\figsetgrpstart
\figsetgrpnum{7.161}
\figsetgrptitle{Total and VLBA spectra of J0832+8006}
\figsetplot{J0832+8006_spectra.pdf}
\figsetgrpnote{Single-dish and VLBA broadband spectra of all the sources of our sample. The source name is specified at the top right of each plot. The single-dish spectra are plotted black, the VLBA spectra are plotted by the red color. In cases of the VLBA non-detection, the upper limits on the VLBA flux density are shown by red arrows.}
\figsetgrpend

\figsetgrpstart
\figsetgrpnum{7.162}
\figsetgrptitle{Total and VLBA spectra of J0835+8350}
\figsetplot{J0835+8350_spectra.pdf}
\figsetgrpnote{Single-dish and VLBA broadband spectra of all the sources of our sample. The source name is specified at the top right of each plot. The single-dish spectra are plotted black, the VLBA spectra are plotted by the red color. In cases of the VLBA non-detection, the upper limits on the VLBA flux density are shown by red arrows.}
\figsetgrpend

\figsetgrpstart
\figsetgrpnum{7.163}
\figsetgrptitle{Total and VLBA spectra of J0838+8446}
\figsetplot{J0838+8446_spectra.pdf}
\figsetgrpnote{Single-dish and VLBA broadband spectra of all the sources of our sample. The source name is specified at the top right of each plot. The single-dish spectra are plotted black, the VLBA spectra are plotted by the red color. In cases of the VLBA non-detection, the upper limits on the VLBA flux density are shown by red arrows.}
\figsetgrpend

\figsetgrpstart
\figsetgrpnum{7.164}
\figsetgrptitle{Total and VLBA spectra of J0848+7830}
\figsetplot{J0848+7830_spectra.pdf}
\figsetgrpnote{Single-dish and VLBA broadband spectra of all the sources of our sample. The source name is specified at the top right of each plot. The single-dish spectra are plotted black, the VLBA spectra are plotted by the red color. In cases of the VLBA non-detection, the upper limits on the VLBA flux density are shown by red arrows.}
\figsetgrpend

\figsetgrpstart
\figsetgrpnum{7.165}
\figsetgrptitle{Total and VLBA spectra of J0852+7840}
\figsetplot{J0852+7840_spectra.pdf}
\figsetgrpnote{Single-dish and VLBA broadband spectra of all the sources of our sample. The source name is specified at the top right of each plot. The single-dish spectra are plotted black, the VLBA spectra are plotted by the red color. In cases of the VLBA non-detection, the upper limits on the VLBA flux density are shown by red arrows.}
\figsetgrpend

\figsetgrpstart
\figsetgrpnum{7.166}
\figsetgrptitle{Total and VLBA spectra of J0858+7501}
\figsetplot{J0858+7501_spectra.pdf}
\figsetgrpnote{Single-dish and VLBA broadband spectra of all the sources of our sample. The source name is specified at the top right of each plot. The single-dish spectra are plotted black, the VLBA spectra are plotted by the red color. In cases of the VLBA non-detection, the upper limits on the VLBA flux density are shown by red arrows.}
\figsetgrpend

\figsetgrpstart
\figsetgrpnum{7.167}
\figsetgrptitle{Total and VLBA spectra of J0901+7809}
\figsetplot{J0901+7809_spectra.pdf}
\figsetgrpnote{Single-dish and VLBA broadband spectra of all the sources of our sample. The source name is specified at the top right of each plot. The single-dish spectra are plotted black, the VLBA spectra are plotted by the red color. In cases of the VLBA non-detection, the upper limits on the VLBA flux density are shown by red arrows.}
\figsetgrpend

\figsetgrpstart
\figsetgrpnum{7.168}
\figsetgrptitle{Total and VLBA spectra of J0908+8345}
\figsetplot{J0908+8345_spectra.pdf}
\figsetgrpnote{Single-dish and VLBA broadband spectra of all the sources of our sample. The source name is specified at the top right of each plot. The single-dish spectra are plotted black, the VLBA spectra are plotted by the red color. In cases of the VLBA non-detection, the upper limits on the VLBA flux density are shown by red arrows.}
\figsetgrpend

\figsetgrpstart
\figsetgrpnum{7.169}
\figsetgrptitle{Total and VLBA spectra of J0909+8327}
\figsetplot{J0909+8327_spectra.pdf}
\figsetgrpnote{Single-dish and VLBA broadband spectra of all the sources of our sample. The source name is specified at the top right of each plot. The single-dish spectra are plotted black, the VLBA spectra are plotted by the red color. In cases of the VLBA non-detection, the upper limits on the VLBA flux density are shown by red arrows.}
\figsetgrpend

\figsetgrpstart
\figsetgrpnum{7.170}
\figsetgrptitle{Total and VLBA spectra of J0911+8607}
\figsetplot{J0911+8607_spectra.pdf}
\figsetgrpnote{Single-dish and VLBA broadband spectra of all the sources of our sample. The source name is specified at the top right of each plot. The single-dish spectra are plotted black, the VLBA spectra are plotted by the red color. In cases of the VLBA non-detection, the upper limits on the VLBA flux density are shown by red arrows.}
\figsetgrpend

\figsetgrpstart
\figsetgrpnum{7.171}
\figsetgrptitle{Total and VLBA spectra of J0917+8021}
\figsetplot{J0917+8021_spectra.pdf}
\figsetgrpnote{Single-dish and VLBA broadband spectra of all the sources of our sample. The source name is specified at the top right of each plot. The single-dish spectra are plotted black, the VLBA spectra are plotted by the red color. In cases of the VLBA non-detection, the upper limits on the VLBA flux density are shown by red arrows.}
\figsetgrpend

\figsetgrpstart
\figsetgrpnum{7.172}
\figsetgrptitle{Total and VLBA spectra of J0918+8111}
\figsetplot{J0918+8111_spectra.pdf}
\figsetgrpnote{Single-dish and VLBA broadband spectra of all the sources of our sample. The source name is specified at the top right of each plot. The single-dish spectra are plotted black, the VLBA spectra are plotted by the red color. In cases of the VLBA non-detection, the upper limits on the VLBA flux density are shown by red arrows.}
\figsetgrpend

\figsetgrpstart
\figsetgrpnum{7.173}
\figsetgrptitle{Total and VLBA spectra of J0919+7727}
\figsetplot{J0919+7727_spectra.pdf}
\figsetgrpnote{Single-dish and VLBA broadband spectra of all the sources of our sample. The source name is specified at the top right of each plot. The single-dish spectra are plotted black, the VLBA spectra are plotted by the red color. In cases of the VLBA non-detection, the upper limits on the VLBA flux density are shown by red arrows.}
\figsetgrpend

\figsetgrpstart
\figsetgrpnum{7.174}
\figsetgrptitle{Total and VLBA spectra of J0919+7825}
\figsetplot{J0919+7825_spectra.pdf}
\figsetgrpnote{Single-dish and VLBA broadband spectra of all the sources of our sample. The source name is specified at the top right of each plot. The single-dish spectra are plotted black, the VLBA spectra are plotted by the red color. In cases of the VLBA non-detection, the upper limits on the VLBA flux density are shown by red arrows.}
\figsetgrpend

\figsetgrpstart
\figsetgrpnum{7.175}
\figsetgrptitle{Total and VLBA spectra of J0920+7636}
\figsetplot{J0920+7636_spectra.pdf}
\figsetgrpnote{Single-dish and VLBA broadband spectra of all the sources of our sample. The source name is specified at the top right of each plot. The single-dish spectra are plotted black, the VLBA spectra are plotted by the red color. In cases of the VLBA non-detection, the upper limits on the VLBA flux density are shown by red arrows.}
\figsetgrpend

\figsetgrpstart
\figsetgrpnum{7.176}
\figsetgrptitle{Total and VLBA spectra of J0920+8628}
\figsetplot{J0920+8628_spectra.pdf}
\figsetgrpnote{Single-dish and VLBA broadband spectra of all the sources of our sample. The source name is specified at the top right of each plot. The single-dish spectra are plotted black, the VLBA spectra are plotted by the red color. In cases of the VLBA non-detection, the upper limits on the VLBA flux density are shown by red arrows.}
\figsetgrpend

\figsetgrpstart
\figsetgrpnum{7.177}
\figsetgrptitle{Total and VLBA spectra of J0923+8116}
\figsetplot{J0923+8116_spectra.pdf}
\figsetgrpnote{Single-dish and VLBA broadband spectra of all the sources of our sample. The source name is specified at the top right of each plot. The single-dish spectra are plotted black, the VLBA spectra are plotted by the red color. In cases of the VLBA non-detection, the upper limits on the VLBA flux density are shown by red arrows.}
\figsetgrpend

\figsetgrpstart
\figsetgrpnum{7.178}
\figsetgrptitle{Total and VLBA spectra of J0932+7906}
\figsetplot{J0932+7906_spectra.pdf}
\figsetgrpnote{Single-dish and VLBA broadband spectra of all the sources of our sample. The source name is specified at the top right of each plot. The single-dish spectra are plotted black, the VLBA spectra are plotted by the red color. In cases of the VLBA non-detection, the upper limits on the VLBA flux density are shown by red arrows.}
\figsetgrpend

\figsetgrpstart
\figsetgrpnum{7.179}
\figsetgrptitle{Total and VLBA spectra of J0938+7815}
\figsetplot{J0938+7815_spectra.pdf}
\figsetgrpnote{Single-dish and VLBA broadband spectra of all the sources of our sample. The source name is specified at the top right of each plot. The single-dish spectra are plotted black, the VLBA spectra are plotted by the red color. In cases of the VLBA non-detection, the upper limits on the VLBA flux density are shown by red arrows.}
\figsetgrpend

\figsetgrpstart
\figsetgrpnum{7.180}
\figsetgrptitle{Total and VLBA spectra of J0939+8315}
\figsetplot{J0939+8315_spectra.pdf}
\figsetgrpnote{Single-dish and VLBA broadband spectra of all the sources of our sample. The source name is specified at the top right of each plot. The single-dish spectra are plotted black, the VLBA spectra are plotted by the red color. In cases of the VLBA non-detection, the upper limits on the VLBA flux density are shown by red arrows.}
\figsetgrpend

\figsetgrpstart
\figsetgrpnum{7.181}
\figsetgrptitle{Total and VLBA spectra of J0944+8254}
\figsetplot{J0944+8254_spectra.pdf}
\figsetgrpnote{Single-dish and VLBA broadband spectra of all the sources of our sample. The source name is specified at the top right of each plot. The single-dish spectra are plotted black, the VLBA spectra are plotted by the red color. In cases of the VLBA non-detection, the upper limits on the VLBA flux density are shown by red arrows.}
\figsetgrpend

\figsetgrpstart
\figsetgrpnum{7.182}
\figsetgrptitle{Total and VLBA spectra of J0945+7932}
\figsetplot{J0945+7932_spectra.pdf}
\figsetgrpnote{Single-dish and VLBA broadband spectra of all the sources of our sample. The source name is specified at the top right of each plot. The single-dish spectra are plotted black, the VLBA spectra are plotted by the red color. In cases of the VLBA non-detection, the upper limits on the VLBA flux density are shown by red arrows.}
\figsetgrpend

\figsetgrpstart
\figsetgrpnum{7.183}
\figsetgrptitle{Total and VLBA spectra of J0953+8034}
\figsetplot{J0953+8034_spectra.pdf}
\figsetgrpnote{Single-dish and VLBA broadband spectra of all the sources of our sample. The source name is specified at the top right of each plot. The single-dish spectra are plotted black, the VLBA spectra are plotted by the red color. In cases of the VLBA non-detection, the upper limits on the VLBA flux density are shown by red arrows.}
\figsetgrpend

\figsetgrpstart
\figsetgrpnum{7.184}
\figsetgrptitle{Total and VLBA spectra of J0956+7911}
\figsetplot{J0956+7911_spectra.pdf}
\figsetgrpnote{Single-dish and VLBA broadband spectra of all the sources of our sample. The source name is specified at the top right of each plot. The single-dish spectra are plotted black, the VLBA spectra are plotted by the red color. In cases of the VLBA non-detection, the upper limits on the VLBA flux density are shown by red arrows.}
\figsetgrpend

\figsetgrpstart
\figsetgrpnum{7.185}
\figsetgrptitle{Total and VLBA spectra of J1000+8127}
\figsetplot{J1000+8127_spectra.pdf}
\figsetgrpnote{Single-dish and VLBA broadband spectra of all the sources of our sample. The source name is specified at the top right of each plot. The single-dish spectra are plotted black, the VLBA spectra are plotted by the red color. In cases of the VLBA non-detection, the upper limits on the VLBA flux density are shown by red arrows.}
\figsetgrpend

\figsetgrpstart
\figsetgrpnum{7.186}
\figsetgrptitle{Total and VLBA spectra of J1003+7908}
\figsetplot{J1003+7908_spectra.pdf}
\figsetgrpnote{Single-dish and VLBA broadband spectra of all the sources of our sample. The source name is specified at the top right of each plot. The single-dish spectra are plotted black, the VLBA spectra are plotted by the red color. In cases of the VLBA non-detection, the upper limits on the VLBA flux density are shown by red arrows.}
\figsetgrpend

\figsetgrpstart
\figsetgrpnum{7.187}
\figsetgrptitle{Total and VLBA spectra of J1005+7739}
\figsetplot{J1005+7739_spectra.pdf}
\figsetgrpnote{Single-dish and VLBA broadband spectra of all the sources of our sample. The source name is specified at the top right of each plot. The single-dish spectra are plotted black, the VLBA spectra are plotted by the red color. In cases of the VLBA non-detection, the upper limits on the VLBA flux density are shown by red arrows.}
\figsetgrpend

\figsetgrpstart
\figsetgrpnum{7.188}
\figsetgrptitle{Total and VLBA spectra of J1007+8131}
\figsetplot{J1007+8131_spectra.pdf}
\figsetgrpnote{Single-dish and VLBA broadband spectra of all the sources of our sample. The source name is specified at the top right of each plot. The single-dish spectra are plotted black, the VLBA spectra are plotted by the red color. In cases of the VLBA non-detection, the upper limits on the VLBA flux density are shown by red arrows.}
\figsetgrpend

\figsetgrpstart
\figsetgrpnum{7.189}
\figsetgrptitle{Total and VLBA spectra of J1009+8107}
\figsetplot{J1009+8107_spectra.pdf}
\figsetgrpnote{Single-dish and VLBA broadband spectra of all the sources of our sample. The source name is specified at the top right of each plot. The single-dish spectra are plotted black, the VLBA spectra are plotted by the red color. In cases of the VLBA non-detection, the upper limits on the VLBA flux density are shown by red arrows.}
\figsetgrpend

\figsetgrpstart
\figsetgrpnum{7.190}
\figsetgrptitle{Total and VLBA spectra of J1010+7650}
\figsetplot{J1010+7650_spectra.pdf}
\figsetgrpnote{Single-dish and VLBA broadband spectra of all the sources of our sample. The source name is specified at the top right of each plot. The single-dish spectra are plotted black, the VLBA spectra are plotted by the red color. In cases of the VLBA non-detection, the upper limits on the VLBA flux density are shown by red arrows.}
\figsetgrpend

\figsetgrpstart
\figsetgrpnum{7.191}
\figsetgrptitle{Total and VLBA spectra of J1010+8250}
\figsetplot{J1010+8250_spectra.pdf}
\figsetgrpnote{Single-dish and VLBA broadband spectra of all the sources of our sample. The source name is specified at the top right of each plot. The single-dish spectra are plotted black, the VLBA spectra are plotted by the red color. In cases of the VLBA non-detection, the upper limits on the VLBA flux density are shown by red arrows.}
\figsetgrpend

\figsetgrpstart
\figsetgrpnum{7.192}
\figsetgrptitle{Total and VLBA spectra of J1014+8553}
\figsetplot{J1014+8553_spectra.pdf}
\figsetgrpnote{Single-dish and VLBA broadband spectra of all the sources of our sample. The source name is specified at the top right of each plot. The single-dish spectra are plotted black, the VLBA spectra are plotted by the red color. In cases of the VLBA non-detection, the upper limits on the VLBA flux density are shown by red arrows.}
\figsetgrpend

\figsetgrpstart
\figsetgrpnum{7.193}
\figsetgrptitle{Total and VLBA spectra of J1016+7617}
\figsetplot{J1016+7617_spectra.pdf}
\figsetgrpnote{Single-dish and VLBA broadband spectra of all the sources of our sample. The source name is specified at the top right of each plot. The single-dish spectra are plotted black, the VLBA spectra are plotted by the red color. In cases of the VLBA non-detection, the upper limits on the VLBA flux density are shown by red arrows.}
\figsetgrpend

\figsetgrpstart
\figsetgrpnum{7.194}
\figsetgrptitle{Total and VLBA spectra of J1017+8105}
\figsetplot{J1017+8105_spectra.pdf}
\figsetgrpnote{Single-dish and VLBA broadband spectra of all the sources of our sample. The source name is specified at the top right of each plot. The single-dish spectra are plotted black, the VLBA spectra are plotted by the red color. In cases of the VLBA non-detection, the upper limits on the VLBA flux density are shown by red arrows.}
\figsetgrpend

\figsetgrpstart
\figsetgrpnum{7.195}
\figsetgrptitle{Total and VLBA spectra of J1017+8756}
\figsetplot{J1017+8756_spectra.pdf}
\figsetgrpnote{Single-dish and VLBA broadband spectra of all the sources of our sample. The source name is specified at the top right of each plot. The single-dish spectra are plotted black, the VLBA spectra are plotted by the red color. In cases of the VLBA non-detection, the upper limits on the VLBA flux density are shown by red arrows.}
\figsetgrpend

\figsetgrpstart
\figsetgrpnum{7.196}
\figsetgrptitle{Total and VLBA spectra of J1021+7655}
\figsetplot{J1021+7655_spectra.pdf}
\figsetgrpnote{Single-dish and VLBA broadband spectra of all the sources of our sample. The source name is specified at the top right of each plot. The single-dish spectra are plotted black, the VLBA spectra are plotted by the red color. In cases of the VLBA non-detection, the upper limits on the VLBA flux density are shown by red arrows.}
\figsetgrpend

\figsetgrpstart
\figsetgrpnum{7.197}
\figsetgrptitle{Total and VLBA spectra of J1023+8032}
\figsetplot{J1023+8032_spectra.pdf}
\figsetgrpnote{Single-dish and VLBA broadband spectra of all the sources of our sample. The source name is specified at the top right of each plot. The single-dish spectra are plotted black, the VLBA spectra are plotted by the red color. In cases of the VLBA non-detection, the upper limits on the VLBA flux density are shown by red arrows.}
\figsetgrpend

\figsetgrpstart
\figsetgrpnum{7.198}
\figsetgrptitle{Total and VLBA spectra of J1025+8144}
\figsetplot{J1025+8144_spectra.pdf}
\figsetgrpnote{Single-dish and VLBA broadband spectra of all the sources of our sample. The source name is specified at the top right of each plot. The single-dish spectra are plotted black, the VLBA spectra are plotted by the red color. In cases of the VLBA non-detection, the upper limits on the VLBA flux density are shown by red arrows.}
\figsetgrpend

\figsetgrpstart
\figsetgrpnum{7.199}
\figsetgrptitle{Total and VLBA spectra of J1029+7852}
\figsetplot{J1029+7852_spectra.pdf}
\figsetgrpnote{Single-dish and VLBA broadband spectra of all the sources of our sample. The source name is specified at the top right of each plot. The single-dish spectra are plotted black, the VLBA spectra are plotted by the red color. In cases of the VLBA non-detection, the upper limits on the VLBA flux density are shown by red arrows.}
\figsetgrpend

\figsetgrpstart
\figsetgrpnum{7.200}
\figsetgrptitle{Total and VLBA spectra of J1031+7845}
\figsetplot{J1031+7845_spectra.pdf}
\figsetgrpnote{Single-dish and VLBA broadband spectra of all the sources of our sample. The source name is specified at the top right of each plot. The single-dish spectra are plotted black, the VLBA spectra are plotted by the red color. In cases of the VLBA non-detection, the upper limits on the VLBA flux density are shown by red arrows.}
\figsetgrpend

\figsetgrpstart
\figsetgrpnum{7.201}
\figsetgrptitle{Total and VLBA spectra of J1037+7632}
\figsetplot{J1037+7632_spectra.pdf}
\figsetgrpnote{Single-dish and VLBA broadband spectra of all the sources of our sample. The source name is specified at the top right of each plot. The single-dish spectra are plotted black, the VLBA spectra are plotted by the red color. In cases of the VLBA non-detection, the upper limits on the VLBA flux density are shown by red arrows.}
\figsetgrpend

\figsetgrpstart
\figsetgrpnum{7.202}
\figsetgrptitle{Total and VLBA spectra of J1042+7545}
\figsetplot{J1042+7545_spectra.pdf}
\figsetgrpnote{Single-dish and VLBA broadband spectra of all the sources of our sample. The source name is specified at the top right of each plot. The single-dish spectra are plotted black, the VLBA spectra are plotted by the red color. In cases of the VLBA non-detection, the upper limits on the VLBA flux density are shown by red arrows.}
\figsetgrpend

\figsetgrpstart
\figsetgrpnum{7.203}
\figsetgrptitle{Total and VLBA spectra of J1044+8054}
\figsetplot{J1044+8054_spectra.pdf}
\figsetgrpnote{Single-dish and VLBA broadband spectra of all the sources of our sample. The source name is specified at the top right of each plot. The single-dish spectra are plotted black, the VLBA spectra are plotted by the red color. In cases of the VLBA non-detection, the upper limits on the VLBA flux density are shown by red arrows.}
\figsetgrpend

\figsetgrpstart
\figsetgrpnum{7.204}
\figsetgrptitle{Total and VLBA spectra of J1049+8121}
\figsetplot{J1049+8121_spectra.pdf}
\figsetgrpnote{Single-dish and VLBA broadband spectra of all the sources of our sample. The source name is specified at the top right of each plot. The single-dish spectra are plotted black, the VLBA spectra are plotted by the red color. In cases of the VLBA non-detection, the upper limits on the VLBA flux density are shown by red arrows.}
\figsetgrpend

\figsetgrpstart
\figsetgrpnum{7.205}
\figsetgrptitle{Total and VLBA spectra of J1051+7913}
\figsetplot{J1051+7913_spectra.pdf}
\figsetgrpnote{Single-dish and VLBA broadband spectra of all the sources of our sample. The source name is specified at the top right of each plot. The single-dish spectra are plotted black, the VLBA spectra are plotted by the red color. In cases of the VLBA non-detection, the upper limits on the VLBA flux density are shown by red arrows.}
\figsetgrpend

\figsetgrpstart
\figsetgrpnum{7.206}
\figsetgrptitle{Total and VLBA spectra of J1052+8317}
\figsetplot{J1052+8317_spectra.pdf}
\figsetgrpnote{Single-dish and VLBA broadband spectra of all the sources of our sample. The source name is specified at the top right of each plot. The single-dish spectra are plotted black, the VLBA spectra are plotted by the red color. In cases of the VLBA non-detection, the upper limits on the VLBA flux density are shown by red arrows.}
\figsetgrpend

\figsetgrpstart
\figsetgrpnum{7.207}
\figsetgrptitle{Total and VLBA spectra of J1054+8629}
\figsetplot{J1054+8629_spectra.pdf}
\figsetgrpnote{Single-dish and VLBA broadband spectra of all the sources of our sample. The source name is specified at the top right of each plot. The single-dish spectra are plotted black, the VLBA spectra are plotted by the red color. In cases of the VLBA non-detection, the upper limits on the VLBA flux density are shown by red arrows.}
\figsetgrpend

\figsetgrpstart
\figsetgrpnum{7.208}
\figsetgrptitle{Total and VLBA spectra of J1057+8858}
\figsetplot{J1057+8858_spectra.pdf}
\figsetgrpnote{Single-dish and VLBA broadband spectra of all the sources of our sample. The source name is specified at the top right of each plot. The single-dish spectra are plotted black, the VLBA spectra are plotted by the red color. In cases of the VLBA non-detection, the upper limits on the VLBA flux density are shown by red arrows.}
\figsetgrpend

\figsetgrpstart
\figsetgrpnum{7.209}
\figsetgrptitle{Total and VLBA spectra of J1058+8114}
\figsetplot{J1058+8114_spectra.pdf}
\figsetgrpnote{Single-dish and VLBA broadband spectra of all the sources of our sample. The source name is specified at the top right of each plot. The single-dish spectra are plotted black, the VLBA spectra are plotted by the red color. In cases of the VLBA non-detection, the upper limits on the VLBA flux density are shown by red arrows.}
\figsetgrpend

\figsetgrpstart
\figsetgrpnum{7.210}
\figsetgrptitle{Total and VLBA spectra of J1102+7905}
\figsetplot{J1102+7905_spectra.pdf}
\figsetgrpnote{Single-dish and VLBA broadband spectra of all the sources of our sample. The source name is specified at the top right of each plot. The single-dish spectra are plotted black, the VLBA spectra are plotted by the red color. In cases of the VLBA non-detection, the upper limits on the VLBA flux density are shown by red arrows.}
\figsetgrpend

\figsetgrpstart
\figsetgrpnum{7.211}
\figsetgrptitle{Total and VLBA spectra of J1104+7658}
\figsetplot{J1104+7658_spectra.pdf}
\figsetgrpnote{Single-dish and VLBA broadband spectra of all the sources of our sample. The source name is specified at the top right of each plot. The single-dish spectra are plotted black, the VLBA spectra are plotted by the red color. In cases of the VLBA non-detection, the upper limits on the VLBA flux density are shown by red arrows.}
\figsetgrpend

\figsetgrpstart
\figsetgrpnum{7.212}
\figsetgrptitle{Total and VLBA spectra of J1104+7932}
\figsetplot{J1104+7932_spectra.pdf}
\figsetgrpnote{Single-dish and VLBA broadband spectra of all the sources of our sample. The source name is specified at the top right of each plot. The single-dish spectra are plotted black, the VLBA spectra are plotted by the red color. In cases of the VLBA non-detection, the upper limits on the VLBA flux density are shown by red arrows.}
\figsetgrpend

\figsetgrpstart
\figsetgrpnum{7.213}
\figsetgrptitle{Total and VLBA spectra of J1105+8328}
\figsetplot{J1105+8328_spectra.pdf}
\figsetgrpnote{Single-dish and VLBA broadband spectra of all the sources of our sample. The source name is specified at the top right of each plot. The single-dish spectra are plotted black, the VLBA spectra are plotted by the red color. In cases of the VLBA non-detection, the upper limits on the VLBA flux density are shown by red arrows.}
\figsetgrpend

\figsetgrpstart
\figsetgrpnum{7.214}
\figsetgrptitle{Total and VLBA spectra of J1113+7654}
\figsetplot{J1113+7654_spectra.pdf}
\figsetgrpnote{Single-dish and VLBA broadband spectra of all the sources of our sample. The source name is specified at the top right of each plot. The single-dish spectra are plotted black, the VLBA spectra are plotted by the red color. In cases of the VLBA non-detection, the upper limits on the VLBA flux density are shown by red arrows.}
\figsetgrpend

\figsetgrpstart
\figsetgrpnum{7.215}
\figsetgrptitle{Total and VLBA spectra of J1113+8332}
\figsetplot{J1113+8332_spectra.pdf}
\figsetgrpnote{Single-dish and VLBA broadband spectra of all the sources of our sample. The source name is specified at the top right of each plot. The single-dish spectra are plotted black, the VLBA spectra are plotted by the red color. In cases of the VLBA non-detection, the upper limits on the VLBA flux density are shown by red arrows.}
\figsetgrpend

\figsetgrpstart
\figsetgrpnum{7.216}
\figsetgrptitle{Total and VLBA spectra of J1116+7658}
\figsetplot{J1116+7658_spectra.pdf}
\figsetgrpnote{Single-dish and VLBA broadband spectra of all the sources of our sample. The source name is specified at the top right of each plot. The single-dish spectra are plotted black, the VLBA spectra are plotted by the red color. In cases of the VLBA non-detection, the upper limits on the VLBA flux density are shown by red arrows.}
\figsetgrpend

\figsetgrpstart
\figsetgrpnum{7.217}
\figsetgrptitle{Total and VLBA spectra of J1116+8335}
\figsetplot{J1116+8335_spectra.pdf}
\figsetgrpnote{Single-dish and VLBA broadband spectra of all the sources of our sample. The source name is specified at the top right of each plot. The single-dish spectra are plotted black, the VLBA spectra are plotted by the red color. In cases of the VLBA non-detection, the upper limits on the VLBA flux density are shown by red arrows.}
\figsetgrpend

\figsetgrpstart
\figsetgrpnum{7.218}
\figsetgrptitle{Total and VLBA spectra of J1119+8048}
\figsetplot{J1119+8048_spectra.pdf}
\figsetgrpnote{Single-dish and VLBA broadband spectra of all the sources of our sample. The source name is specified at the top right of each plot. The single-dish spectra are plotted black, the VLBA spectra are plotted by the red color. In cases of the VLBA non-detection, the upper limits on the VLBA flux density are shown by red arrows.}
\figsetgrpend

\figsetgrpstart
\figsetgrpnum{7.219}
\figsetgrptitle{Total and VLBA spectra of J1123+7731}
\figsetplot{J1123+7731_spectra.pdf}
\figsetgrpnote{Single-dish and VLBA broadband spectra of all the sources of our sample. The source name is specified at the top right of each plot. The single-dish spectra are plotted black, the VLBA spectra are plotted by the red color. In cases of the VLBA non-detection, the upper limits on the VLBA flux density are shown by red arrows.}
\figsetgrpend

\figsetgrpstart
\figsetgrpnum{7.220}
\figsetgrptitle{Total and VLBA spectra of J1133+7831}
\figsetplot{J1133+7831_spectra.pdf}
\figsetgrpnote{Single-dish and VLBA broadband spectra of all the sources of our sample. The source name is specified at the top right of each plot. The single-dish spectra are plotted black, the VLBA spectra are plotted by the red color. In cases of the VLBA non-detection, the upper limits on the VLBA flux density are shown by red arrows.}
\figsetgrpend

\figsetgrpstart
\figsetgrpnum{7.221}
\figsetgrptitle{Total and VLBA spectra of J1139+8321}
\figsetplot{J1139+8321_spectra.pdf}
\figsetgrpnote{Single-dish and VLBA broadband spectra of all the sources of our sample. The source name is specified at the top right of each plot. The single-dish spectra are plotted black, the VLBA spectra are plotted by the red color. In cases of the VLBA non-detection, the upper limits on the VLBA flux density are shown by red arrows.}
\figsetgrpend

\figsetgrpstart
\figsetgrpnum{7.222}
\figsetgrptitle{Total and VLBA spectra of J1147+8346}
\figsetplot{J1147+8346_spectra.pdf}
\figsetgrpnote{Single-dish and VLBA broadband spectra of all the sources of our sample. The source name is specified at the top right of each plot. The single-dish spectra are plotted black, the VLBA spectra are plotted by the red color. In cases of the VLBA non-detection, the upper limits on the VLBA flux density are shown by red arrows.}
\figsetgrpend

\figsetgrpstart
\figsetgrpnum{7.223}
\figsetgrptitle{Total and VLBA spectra of J1148+7827}
\figsetplot{J1148+7827_spectra.pdf}
\figsetgrpnote{Single-dish and VLBA broadband spectra of all the sources of our sample. The source name is specified at the top right of each plot. The single-dish spectra are plotted black, the VLBA spectra are plotted by the red color. In cases of the VLBA non-detection, the upper limits on the VLBA flux density are shown by red arrows.}
\figsetgrpend

\figsetgrpstart
\figsetgrpnum{7.224}
\figsetgrptitle{Total and VLBA spectra of J1149+7645}
\figsetplot{J1149+7645_spectra.pdf}
\figsetgrpnote{Single-dish and VLBA broadband spectra of all the sources of our sample. The source name is specified at the top right of each plot. The single-dish spectra are plotted black, the VLBA spectra are plotted by the red color. In cases of the VLBA non-detection, the upper limits on the VLBA flux density are shown by red arrows.}
\figsetgrpend

\figsetgrpstart
\figsetgrpnum{7.225}
\figsetgrptitle{Total and VLBA spectra of J1153+8058}
\figsetplot{J1153+8058_spectra.pdf}
\figsetgrpnote{Single-dish and VLBA broadband spectra of all the sources of our sample. The source name is specified at the top right of each plot. The single-dish spectra are plotted black, the VLBA spectra are plotted by the red color. In cases of the VLBA non-detection, the upper limits on the VLBA flux density are shown by red arrows.}
\figsetgrpend

\figsetgrpstart
\figsetgrpnum{7.226}
\figsetgrptitle{Total and VLBA spectra of J1153+8256}
\figsetplot{J1153+8256_spectra.pdf}
\figsetgrpnote{Single-dish and VLBA broadband spectra of all the sources of our sample. The source name is specified at the top right of each plot. The single-dish spectra are plotted black, the VLBA spectra are plotted by the red color. In cases of the VLBA non-detection, the upper limits on the VLBA flux density are shown by red arrows.}
\figsetgrpend

\figsetgrpstart
\figsetgrpnum{7.227}
\figsetgrptitle{Total and VLBA spectra of J1155+7534}
\figsetplot{J1155+7534_spectra.pdf}
\figsetgrpnote{Single-dish and VLBA broadband spectra of all the sources of our sample. The source name is specified at the top right of each plot. The single-dish spectra are plotted black, the VLBA spectra are plotted by the red color. In cases of the VLBA non-detection, the upper limits on the VLBA flux density are shown by red arrows.}
\figsetgrpend

\figsetgrpstart
\figsetgrpnum{7.228}
\figsetgrptitle{Total and VLBA spectra of J1155+8157}
\figsetplot{J1155+8157_spectra.pdf}
\figsetgrpnote{Single-dish and VLBA broadband spectra of all the sources of our sample. The source name is specified at the top right of each plot. The single-dish spectra are plotted black, the VLBA spectra are plotted by the red color. In cases of the VLBA non-detection, the upper limits on the VLBA flux density are shown by red arrows.}
\figsetgrpend

\figsetgrpstart
\figsetgrpnum{7.229}
\figsetgrptitle{Total and VLBA spectra of J1156+8235}
\figsetplot{J1156+8235_spectra.pdf}
\figsetgrpnote{Single-dish and VLBA broadband spectra of all the sources of our sample. The source name is specified at the top right of each plot. The single-dish spectra are plotted black, the VLBA spectra are plotted by the red color. In cases of the VLBA non-detection, the upper limits on the VLBA flux density are shown by red arrows.}
\figsetgrpend

\figsetgrpstart
\figsetgrpnum{7.230}
\figsetgrptitle{Total and VLBA spectra of J1157+7526}
\figsetplot{J1157+7526_spectra.pdf}
\figsetgrpnote{Single-dish and VLBA broadband spectra of all the sources of our sample. The source name is specified at the top right of each plot. The single-dish spectra are plotted black, the VLBA spectra are plotted by the red color. In cases of the VLBA non-detection, the upper limits on the VLBA flux density are shown by red arrows.}
\figsetgrpend

\figsetgrpstart
\figsetgrpnum{7.231}
\figsetgrptitle{Total and VLBA spectra of J1157+8118}
\figsetplot{J1157+8118_spectra.pdf}
\figsetgrpnote{Single-dish and VLBA broadband spectra of all the sources of our sample. The source name is specified at the top right of each plot. The single-dish spectra are plotted black, the VLBA spectra are plotted by the red color. In cases of the VLBA non-detection, the upper limits on the VLBA flux density are shown by red arrows.}
\figsetgrpend

\figsetgrpstart
\figsetgrpnum{7.232}
\figsetgrptitle{Total and VLBA spectra of J1203+8206}
\figsetplot{J1203+8206_spectra.pdf}
\figsetgrpnote{Single-dish and VLBA broadband spectra of all the sources of our sample. The source name is specified at the top right of each plot. The single-dish spectra are plotted black, the VLBA spectra are plotted by the red color. In cases of the VLBA non-detection, the upper limits on the VLBA flux density are shown by red arrows.}
\figsetgrpend

\figsetgrpstart
\figsetgrpnum{7.233}
\figsetgrptitle{Total and VLBA spectra of J1203+8401}
\figsetplot{J1203+8401_spectra.pdf}
\figsetgrpnote{Single-dish and VLBA broadband spectra of all the sources of our sample. The source name is specified at the top right of each plot. The single-dish spectra are plotted black, the VLBA spectra are plotted by the red color. In cases of the VLBA non-detection, the upper limits on the VLBA flux density are shown by red arrows.}
\figsetgrpend

\figsetgrpstart
\figsetgrpnum{7.234}
\figsetgrptitle{Total and VLBA spectra of J1203+8534}
\figsetplot{J1203+8534_spectra.pdf}
\figsetgrpnote{Single-dish and VLBA broadband spectra of all the sources of our sample. The source name is specified at the top right of each plot. The single-dish spectra are plotted black, the VLBA spectra are plotted by the red color. In cases of the VLBA non-detection, the upper limits on the VLBA flux density are shown by red arrows.}
\figsetgrpend

\figsetgrpstart
\figsetgrpnum{7.235}
\figsetgrptitle{Total and VLBA spectra of J1204+8354}
\figsetplot{J1204+8354_spectra.pdf}
\figsetgrpnote{Single-dish and VLBA broadband spectra of all the sources of our sample. The source name is specified at the top right of each plot. The single-dish spectra are plotted black, the VLBA spectra are plotted by the red color. In cases of the VLBA non-detection, the upper limits on the VLBA flux density are shown by red arrows.}
\figsetgrpend

\figsetgrpstart
\figsetgrpnum{7.236}
\figsetgrptitle{Total and VLBA spectra of J1207+7756}
\figsetplot{J1207+7756_spectra.pdf}
\figsetgrpnote{Single-dish and VLBA broadband spectra of all the sources of our sample. The source name is specified at the top right of each plot. The single-dish spectra are plotted black, the VLBA spectra are plotted by the red color. In cases of the VLBA non-detection, the upper limits on the VLBA flux density are shown by red arrows.}
\figsetgrpend

\figsetgrpstart
\figsetgrpnum{7.237}
\figsetgrptitle{Total and VLBA spectra of J1212+7539}
\figsetplot{J1212+7539_spectra.pdf}
\figsetgrpnote{Single-dish and VLBA broadband spectra of all the sources of our sample. The source name is specified at the top right of each plot. The single-dish spectra are plotted black, the VLBA spectra are plotted by the red color. In cases of the VLBA non-detection, the upper limits on the VLBA flux density are shown by red arrows.}
\figsetgrpend

\figsetgrpstart
\figsetgrpnum{7.238}
\figsetgrptitle{Total and VLBA spectra of J1218+7934}
\figsetplot{J1218+7934_spectra.pdf}
\figsetgrpnote{Single-dish and VLBA broadband spectra of all the sources of our sample. The source name is specified at the top right of each plot. The single-dish spectra are plotted black, the VLBA spectra are plotted by the red color. In cases of the VLBA non-detection, the upper limits on the VLBA flux density are shown by red arrows.}
\figsetgrpend

\figsetgrpstart
\figsetgrpnum{7.239}
\figsetgrptitle{Total and VLBA spectra of J1220+7927}
\figsetplot{J1220+7927_spectra.pdf}
\figsetgrpnote{Single-dish and VLBA broadband spectra of all the sources of our sample. The source name is specified at the top right of each plot. The single-dish spectra are plotted black, the VLBA spectra are plotted by the red color. In cases of the VLBA non-detection, the upper limits on the VLBA flux density are shown by red arrows.}
\figsetgrpend

\figsetgrpstart
\figsetgrpnum{7.240}
\figsetgrptitle{Total and VLBA spectra of J1223+8040}
\figsetplot{J1223+8040_spectra.pdf}
\figsetgrpnote{Single-dish and VLBA broadband spectra of all the sources of our sample. The source name is specified at the top right of each plot. The single-dish spectra are plotted black, the VLBA spectra are plotted by the red color. In cases of the VLBA non-detection, the upper limits on the VLBA flux density are shown by red arrows.}
\figsetgrpend

\figsetgrpstart
\figsetgrpnum{7.241}
\figsetgrptitle{Total and VLBA spectra of J1223+8436}
\figsetplot{J1223+8436_spectra.pdf}
\figsetgrpnote{Single-dish and VLBA broadband spectra of all the sources of our sample. The source name is specified at the top right of each plot. The single-dish spectra are plotted black, the VLBA spectra are plotted by the red color. In cases of the VLBA non-detection, the upper limits on the VLBA flux density are shown by red arrows.}
\figsetgrpend

\figsetgrpstart
\figsetgrpnum{7.242}
\figsetgrptitle{Total and VLBA spectra of J1225+8608}
\figsetplot{J1225+8608_spectra.pdf}
\figsetgrpnote{Single-dish and VLBA broadband spectra of all the sources of our sample. The source name is specified at the top right of each plot. The single-dish spectra are plotted black, the VLBA spectra are plotted by the red color. In cases of the VLBA non-detection, the upper limits on the VLBA flux density are shown by red arrows.}
\figsetgrpend

\figsetgrpstart
\figsetgrpnum{7.243}
\figsetgrptitle{Total and VLBA spectra of J1233+8054}
\figsetplot{J1233+8054_spectra.pdf}
\figsetgrpnote{Single-dish and VLBA broadband spectra of all the sources of our sample. The source name is specified at the top right of each plot. The single-dish spectra are plotted black, the VLBA spectra are plotted by the red color. In cases of the VLBA non-detection, the upper limits on the VLBA flux density are shown by red arrows.}
\figsetgrpend

\figsetgrpstart
\figsetgrpnum{7.244}
\figsetgrptitle{Total and VLBA spectra of J1234+7730}
\figsetplot{J1234+7730_spectra.pdf}
\figsetgrpnote{Single-dish and VLBA broadband spectra of all the sources of our sample. The source name is specified at the top right of each plot. The single-dish spectra are plotted black, the VLBA spectra are plotted by the red color. In cases of the VLBA non-detection, the upper limits on the VLBA flux density are shown by red arrows.}
\figsetgrpend

\figsetgrpstart
\figsetgrpnum{7.245}
\figsetgrptitle{Total and VLBA spectra of J1237+7626}
\figsetplot{J1237+7626_spectra.pdf}
\figsetgrpnote{Single-dish and VLBA broadband spectra of all the sources of our sample. The source name is specified at the top right of each plot. The single-dish spectra are plotted black, the VLBA spectra are plotted by the red color. In cases of the VLBA non-detection, the upper limits on the VLBA flux density are shown by red arrows.}
\figsetgrpend

\figsetgrpstart
\figsetgrpnum{7.246}
\figsetgrptitle{Total and VLBA spectra of J1237+8357}
\figsetplot{J1237+8357_spectra.pdf}
\figsetgrpnote{Single-dish and VLBA broadband spectra of all the sources of our sample. The source name is specified at the top right of each plot. The single-dish spectra are plotted black, the VLBA spectra are plotted by the red color. In cases of the VLBA non-detection, the upper limits on the VLBA flux density are shown by red arrows.}
\figsetgrpend

\figsetgrpstart
\figsetgrpnum{7.247}
\figsetgrptitle{Total and VLBA spectra of J1241+8612}
\figsetplot{J1241+8612_spectra.pdf}
\figsetgrpnote{Single-dish and VLBA broadband spectra of all the sources of our sample. The source name is specified at the top right of each plot. The single-dish spectra are plotted black, the VLBA spectra are plotted by the red color. In cases of the VLBA non-detection, the upper limits on the VLBA flux density are shown by red arrows.}
\figsetgrpend

\figsetgrpstart
\figsetgrpnum{7.248}
\figsetgrptitle{Total and VLBA spectra of J1242+8000}
\figsetplot{J1242+8000_spectra.pdf}
\figsetgrpnote{Single-dish and VLBA broadband spectra of all the sources of our sample. The source name is specified at the top right of each plot. The single-dish spectra are plotted black, the VLBA spectra are plotted by the red color. In cases of the VLBA non-detection, the upper limits on the VLBA flux density are shown by red arrows.}
\figsetgrpend

\figsetgrpstart
\figsetgrpnum{7.249}
\figsetgrptitle{Total and VLBA spectra of J1244+8755}
\figsetplot{J1244+8755_spectra.pdf}
\figsetgrpnote{Single-dish and VLBA broadband spectra of all the sources of our sample. The source name is specified at the top right of each plot. The single-dish spectra are plotted black, the VLBA spectra are plotted by the red color. In cases of the VLBA non-detection, the upper limits on the VLBA flux density are shown by red arrows.}
\figsetgrpend

\figsetgrpstart
\figsetgrpnum{7.250}
\figsetgrptitle{Total and VLBA spectra of J1245+7613}
\figsetplot{J1245+7613_spectra.pdf}
\figsetgrpnote{Single-dish and VLBA broadband spectra of all the sources of our sample. The source name is specified at the top right of each plot. The single-dish spectra are plotted black, the VLBA spectra are plotted by the red color. In cases of the VLBA non-detection, the upper limits on the VLBA flux density are shown by red arrows.}
\figsetgrpend

\figsetgrpstart
\figsetgrpnum{7.251}
\figsetgrptitle{Total and VLBA spectra of J1250+7557}
\figsetplot{J1250+7557_spectra.pdf}
\figsetgrpnote{Single-dish and VLBA broadband spectra of all the sources of our sample. The source name is specified at the top right of each plot. The single-dish spectra are plotted black, the VLBA spectra are plotted by the red color. In cases of the VLBA non-detection, the upper limits on the VLBA flux density are shown by red arrows.}
\figsetgrpend

\figsetgrpstart
\figsetgrpnum{7.252}
\figsetgrptitle{Total and VLBA spectra of J1257+7958}
\figsetplot{J1257+7958_spectra.pdf}
\figsetgrpnote{Single-dish and VLBA broadband spectra of all the sources of our sample. The source name is specified at the top right of each plot. The single-dish spectra are plotted black, the VLBA spectra are plotted by the red color. In cases of the VLBA non-detection, the upper limits on the VLBA flux density are shown by red arrows.}
\figsetgrpend

\figsetgrpstart
\figsetgrpnum{7.253}
\figsetgrptitle{Total and VLBA spectra of J1257+8342}
\figsetplot{J1257+8342_spectra.pdf}
\figsetgrpnote{Single-dish and VLBA broadband spectra of all the sources of our sample. The source name is specified at the top right of each plot. The single-dish spectra are plotted black, the VLBA spectra are plotted by the red color. In cases of the VLBA non-detection, the upper limits on the VLBA flux density are shown by red arrows.}
\figsetgrpend

\figsetgrpstart
\figsetgrpnum{7.254}
\figsetgrptitle{Total and VLBA spectra of J1300+8054}
\figsetplot{J1300+8054_spectra.pdf}
\figsetgrpnote{Single-dish and VLBA broadband spectra of all the sources of our sample. The source name is specified at the top right of each plot. The single-dish spectra are plotted black, the VLBA spectra are plotted by the red color. In cases of the VLBA non-detection, the upper limits on the VLBA flux density are shown by red arrows.}
\figsetgrpend

\figsetgrpstart
\figsetgrpnum{7.255}
\figsetgrptitle{Total and VLBA spectra of J1305+7854}
\figsetplot{J1305+7854_spectra.pdf}
\figsetgrpnote{Single-dish and VLBA broadband spectra of all the sources of our sample. The source name is specified at the top right of each plot. The single-dish spectra are plotted black, the VLBA spectra are plotted by the red color. In cases of the VLBA non-detection, the upper limits on the VLBA flux density are shown by red arrows.}
\figsetgrpend

\figsetgrpstart
\figsetgrpnum{7.256}
\figsetgrptitle{Total and VLBA spectra of J1305+8156}
\figsetplot{J1305+8156_spectra.pdf}
\figsetgrpnote{Single-dish and VLBA broadband spectra of all the sources of our sample. The source name is specified at the top right of each plot. The single-dish spectra are plotted black, the VLBA spectra are plotted by the red color. In cases of the VLBA non-detection, the upper limits on the VLBA flux density are shown by red arrows.}
\figsetgrpend

\figsetgrpstart
\figsetgrpnum{7.257}
\figsetgrptitle{Total and VLBA spectra of J1305+8216}
\figsetplot{J1305+8216_spectra.pdf}
\figsetgrpnote{Single-dish and VLBA broadband spectra of all the sources of our sample. The source name is specified at the top right of each plot. The single-dish spectra are plotted black, the VLBA spectra are plotted by the red color. In cases of the VLBA non-detection, the upper limits on the VLBA flux density are shown by red arrows.}
\figsetgrpend

\figsetgrpstart
\figsetgrpnum{7.258}
\figsetgrptitle{Total and VLBA spectra of J1306+8008}
\figsetplot{J1306+8008_spectra.pdf}
\figsetgrpnote{Single-dish and VLBA broadband spectra of all the sources of our sample. The source name is specified at the top right of each plot. The single-dish spectra are plotted black, the VLBA spectra are plotted by the red color. In cases of the VLBA non-detection, the upper limits on the VLBA flux density are shown by red arrows.}
\figsetgrpend

\figsetgrpstart
\figsetgrpnum{7.259}
\figsetgrptitle{Total and VLBA spectra of J1306+8019}
\figsetplot{J1306+8019_spectra.pdf}
\figsetgrpnote{Single-dish and VLBA broadband spectra of all the sources of our sample. The source name is specified at the top right of each plot. The single-dish spectra are plotted black, the VLBA spectra are plotted by the red color. In cases of the VLBA non-detection, the upper limits on the VLBA flux density are shown by red arrows.}
\figsetgrpend

\figsetgrpstart
\figsetgrpnum{7.260}
\figsetgrptitle{Total and VLBA spectra of J1307+7649}
\figsetplot{J1307+7649_spectra.pdf}
\figsetgrpnote{Single-dish and VLBA broadband spectra of all the sources of our sample. The source name is specified at the top right of each plot. The single-dish spectra are plotted black, the VLBA spectra are plotted by the red color. In cases of the VLBA non-detection, the upper limits on the VLBA flux density are shown by red arrows.}
\figsetgrpend

\figsetgrpstart
\figsetgrpnum{7.261}
\figsetgrptitle{Total and VLBA spectra of J1308+8544}
\figsetplot{J1308+8544_spectra.pdf}
\figsetgrpnote{Single-dish and VLBA broadband spectra of all the sources of our sample. The source name is specified at the top right of each plot. The single-dish spectra are plotted black, the VLBA spectra are plotted by the red color. In cases of the VLBA non-detection, the upper limits on the VLBA flux density are shown by red arrows.}
\figsetgrpend

\figsetgrpstart
\figsetgrpnum{7.262}
\figsetgrptitle{Total and VLBA spectra of J1311+7809}
\figsetplot{J1311+7809_spectra.pdf}
\figsetgrpnote{Single-dish and VLBA broadband spectra of all the sources of our sample. The source name is specified at the top right of each plot. The single-dish spectra are plotted black, the VLBA spectra are plotted by the red color. In cases of the VLBA non-detection, the upper limits on the VLBA flux density are shown by red arrows.}
\figsetgrpend

\figsetgrpstart
\figsetgrpnum{7.263}
\figsetgrptitle{Total and VLBA spectra of J1317+8219}
\figsetplot{J1317+8219_spectra.pdf}
\figsetgrpnote{Single-dish and VLBA broadband spectra of all the sources of our sample. The source name is specified at the top right of each plot. The single-dish spectra are plotted black, the VLBA spectra are plotted by the red color. In cases of the VLBA non-detection, the upper limits on the VLBA flux density are shown by red arrows.}
\figsetgrpend

\figsetgrpstart
\figsetgrpnum{7.264}
\figsetgrptitle{Total and VLBA spectra of J1320+8450}
\figsetplot{J1320+8450_spectra.pdf}
\figsetgrpnote{Single-dish and VLBA broadband spectra of all the sources of our sample. The source name is specified at the top right of each plot. The single-dish spectra are plotted black, the VLBA spectra are plotted by the red color. In cases of the VLBA non-detection, the upper limits on the VLBA flux density are shown by red arrows.}
\figsetgrpend

\figsetgrpstart
\figsetgrpnum{7.265}
\figsetgrptitle{Total and VLBA spectra of J1321+8316}
\figsetplot{J1321+8316_spectra.pdf}
\figsetgrpnote{Single-dish and VLBA broadband spectra of all the sources of our sample. The source name is specified at the top right of each plot. The single-dish spectra are plotted black, the VLBA spectra are plotted by the red color. In cases of the VLBA non-detection, the upper limits on the VLBA flux density are shown by red arrows.}
\figsetgrpend

\figsetgrpstart
\figsetgrpnum{7.266}
\figsetgrptitle{Total and VLBA spectra of J1323+7809}
\figsetplot{J1323+7809_spectra.pdf}
\figsetgrpnote{Single-dish and VLBA broadband spectra of all the sources of our sample. The source name is specified at the top right of each plot. The single-dish spectra are plotted black, the VLBA spectra are plotted by the red color. In cases of the VLBA non-detection, the upper limits on the VLBA flux density are shown by red arrows.}
\figsetgrpend

\figsetgrpstart
\figsetgrpnum{7.267}
\figsetgrptitle{Total and VLBA spectra of J1323+7942}
\figsetplot{J1323+7942_spectra.pdf}
\figsetgrpnote{Single-dish and VLBA broadband spectra of all the sources of our sample. The source name is specified at the top right of each plot. The single-dish spectra are plotted black, the VLBA spectra are plotted by the red color. In cases of the VLBA non-detection, the upper limits on the VLBA flux density are shown by red arrows.}
\figsetgrpend

\figsetgrpstart
\figsetgrpnum{7.268}
\figsetgrptitle{Total and VLBA spectra of J1325+7535}
\figsetplot{J1325+7535_spectra.pdf}
\figsetgrpnote{Single-dish and VLBA broadband spectra of all the sources of our sample. The source name is specified at the top right of each plot. The single-dish spectra are plotted black, the VLBA spectra are plotted by the red color. In cases of the VLBA non-detection, the upper limits on the VLBA flux density are shown by red arrows.}
\figsetgrpend

\figsetgrpstart
\figsetgrpnum{7.269}
\figsetgrptitle{Total and VLBA spectra of J1327+8221}
\figsetplot{J1327+8221_spectra.pdf}
\figsetgrpnote{Single-dish and VLBA broadband spectra of all the sources of our sample. The source name is specified at the top right of each plot. The single-dish spectra are plotted black, the VLBA spectra are plotted by the red color. In cases of the VLBA non-detection, the upper limits on the VLBA flux density are shown by red arrows.}
\figsetgrpend

\figsetgrpstart
\figsetgrpnum{7.270}
\figsetgrptitle{Total and VLBA spectra of J1332+8650}
\figsetplot{J1332+8650_spectra.pdf}
\figsetgrpnote{Single-dish and VLBA broadband spectra of all the sources of our sample. The source name is specified at the top right of each plot. The single-dish spectra are plotted black, the VLBA spectra are plotted by the red color. In cases of the VLBA non-detection, the upper limits on the VLBA flux density are shown by red arrows.}
\figsetgrpend

\figsetgrpstart
\figsetgrpnum{7.271}
\figsetgrptitle{Total and VLBA spectra of J1337+7914}
\figsetplot{J1337+7914_spectra.pdf}
\figsetgrpnote{Single-dish and VLBA broadband spectra of all the sources of our sample. The source name is specified at the top right of each plot. The single-dish spectra are plotted black, the VLBA spectra are plotted by the red color. In cases of the VLBA non-detection, the upper limits on the VLBA flux density are shown by red arrows.}
\figsetgrpend

\figsetgrpstart
\figsetgrpnum{7.272}
\figsetgrptitle{Total and VLBA spectra of J1337+8033}
\figsetplot{J1337+8033_spectra.pdf}
\figsetgrpnote{Single-dish and VLBA broadband spectra of all the sources of our sample. The source name is specified at the top right of each plot. The single-dish spectra are plotted black, the VLBA spectra are plotted by the red color. In cases of the VLBA non-detection, the upper limits on the VLBA flux density are shown by red arrows.}
\figsetgrpend

\figsetgrpstart
\figsetgrpnum{7.273}
\figsetgrptitle{Total and VLBA spectra of J1344+7907}
\figsetplot{J1344+7907_spectra.pdf}
\figsetgrpnote{Single-dish and VLBA broadband spectra of all the sources of our sample. The source name is specified at the top right of each plot. The single-dish spectra are plotted black, the VLBA spectra are plotted by the red color. In cases of the VLBA non-detection, the upper limits on the VLBA flux density are shown by red arrows.}
\figsetgrpend

\figsetgrpstart
\figsetgrpnum{7.274}
\figsetgrptitle{Total and VLBA spectra of J1351+7556}
\figsetplot{J1351+7556_spectra.pdf}
\figsetgrpnote{Single-dish and VLBA broadband spectra of all the sources of our sample. The source name is specified at the top right of each plot. The single-dish spectra are plotted black, the VLBA spectra are plotted by the red color. In cases of the VLBA non-detection, the upper limits on the VLBA flux density are shown by red arrows.}
\figsetgrpend

\figsetgrpstart
\figsetgrpnum{7.275}
\figsetgrptitle{Total and VLBA spectra of J1356+7943}
\figsetplot{J1356+7943_spectra.pdf}
\figsetgrpnote{Single-dish and VLBA broadband spectra of all the sources of our sample. The source name is specified at the top right of each plot. The single-dish spectra are plotted black, the VLBA spectra are plotted by the red color. In cases of the VLBA non-detection, the upper limits on the VLBA flux density are shown by red arrows.}
\figsetgrpend

\figsetgrpstart
\figsetgrpnum{7.276}
\figsetgrptitle{Total and VLBA spectra of J1357+7643}
\figsetplot{J1357+7643_spectra.pdf}
\figsetgrpnote{Single-dish and VLBA broadband spectra of all the sources of our sample. The source name is specified at the top right of each plot. The single-dish spectra are plotted black, the VLBA spectra are plotted by the red color. In cases of the VLBA non-detection, the upper limits on the VLBA flux density are shown by red arrows.}
\figsetgrpend

\figsetgrpstart
\figsetgrpnum{7.277}
\figsetgrptitle{Total and VLBA spectra of J1358+7855}
\figsetplot{J1358+7855_spectra.pdf}
\figsetgrpnote{Single-dish and VLBA broadband spectra of all the sources of our sample. The source name is specified at the top right of each plot. The single-dish spectra are plotted black, the VLBA spectra are plotted by the red color. In cases of the VLBA non-detection, the upper limits on the VLBA flux density are shown by red arrows.}
\figsetgrpend

\figsetgrpstart
\figsetgrpnum{7.278}
\figsetgrptitle{Total and VLBA spectra of J1358+8319}
\figsetplot{J1358+8319_spectra.pdf}
\figsetgrpnote{Single-dish and VLBA broadband spectra of all the sources of our sample. The source name is specified at the top right of each plot. The single-dish spectra are plotted black, the VLBA spectra are plotted by the red color. In cases of the VLBA non-detection, the upper limits on the VLBA flux density are shown by red arrows.}
\figsetgrpend

\figsetgrpstart
\figsetgrpnum{7.279}
\figsetgrptitle{Total and VLBA spectra of J1358+8340}
\figsetplot{J1358+8340_spectra.pdf}
\figsetgrpnote{Single-dish and VLBA broadband spectra of all the sources of our sample. The source name is specified at the top right of each plot. The single-dish spectra are plotted black, the VLBA spectra are plotted by the red color. In cases of the VLBA non-detection, the upper limits on the VLBA flux density are shown by red arrows.}
\figsetgrpend

\figsetgrpstart
\figsetgrpnum{7.280}
\figsetgrptitle{Total and VLBA spectra of J1400+8609}
\figsetplot{J1400+8609_spectra.pdf}
\figsetgrpnote{Single-dish and VLBA broadband spectra of all the sources of our sample. The source name is specified at the top right of each plot. The single-dish spectra are plotted black, the VLBA spectra are plotted by the red color. In cases of the VLBA non-detection, the upper limits on the VLBA flux density are shown by red arrows.}
\figsetgrpend

\figsetgrpstart
\figsetgrpnum{7.281}
\figsetgrptitle{Total and VLBA spectra of J1406+7828}
\figsetplot{J1406+7828_spectra.pdf}
\figsetgrpnote{Single-dish and VLBA broadband spectra of all the sources of our sample. The source name is specified at the top right of each plot. The single-dish spectra are plotted black, the VLBA spectra are plotted by the red color. In cases of the VLBA non-detection, the upper limits on the VLBA flux density are shown by red arrows.}
\figsetgrpend

\figsetgrpstart
\figsetgrpnum{7.282}
\figsetgrptitle{Total and VLBA spectra of J1414+7905}
\figsetplot{J1414+7905_spectra.pdf}
\figsetgrpnote{Single-dish and VLBA broadband spectra of all the sources of our sample. The source name is specified at the top right of each plot. The single-dish spectra are plotted black, the VLBA spectra are plotted by the red color. In cases of the VLBA non-detection, the upper limits on the VLBA flux density are shown by red arrows.}
\figsetgrpend

\figsetgrpstart
\figsetgrpnum{7.283}
\figsetgrptitle{Total and VLBA spectra of J1417+8059}
\figsetplot{J1417+8059_spectra.pdf}
\figsetgrpnote{Single-dish and VLBA broadband spectra of all the sources of our sample. The source name is specified at the top right of each plot. The single-dish spectra are plotted black, the VLBA spectra are plotted by the red color. In cases of the VLBA non-detection, the upper limits on the VLBA flux density are shown by red arrows.}
\figsetgrpend

\figsetgrpstart
\figsetgrpnum{7.284}
\figsetgrptitle{Total and VLBA spectra of J1417+8529}
\figsetplot{J1417+8529_spectra.pdf}
\figsetgrpnote{Single-dish and VLBA broadband spectra of all the sources of our sample. The source name is specified at the top right of each plot. The single-dish spectra are plotted black, the VLBA spectra are plotted by the red color. In cases of the VLBA non-detection, the upper limits on the VLBA flux density are shown by red arrows.}
\figsetgrpend

\figsetgrpstart
\figsetgrpnum{7.285}
\figsetgrptitle{Total and VLBA spectra of J1418+7810}
\figsetplot{J1418+7810_spectra.pdf}
\figsetgrpnote{Single-dish and VLBA broadband spectra of all the sources of our sample. The source name is specified at the top right of each plot. The single-dish spectra are plotted black, the VLBA spectra are plotted by the red color. In cases of the VLBA non-detection, the upper limits on the VLBA flux density are shown by red arrows.}
\figsetgrpend

\figsetgrpstart
\figsetgrpnum{7.286}
\figsetgrptitle{Total and VLBA spectra of J1419+7600}
\figsetplot{J1419+7600_spectra.pdf}
\figsetgrpnote{Single-dish and VLBA broadband spectra of all the sources of our sample. The source name is specified at the top right of each plot. The single-dish spectra are plotted black, the VLBA spectra are plotted by the red color. In cases of the VLBA non-detection, the upper limits on the VLBA flux density are shown by red arrows.}
\figsetgrpend

\figsetgrpstart
\figsetgrpnum{7.287}
\figsetgrptitle{Total and VLBA spectra of J1421+7513}
\figsetplot{J1421+7513_spectra.pdf}
\figsetgrpnote{Single-dish and VLBA broadband spectra of all the sources of our sample. The source name is specified at the top right of each plot. The single-dish spectra are plotted black, the VLBA spectra are plotted by the red color. In cases of the VLBA non-detection, the upper limits on the VLBA flux density are shown by red arrows.}
\figsetgrpend

\figsetgrpstart
\figsetgrpnum{7.288}
\figsetgrptitle{Total and VLBA spectra of J1422+7607}
\figsetplot{J1422+7607_spectra.pdf}
\figsetgrpnote{Single-dish and VLBA broadband spectra of all the sources of our sample. The source name is specified at the top right of each plot. The single-dish spectra are plotted black, the VLBA spectra are plotted by the red color. In cases of the VLBA non-detection, the upper limits on the VLBA flux density are shown by red arrows.}
\figsetgrpend

\figsetgrpstart
\figsetgrpnum{7.289}
\figsetgrptitle{Total and VLBA spectra of J1422+7704}
\figsetplot{J1422+7704_spectra.pdf}
\figsetgrpnote{Single-dish and VLBA broadband spectra of all the sources of our sample. The source name is specified at the top right of each plot. The single-dish spectra are plotted black, the VLBA spectra are plotted by the red color. In cases of the VLBA non-detection, the upper limits on the VLBA flux density are shown by red arrows.}
\figsetgrpend

\figsetgrpstart
\figsetgrpnum{7.290}
\figsetgrptitle{Total and VLBA spectra of J1426+7946}
\figsetplot{J1426+7946_spectra.pdf}
\figsetgrpnote{Single-dish and VLBA broadband spectra of all the sources of our sample. The source name is specified at the top right of each plot. The single-dish spectra are plotted black, the VLBA spectra are plotted by the red color. In cases of the VLBA non-detection, the upper limits on the VLBA flux density are shown by red arrows.}
\figsetgrpend

\figsetgrpstart
\figsetgrpnum{7.291}
\figsetgrptitle{Total and VLBA spectra of J1435+7605}
\figsetplot{J1435+7605_spectra.pdf}
\figsetgrpnote{Single-dish and VLBA broadband spectra of all the sources of our sample. The source name is specified at the top right of each plot. The single-dish spectra are plotted black, the VLBA spectra are plotted by the red color. In cases of the VLBA non-detection, the upper limits on the VLBA flux density are shown by red arrows.}
\figsetgrpend

\figsetgrpstart
\figsetgrpnum{7.292}
\figsetgrptitle{Total and VLBA spectra of J1436+8003}
\figsetplot{J1436+8003_spectra.pdf}
\figsetgrpnote{Single-dish and VLBA broadband spectra of all the sources of our sample. The source name is specified at the top right of each plot. The single-dish spectra are plotted black, the VLBA spectra are plotted by the red color. In cases of the VLBA non-detection, the upper limits on the VLBA flux density are shown by red arrows.}
\figsetgrpend

\figsetgrpstart
\figsetgrpnum{7.293}
\figsetgrptitle{Total and VLBA spectra of J1443+7707}
\figsetplot{J1443+7707_spectra.pdf}
\figsetgrpnote{Single-dish and VLBA broadband spectra of all the sources of our sample. The source name is specified at the top right of each plot. The single-dish spectra are plotted black, the VLBA spectra are plotted by the red color. In cases of the VLBA non-detection, the upper limits on the VLBA flux density are shown by red arrows.}
\figsetgrpend

\figsetgrpstart
\figsetgrpnum{7.294}
\figsetgrptitle{Total and VLBA spectra of J1444+7501}
\figsetplot{J1444+7501_spectra.pdf}
\figsetgrpnote{Single-dish and VLBA broadband spectra of all the sources of our sample. The source name is specified at the top right of each plot. The single-dish spectra are plotted black, the VLBA spectra are plotted by the red color. In cases of the VLBA non-detection, the upper limits on the VLBA flux density are shown by red arrows.}
\figsetgrpend

\figsetgrpstart
\figsetgrpnum{7.295}
\figsetgrptitle{Total and VLBA spectra of J1447+7656}
\figsetplot{J1447+7656_spectra.pdf}
\figsetgrpnote{Single-dish and VLBA broadband spectra of all the sources of our sample. The source name is specified at the top right of each plot. The single-dish spectra are plotted black, the VLBA spectra are plotted by the red color. In cases of the VLBA non-detection, the upper limits on the VLBA flux density are shown by red arrows.}
\figsetgrpend

\figsetgrpstart
\figsetgrpnum{7.296}
\figsetgrptitle{Total and VLBA spectra of J1453+8016}
\figsetplot{J1453+8016_spectra.pdf}
\figsetgrpnote{Single-dish and VLBA broadband spectra of all the sources of our sample. The source name is specified at the top right of each plot. The single-dish spectra are plotted black, the VLBA spectra are plotted by the red color. In cases of the VLBA non-detection, the upper limits on the VLBA flux density are shown by red arrows.}
\figsetgrpend

\figsetgrpstart
\figsetgrpnum{7.297}
\figsetgrptitle{Total and VLBA spectra of J1500+7518}
\figsetplot{J1500+7518_spectra.pdf}
\figsetgrpnote{Single-dish and VLBA broadband spectra of all the sources of our sample. The source name is specified at the top right of each plot. The single-dish spectra are plotted black, the VLBA spectra are plotted by the red color. In cases of the VLBA non-detection, the upper limits on the VLBA flux density are shown by red arrows.}
\figsetgrpend

\figsetgrpstart
\figsetgrpnum{7.298}
\figsetgrptitle{Total and VLBA spectra of J1502+8608}
\figsetplot{J1502+8608_spectra.pdf}
\figsetgrpnote{Single-dish and VLBA broadband spectra of all the sources of our sample. The source name is specified at the top right of each plot. The single-dish spectra are plotted black, the VLBA spectra are plotted by the red color. In cases of the VLBA non-detection, the upper limits on the VLBA flux density are shown by red arrows.}
\figsetgrpend

\figsetgrpstart
\figsetgrpnum{7.299}
\figsetgrptitle{Total and VLBA spectra of J1503+8542}
\figsetplot{J1503+8542_spectra.pdf}
\figsetgrpnote{Single-dish and VLBA broadband spectra of all the sources of our sample. The source name is specified at the top right of each plot. The single-dish spectra are plotted black, the VLBA spectra are plotted by the red color. In cases of the VLBA non-detection, the upper limits on the VLBA flux density are shown by red arrows.}
\figsetgrpend

\figsetgrpstart
\figsetgrpnum{7.300}
\figsetgrptitle{Total and VLBA spectra of J1504+8022}
\figsetplot{J1504+8022_spectra.pdf}
\figsetgrpnote{Single-dish and VLBA broadband spectra of all the sources of our sample. The source name is specified at the top right of each plot. The single-dish spectra are plotted black, the VLBA spectra are plotted by the red color. In cases of the VLBA non-detection, the upper limits on the VLBA flux density are shown by red arrows.}
\figsetgrpend

\figsetgrpstart
\figsetgrpnum{7.301}
\figsetgrptitle{Total and VLBA spectra of J1506+8319}
\figsetplot{J1506+8319_spectra.pdf}
\figsetgrpnote{Single-dish and VLBA broadband spectra of all the sources of our sample. The source name is specified at the top right of each plot. The single-dish spectra are plotted black, the VLBA spectra are plotted by the red color. In cases of the VLBA non-detection, the upper limits on the VLBA flux density are shown by red arrows.}
\figsetgrpend

\figsetgrpstart
\figsetgrpnum{7.302}
\figsetgrptitle{Total and VLBA spectra of J1508+7528}
\figsetplot{J1508+7528_spectra.pdf}
\figsetgrpnote{Single-dish and VLBA broadband spectra of all the sources of our sample. The source name is specified at the top right of each plot. The single-dish spectra are plotted black, the VLBA spectra are plotted by the red color. In cases of the VLBA non-detection, the upper limits on the VLBA flux density are shown by red arrows.}
\figsetgrpend

\figsetgrpstart
\figsetgrpnum{7.303}
\figsetgrptitle{Total and VLBA spectra of J1513+8143}
\figsetplot{J1513+8143_spectra.pdf}
\figsetgrpnote{Single-dish and VLBA broadband spectra of all the sources of our sample. The source name is specified at the top right of each plot. The single-dish spectra are plotted black, the VLBA spectra are plotted by the red color. In cases of the VLBA non-detection, the upper limits on the VLBA flux density are shown by red arrows.}
\figsetgrpend

\figsetgrpstart
\figsetgrpnum{7.304}
\figsetgrptitle{Total and VLBA spectra of J1523+7645}
\figsetplot{J1523+7645_spectra.pdf}
\figsetgrpnote{Single-dish and VLBA broadband spectra of all the sources of our sample. The source name is specified at the top right of each plot. The single-dish spectra are plotted black, the VLBA spectra are plotted by the red color. In cases of the VLBA non-detection, the upper limits on the VLBA flux density are shown by red arrows.}
\figsetgrpend

\figsetgrpstart
\figsetgrpnum{7.305}
\figsetgrptitle{Total and VLBA spectra of J1527+7850}
\figsetplot{J1527+7850_spectra.pdf}
\figsetgrpnote{Single-dish and VLBA broadband spectra of all the sources of our sample. The source name is specified at the top right of each plot. The single-dish spectra are plotted black, the VLBA spectra are plotted by the red color. In cases of the VLBA non-detection, the upper limits on the VLBA flux density are shown by red arrows.}
\figsetgrpend

\figsetgrpstart
\figsetgrpnum{7.306}
\figsetgrptitle{Total and VLBA spectra of J1528+8217}
\figsetplot{J1528+8217_spectra.pdf}
\figsetgrpnote{Single-dish and VLBA broadband spectra of all the sources of our sample. The source name is specified at the top right of each plot. The single-dish spectra are plotted black, the VLBA spectra are plotted by the red color. In cases of the VLBA non-detection, the upper limits on the VLBA flux density are shown by red arrows.}
\figsetgrpend

\figsetgrpstart
\figsetgrpnum{7.307}
\figsetgrptitle{Total and VLBA spectra of J1530+7844}
\figsetplot{J1530+7844_spectra.pdf}
\figsetgrpnote{Single-dish and VLBA broadband spectra of all the sources of our sample. The source name is specified at the top right of each plot. The single-dish spectra are plotted black, the VLBA spectra are plotted by the red color. In cases of the VLBA non-detection, the upper limits on the VLBA flux density are shown by red arrows.}
\figsetgrpend

\figsetgrpstart
\figsetgrpnum{7.308}
\figsetgrptitle{Total and VLBA spectra of J1530+8251}
\figsetplot{J1530+8251_spectra.pdf}
\figsetgrpnote{Single-dish and VLBA broadband spectra of all the sources of our sample. The source name is specified at the top right of each plot. The single-dish spectra are plotted black, the VLBA spectra are plotted by the red color. In cases of the VLBA non-detection, the upper limits on the VLBA flux density are shown by red arrows.}
\figsetgrpend

\figsetgrpstart
\figsetgrpnum{7.309}
\figsetgrptitle{Total and VLBA spectra of J1531+7706}
\figsetplot{J1531+7706_spectra.pdf}
\figsetgrpnote{Single-dish and VLBA broadband spectra of all the sources of our sample. The source name is specified at the top right of each plot. The single-dish spectra are plotted black, the VLBA spectra are plotted by the red color. In cases of the VLBA non-detection, the upper limits on the VLBA flux density are shown by red arrows.}
\figsetgrpend

\figsetgrpstart
\figsetgrpnum{7.310}
\figsetgrptitle{Total and VLBA spectra of J1537+8154}
\figsetplot{J1537+8154_spectra.pdf}
\figsetgrpnote{Single-dish and VLBA broadband spectra of all the sources of our sample. The source name is specified at the top right of each plot. The single-dish spectra are plotted black, the VLBA spectra are plotted by the red color. In cases of the VLBA non-detection, the upper limits on the VLBA flux density are shown by red arrows.}
\figsetgrpend

\figsetgrpstart
\figsetgrpnum{7.311}
\figsetgrptitle{Total and VLBA spectra of J1539+7814}
\figsetplot{J1539+7814_spectra.pdf}
\figsetgrpnote{Single-dish and VLBA broadband spectra of all the sources of our sample. The source name is specified at the top right of each plot. The single-dish spectra are plotted black, the VLBA spectra are plotted by the red color. In cases of the VLBA non-detection, the upper limits on the VLBA flux density are shown by red arrows.}
\figsetgrpend

\figsetgrpstart
\figsetgrpnum{7.312}
\figsetgrptitle{Total and VLBA spectra of J1553+7517}
\figsetplot{J1553+7517_spectra.pdf}
\figsetgrpnote{Single-dish and VLBA broadband spectra of all the sources of our sample. The source name is specified at the top right of each plot. The single-dish spectra are plotted black, the VLBA spectra are plotted by the red color. In cases of the VLBA non-detection, the upper limits on the VLBA flux density are shown by red arrows.}
\figsetgrpend

\figsetgrpstart
\figsetgrpnum{7.313}
\figsetgrptitle{Total and VLBA spectra of J1554+7938}
\figsetplot{J1554+7938_spectra.pdf}
\figsetgrpnote{Single-dish and VLBA broadband spectra of all the sources of our sample. The source name is specified at the top right of each plot. The single-dish spectra are plotted black, the VLBA spectra are plotted by the red color. In cases of the VLBA non-detection, the upper limits on the VLBA flux density are shown by red arrows.}
\figsetgrpend

\figsetgrpstart
\figsetgrpnum{7.314}
\figsetgrptitle{Total and VLBA spectra of J1559+7816}
\figsetplot{J1559+7816_spectra.pdf}
\figsetgrpnote{Single-dish and VLBA broadband spectra of all the sources of our sample. The source name is specified at the top right of each plot. The single-dish spectra are plotted black, the VLBA spectra are plotted by the red color. In cases of the VLBA non-detection, the upper limits on the VLBA flux density are shown by red arrows.}
\figsetgrpend

\figsetgrpstart
\figsetgrpnum{7.315}
\figsetgrptitle{Total and VLBA spectra of J1600+8747}
\figsetplot{J1600+8747_spectra.pdf}
\figsetgrpnote{Single-dish and VLBA broadband spectra of all the sources of our sample. The source name is specified at the top right of each plot. The single-dish spectra are plotted black, the VLBA spectra are plotted by the red color. In cases of the VLBA non-detection, the upper limits on the VLBA flux density are shown by red arrows.}
\figsetgrpend

\figsetgrpstart
\figsetgrpnum{7.316}
\figsetgrptitle{Total and VLBA spectra of J1602+8015}
\figsetplot{J1602+8015_spectra.pdf}
\figsetgrpnote{Single-dish and VLBA broadband spectra of all the sources of our sample. The source name is specified at the top right of each plot. The single-dish spectra are plotted black, the VLBA spectra are plotted by the red color. In cases of the VLBA non-detection, the upper limits on the VLBA flux density are shown by red arrows.}
\figsetgrpend

\figsetgrpstart
\figsetgrpnum{7.317}
\figsetgrptitle{Total and VLBA spectra of J1606+7658}
\figsetplot{J1606+7658_spectra.pdf}
\figsetgrpnote{Single-dish and VLBA broadband spectra of all the sources of our sample. The source name is specified at the top right of each plot. The single-dish spectra are plotted black, the VLBA spectra are plotted by the red color. In cases of the VLBA non-detection, the upper limits on the VLBA flux density are shown by red arrows.}
\figsetgrpend

\figsetgrpstart
\figsetgrpnum{7.318}
\figsetgrptitle{Total and VLBA spectra of J1607+8201}
\figsetplot{J1607+8201_spectra.pdf}
\figsetgrpnote{Single-dish and VLBA broadband spectra of all the sources of our sample. The source name is specified at the top right of each plot. The single-dish spectra are plotted black, the VLBA spectra are plotted by the red color. In cases of the VLBA non-detection, the upper limits on the VLBA flux density are shown by red arrows.}
\figsetgrpend

\figsetgrpstart
\figsetgrpnum{7.319}
\figsetgrptitle{Total and VLBA spectra of J1607+8501}
\figsetplot{J1607+8501_spectra.pdf}
\figsetgrpnote{Single-dish and VLBA broadband spectra of all the sources of our sample. The source name is specified at the top right of each plot. The single-dish spectra are plotted black, the VLBA spectra are plotted by the red color. In cases of the VLBA non-detection, the upper limits on the VLBA flux density are shown by red arrows.}
\figsetgrpend

\figsetgrpstart
\figsetgrpnum{7.320}
\figsetgrptitle{Total and VLBA spectra of J1609+7939}
\figsetplot{J1609+7939_spectra.pdf}
\figsetgrpnote{Single-dish and VLBA broadband spectra of all the sources of our sample. The source name is specified at the top right of each plot. The single-dish spectra are plotted black, the VLBA spectra are plotted by the red color. In cases of the VLBA non-detection, the upper limits on the VLBA flux density are shown by red arrows.}
\figsetgrpend

\figsetgrpstart
\figsetgrpnum{7.321}
\figsetgrptitle{Total and VLBA spectra of J1611+7911}
\figsetplot{J1611+7911_spectra.pdf}
\figsetgrpnote{Single-dish and VLBA broadband spectra of all the sources of our sample. The source name is specified at the top right of each plot. The single-dish spectra are plotted black, the VLBA spectra are plotted by the red color. In cases of the VLBA non-detection, the upper limits on the VLBA flux density are shown by red arrows.}
\figsetgrpend

\figsetgrpstart
\figsetgrpnum{7.322}
\figsetgrptitle{Total and VLBA spectra of J1617+7624}
\figsetplot{J1617+7624_spectra.pdf}
\figsetgrpnote{Single-dish and VLBA broadband spectra of all the sources of our sample. The source name is specified at the top right of each plot. The single-dish spectra are plotted black, the VLBA spectra are plotted by the red color. In cases of the VLBA non-detection, the upper limits on the VLBA flux density are shown by red arrows.}
\figsetgrpend

\figsetgrpstart
\figsetgrpnum{7.323}
\figsetgrptitle{Total and VLBA spectra of J1619+8549}
\figsetplot{J1619+8549_spectra.pdf}
\figsetgrpnote{Single-dish and VLBA broadband spectra of all the sources of our sample. The source name is specified at the top right of each plot. The single-dish spectra are plotted black, the VLBA spectra are plotted by the red color. In cases of the VLBA non-detection, the upper limits on the VLBA flux density are shown by red arrows.}
\figsetgrpend

\figsetgrpstart
\figsetgrpnum{7.324}
\figsetgrptitle{Total and VLBA spectra of J1624+8116}
\figsetplot{J1624+8116_spectra.pdf}
\figsetgrpnote{Single-dish and VLBA broadband spectra of all the sources of our sample. The source name is specified at the top right of each plot. The single-dish spectra are plotted black, the VLBA spectra are plotted by the red color. In cases of the VLBA non-detection, the upper limits on the VLBA flux density are shown by red arrows.}
\figsetgrpend

\figsetgrpstart
\figsetgrpnum{7.325}
\figsetgrptitle{Total and VLBA spectra of J1624+8209}
\figsetplot{J1624+8209_spectra.pdf}
\figsetgrpnote{Single-dish and VLBA broadband spectra of all the sources of our sample. The source name is specified at the top right of each plot. The single-dish spectra are plotted black, the VLBA spectra are plotted by the red color. In cases of the VLBA non-detection, the upper limits on the VLBA flux density are shown by red arrows.}
\figsetgrpend

\figsetgrpstart
\figsetgrpnum{7.326}
\figsetgrptitle{Total and VLBA spectra of J1625+7526}
\figsetplot{J1625+7526_spectra.pdf}
\figsetgrpnote{Single-dish and VLBA broadband spectra of all the sources of our sample. The source name is specified at the top right of each plot. The single-dish spectra are plotted black, the VLBA spectra are plotted by the red color. In cases of the VLBA non-detection, the upper limits on the VLBA flux density are shown by red arrows.}
\figsetgrpend

\figsetgrpstart
\figsetgrpnum{7.327}
\figsetgrptitle{Total and VLBA spectra of J1632+8232}
\figsetplot{J1632+8232_spectra.pdf}
\figsetgrpnote{Single-dish and VLBA broadband spectra of all the sources of our sample. The source name is specified at the top right of each plot. The single-dish spectra are plotted black, the VLBA spectra are plotted by the red color. In cases of the VLBA non-detection, the upper limits on the VLBA flux density are shown by red arrows.}
\figsetgrpend

\figsetgrpstart
\figsetgrpnum{7.328}
\figsetgrptitle{Total and VLBA spectra of J1639+8631}
\figsetplot{J1639+8631_spectra.pdf}
\figsetgrpnote{Single-dish and VLBA broadband spectra of all the sources of our sample. The source name is specified at the top right of each plot. The single-dish spectra are plotted black, the VLBA spectra are plotted by the red color. In cases of the VLBA non-detection, the upper limits on the VLBA flux density are shown by red arrows.}
\figsetgrpend

\figsetgrpstart
\figsetgrpnum{7.329}
\figsetgrptitle{Total and VLBA spectra of J1639+8805}
\figsetplot{J1639+8805_spectra.pdf}
\figsetgrpnote{Single-dish and VLBA broadband spectra of all the sources of our sample. The source name is specified at the top right of each plot. The single-dish spectra are plotted black, the VLBA spectra are plotted by the red color. In cases of the VLBA non-detection, the upper limits on the VLBA flux density are shown by red arrows.}
\figsetgrpend

\figsetgrpstart
\figsetgrpnum{7.330}
\figsetgrptitle{Total and VLBA spectra of J1644+7547}
\figsetplot{J1644+7547_spectra.pdf}
\figsetgrpnote{Single-dish and VLBA broadband spectra of all the sources of our sample. The source name is specified at the top right of each plot. The single-dish spectra are plotted black, the VLBA spectra are plotted by the red color. In cases of the VLBA non-detection, the upper limits on the VLBA flux density are shown by red arrows.}
\figsetgrpend

\figsetgrpstart
\figsetgrpnum{7.331}
\figsetgrptitle{Total and VLBA spectra of J1648+7546}
\figsetplot{J1648+7546_spectra.pdf}
\figsetgrpnote{Single-dish and VLBA broadband spectra of all the sources of our sample. The source name is specified at the top right of each plot. The single-dish spectra are plotted black, the VLBA spectra are plotted by the red color. In cases of the VLBA non-detection, the upper limits on the VLBA flux density are shown by red arrows.}
\figsetgrpend

\figsetgrpstart
\figsetgrpnum{7.332}
\figsetgrptitle{Total and VLBA spectra of J1651+7828}
\figsetplot{J1651+7828_spectra.pdf}
\figsetgrpnote{Single-dish and VLBA broadband spectra of all the sources of our sample. The source name is specified at the top right of each plot. The single-dish spectra are plotted black, the VLBA spectra are plotted by the red color. In cases of the VLBA non-detection, the upper limits on the VLBA flux density are shown by red arrows.}
\figsetgrpend

\figsetgrpstart
\figsetgrpnum{7.333}
\figsetgrptitle{Total and VLBA spectra of J1654+7628}
\figsetplot{J1654+7628_spectra.pdf}
\figsetgrpnote{Single-dish and VLBA broadband spectra of all the sources of our sample. The source name is specified at the top right of each plot. The single-dish spectra are plotted black, the VLBA spectra are plotted by the red color. In cases of the VLBA non-detection, the upper limits on the VLBA flux density are shown by red arrows.}
\figsetgrpend

\figsetgrpstart
\figsetgrpnum{7.334}
\figsetgrptitle{Total and VLBA spectra of J1655+7758}
\figsetplot{J1655+7758_spectra.pdf}
\figsetgrpnote{Single-dish and VLBA broadband spectra of all the sources of our sample. The source name is specified at the top right of each plot. The single-dish spectra are plotted black, the VLBA spectra are plotted by the red color. In cases of the VLBA non-detection, the upper limits on the VLBA flux density are shown by red arrows.}
\figsetgrpend

\figsetgrpstart
\figsetgrpnum{7.335}
\figsetgrptitle{Total and VLBA spectra of J1657+7928}
\figsetplot{J1657+7928_spectra.pdf}
\figsetgrpnote{Single-dish and VLBA broadband spectra of all the sources of our sample. The source name is specified at the top right of each plot. The single-dish spectra are plotted black, the VLBA spectra are plotted by the red color. In cases of the VLBA non-detection, the upper limits on the VLBA flux density are shown by red arrows.}
\figsetgrpend

\figsetgrpstart
\figsetgrpnum{7.336}
\figsetgrptitle{Total and VLBA spectra of J1701+7652}
\figsetplot{J1701+7652_spectra.pdf}
\figsetgrpnote{Single-dish and VLBA broadband spectra of all the sources of our sample. The source name is specified at the top right of each plot. The single-dish spectra are plotted black, the VLBA spectra are plotted by the red color. In cases of the VLBA non-detection, the upper limits on the VLBA flux density are shown by red arrows.}
\figsetgrpend

\figsetgrpstart
\figsetgrpnum{7.337}
\figsetgrptitle{Total and VLBA spectra of J1705+7756}
\figsetplot{J1705+7756_spectra.pdf}
\figsetgrpnote{Single-dish and VLBA broadband spectra of all the sources of our sample. The source name is specified at the top right of each plot. The single-dish spectra are plotted black, the VLBA spectra are plotted by the red color. In cases of the VLBA non-detection, the upper limits on the VLBA flux density are shown by red arrows.}
\figsetgrpend

\figsetgrpstart
\figsetgrpnum{7.338}
\figsetgrptitle{Total and VLBA spectra of J1705+8316}
\figsetplot{J1705+8316_spectra.pdf}
\figsetgrpnote{Single-dish and VLBA broadband spectra of all the sources of our sample. The source name is specified at the top right of each plot. The single-dish spectra are plotted black, the VLBA spectra are plotted by the red color. In cases of the VLBA non-detection, the upper limits on the VLBA flux density are shown by red arrows.}
\figsetgrpend

\figsetgrpstart
\figsetgrpnum{7.339}
\figsetgrptitle{Total and VLBA spectra of J1706+7707}
\figsetplot{J1706+7707_spectra.pdf}
\figsetgrpnote{Single-dish and VLBA broadband spectra of all the sources of our sample. The source name is specified at the top right of each plot. The single-dish spectra are plotted black, the VLBA spectra are plotted by the red color. In cases of the VLBA non-detection, the upper limits on the VLBA flux density are shown by red arrows.}
\figsetgrpend

\figsetgrpstart
\figsetgrpnum{7.340}
\figsetgrptitle{Total and VLBA spectra of J1708+8006}
\figsetplot{J1708+8006_spectra.pdf}
\figsetgrpnote{Single-dish and VLBA broadband spectra of all the sources of our sample. The source name is specified at the top right of each plot. The single-dish spectra are plotted black, the VLBA spectra are plotted by the red color. In cases of the VLBA non-detection, the upper limits on the VLBA flux density are shown by red arrows.}
\figsetgrpend

\figsetgrpstart
\figsetgrpnum{7.341}
\figsetgrptitle{Total and VLBA spectra of J1713+8657}
\figsetplot{J1713+8657_spectra.pdf}
\figsetgrpnote{Single-dish and VLBA broadband spectra of all the sources of our sample. The source name is specified at the top right of each plot. The single-dish spectra are plotted black, the VLBA spectra are plotted by the red color. In cases of the VLBA non-detection, the upper limits on the VLBA flux density are shown by red arrows.}
\figsetgrpend

\figsetgrpstart
\figsetgrpnum{7.342}
\figsetgrptitle{Total and VLBA spectra of J1714+7612}
\figsetplot{J1714+7612_spectra.pdf}
\figsetgrpnote{Single-dish and VLBA broadband spectra of all the sources of our sample. The source name is specified at the top right of each plot. The single-dish spectra are plotted black, the VLBA spectra are plotted by the red color. In cases of the VLBA non-detection, the upper limits on the VLBA flux density are shown by red arrows.}
\figsetgrpend

\figsetgrpstart
\figsetgrpnum{7.343}
\figsetgrptitle{Total and VLBA spectra of J1723+7533}
\figsetplot{J1723+7533_spectra.pdf}
\figsetgrpnote{Single-dish and VLBA broadband spectra of all the sources of our sample. The source name is specified at the top right of each plot. The single-dish spectra are plotted black, the VLBA spectra are plotted by the red color. In cases of the VLBA non-detection, the upper limits on the VLBA flux density are shown by red arrows.}
\figsetgrpend

\figsetgrpstart
\figsetgrpnum{7.344}
\figsetgrptitle{Total and VLBA spectra of J1723+7653}
\figsetplot{J1723+7653_spectra.pdf}
\figsetgrpnote{Single-dish and VLBA broadband spectra of all the sources of our sample. The source name is specified at the top right of each plot. The single-dish spectra are plotted black, the VLBA spectra are plotted by the red color. In cases of the VLBA non-detection, the upper limits on the VLBA flux density are shown by red arrows.}
\figsetgrpend

\figsetgrpstart
\figsetgrpnum{7.345}
\figsetgrptitle{Total and VLBA spectra of J1725+7708}
\figsetplot{J1725+7708_spectra.pdf}
\figsetgrpnote{Single-dish and VLBA broadband spectra of all the sources of our sample. The source name is specified at the top right of each plot. The single-dish spectra are plotted black, the VLBA spectra are plotted by the red color. In cases of the VLBA non-detection, the upper limits on the VLBA flux density are shown by red arrows.}
\figsetgrpend

\figsetgrpstart
\figsetgrpnum{7.346}
\figsetgrptitle{Total and VLBA spectra of J1725+7726}
\figsetplot{J1725+7726_spectra.pdf}
\figsetgrpnote{Single-dish and VLBA broadband spectra of all the sources of our sample. The source name is specified at the top right of each plot. The single-dish spectra are plotted black, the VLBA spectra are plotted by the red color. In cases of the VLBA non-detection, the upper limits on the VLBA flux density are shown by red arrows.}
\figsetgrpend

\figsetgrpstart
\figsetgrpnum{7.347}
\figsetgrptitle{Total and VLBA spectra of J1730+7848}
\figsetplot{J1730+7848_spectra.pdf}
\figsetgrpnote{Single-dish and VLBA broadband spectra of all the sources of our sample. The source name is specified at the top right of each plot. The single-dish spectra are plotted black, the VLBA spectra are plotted by the red color. In cases of the VLBA non-detection, the upper limits on the VLBA flux density are shown by red arrows.}
\figsetgrpend

\figsetgrpstart
\figsetgrpnum{7.348}
\figsetgrptitle{Total and VLBA spectra of J1730+7949}
\figsetplot{J1730+7949_spectra.pdf}
\figsetgrpnote{Single-dish and VLBA broadband spectra of all the sources of our sample. The source name is specified at the top right of each plot. The single-dish spectra are plotted black, the VLBA spectra are plotted by the red color. In cases of the VLBA non-detection, the upper limits on the VLBA flux density are shown by red arrows.}
\figsetgrpend

\figsetgrpstart
\figsetgrpnum{7.349}
\figsetgrptitle{Total and VLBA spectra of J1733+7758}
\figsetplot{J1733+7758_spectra.pdf}
\figsetgrpnote{Single-dish and VLBA broadband spectra of all the sources of our sample. The source name is specified at the top right of each plot. The single-dish spectra are plotted black, the VLBA spectra are plotted by the red color. In cases of the VLBA non-detection, the upper limits on the VLBA flux density are shown by red arrows.}
\figsetgrpend

\figsetgrpstart
\figsetgrpnum{7.350}
\figsetgrptitle{Total and VLBA spectra of J1737+8445}
\figsetplot{J1737+8445_spectra.pdf}
\figsetgrpnote{Single-dish and VLBA broadband spectra of all the sources of our sample. The source name is specified at the top right of each plot. The single-dish spectra are plotted black, the VLBA spectra are plotted by the red color. In cases of the VLBA non-detection, the upper limits on the VLBA flux density are shown by red arrows.}
\figsetgrpend

\figsetgrpstart
\figsetgrpnum{7.351}
\figsetgrptitle{Total and VLBA spectra of J1739+7553}
\figsetplot{J1739+7553_spectra.pdf}
\figsetgrpnote{Single-dish and VLBA broadband spectra of all the sources of our sample. The source name is specified at the top right of each plot. The single-dish spectra are plotted black, the VLBA spectra are plotted by the red color. In cases of the VLBA non-detection, the upper limits on the VLBA flux density are shown by red arrows.}
\figsetgrpend

\figsetgrpstart
\figsetgrpnum{7.352}
\figsetgrptitle{Total and VLBA spectra of J1743+8004}
\figsetplot{J1743+8004_spectra.pdf}
\figsetgrpnote{Single-dish and VLBA broadband spectra of all the sources of our sample. The source name is specified at the top right of each plot. The single-dish spectra are plotted black, the VLBA spectra are plotted by the red color. In cases of the VLBA non-detection, the upper limits on the VLBA flux density are shown by red arrows.}
\figsetgrpend

\figsetgrpstart
\figsetgrpnum{7.353}
\figsetgrptitle{Total and VLBA spectra of J1750+8147}
\figsetplot{J1750+8147_spectra.pdf}
\figsetgrpnote{Single-dish and VLBA broadband spectra of all the sources of our sample. The source name is specified at the top right of each plot. The single-dish spectra are plotted black, the VLBA spectra are plotted by the red color. In cases of the VLBA non-detection, the upper limits on the VLBA flux density are shown by red arrows.}
\figsetgrpend

\figsetgrpstart
\figsetgrpnum{7.354}
\figsetgrptitle{Total and VLBA spectra of J1759+7507}
\figsetplot{J1759+7507_spectra.pdf}
\figsetgrpnote{Single-dish and VLBA broadband spectra of all the sources of our sample. The source name is specified at the top right of each plot. The single-dish spectra are plotted black, the VLBA spectra are plotted by the red color. In cases of the VLBA non-detection, the upper limits on the VLBA flux density are shown by red arrows.}
\figsetgrpend

\figsetgrpstart
\figsetgrpnum{7.355}
\figsetgrptitle{Total and VLBA spectra of J1800+7828}
\figsetplot{J1800+7828_spectra.pdf}
\figsetgrpnote{Single-dish and VLBA broadband spectra of all the sources of our sample. The source name is specified at the top right of each plot. The single-dish spectra are plotted black, the VLBA spectra are plotted by the red color. In cases of the VLBA non-detection, the upper limits on the VLBA flux density are shown by red arrows.}
\figsetgrpend

\figsetgrpstart
\figsetgrpnum{7.356}
\figsetgrptitle{Total and VLBA spectra of J1803+7601}
\figsetplot{J1803+7601_spectra.pdf}
\figsetgrpnote{Single-dish and VLBA broadband spectra of all the sources of our sample. The source name is specified at the top right of each plot. The single-dish spectra are plotted black, the VLBA spectra are plotted by the red color. In cases of the VLBA non-detection, the upper limits on the VLBA flux density are shown by red arrows.}
\figsetgrpend

\figsetgrpstart
\figsetgrpnum{7.357}
\figsetgrptitle{Total and VLBA spectra of J1803+8219}
\figsetplot{J1803+8219_spectra.pdf}
\figsetgrpnote{Single-dish and VLBA broadband spectra of all the sources of our sample. The source name is specified at the top right of each plot. The single-dish spectra are plotted black, the VLBA spectra are plotted by the red color. In cases of the VLBA non-detection, the upper limits on the VLBA flux density are shown by red arrows.}
\figsetgrpend

\figsetgrpstart
\figsetgrpnum{7.358}
\figsetgrptitle{Total and VLBA spectra of J1809+7503}
\figsetplot{J1809+7503_spectra.pdf}
\figsetgrpnote{Single-dish and VLBA broadband spectra of all the sources of our sample. The source name is specified at the top right of each plot. The single-dish spectra are plotted black, the VLBA spectra are plotted by the red color. In cases of the VLBA non-detection, the upper limits on the VLBA flux density are shown by red arrows.}
\figsetgrpend

\figsetgrpstart
\figsetgrpnum{7.359}
\figsetgrptitle{Total and VLBA spectra of J1822+8000}
\figsetplot{J1822+8000_spectra.pdf}
\figsetgrpnote{Single-dish and VLBA broadband spectra of all the sources of our sample. The source name is specified at the top right of each plot. The single-dish spectra are plotted black, the VLBA spectra are plotted by the red color. In cases of the VLBA non-detection, the upper limits on the VLBA flux density are shown by red arrows.}
\figsetgrpend

\figsetgrpstart
\figsetgrpnum{7.360}
\figsetgrptitle{Total and VLBA spectra of J1822+8053}
\figsetplot{J1822+8053_spectra.pdf}
\figsetgrpnote{Single-dish and VLBA broadband spectra of all the sources of our sample. The source name is specified at the top right of each plot. The single-dish spectra are plotted black, the VLBA spectra are plotted by the red color. In cases of the VLBA non-detection, the upper limits on the VLBA flux density are shown by red arrows.}
\figsetgrpend

\figsetgrpstart
\figsetgrpnum{7.361}
\figsetgrptitle{Total and VLBA spectra of J1822+8257}
\figsetplot{J1822+8257_spectra.pdf}
\figsetgrpnote{Single-dish and VLBA broadband spectra of all the sources of our sample. The source name is specified at the top right of each plot. The single-dish spectra are plotted black, the VLBA spectra are plotted by the red color. In cases of the VLBA non-detection, the upper limits on the VLBA flux density are shown by red arrows.}
\figsetgrpend

\figsetgrpstart
\figsetgrpnum{7.362}
\figsetgrptitle{Total and VLBA spectra of J1823+7938}
\figsetplot{J1823+7938_spectra.pdf}
\figsetgrpnote{Single-dish and VLBA broadband spectra of all the sources of our sample. The source name is specified at the top right of each plot. The single-dish spectra are plotted black, the VLBA spectra are plotted by the red color. In cases of the VLBA non-detection, the upper limits on the VLBA flux density are shown by red arrows.}
\figsetgrpend

\figsetgrpstart
\figsetgrpnum{7.363}
\figsetgrptitle{Total and VLBA spectra of J1823+8212}
\figsetplot{J1823+8212_spectra.pdf}
\figsetgrpnote{Single-dish and VLBA broadband spectra of all the sources of our sample. The source name is specified at the top right of each plot. The single-dish spectra are plotted black, the VLBA spectra are plotted by the red color. In cases of the VLBA non-detection, the upper limits on the VLBA flux density are shown by red arrows.}
\figsetgrpend

\figsetgrpstart
\figsetgrpnum{7.364}
\figsetgrptitle{Total and VLBA spectra of J1824+7630}
\figsetplot{J1824+7630_spectra.pdf}
\figsetgrpnote{Single-dish and VLBA broadband spectra of all the sources of our sample. The source name is specified at the top right of each plot. The single-dish spectra are plotted black, the VLBA spectra are plotted by the red color. In cases of the VLBA non-detection, the upper limits on the VLBA flux density are shown by red arrows.}
\figsetgrpend

\figsetgrpstart
\figsetgrpnum{7.365}
\figsetgrptitle{Total and VLBA spectra of J1826+7944}
\figsetplot{J1826+7944_spectra.pdf}
\figsetgrpnote{Single-dish and VLBA broadband spectra of all the sources of our sample. The source name is specified at the top right of each plot. The single-dish spectra are plotted black, the VLBA spectra are plotted by the red color. In cases of the VLBA non-detection, the upper limits on the VLBA flux density are shown by red arrows.}
\figsetgrpend

\figsetgrpstart
\figsetgrpnum{7.366}
\figsetgrptitle{Total and VLBA spectra of J1829+7801}
\figsetplot{J1829+7801_spectra.pdf}
\figsetgrpnote{Single-dish and VLBA broadband spectra of all the sources of our sample. The source name is specified at the top right of each plot. The single-dish spectra are plotted black, the VLBA spectra are plotted by the red color. In cases of the VLBA non-detection, the upper limits on the VLBA flux density are shown by red arrows.}
\figsetgrpend

\figsetgrpstart
\figsetgrpnum{7.367}
\figsetgrptitle{Total and VLBA spectra of J1831+7901}
\figsetplot{J1831+7901_spectra.pdf}
\figsetgrpnote{Single-dish and VLBA broadband spectra of all the sources of our sample. The source name is specified at the top right of each plot. The single-dish spectra are plotted black, the VLBA spectra are plotted by the red color. In cases of the VLBA non-detection, the upper limits on the VLBA flux density are shown by red arrows.}
\figsetgrpend

\figsetgrpstart
\figsetgrpnum{7.368}
\figsetgrptitle{Total and VLBA spectra of J1831+8416}
\figsetplot{J1831+8416_spectra.pdf}
\figsetgrpnote{Single-dish and VLBA broadband spectra of all the sources of our sample. The source name is specified at the top right of each plot. The single-dish spectra are plotted black, the VLBA spectra are plotted by the red color. In cases of the VLBA non-detection, the upper limits on the VLBA flux density are shown by red arrows.}
\figsetgrpend

\figsetgrpstart
\figsetgrpnum{7.369}
\figsetgrptitle{Total and VLBA spectra of J1832+7515}
\figsetplot{J1832+7515_spectra.pdf}
\figsetgrpnote{Single-dish and VLBA broadband spectra of all the sources of our sample. The source name is specified at the top right of each plot. The single-dish spectra are plotted black, the VLBA spectra are plotted by the red color. In cases of the VLBA non-detection, the upper limits on the VLBA flux density are shown by red arrows.}
\figsetgrpend

\figsetgrpstart
\figsetgrpnum{7.370}
\figsetgrptitle{Total and VLBA spectra of J1832+8049}
\figsetplot{J1832+8049_spectra.pdf}
\figsetgrpnote{Single-dish and VLBA broadband spectra of all the sources of our sample. The source name is specified at the top right of each plot. The single-dish spectra are plotted black, the VLBA spectra are plotted by the red color. In cases of the VLBA non-detection, the upper limits on the VLBA flux density are shown by red arrows.}
\figsetgrpend

\figsetgrpstart
\figsetgrpnum{7.371}
\figsetgrptitle{Total and VLBA spectra of J1837+8514}
\figsetplot{J1837+8514_spectra.pdf}
\figsetgrpnote{Single-dish and VLBA broadband spectra of all the sources of our sample. The source name is specified at the top right of each plot. The single-dish spectra are plotted black, the VLBA spectra are plotted by the red color. In cases of the VLBA non-detection, the upper limits on the VLBA flux density are shown by red arrows.}
\figsetgrpend

\figsetgrpstart
\figsetgrpnum{7.372}
\figsetgrptitle{Total and VLBA spectra of J1841+7648}
\figsetplot{J1841+7648_spectra.pdf}
\figsetgrpnote{Single-dish and VLBA broadband spectra of all the sources of our sample. The source name is specified at the top right of each plot. The single-dish spectra are plotted black, the VLBA spectra are plotted by the red color. In cases of the VLBA non-detection, the upper limits on the VLBA flux density are shown by red arrows.}
\figsetgrpend

\figsetgrpstart
\figsetgrpnum{7.373}
\figsetgrptitle{Total and VLBA spectra of J1842+7946}
\figsetplot{J1842+7946_spectra.pdf}
\figsetgrpnote{Single-dish and VLBA broadband spectra of all the sources of our sample. The source name is specified at the top right of each plot. The single-dish spectra are plotted black, the VLBA spectra are plotted by the red color. In cases of the VLBA non-detection, the upper limits on the VLBA flux density are shown by red arrows.}
\figsetgrpend

\figsetgrpstart
\figsetgrpnum{7.374}
\figsetgrptitle{Total and VLBA spectra of J1845+7652}
\figsetplot{J1845+7652_spectra.pdf}
\figsetgrpnote{Single-dish and VLBA broadband spectra of all the sources of our sample. The source name is specified at the top right of each plot. The single-dish spectra are plotted black, the VLBA spectra are plotted by the red color. In cases of the VLBA non-detection, the upper limits on the VLBA flux density are shown by red arrows.}
\figsetgrpend

\figsetgrpstart
\figsetgrpnum{7.375}
\figsetgrptitle{Total and VLBA spectra of J1845+8150}
\figsetplot{J1845+8150_spectra.pdf}
\figsetgrpnote{Single-dish and VLBA broadband spectra of all the sources of our sample. The source name is specified at the top right of each plot. The single-dish spectra are plotted black, the VLBA spectra are plotted by the red color. In cases of the VLBA non-detection, the upper limits on the VLBA flux density are shown by red arrows.}
\figsetgrpend

\figsetgrpstart
\figsetgrpnum{7.376}
\figsetgrptitle{Total and VLBA spectra of J1849+7519}
\figsetplot{J1849+7519_spectra.pdf}
\figsetgrpnote{Single-dish and VLBA broadband spectra of all the sources of our sample. The source name is specified at the top right of each plot. The single-dish spectra are plotted black, the VLBA spectra are plotted by the red color. In cases of the VLBA non-detection, the upper limits on the VLBA flux density are shown by red arrows.}
\figsetgrpend

\figsetgrpstart
\figsetgrpnum{7.377}
\figsetgrptitle{Total and VLBA spectra of J1853+8437}
\figsetplot{J1853+8437_spectra.pdf}
\figsetgrpnote{Single-dish and VLBA broadband spectra of all the sources of our sample. The source name is specified at the top right of each plot. The single-dish spectra are plotted black, the VLBA spectra are plotted by the red color. In cases of the VLBA non-detection, the upper limits on the VLBA flux density are shown by red arrows.}
\figsetgrpend

\figsetgrpstart
\figsetgrpnum{7.378}
\figsetgrptitle{Total and VLBA spectra of J1853+8610}
\figsetplot{J1853+8610_spectra.pdf}
\figsetgrpnote{Single-dish and VLBA broadband spectra of all the sources of our sample. The source name is specified at the top right of each plot. The single-dish spectra are plotted black, the VLBA spectra are plotted by the red color. In cases of the VLBA non-detection, the upper limits on the VLBA flux density are shown by red arrows.}
\figsetgrpend

\figsetgrpstart
\figsetgrpnum{7.379}
\figsetgrptitle{Total and VLBA spectra of J1857+7746}
\figsetplot{J1857+7746_spectra.pdf}
\figsetgrpnote{Single-dish and VLBA broadband spectra of all the sources of our sample. The source name is specified at the top right of each plot. The single-dish spectra are plotted black, the VLBA spectra are plotted by the red color. In cases of the VLBA non-detection, the upper limits on the VLBA flux density are shown by red arrows.}
\figsetgrpend

\figsetgrpstart
\figsetgrpnum{7.380}
\figsetgrptitle{Total and VLBA spectra of J1858+7710}
\figsetplot{J1858+7710_spectra.pdf}
\figsetgrpnote{Single-dish and VLBA broadband spectra of all the sources of our sample. The source name is specified at the top right of each plot. The single-dish spectra are plotted black, the VLBA spectra are plotted by the red color. In cases of the VLBA non-detection, the upper limits on the VLBA flux density are shown by red arrows.}
\figsetgrpend

\figsetgrpstart
\figsetgrpnum{7.381}
\figsetgrptitle{Total and VLBA spectra of J1901+8623}
\figsetplot{J1901+8623_spectra.pdf}
\figsetgrpnote{Single-dish and VLBA broadband spectra of all the sources of our sample. The source name is specified at the top right of each plot. The single-dish spectra are plotted black, the VLBA spectra are plotted by the red color. In cases of the VLBA non-detection, the upper limits on the VLBA flux density are shown by red arrows.}
\figsetgrpend

\figsetgrpstart
\figsetgrpnum{7.382}
\figsetgrptitle{Total and VLBA spectra of J1903+8536}
\figsetplot{J1903+8536_spectra.pdf}
\figsetgrpnote{Single-dish and VLBA broadband spectra of all the sources of our sample. The source name is specified at the top right of each plot. The single-dish spectra are plotted black, the VLBA spectra are plotted by the red color. In cases of the VLBA non-detection, the upper limits on the VLBA flux density are shown by red arrows.}
\figsetgrpend

\figsetgrpstart
\figsetgrpnum{7.383}
\figsetgrptitle{Total and VLBA spectra of J1904+7632}
\figsetplot{J1904+7632_spectra.pdf}
\figsetgrpnote{Single-dish and VLBA broadband spectra of all the sources of our sample. The source name is specified at the top right of each plot. The single-dish spectra are plotted black, the VLBA spectra are plotted by the red color. In cases of the VLBA non-detection, the upper limits on the VLBA flux density are shown by red arrows.}
\figsetgrpend

\figsetgrpstart
\figsetgrpnum{7.384}
\figsetgrptitle{Total and VLBA spectra of J1904+7648}
\figsetplot{J1904+7648_spectra.pdf}
\figsetgrpnote{Single-dish and VLBA broadband spectra of all the sources of our sample. The source name is specified at the top right of each plot. The single-dish spectra are plotted black, the VLBA spectra are plotted by the red color. In cases of the VLBA non-detection, the upper limits on the VLBA flux density are shown by red arrows.}
\figsetgrpend

\figsetgrpstart
\figsetgrpnum{7.385}
\figsetgrptitle{Total and VLBA spectra of J1906+8100}
\figsetplot{J1906+8100_spectra.pdf}
\figsetgrpnote{Single-dish and VLBA broadband spectra of all the sources of our sample. The source name is specified at the top right of each plot. The single-dish spectra are plotted black, the VLBA spectra are plotted by the red color. In cases of the VLBA non-detection, the upper limits on the VLBA flux density are shown by red arrows.}
\figsetgrpend

\figsetgrpstart
\figsetgrpnum{7.386}
\figsetgrptitle{Total and VLBA spectra of J1909+7813}
\figsetplot{J1909+7813_spectra.pdf}
\figsetgrpnote{Single-dish and VLBA broadband spectra of all the sources of our sample. The source name is specified at the top right of each plot. The single-dish spectra are plotted black, the VLBA spectra are plotted by the red color. In cases of the VLBA non-detection, the upper limits on the VLBA flux density are shown by red arrows.}
\figsetgrpend

\figsetgrpstart
\figsetgrpnum{7.387}
\figsetgrptitle{Total and VLBA spectra of J1912+8026}
\figsetplot{J1912+8026_spectra.pdf}
\figsetgrpnote{Single-dish and VLBA broadband spectra of all the sources of our sample. The source name is specified at the top right of each plot. The single-dish spectra are plotted black, the VLBA spectra are plotted by the red color. In cases of the VLBA non-detection, the upper limits on the VLBA flux density are shown by red arrows.}
\figsetgrpend

\figsetgrpstart
\figsetgrpnum{7.388}
\figsetgrptitle{Total and VLBA spectra of J1922+7755}
\figsetplot{J1922+7755_spectra.pdf}
\figsetgrpnote{Single-dish and VLBA broadband spectra of all the sources of our sample. The source name is specified at the top right of each plot. The single-dish spectra are plotted black, the VLBA spectra are plotted by the red color. In cases of the VLBA non-detection, the upper limits on the VLBA flux density are shown by red arrows.}
\figsetgrpend

\figsetgrpstart
\figsetgrpnum{7.389}
\figsetgrptitle{Total and VLBA spectra of J1922+7918}
\figsetplot{J1922+7918_spectra.pdf}
\figsetgrpnote{Single-dish and VLBA broadband spectra of all the sources of our sample. The source name is specified at the top right of each plot. The single-dish spectra are plotted black, the VLBA spectra are plotted by the red color. In cases of the VLBA non-detection, the upper limits on the VLBA flux density are shown by red arrows.}
\figsetgrpend

\figsetgrpstart
\figsetgrpnum{7.390}
\figsetgrptitle{Total and VLBA spectra of J1934+7956}
\figsetplot{J1934+7956_spectra.pdf}
\figsetgrpnote{Single-dish and VLBA broadband spectra of all the sources of our sample. The source name is specified at the top right of each plot. The single-dish spectra are plotted black, the VLBA spectra are plotted by the red color. In cases of the VLBA non-detection, the upper limits on the VLBA flux density are shown by red arrows.}
\figsetgrpend

\figsetgrpstart
\figsetgrpnum{7.391}
\figsetgrptitle{Total and VLBA spectra of J1935+8130}
\figsetplot{J1935+8130_spectra.pdf}
\figsetgrpnote{Single-dish and VLBA broadband spectra of all the sources of our sample. The source name is specified at the top right of each plot. The single-dish spectra are plotted black, the VLBA spectra are plotted by the red color. In cases of the VLBA non-detection, the upper limits on the VLBA flux density are shown by red arrows.}
\figsetgrpend

\figsetgrpstart
\figsetgrpnum{7.392}
\figsetgrptitle{Total and VLBA spectra of J1936+7516}
\figsetplot{J1936+7516_spectra.pdf}
\figsetgrpnote{Single-dish and VLBA broadband spectra of all the sources of our sample. The source name is specified at the top right of each plot. The single-dish spectra are plotted black, the VLBA spectra are plotted by the red color. In cases of the VLBA non-detection, the upper limits on the VLBA flux density are shown by red arrows.}
\figsetgrpend

\figsetgrpstart
\figsetgrpnum{7.393}
\figsetgrptitle{Total and VLBA spectra of J1937+8356}
\figsetplot{J1937+8356_spectra.pdf}
\figsetgrpnote{Single-dish and VLBA broadband spectra of all the sources of our sample. The source name is specified at the top right of each plot. The single-dish spectra are plotted black, the VLBA spectra are plotted by the red color. In cases of the VLBA non-detection, the upper limits on the VLBA flux density are shown by red arrows.}
\figsetgrpend

\figsetgrpstart
\figsetgrpnum{7.394}
\figsetgrptitle{Total and VLBA spectra of J1939+7737}
\figsetplot{J1939+7737_spectra.pdf}
\figsetgrpnote{Single-dish and VLBA broadband spectra of all the sources of our sample. The source name is specified at the top right of each plot. The single-dish spectra are plotted black, the VLBA spectra are plotted by the red color. In cases of the VLBA non-detection, the upper limits on the VLBA flux density are shown by red arrows.}
\figsetgrpend

\figsetgrpstart
\figsetgrpnum{7.395}
\figsetgrptitle{Total and VLBA spectra of J1941+8501}
\figsetplot{J1941+8501_spectra.pdf}
\figsetgrpnote{Single-dish and VLBA broadband spectra of all the sources of our sample. The source name is specified at the top right of each plot. The single-dish spectra are plotted black, the VLBA spectra are plotted by the red color. In cases of the VLBA non-detection, the upper limits on the VLBA flux density are shown by red arrows.}
\figsetgrpend

\figsetgrpstart
\figsetgrpnum{7.396}
\figsetgrptitle{Total and VLBA spectra of J1943+7632}
\figsetplot{J1943+7632_spectra.pdf}
\figsetgrpnote{Single-dish and VLBA broadband spectra of all the sources of our sample. The source name is specified at the top right of each plot. The single-dish spectra are plotted black, the VLBA spectra are plotted by the red color. In cases of the VLBA non-detection, the upper limits on the VLBA flux density are shown by red arrows.}
\figsetgrpend

\figsetgrpstart
\figsetgrpnum{7.397}
\figsetgrptitle{Total and VLBA spectra of J1943+7841}
\figsetplot{J1943+7841_spectra.pdf}
\figsetgrpnote{Single-dish and VLBA broadband spectra of all the sources of our sample. The source name is specified at the top right of each plot. The single-dish spectra are plotted black, the VLBA spectra are plotted by the red color. In cases of the VLBA non-detection, the upper limits on the VLBA flux density are shown by red arrows.}
\figsetgrpend

\figsetgrpstart
\figsetgrpnum{7.398}
\figsetgrptitle{Total and VLBA spectra of J1943+7858}
\figsetplot{J1943+7858_spectra.pdf}
\figsetgrpnote{Single-dish and VLBA broadband spectra of all the sources of our sample. The source name is specified at the top right of each plot. The single-dish spectra are plotted black, the VLBA spectra are plotted by the red color. In cases of the VLBA non-detection, the upper limits on the VLBA flux density are shown by red arrows.}
\figsetgrpend

\figsetgrpstart
\figsetgrpnum{7.399}
\figsetgrptitle{Total and VLBA spectra of J1944+7816}
\figsetplot{J1944+7816_spectra.pdf}
\figsetgrpnote{Single-dish and VLBA broadband spectra of all the sources of our sample. The source name is specified at the top right of each plot. The single-dish spectra are plotted black, the VLBA spectra are plotted by the red color. In cases of the VLBA non-detection, the upper limits on the VLBA flux density are shown by red arrows.}
\figsetgrpend

\figsetgrpstart
\figsetgrpnum{7.400}
\figsetgrptitle{Total and VLBA spectra of J1945+7515}
\figsetplot{J1945+7515_spectra.pdf}
\figsetgrpnote{Single-dish and VLBA broadband spectra of all the sources of our sample. The source name is specified at the top right of each plot. The single-dish spectra are plotted black, the VLBA spectra are plotted by the red color. In cases of the VLBA non-detection, the upper limits on the VLBA flux density are shown by red arrows.}
\figsetgrpend

\figsetgrpstart
\figsetgrpnum{7.401}
\figsetgrptitle{Total and VLBA spectra of J1945+8257}
\figsetplot{J1945+8257_spectra.pdf}
\figsetgrpnote{Single-dish and VLBA broadband spectra of all the sources of our sample. The source name is specified at the top right of each plot. The single-dish spectra are plotted black, the VLBA spectra are plotted by the red color. In cases of the VLBA non-detection, the upper limits on the VLBA flux density are shown by red arrows.}
\figsetgrpend

\figsetgrpstart
\figsetgrpnum{7.402}
\figsetgrptitle{Total and VLBA spectra of J1949+7654}
\figsetplot{J1949+7654_spectra.pdf}
\figsetgrpnote{Single-dish and VLBA broadband spectra of all the sources of our sample. The source name is specified at the top right of each plot. The single-dish spectra are plotted black, the VLBA spectra are plotted by the red color. In cases of the VLBA non-detection, the upper limits on the VLBA flux density are shown by red arrows.}
\figsetgrpend

\figsetgrpstart
\figsetgrpnum{7.403}
\figsetgrptitle{Total and VLBA spectra of J1957+8417}
\figsetplot{J1957+8417_spectra.pdf}
\figsetgrpnote{Single-dish and VLBA broadband spectra of all the sources of our sample. The source name is specified at the top right of each plot. The single-dish spectra are plotted black, the VLBA spectra are plotted by the red color. In cases of the VLBA non-detection, the upper limits on the VLBA flux density are shown by red arrows.}
\figsetgrpend

\figsetgrpstart
\figsetgrpnum{7.404}
\figsetgrptitle{Total and VLBA spectra of J1958+7711}
\figsetplot{J1958+7711_spectra.pdf}
\figsetgrpnote{Single-dish and VLBA broadband spectra of all the sources of our sample. The source name is specified at the top right of each plot. The single-dish spectra are plotted black, the VLBA spectra are plotted by the red color. In cases of the VLBA non-detection, the upper limits on the VLBA flux density are shown by red arrows.}
\figsetgrpend

\figsetgrpstart
\figsetgrpnum{7.405}
\figsetgrptitle{Total and VLBA spectra of J2000+8213}
\figsetplot{J2000+8213_spectra.pdf}
\figsetgrpnote{Single-dish and VLBA broadband spectra of all the sources of our sample. The source name is specified at the top right of each plot. The single-dish spectra are plotted black, the VLBA spectra are plotted by the red color. In cases of the VLBA non-detection, the upper limits on the VLBA flux density are shown by red arrows.}
\figsetgrpend

\figsetgrpstart
\figsetgrpnum{7.406}
\figsetgrptitle{Total and VLBA spectra of J2003+7814}
\figsetplot{J2003+7814_spectra.pdf}
\figsetgrpnote{Single-dish and VLBA broadband spectra of all the sources of our sample. The source name is specified at the top right of each plot. The single-dish spectra are plotted black, the VLBA spectra are plotted by the red color. In cases of the VLBA non-detection, the upper limits on the VLBA flux density are shown by red arrows.}
\figsetgrpend

\figsetgrpstart
\figsetgrpnum{7.407}
\figsetgrptitle{Total and VLBA spectra of J2004+7623}
\figsetplot{J2004+7623_spectra.pdf}
\figsetgrpnote{Single-dish and VLBA broadband spectra of all the sources of our sample. The source name is specified at the top right of each plot. The single-dish spectra are plotted black, the VLBA spectra are plotted by the red color. In cases of the VLBA non-detection, the upper limits on the VLBA flux density are shown by red arrows.}
\figsetgrpend

\figsetgrpstart
\figsetgrpnum{7.408}
\figsetgrptitle{Total and VLBA spectra of J2005+7752}
\figsetplot{J2005+7752_spectra.pdf}
\figsetgrpnote{Single-dish and VLBA broadband spectra of all the sources of our sample. The source name is specified at the top right of each plot. The single-dish spectra are plotted black, the VLBA spectra are plotted by the red color. In cases of the VLBA non-detection, the upper limits on the VLBA flux density are shown by red arrows.}
\figsetgrpend

\figsetgrpstart
\figsetgrpnum{7.409}
\figsetgrptitle{Total and VLBA spectra of J2007+7942}
\figsetplot{J2007+7942_spectra.pdf}
\figsetgrpnote{Single-dish and VLBA broadband spectra of all the sources of our sample. The source name is specified at the top right of each plot. The single-dish spectra are plotted black, the VLBA spectra are plotted by the red color. In cases of the VLBA non-detection, the upper limits on the VLBA flux density are shown by red arrows.}
\figsetgrpend

\figsetgrpstart
\figsetgrpnum{7.410}
\figsetgrptitle{Total and VLBA spectra of J2011+8811}
\figsetplot{J2011+8811_spectra.pdf}
\figsetgrpnote{Single-dish and VLBA broadband spectra of all the sources of our sample. The source name is specified at the top right of each plot. The single-dish spectra are plotted black, the VLBA spectra are plotted by the red color. In cases of the VLBA non-detection, the upper limits on the VLBA flux density are shown by red arrows.}
\figsetgrpend

\figsetgrpstart
\figsetgrpnum{7.411}
\figsetgrptitle{Total and VLBA spectra of J2013+7646}
\figsetplot{J2013+7646_spectra.pdf}
\figsetgrpnote{Single-dish and VLBA broadband spectra of all the sources of our sample. The source name is specified at the top right of each plot. The single-dish spectra are plotted black, the VLBA spectra are plotted by the red color. In cases of the VLBA non-detection, the upper limits on the VLBA flux density are shown by red arrows.}
\figsetgrpend

\figsetgrpstart
\figsetgrpnum{7.412}
\figsetgrptitle{Total and VLBA spectra of J2021+7833}
\figsetplot{J2021+7833_spectra.pdf}
\figsetgrpnote{Single-dish and VLBA broadband spectra of all the sources of our sample. The source name is specified at the top right of each plot. The single-dish spectra are plotted black, the VLBA spectra are plotted by the red color. In cases of the VLBA non-detection, the upper limits on the VLBA flux density are shown by red arrows.}
\figsetgrpend

\figsetgrpstart
\figsetgrpnum{7.413}
\figsetgrptitle{Total and VLBA spectra of J2022+7611}
\figsetplot{J2022+7611_spectra.pdf}
\figsetgrpnote{Single-dish and VLBA broadband spectra of all the sources of our sample. The source name is specified at the top right of each plot. The single-dish spectra are plotted black, the VLBA spectra are plotted by the red color. In cases of the VLBA non-detection, the upper limits on the VLBA flux density are shown by red arrows.}
\figsetgrpend

\figsetgrpstart
\figsetgrpnum{7.414}
\figsetgrptitle{Total and VLBA spectra of J2023+8354}
\figsetplot{J2023+8354_spectra.pdf}
\figsetgrpnote{Single-dish and VLBA broadband spectra of all the sources of our sample. The source name is specified at the top right of each plot. The single-dish spectra are plotted black, the VLBA spectra are plotted by the red color. In cases of the VLBA non-detection, the upper limits on the VLBA flux density are shown by red arrows.}
\figsetgrpend

\figsetgrpstart
\figsetgrpnum{7.415}
\figsetgrptitle{Total and VLBA spectra of J2024+8407}
\figsetplot{J2024+8407_spectra.pdf}
\figsetgrpnote{Single-dish and VLBA broadband spectra of all the sources of our sample. The source name is specified at the top right of each plot. The single-dish spectra are plotted black, the VLBA spectra are plotted by the red color. In cases of the VLBA non-detection, the upper limits on the VLBA flux density are shown by red arrows.}
\figsetgrpend

\figsetgrpstart
\figsetgrpnum{7.416}
\figsetgrptitle{Total and VLBA spectra of J2039+7612}
\figsetplot{J2039+7612_spectra.pdf}
\figsetgrpnote{Single-dish and VLBA broadband spectra of all the sources of our sample. The source name is specified at the top right of each plot. The single-dish spectra are plotted black, the VLBA spectra are plotted by the red color. In cases of the VLBA non-detection, the upper limits on the VLBA flux density are shown by red arrows.}
\figsetgrpend

\figsetgrpstart
\figsetgrpnum{7.417}
\figsetgrptitle{Total and VLBA spectra of J2042+7508}
\figsetplot{J2042+7508_spectra.pdf}
\figsetgrpnote{Single-dish and VLBA broadband spectra of all the sources of our sample. The source name is specified at the top right of each plot. The single-dish spectra are plotted black, the VLBA spectra are plotted by the red color. In cases of the VLBA non-detection, the upper limits on the VLBA flux density are shown by red arrows.}
\figsetgrpend

\figsetgrpstart
\figsetgrpnum{7.418}
\figsetgrptitle{Total and VLBA spectra of J2043+8635}
\figsetplot{J2043+8635_spectra.pdf}
\figsetgrpnote{Single-dish and VLBA broadband spectra of all the sources of our sample. The source name is specified at the top right of each plot. The single-dish spectra are plotted black, the VLBA spectra are plotted by the red color. In cases of the VLBA non-detection, the upper limits on the VLBA flux density are shown by red arrows.}
\figsetgrpend

\figsetgrpstart
\figsetgrpnum{7.419}
\figsetgrptitle{Total and VLBA spectra of J2045+7625}
\figsetplot{J2045+7625_spectra.pdf}
\figsetgrpnote{Single-dish and VLBA broadband spectra of all the sources of our sample. The source name is specified at the top right of each plot. The single-dish spectra are plotted black, the VLBA spectra are plotted by the red color. In cases of the VLBA non-detection, the upper limits on the VLBA flux density are shown by red arrows.}
\figsetgrpend

\figsetgrpstart
\figsetgrpnum{7.420}
\figsetgrptitle{Total and VLBA spectra of J2050+7526}
\figsetplot{J2050+7526_spectra.pdf}
\figsetgrpnote{Single-dish and VLBA broadband spectra of all the sources of our sample. The source name is specified at the top right of each plot. The single-dish spectra are plotted black, the VLBA spectra are plotted by the red color. In cases of the VLBA non-detection, the upper limits on the VLBA flux density are shown by red arrows.}
\figsetgrpend

\figsetgrpstart
\figsetgrpnum{7.421}
\figsetgrptitle{Total and VLBA spectra of J2056+7514}
\figsetplot{J2056+7514_spectra.pdf}
\figsetgrpnote{Single-dish and VLBA broadband spectra of all the sources of our sample. The source name is specified at the top right of each plot. The single-dish spectra are plotted black, the VLBA spectra are plotted by the red color. In cases of the VLBA non-detection, the upper limits on the VLBA flux density are shown by red arrows.}
\figsetgrpend

\figsetgrpstart
\figsetgrpnum{7.422}
\figsetgrptitle{Total and VLBA spectra of J2102+7653}
\figsetplot{J2102+7653_spectra.pdf}
\figsetgrpnote{Single-dish and VLBA broadband spectra of all the sources of our sample. The source name is specified at the top right of each plot. The single-dish spectra are plotted black, the VLBA spectra are plotted by the red color. In cases of the VLBA non-detection, the upper limits on the VLBA flux density are shown by red arrows.}
\figsetgrpend

\figsetgrpstart
\figsetgrpnum{7.423}
\figsetgrptitle{Total and VLBA spectra of J2104+7633}
\figsetplot{J2104+7633_spectra.pdf}
\figsetgrpnote{Single-dish and VLBA broadband spectra of all the sources of our sample. The source name is specified at the top right of each plot. The single-dish spectra are plotted black, the VLBA spectra are plotted by the red color. In cases of the VLBA non-detection, the upper limits on the VLBA flux density are shown by red arrows.}
\figsetgrpend

\figsetgrpstart
\figsetgrpnum{7.424}
\figsetgrptitle{Total and VLBA spectra of J2108+7752}
\figsetplot{J2108+7752_spectra.pdf}
\figsetgrpnote{Single-dish and VLBA broadband spectra of all the sources of our sample. The source name is specified at the top right of each plot. The single-dish spectra are plotted black, the VLBA spectra are plotted by the red color. In cases of the VLBA non-detection, the upper limits on the VLBA flux density are shown by red arrows.}
\figsetgrpend

\figsetgrpstart
\figsetgrpnum{7.425}
\figsetgrptitle{Total and VLBA spectra of J2109+8021}
\figsetplot{J2109+8021_spectra.pdf}
\figsetgrpnote{Single-dish and VLBA broadband spectra of all the sources of our sample. The source name is specified at the top right of each plot. The single-dish spectra are plotted black, the VLBA spectra are plotted by the red color. In cases of the VLBA non-detection, the upper limits on the VLBA flux density are shown by red arrows.}
\figsetgrpend

\figsetgrpstart
\figsetgrpnum{7.426}
\figsetgrptitle{Total and VLBA spectra of J2109+8255}
\figsetplot{J2109+8255_spectra.pdf}
\figsetgrpnote{Single-dish and VLBA broadband spectra of all the sources of our sample. The source name is specified at the top right of each plot. The single-dish spectra are plotted black, the VLBA spectra are plotted by the red color. In cases of the VLBA non-detection, the upper limits on the VLBA flux density are shown by red arrows.}
\figsetgrpend

\figsetgrpstart
\figsetgrpnum{7.427}
\figsetgrptitle{Total and VLBA spectra of J2113+7720}
\figsetplot{J2113+7720_spectra.pdf}
\figsetgrpnote{Single-dish and VLBA broadband spectra of all the sources of our sample. The source name is specified at the top right of each plot. The single-dish spectra are plotted black, the VLBA spectra are plotted by the red color. In cases of the VLBA non-detection, the upper limits on the VLBA flux density are shown by red arrows.}
\figsetgrpend

\figsetgrpstart
\figsetgrpnum{7.428}
\figsetgrptitle{Total and VLBA spectra of J2114+8204}
\figsetplot{J2114+8204_spectra.pdf}
\figsetgrpnote{Single-dish and VLBA broadband spectra of all the sources of our sample. The source name is specified at the top right of each plot. The single-dish spectra are plotted black, the VLBA spectra are plotted by the red color. In cases of the VLBA non-detection, the upper limits on the VLBA flux density are shown by red arrows.}
\figsetgrpend

\figsetgrpstart
\figsetgrpnum{7.429}
\figsetgrptitle{Total and VLBA spectra of J2118+7511}
\figsetplot{J2118+7511_spectra.pdf}
\figsetgrpnote{Single-dish and VLBA broadband spectra of all the sources of our sample. The source name is specified at the top right of each plot. The single-dish spectra are plotted black, the VLBA spectra are plotted by the red color. In cases of the VLBA non-detection, the upper limits on the VLBA flux density are shown by red arrows.}
\figsetgrpend

\figsetgrpstart
\figsetgrpnum{7.430}
\figsetgrptitle{Total and VLBA spectra of J2119+7657}
\figsetplot{J2119+7657_spectra.pdf}
\figsetgrpnote{Single-dish and VLBA broadband spectra of all the sources of our sample. The source name is specified at the top right of each plot. The single-dish spectra are plotted black, the VLBA spectra are plotted by the red color. In cases of the VLBA non-detection, the upper limits on the VLBA flux density are shown by red arrows.}
\figsetgrpend

\figsetgrpstart
\figsetgrpnum{7.431}
\figsetgrptitle{Total and VLBA spectra of J2129+8453}
\figsetplot{J2129+8453_spectra.pdf}
\figsetgrpnote{Single-dish and VLBA broadband spectra of all the sources of our sample. The source name is specified at the top right of each plot. The single-dish spectra are plotted black, the VLBA spectra are plotted by the red color. In cases of the VLBA non-detection, the upper limits on the VLBA flux density are shown by red arrows.}
\figsetgrpend

\figsetgrpstart
\figsetgrpnum{7.432}
\figsetgrptitle{Total and VLBA spectra of J2130+8357}
\figsetplot{J2130+8357_spectra.pdf}
\figsetgrpnote{Single-dish and VLBA broadband spectra of all the sources of our sample. The source name is specified at the top right of each plot. The single-dish spectra are plotted black, the VLBA spectra are plotted by the red color. In cases of the VLBA non-detection, the upper limits on the VLBA flux density are shown by red arrows.}
\figsetgrpend

\figsetgrpstart
\figsetgrpnum{7.433}
\figsetgrptitle{Total and VLBA spectra of J2131+8430}
\figsetplot{J2131+8430_spectra.pdf}
\figsetgrpnote{Single-dish and VLBA broadband spectra of all the sources of our sample. The source name is specified at the top right of each plot. The single-dish spectra are plotted black, the VLBA spectra are plotted by the red color. In cases of the VLBA non-detection, the upper limits on the VLBA flux density are shown by red arrows.}
\figsetgrpend

\figsetgrpstart
\figsetgrpnum{7.434}
\figsetgrptitle{Total and VLBA spectra of J2132+7930}
\figsetplot{J2132+7930_spectra.pdf}
\figsetgrpnote{Single-dish and VLBA broadband spectra of all the sources of our sample. The source name is specified at the top right of each plot. The single-dish spectra are plotted black, the VLBA spectra are plotted by the red color. In cases of the VLBA non-detection, the upper limits on the VLBA flux density are shown by red arrows.}
\figsetgrpend

\figsetgrpstart
\figsetgrpnum{7.435}
\figsetgrptitle{Total and VLBA spectra of J2133+7613}
\figsetplot{J2133+7613_spectra.pdf}
\figsetgrpnote{Single-dish and VLBA broadband spectra of all the sources of our sample. The source name is specified at the top right of each plot. The single-dish spectra are plotted black, the VLBA spectra are plotted by the red color. In cases of the VLBA non-detection, the upper limits on the VLBA flux density are shown by red arrows.}
\figsetgrpend

\figsetgrpstart
\figsetgrpnum{7.436}
\figsetgrptitle{Total and VLBA spectra of J2133+8239}
\figsetplot{J2133+8239_spectra.pdf}
\figsetgrpnote{Single-dish and VLBA broadband spectra of all the sources of our sample. The source name is specified at the top right of each plot. The single-dish spectra are plotted black, the VLBA spectra are plotted by the red color. In cases of the VLBA non-detection, the upper limits on the VLBA flux density are shown by red arrows.}
\figsetgrpend

\figsetgrpstart
\figsetgrpnum{7.437}
\figsetgrptitle{Total and VLBA spectra of J2136+7638}
\figsetplot{J2136+7638_spectra.pdf}
\figsetgrpnote{Single-dish and VLBA broadband spectra of all the sources of our sample. The source name is specified at the top right of each plot. The single-dish spectra are plotted black, the VLBA spectra are plotted by the red color. In cases of the VLBA non-detection, the upper limits on the VLBA flux density are shown by red arrows.}
\figsetgrpend

\figsetgrpstart
\figsetgrpnum{7.438}
\figsetgrptitle{Total and VLBA spectra of J2139+8339}
\figsetplot{J2139+8339_spectra.pdf}
\figsetgrpnote{Single-dish and VLBA broadband spectra of all the sources of our sample. The source name is specified at the top right of each plot. The single-dish spectra are plotted black, the VLBA spectra are plotted by the red color. In cases of the VLBA non-detection, the upper limits on the VLBA flux density are shown by red arrows.}
\figsetgrpend

\figsetgrpstart
\figsetgrpnum{7.439}
\figsetgrptitle{Total and VLBA spectra of J2139+8658}
\figsetplot{J2139+8658_spectra.pdf}
\figsetgrpnote{Single-dish and VLBA broadband spectra of all the sources of our sample. The source name is specified at the top right of each plot. The single-dish spectra are plotted black, the VLBA spectra are plotted by the red color. In cases of the VLBA non-detection, the upper limits on the VLBA flux density are shown by red arrows.}
\figsetgrpend

\figsetgrpstart
\figsetgrpnum{7.440}
\figsetgrptitle{Total and VLBA spectra of J2140+7505}
\figsetplot{J2140+7505_spectra.pdf}
\figsetgrpnote{Single-dish and VLBA broadband spectra of all the sources of our sample. The source name is specified at the top right of each plot. The single-dish spectra are plotted black, the VLBA spectra are plotted by the red color. In cases of the VLBA non-detection, the upper limits on the VLBA flux density are shown by red arrows.}
\figsetgrpend

\figsetgrpstart
\figsetgrpnum{7.441}
\figsetgrptitle{Total and VLBA spectra of J2144+8147}
\figsetplot{J2144+8147_spectra.pdf}
\figsetgrpnote{Single-dish and VLBA broadband spectra of all the sources of our sample. The source name is specified at the top right of each plot. The single-dish spectra are plotted black, the VLBA spectra are plotted by the red color. In cases of the VLBA non-detection, the upper limits on the VLBA flux density are shown by red arrows.}
\figsetgrpend

\figsetgrpstart
\figsetgrpnum{7.442}
\figsetgrptitle{Total and VLBA spectra of J2146+7536}
\figsetplot{J2146+7536_spectra.pdf}
\figsetgrpnote{Single-dish and VLBA broadband spectra of all the sources of our sample. The source name is specified at the top right of each plot. The single-dish spectra are plotted black, the VLBA spectra are plotted by the red color. In cases of the VLBA non-detection, the upper limits on the VLBA flux density are shown by red arrows.}
\figsetgrpend

\figsetgrpstart
\figsetgrpnum{7.443}
\figsetgrptitle{Total and VLBA spectra of J2149+7540}
\figsetplot{J2149+7540_spectra.pdf}
\figsetgrpnote{Single-dish and VLBA broadband spectra of all the sources of our sample. The source name is specified at the top right of each plot. The single-dish spectra are plotted black, the VLBA spectra are plotted by the red color. In cases of the VLBA non-detection, the upper limits on the VLBA flux density are shown by red arrows.}
\figsetgrpend

\figsetgrpstart
\figsetgrpnum{7.444}
\figsetgrptitle{Total and VLBA spectra of J2156+8337}
\figsetplot{J2156+8337_spectra.pdf}
\figsetgrpnote{Single-dish and VLBA broadband spectra of all the sources of our sample. The source name is specified at the top right of each plot. The single-dish spectra are plotted black, the VLBA spectra are plotted by the red color. In cases of the VLBA non-detection, the upper limits on the VLBA flux density are shown by red arrows.}
\figsetgrpend

\figsetgrpstart
\figsetgrpnum{7.445}
\figsetgrptitle{Total and VLBA spectra of J2157+7646}
\figsetplot{J2157+7646_spectra.pdf}
\figsetgrpnote{Single-dish and VLBA broadband spectra of all the sources of our sample. The source name is specified at the top right of each plot. The single-dish spectra are plotted black, the VLBA spectra are plotted by the red color. In cases of the VLBA non-detection, the upper limits on the VLBA flux density are shown by red arrows.}
\figsetgrpend

\figsetgrpstart
\figsetgrpnum{7.446}
\figsetgrptitle{Total and VLBA spectra of J2158+7722}
\figsetplot{J2158+7722_spectra.pdf}
\figsetgrpnote{Single-dish and VLBA broadband spectra of all the sources of our sample. The source name is specified at the top right of each plot. The single-dish spectra are plotted black, the VLBA spectra are plotted by the red color. In cases of the VLBA non-detection, the upper limits on the VLBA flux density are shown by red arrows.}
\figsetgrpend

\figsetgrpstart
\figsetgrpnum{7.447}
\figsetgrptitle{Total and VLBA spectra of J2200+7535}
\figsetplot{J2200+7535_spectra.pdf}
\figsetgrpnote{Single-dish and VLBA broadband spectra of all the sources of our sample. The source name is specified at the top right of each plot. The single-dish spectra are plotted black, the VLBA spectra are plotted by the red color. In cases of the VLBA non-detection, the upper limits on the VLBA flux density are shown by red arrows.}
\figsetgrpend

\figsetgrpstart
\figsetgrpnum{7.448}
\figsetgrptitle{Total and VLBA spectra of J2209+8353}
\figsetplot{J2209+8353_spectra.pdf}
\figsetgrpnote{Single-dish and VLBA broadband spectra of all the sources of our sample. The source name is specified at the top right of each plot. The single-dish spectra are plotted black, the VLBA spectra are plotted by the red color. In cases of the VLBA non-detection, the upper limits on the VLBA flux density are shown by red arrows.}
\figsetgrpend

\figsetgrpstart
\figsetgrpnum{7.449}
\figsetgrptitle{Total and VLBA spectra of J2217+7858}
\figsetplot{J2217+7858_spectra.pdf}
\figsetgrpnote{Single-dish and VLBA broadband spectra of all the sources of our sample. The source name is specified at the top right of each plot. The single-dish spectra are plotted black, the VLBA spectra are plotted by the red color. In cases of the VLBA non-detection, the upper limits on the VLBA flux density are shown by red arrows.}
\figsetgrpend

\figsetgrpstart
\figsetgrpnum{7.450}
\figsetgrptitle{Total and VLBA spectra of J2223+7631}
\figsetplot{J2223+7631_spectra.pdf}
\figsetgrpnote{Single-dish and VLBA broadband spectra of all the sources of our sample. The source name is specified at the top right of each plot. The single-dish spectra are plotted black, the VLBA spectra are plotted by the red color. In cases of the VLBA non-detection, the upper limits on the VLBA flux density are shown by red arrows.}
\figsetgrpend

\figsetgrpstart
\figsetgrpnum{7.451}
\figsetgrptitle{Total and VLBA spectra of J2228+7532}
\figsetplot{J2228+7532_spectra.pdf}
\figsetgrpnote{Single-dish and VLBA broadband spectra of all the sources of our sample. The source name is specified at the top right of each plot. The single-dish spectra are plotted black, the VLBA spectra are plotted by the red color. In cases of the VLBA non-detection, the upper limits on the VLBA flux density are shown by red arrows.}
\figsetgrpend

\figsetgrpstart
\figsetgrpnum{7.452}
\figsetgrptitle{Total and VLBA spectra of J2232+8039}
\figsetplot{J2232+8039_spectra.pdf}
\figsetgrpnote{Single-dish and VLBA broadband spectra of all the sources of our sample. The source name is specified at the top right of each plot. The single-dish spectra are plotted black, the VLBA spectra are plotted by the red color. In cases of the VLBA non-detection, the upper limits on the VLBA flux density are shown by red arrows.}
\figsetgrpend

\figsetgrpstart
\figsetgrpnum{7.453}
\figsetgrptitle{Total and VLBA spectra of J2236+7648}
\figsetplot{J2236+7648_spectra.pdf}
\figsetgrpnote{Single-dish and VLBA broadband spectra of all the sources of our sample. The source name is specified at the top right of each plot. The single-dish spectra are plotted black, the VLBA spectra are plotted by the red color. In cases of the VLBA non-detection, the upper limits on the VLBA flux density are shown by red arrows.}
\figsetgrpend

\figsetgrpstart
\figsetgrpnum{7.454}
\figsetgrptitle{Total and VLBA spectra of J2236+8815}
\figsetplot{J2236+8815_spectra.pdf}
\figsetgrpnote{Single-dish and VLBA broadband spectra of all the sources of our sample. The source name is specified at the top right of each plot. The single-dish spectra are plotted black, the VLBA spectra are plotted by the red color. In cases of the VLBA non-detection, the upper limits on the VLBA flux density are shown by red arrows.}
\figsetgrpend

\figsetgrpstart
\figsetgrpnum{7.455}
\figsetgrptitle{Total and VLBA spectra of J2238+8148}
\figsetplot{J2238+8148_spectra.pdf}
\figsetgrpnote{Single-dish and VLBA broadband spectra of all the sources of our sample. The source name is specified at the top right of each plot. The single-dish spectra are plotted black, the VLBA spectra are plotted by the red color. In cases of the VLBA non-detection, the upper limits on the VLBA flux density are shown by red arrows.}
\figsetgrpend

\figsetgrpstart
\figsetgrpnum{7.456}
\figsetgrptitle{Total and VLBA spectra of J2242+8224}
\figsetplot{J2242+8224_spectra.pdf}
\figsetgrpnote{Single-dish and VLBA broadband spectra of all the sources of our sample. The source name is specified at the top right of each plot. The single-dish spectra are plotted black, the VLBA spectra are plotted by the red color. In cases of the VLBA non-detection, the upper limits on the VLBA flux density are shown by red arrows.}
\figsetgrpend

\figsetgrpstart
\figsetgrpnum{7.457}
\figsetgrptitle{Total and VLBA spectra of J2247+8555}
\figsetplot{J2247+8555_spectra.pdf}
\figsetgrpnote{Single-dish and VLBA broadband spectra of all the sources of our sample. The source name is specified at the top right of each plot. The single-dish spectra are plotted black, the VLBA spectra are plotted by the red color. In cases of the VLBA non-detection, the upper limits on the VLBA flux density are shown by red arrows.}
\figsetgrpend

\figsetgrpstart
\figsetgrpnum{7.458}
\figsetgrptitle{Total and VLBA spectra of J2248+7718}
\figsetplot{J2248+7718_spectra.pdf}
\figsetgrpnote{Single-dish and VLBA broadband spectra of all the sources of our sample. The source name is specified at the top right of each plot. The single-dish spectra are plotted black, the VLBA spectra are plotted by the red color. In cases of the VLBA non-detection, the upper limits on the VLBA flux density are shown by red arrows.}
\figsetgrpend

\figsetgrpstart
\figsetgrpnum{7.459}
\figsetgrptitle{Total and VLBA spectra of J2249+7601}
\figsetplot{J2249+7601_spectra.pdf}
\figsetgrpnote{Single-dish and VLBA broadband spectra of all the sources of our sample. The source name is specified at the top right of each plot. The single-dish spectra are plotted black, the VLBA spectra are plotted by the red color. In cases of the VLBA non-detection, the upper limits on the VLBA flux density are shown by red arrows.}
\figsetgrpend

\figsetgrpstart
\figsetgrpnum{7.460}
\figsetgrptitle{Total and VLBA spectra of J2250+7619}
\figsetplot{J2250+7619_spectra.pdf}
\figsetgrpnote{Single-dish and VLBA broadband spectra of all the sources of our sample. The source name is specified at the top right of each plot. The single-dish spectra are plotted black, the VLBA spectra are plotted by the red color. In cases of the VLBA non-detection, the upper limits on the VLBA flux density are shown by red arrows.}
\figsetgrpend

\figsetgrpstart
\figsetgrpnum{7.461}
\figsetgrptitle{Total and VLBA spectra of J2251+8013}
\figsetplot{J2251+8013_spectra.pdf}
\figsetgrpnote{Single-dish and VLBA broadband spectra of all the sources of our sample. The source name is specified at the top right of each plot. The single-dish spectra are plotted black, the VLBA spectra are plotted by the red color. In cases of the VLBA non-detection, the upper limits on the VLBA flux density are shown by red arrows.}
\figsetgrpend

\figsetgrpstart
\figsetgrpnum{7.462}
\figsetgrptitle{Total and VLBA spectra of J2301+7954}
\figsetplot{J2301+7954_spectra.pdf}
\figsetgrpnote{Single-dish and VLBA broadband spectra of all the sources of our sample. The source name is specified at the top right of each plot. The single-dish spectra are plotted black, the VLBA spectra are plotted by the red color. In cases of the VLBA non-detection, the upper limits on the VLBA flux density are shown by red arrows.}
\figsetgrpend

\figsetgrpstart
\figsetgrpnum{7.463}
\figsetgrptitle{Total and VLBA spectra of J2301+8200}
\figsetplot{J2301+8200_spectra.pdf}
\figsetgrpnote{Single-dish and VLBA broadband spectra of all the sources of our sample. The source name is specified at the top right of each plot. The single-dish spectra are plotted black, the VLBA spectra are plotted by the red color. In cases of the VLBA non-detection, the upper limits on the VLBA flux density are shown by red arrows.}
\figsetgrpend

\figsetgrpstart
\figsetgrpnum{7.464}
\figsetgrptitle{Total and VLBA spectra of J2310+8857}
\figsetplot{J2310+8857_spectra.pdf}
\figsetgrpnote{Single-dish and VLBA broadband spectra of all the sources of our sample. The source name is specified at the top right of each plot. The single-dish spectra are plotted black, the VLBA spectra are plotted by the red color. In cases of the VLBA non-detection, the upper limits on the VLBA flux density are shown by red arrows.}
\figsetgrpend

\figsetgrpstart
\figsetgrpnum{7.465}
\figsetgrptitle{Total and VLBA spectra of J2317+7554}
\figsetplot{J2317+7554_spectra.pdf}
\figsetgrpnote{Single-dish and VLBA broadband spectra of all the sources of our sample. The source name is specified at the top right of each plot. The single-dish spectra are plotted black, the VLBA spectra are plotted by the red color. In cases of the VLBA non-detection, the upper limits on the VLBA flux density are shown by red arrows.}
\figsetgrpend

\figsetgrpstart
\figsetgrpnum{7.466}
\figsetgrptitle{Total and VLBA spectra of J2322+7523}
\figsetplot{J2322+7523_spectra.pdf}
\figsetgrpnote{Single-dish and VLBA broadband spectra of all the sources of our sample. The source name is specified at the top right of each plot. The single-dish spectra are plotted black, the VLBA spectra are plotted by the red color. In cases of the VLBA non-detection, the upper limits on the VLBA flux density are shown by red arrows.}
\figsetgrpend

\figsetgrpstart
\figsetgrpnum{7.467}
\figsetgrptitle{Total and VLBA spectra of J2325+7917}
\figsetplot{J2325+7917_spectra.pdf}
\figsetgrpnote{Single-dish and VLBA broadband spectra of all the sources of our sample. The source name is specified at the top right of each plot. The single-dish spectra are plotted black, the VLBA spectra are plotted by the red color. In cases of the VLBA non-detection, the upper limits on the VLBA flux density are shown by red arrows.}
\figsetgrpend

\figsetgrpstart
\figsetgrpnum{7.468}
\figsetgrptitle{Total and VLBA spectra of J2326+7813}
\figsetplot{J2326+7813_spectra.pdf}
\figsetgrpnote{Single-dish and VLBA broadband spectra of all the sources of our sample. The source name is specified at the top right of each plot. The single-dish spectra are plotted black, the VLBA spectra are plotted by the red color. In cases of the VLBA non-detection, the upper limits on the VLBA flux density are shown by red arrows.}
\figsetgrpend

\figsetgrpstart
\figsetgrpnum{7.469}
\figsetgrptitle{Total and VLBA spectra of J2326+8231}
\figsetplot{J2326+8231_spectra.pdf}
\figsetgrpnote{Single-dish and VLBA broadband spectra of all the sources of our sample. The source name is specified at the top right of each plot. The single-dish spectra are plotted black, the VLBA spectra are plotted by the red color. In cases of the VLBA non-detection, the upper limits on the VLBA flux density are shown by red arrows.}
\figsetgrpend

\figsetgrpstart
\figsetgrpnum{7.470}
\figsetgrptitle{Total and VLBA spectra of J2328+7617}
\figsetplot{J2328+7617_spectra.pdf}
\figsetgrpnote{Single-dish and VLBA broadband spectra of all the sources of our sample. The source name is specified at the top right of each plot. The single-dish spectra are plotted black, the VLBA spectra are plotted by the red color. In cases of the VLBA non-detection, the upper limits on the VLBA flux density are shown by red arrows.}
\figsetgrpend

\figsetgrpstart
\figsetgrpnum{7.471}
\figsetgrptitle{Total and VLBA spectra of J2329+8131}
\figsetplot{J2329+8131_spectra.pdf}
\figsetgrpnote{Single-dish and VLBA broadband spectra of all the sources of our sample. The source name is specified at the top right of each plot. The single-dish spectra are plotted black, the VLBA spectra are plotted by the red color. In cases of the VLBA non-detection, the upper limits on the VLBA flux density are shown by red arrows.}
\figsetgrpend

\figsetgrpstart
\figsetgrpnum{7.472}
\figsetgrptitle{Total and VLBA spectra of J2330+7742}
\figsetplot{J2330+7742_spectra.pdf}
\figsetgrpnote{Single-dish and VLBA broadband spectra of all the sources of our sample. The source name is specified at the top right of each plot. The single-dish spectra are plotted black, the VLBA spectra are plotted by the red color. In cases of the VLBA non-detection, the upper limits on the VLBA flux density are shown by red arrows.}
\figsetgrpend

\figsetgrpstart
\figsetgrpnum{7.473}
\figsetgrptitle{Total and VLBA spectra of J2332+7801}
\figsetplot{J2332+7801_spectra.pdf}
\figsetgrpnote{Single-dish and VLBA broadband spectra of all the sources of our sample. The source name is specified at the top right of each plot. The single-dish spectra are plotted black, the VLBA spectra are plotted by the red color. In cases of the VLBA non-detection, the upper limits on the VLBA flux density are shown by red arrows.}
\figsetgrpend

\figsetgrpstart
\figsetgrpnum{7.474}
\figsetgrptitle{Total and VLBA spectra of J2333+8623}
\figsetplot{J2333+8623_spectra.pdf}
\figsetgrpnote{Single-dish and VLBA broadband spectra of all the sources of our sample. The source name is specified at the top right of each plot. The single-dish spectra are plotted black, the VLBA spectra are plotted by the red color. In cases of the VLBA non-detection, the upper limits on the VLBA flux density are shown by red arrows.}
\figsetgrpend

\figsetgrpstart
\figsetgrpnum{7.475}
\figsetgrptitle{Total and VLBA spectra of J2344+8226}
\figsetplot{J2344+8226_spectra.pdf}
\figsetgrpnote{Single-dish and VLBA broadband spectra of all the sources of our sample. The source name is specified at the top right of each plot. The single-dish spectra are plotted black, the VLBA spectra are plotted by the red color. In cases of the VLBA non-detection, the upper limits on the VLBA flux density are shown by red arrows.}
\figsetgrpend

\figsetgrpstart
\figsetgrpnum{7.476}
\figsetgrptitle{Total and VLBA spectra of J2348+8038}
\figsetplot{J2348+8038_spectra.pdf}
\figsetgrpnote{Single-dish and VLBA broadband spectra of all the sources of our sample. The source name is specified at the top right of each plot. The single-dish spectra are plotted black, the VLBA spectra are plotted by the red color. In cases of the VLBA non-detection, the upper limits on the VLBA flux density are shown by red arrows.}
\figsetgrpend

\figsetgrpstart
\figsetgrpnum{7.477}
\figsetgrptitle{Total and VLBA spectra of J2349+7517}
\figsetplot{J2349+7517_spectra.pdf}
\figsetgrpnote{Single-dish and VLBA broadband spectra of all the sources of our sample. The source name is specified at the top right of each plot. The single-dish spectra are plotted black, the VLBA spectra are plotted by the red color. In cases of the VLBA non-detection, the upper limits on the VLBA flux density are shown by red arrows.}
\figsetgrpend

\figsetgrpstart
\figsetgrpnum{7.478}
\figsetgrptitle{Total and VLBA spectra of J2349+7913}
\figsetplot{J2349+7913_spectra.pdf}
\figsetgrpnote{Single-dish and VLBA broadband spectra of all the sources of our sample. The source name is specified at the top right of each plot. The single-dish spectra are plotted black, the VLBA spectra are plotted by the red color. In cases of the VLBA non-detection, the upper limits on the VLBA flux density are shown by red arrows.}
\figsetgrpend

\figsetgrpstart
\figsetgrpnum{7.479}
\figsetgrptitle{Total and VLBA spectra of J2354+8047}
\figsetplot{J2354+8047_spectra.pdf}
\figsetgrpnote{Single-dish and VLBA broadband spectra of all the sources of our sample. The source name is specified at the top right of each plot. The single-dish spectra are plotted black, the VLBA spectra are plotted by the red color. In cases of the VLBA non-detection, the upper limits on the VLBA flux density are shown by red arrows.}
\figsetgrpend

\figsetgrpstart
\figsetgrpnum{7.480}
\figsetgrptitle{Total and VLBA spectra of J2355+7954}
\figsetplot{J2355+7954_spectra.pdf}
\figsetgrpnote{Single-dish and VLBA broadband spectra of all the sources of our sample. The source name is specified at the top right of each plot. The single-dish spectra are plotted black, the VLBA spectra are plotted by the red color. In cases of the VLBA non-detection, the upper limits on the VLBA flux density are shown by red arrows.}
\figsetgrpend

\figsetgrpstart
\figsetgrpnum{7.481}
\figsetgrptitle{Total and VLBA spectra of J2356+8152}
\figsetplot{J2356+8152_spectra.pdf}
\figsetgrpnote{Single-dish and VLBA broadband spectra of all the sources of our sample. The source name is specified at the top right of each plot. The single-dish spectra are plotted black, the VLBA spectra are plotted by the red color. In cases of the VLBA non-detection, the upper limits on the VLBA flux density are shown by red arrows.}
\figsetgrpend

\figsetgrpstart
\figsetgrpnum{7.482}
\figsetgrptitle{Total and VLBA spectra of J2358+8142}
\figsetplot{J2358+8142_spectra.pdf}
\figsetgrpnote{Single-dish and VLBA broadband spectra of all the sources of our sample. The source name is specified at the top right of each plot. The single-dish spectra are plotted black, the VLBA spectra are plotted by the red color. In cases of the VLBA non-detection, the upper limits on the VLBA flux density are shown by red arrows.}
\figsetgrpend

\figsetend

\end{document}